\documentclass[preprint,showpacs,showkeys,preprintnumbers,amsmath,amssymb]{revtex4-2}
\DeclareMathOperator{\diag}{diag}

\usepackage{graphicx}% Include figure files
\usepackage{dcolumn}% Align table columns on decimal point
\usepackage{bm}% bold math
\begin{document}

\title{Universal vulnerability in strong modular networks with
  various degree distributions between inequality and equality
}% Force line breaks with \\

\author{Yukio Hayashi, and Taishi Ogawa}
%\email{yhayashi@jaist.ac.jp}
\affiliation{
%Division of Transdisciplinary Sciences,
%Graduate School of Advanced Science and Technology, 
Japan Advanced Institute of Science and Technology,\\
Ishikawa, 923-1292, Japan}

\date{\today}% It is always \today, today,

\begin{abstract}
Generally,
networks are classified into two sides of inequality and
equality with respect to the number of links at nodes by the types of 
degree distributions.
One side includes many social, technological, and biological networks 
which consist of a few nodes with many links, and many nodes with a few links, 
whereas the other side 
consists of all nodes with an equal number of links.
In comprehensive investigations between them,
we have found that, as a more equal network,
the tolerance of whole connectivity is stronger without fragmentation
against the malfunction of nodes in a wide class of randomized networks.
However, we newly find that 
all networks which include typical well-known network structures 
between them become extremely vulnerable, 
if a strong modular (or community) structure is added with commonalities
of areas, interests, religions, purpose, and so on.
These results will encourage avoiding too dense unions by connecting nodes
and taking into account the balanced resource allocation between 
intra- and inter-links of weak communities.
We must reconsider 
not only efficiency but also tolerance against attacks or disasters,
unless no community that is really impossible.  
\end{abstract}

%\pacs{89.20.-a, 02.60.-x, 02.60.Cb, 05.90.+m}
% Systems obeying scaling law, Markov process, 
% Random walks and Levy flights, Self-organized systems
%05.65.+b Self-organized systems
%89.20.-a Interdisciplinary applications of physics
%89.40-a Transportation
%89.70.-a Information and communication theory
%89.70.Eg Computartional complexity
%89.75.Fb Structures and organization in complex system
%05.45.Df Fractals
%05.90.+m Other topics in statistical physics, thermodynamics, and nonlinear dynamical systems
%02.60.-x Numerical approximation and analysis
%02.60.Cb Numerical simulation; solution of equations
%02.90.+p Other topics in mathematical methods in physics

%arXiv
%primary: Physics and Society;
%secondary: Disordered Systems and Neural Networks;
%           Statistical Mechanics;
%Nonlinear Sciences:
%Computer Science: Machine Learning;
%Mathmatics: Combinatorics; 


\maketitle

%\section{\label{sec:level1}First-level heading:\protect\\ The line
%break was forced \lowercase{via} \textbackslash\textbackslash}
%\subsection{\label{sec:level2}Second-level heading: Formatting}
%\tableofcontents
%\clearpage

\newpage
\section*{Introduction}
The robustness
of connectivity is one of the fundamental issues in network science
as an interdisciplinary research area of statistical physics, computer
science, and biological or social science,
since the disconnections lose the network function
for transferring or delivering information or goods.
The connectivity is therefore essential even abstracted as independent of
what the materials are for linking. 
Despite persistent efforts
\cite{Albert00,Callaway00,Schneider11,Tanizawa12,Holme11,Dorogovtsev04,Cohen10}, 
it is not elucidated what topological structure
is the best or the worst against intentional attacks or large disasters.
At least, 
degree distributions are considered as important.
Because it has been found that 
networks with narrower degree distributions become
more robust as the pure effect of degree distributions \cite{Chujyo22}.
However, the robustness may be affected by modules
with higher interactions among nodes beyond its degree at each node.
In fact 
at three types of SF networks, ER random graphs 
\cite{Shai15,Nguyen21}, and regular networks \cite{Kim23}, 
decreasing the robustness has been numerically shown, 
when a strong community structure is added artificially \cite{Shai15,Nguyen21,Kim23}.
Such vulnerability has also been found 
in spatially modular networks with local concentrations of nodes
connected by short links \cite{Mou24}.

On the above background, through an original approach \cite{Chujyo22,Liao22},
we comprehensively investigate the robustness 
against intentional attacks in networks
with/without modules (or communities as synonym in this paper)
under the condition of continuously changing degree distributions
which interpolate the above three types of well-known network structures.
Moreover,
we find a deep relation between the robustness and
the algebraic connectivity: the second smallest eigenvalue of
Laplacian matrix \cite{Meighem11} of graph under the condition of
a constant total number of links.
Although it has been numerically shown that
the eigenvalue increases as the node or link
connectivity (defined by the minimum number of nodes or links whose
removal causes disconnection) is larger 
in ER random graphs or SF networks
with more (not constant) total links \cite{Jamakovic07},
this analysis is also limited for Poisson and power-law
degree distributions without considering communities and attacks.

\section*{Related Work}
So far it has been discovered about twenty years ago that
many real networks in social, technological, and biological systems
are extremely vulnerable against attacks on hub nodes
\cite{Albert00,Callaway00}
in commonly existing scale-free (SF) structure whose degree distribution
follows a power-law \cite{Barabasi99a,Amaral00}.
It was believed about ten years ago that
the robustness can be optimized as onion-like by increasing positive
degree-degree correlations even from uncorrelated SF networks
\cite{Schneider11,Tanizawa12}.
Instead,
too strong correlations rather weaken the robustness \cite{Tanizawa12},
the whole rewiring \cite{Holme11} to enhance correlations
is not realistic because of discarding the already constructed relations
between nodes.
To retain the existing links,
incrementally growing methods have been proposed in showing the emergence
of onion-like networks \cite{Hayashi14,Hayashi16,Hayashi18a,Hayashi18b}.
Moreover, it has been pointed out that \cite{Hayashi18b} 
the enhancing loops may be more crucial than degree-degree correlations
for improving the robustness.
Just around that time,
the equivalence of dismantling and decycling problems has been suggested
theoretically
in a large class of random networks with light-tailed degree distributions 
\cite{Braunstein16}.
This means that 
enhancing loops leads to the strongest tolerance of connectivity 
for avoiding fragmenting or dismantling as the worst case removal 
of nodes.
Here, the dismantling problem is to find the minimum set of nodes
which removal makes a network fragmented into at most a given size,
while the decycling problem is to find the minimum set of
nodes which are necessary to form loops.
They correspond to
the critical node detection (CND) \cite{Santos18} 
and the minimum feedback vertex set (FVS) \cite{Karp1972} problems,
respectively, 
known as NP-hard in computer science.
Thus, to only measure the size of FVS is intractable,
its maximization by enhancing loops
becomes more difficult.
However, there are several evidences to be solvable somehow as follows.

Recently, 
by several rewiring methods for heuristically enhancing loops,
it has been commonly found that \cite{Chujyo21a}
the widths of minimum and maximum degrees become small
even with negative degree-degree correlations as counterexamples
for onion-like networks \cite{Schneider11,Tanizawa12}.
These networks have a large size of FVS with improving the robustness 
than the conventional optimal \cite{Holme11} against hub attacks.
Moreover, 
the robustness has been comprehensively investigated
for continuously changing degree distributions \cite{Chujyo22}
from power-law to exponential \cite{Krapivsky01}
further to narrower ones \cite{Liao22} with randomization \cite{Boguna05},
although the importance of degree distributions
has been revealed \cite{Albert00,Callaway00,Dorogovtsev04,Cohen10}
for mainly 
power-law and Poisson degree distributions in realistic SF networks 
and classical Erd\"{o}s-R\'{e}nyi (ER) random graphs \cite{ErdosRenyi1959}.
Here, because of eliminating degree-degree correlations and higher
correlations among degrees, 
the randomization is necessary for investigating the pure effect of
degree distribution on the robustness.
From these bases, 
networks with narrower degree distributions are more robust
in randomized networks with continuously changing degree distributions.
It has also been numerically shown by using perturbations \cite{Chujyo23}
that random regular networks with zero width of degrees
are the optimal as suggested by \cite{Ma15}, strictly speaking at least locally,
in the state-of-the-art.

On the other hand,
there widely exist modules or communities based on local areas, interests,
religions, purposes, etc. 
in social, biological \cite{Fortunato10},
and technological networks \cite{Guimera05,Kaluza10}.
Historically community-like structure has been considered with focusing on
internal cohesion among nodes of a subgraph \cite{Wasserman94}.
Such structure is generally characterized by
the mixing of dense intra-links between nodes in a community
and sparse inter-links between nodes in different communities.
Intra- and inter-links correspond to strong-ties or bondings
and weak-ties or bridgings, respectively, related to socialcapital
\cite{Granovetter1973,Claridge18}.

\section*{Methods}
We explain how various networks are generated by using
growing network (GN) \cite{Krapivsky01}
and inverse preferential attachment (IPA) \cite{Liao22} models.
For growing networks, 
there is no other construction method to make 
continuously changing degree distributions $P(k)$ of degree $k$.
From an initial configuration (e.g. a complete graph of $m$ nodes),
a network is generated in prohibiting self-loops and multi-links 
by the attachments of $m$ links from a new node
to the existing nodes at each time-step until the size $N$
as the total number of nodes.
Here, the total number of links is $M \approx m \times N$
except for the initial links, and $1 \leq m \ll N$.
For selecting the attached node $i$,
the attachment probability is proportional to $k_{i}^{\nu}$,
where $k_{i}$ denotes the degree of node $i$,
$\nu$ is a parameter.
Note that a strictly regular network is not created 
even for $\nu \rightarrow - \infty$ in IPA model, 
since there are several degrees $m, m+1, \ldots, 2m$ \cite{Liao22}.

After generating networks with various $P(k)$ as shown later,
to investigate the pure effect of degree distribution on the robustness,
the randomization eliminates degree-degree correlations and other higher
structure such as chain-like \cite{Liao22} 
by the following rewiring in the configuration model \cite{Boguna05}.
Off course,
the generated network with $N$ nodes and $M$ links is different
from an initial one of $m$ nodes.
First, all links are cut to free-ends.
In the initial stage of randomization,
there are $k_{i}$ free-ends of links emanated form each node $i$.
Second, two free-ends are randomly chosen and rewired to a link 
in prohibiting self-loops at a node and multi-links between nodes.
Such process is repeated until constructing a connected network
of $N$ nodes without free-ends of links.
When it is impossible to prohibit self-loops or multi-links,
several rewired links are cut to free-ends.
Then, the above process is continued again.

After the randomization, 
we artificially add modular structure without overlapping among modules.
To simplify the discussion,
we assume that the number of nodes in each module is the same integer
$N/m_{o}$ as an equal size of module \cite{Shai15,Nguyen21},
where $m_{o}$ is a divisor of $N$ as the number of modules.
This case of equal size is the most meaningful but difficult,
since the largest module is dominant, when the sizes are quite different.
In that case of different sizes,
it is sufficient to easily investigate only the largest one in ignoring
the small effects of other modules.
One of $1, 2, \ldots, m_{o}$ is randomly assigned to each node
\cite{Nguyen21} without any bias.
In this stage, the number of intra-links is estimated as
$M/m_{o}$ by the probability that one-end node of link has the same
module number randomly assigned to the other-end node,
while the number of inter-links is estimated as
remaining $(1 - 1/m_{o})M$.

In controlling the strength of modular structure by
a given tentative rewiring rate $w'$, 
some inter-links are rewired to intra-links 
in prohibiting self-loops and multi-links.
The rewiring is slightly modified from the previous method \cite{Nguyen21}
to maintain the whole connectivity under a given $P(k)$,
and the random process is performed 
so as not to have any bias.
First,
$(1 - 1/m_{o})M \times w'$ inter-links are randomly chosen
and cut to free-ends.
%Remember that the initial number of inter-links is
%$(1 - 1/m_{o})M$ by randomly assigned module number $1, 2, \dots, m_{o}$
%to each node
%as mentioned in the third paragraph of Results section.
Second,
when pairs of free-ends exist in each module,
they are randomly chosen and connected to be as many intra-links as possible.
Note that the total numbers of free-ends are even both 
before and after the pairings.
Third,
if there remain free-ends in different modules,
%and other-side's nodes of these links are unconnected,
then the free-ends are randomly chosen and 
connected as returning to inter-links.
The return is repeated if possible.
Finally,
for further remaining two free-ends
(denoted by $i,j$) 
generally belong to different modules,
an intra-link is randomly chosen and cut to new free-ends
(denoted by $k,l$) in other than those modules.
These nodes are connected (as $i$-$k$ and $j$-$l$).
Such final process is repeated until there are no free-ends.
If a constructed network is fragmented,
the sample is discarded.
After the rewirings,
it is trivial to maintain $P(k)$
because there is nothing to add and remove any links at each node.
%Note that links of free-ends are emanated from it originally.
We remark that the actual rewiring rate $w$ is slightly changeable
from a tentative rate $w'$, though it is not essential 
for controlling the strength of modular structure.

For each of the modular networks under the condition of the same $P(k)$ 
by varying the tentative rewiring rate
$w' = 0.0, 0.5, 0.7, 0.9, 0.95$, and $0.98$,
we calculate the modularity \cite{Newman06} defined as
\[
  Q \stackrel{\rm def}{=} \frac{1}{2 M} \sum_{i,j \in V}
  \left( A_{ij} - \frac{k_{i} k_{j}}{2 M} \right) \delta(c_{i}, c_{j}),
\]
where $\delta(c_{i}, c_{j})$ is the Kronecker delta:
$1$ if $c_{i} = c_{j}$ or $0$ otherwise,
$c_{i}$ (or $c_{j}$) is the belonging module number of node $i$ (or $j$).
A larger $Q$ represents stronger modular structure.

\vspace{2mm}
\begin{figure}[htb]
\centering
\includegraphics[width=.62\textwidth]{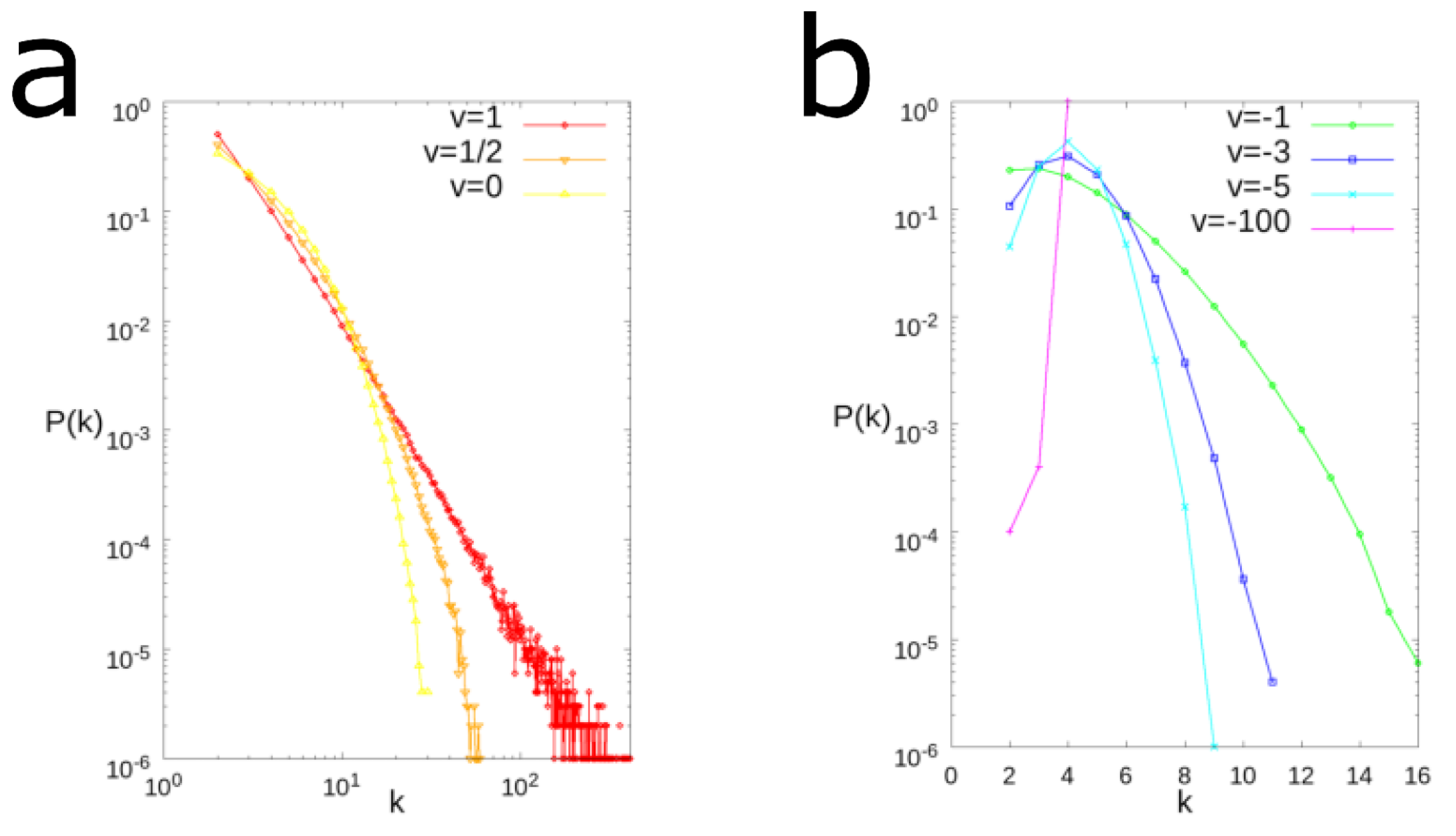}
  \caption{Continuously changing degree distributions from
    (a) power-law as SF networks (red),
    power-law with exponential cut-off (orange), exponential (yellow),
    (b) near Poisson as ER random graphs (green),
    and to narrower ones approaching regular networks
    (blue, light-blue, and purple).
    Note that (a) is a log-log plot, while (b) is a semi-log plot.}
\label{fig_Pk}  
\end{figure}

\begin{figure}[htb]
\centering
\includegraphics[width=.55\linewidth]{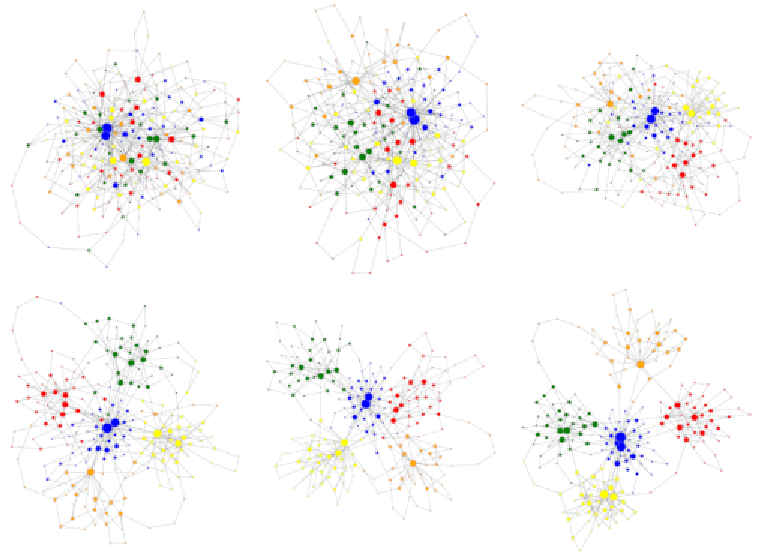}
\caption{Visualization example of SF networks
at $\nu = 1$, $m = 2$, and $N = 200$ with modules
whose strength is increased 
by the tentative rewiring rate
$w' = 0, 0.5, 0.7, 0.9, 0.95$, and $0.98$
from top-left to bottom-right.
Th five colors represent $m_{o} = 5$ modules,
and the circle size of node is proportional to its degree.}
\label{fig1_vis_example}
\end{figure}

\section*{Results}
\subsection*{Various modular networks with continuously changing degree distributions}
We set the parameters
$\nu = 1, 1/2, 0, -1, -3, -5$, and $-100$
in the attachment probabilities proportional to $k_{i}^{\nu}$,
the numbers $m_{o} = 5, 10, 20, 50, 100$, and $200$ of modules,
the number $m = 2$ of attached links per time-step in growing,
and the size $N = 10^{4}$.
Figure \ref{fig_Pk} shows 
power-law (in SF networks),
power-law with exponential cut-off, exponential,
near Poisson \cite{Kusunoki24} (corresponded to ER random graphs),
and narrower degree distributions \cite{Liao22}
(approaching regular networks)
for $\nu = 1, 1/2, 0, -1$, 
and $\nu < -1$, respectively.
These function forms \cite{Krapivsky01} are
  \begin{description}
    \item[power-law] $P(k) = C k^{-\gamma}$, \;\;\; ($\gamma = 3$ for $\nu = 1$), 
    \item[power-law with exponential cut-off]
      $P(k) \approx C' k^{-(1/2 - \mu^{2})} \exp( -2 \mu \sqrt{k})$, 
    \item[exponential] $P(k) = C"  \exp(- a k)$, 
    \item[nearly Poisson] $P(k) \approx \frac{\lambda^{k} e^{- \lambda}}{k !}$,
    \end{description}
and narrower ones with rapidly decreasing tails of 
$\exp(- \beta k \log k)$ \cite{Liao22},
where $C$, $C'$, and $C"$ are normalizing constants,
the values of $\mu$, $a$, $\lambda$ and $\beta$ depends on the values of 
$N$ and $m$ (or $M \approx m \times N$).

\begin{table}[htb]\centering
\caption{Estimated values for the linear relations 
$w = a w' + b$ and $Q = \alpha w + \beta$.}
\begin{footnotesize}
  \begin{tabular}{cc|cc|cc} \hline
    & & $a$ & $b$ 
    & $\alpha$ & $\beta$ \\ \hline
    $\nu=1$    & (SF) & $0.973 \pm 2.21 \times 10^{-3}$ & $0.005 \pm 1.82 \times 10^{-3}$
    & $0.911 \pm 5.73 \times 10^{-3}$ & $0.019 \pm 4.63 \times 10^{-3}$ \\
    $\nu=1/2$  & & $0.998 \pm 3.47 \times 10^{-4}$ & $-0.0008 \pm 2.86 \times 10^{-4}$
    & $0.932 \pm 5.71 \times 10^{-3}$ & $0.003 \pm 4.71 \times 10^{-3}$ \\
    $\nu=0$    & & $0.999 \pm 2.00 \times 10^{-4}$ & $-0.001 \pm 1.65 \times 10^{-4}$
    & $0.934 \pm 5.71 \times 10^{-3}$ & $0.001 \pm 4.71 \times 10^{-3}$ \\
    $\nu=-1$   & (Near ER) & $0.999 \pm 1.67 \times 10^{-4}$ & $-0.001 \pm 1.38 \times 10^{-4}$
    & $0.934 \pm 5.71 \times 10^{-3}$ & $0.001 \pm 4.71 \times 10^{-3}$ \\
    $\nu=-3$   & & $0.999 \pm 1.57 \times 10^{-4}$ & $-0.001 \pm 1.29 \times 10^{-4}$
    & $0.934 \pm 5.70 \times 10^{-3}$ & $0.001 \pm 4.70 \times 10^{-3}$ \\
    $\nu=-5$   & & $0.999 \pm 1.53 \times 10^{-4}$ & $-0.001 \pm 1.27 \times 10^{-4}$
    & $0.934 \pm 5.71 \times 10^{-3}$ & $0.001 \pm 4.71 \times 10^{-3}$ \\
    $\nu=-100$ & (Near Regular) & $0.999 \pm 1.52 \times 10^{-4}$ & $-0.001 \pm 1.26 \times 10^{-4}$
    & $0.934 \pm 5.71 \times 10^{-3}$ & $0.001 \pm 4.71 \times 10^{-3}$ \\\hline
\end{tabular}
\end{footnotesize}
\label{table_estimation}
\end{table}

Figure \ref{fig1_vis_example}
visualizes SF networks with modules by varying
the tentative rewiring rate $w'$.
Similar structural changes are obtained in other networks
without power-law $P(k)$ for $\nu \leq 1/2$.
For the actual rewiring rate $w$ or the corresponded modularity $Q$,
we obtain the linear relations 
$w = a w' + b$ and $Q = \alpha w + \beta$
as estimated in Table \ref{table_estimation}
by using the least squares method over $100$ realizations of
networks with each combination of parameters $\nu$, $w'$, and $m_{o}$.

\subsection*{Robustness of connectivity against attacks}
\begin{figure}[htb]
\centering
\includegraphics[width=.8\linewidth]{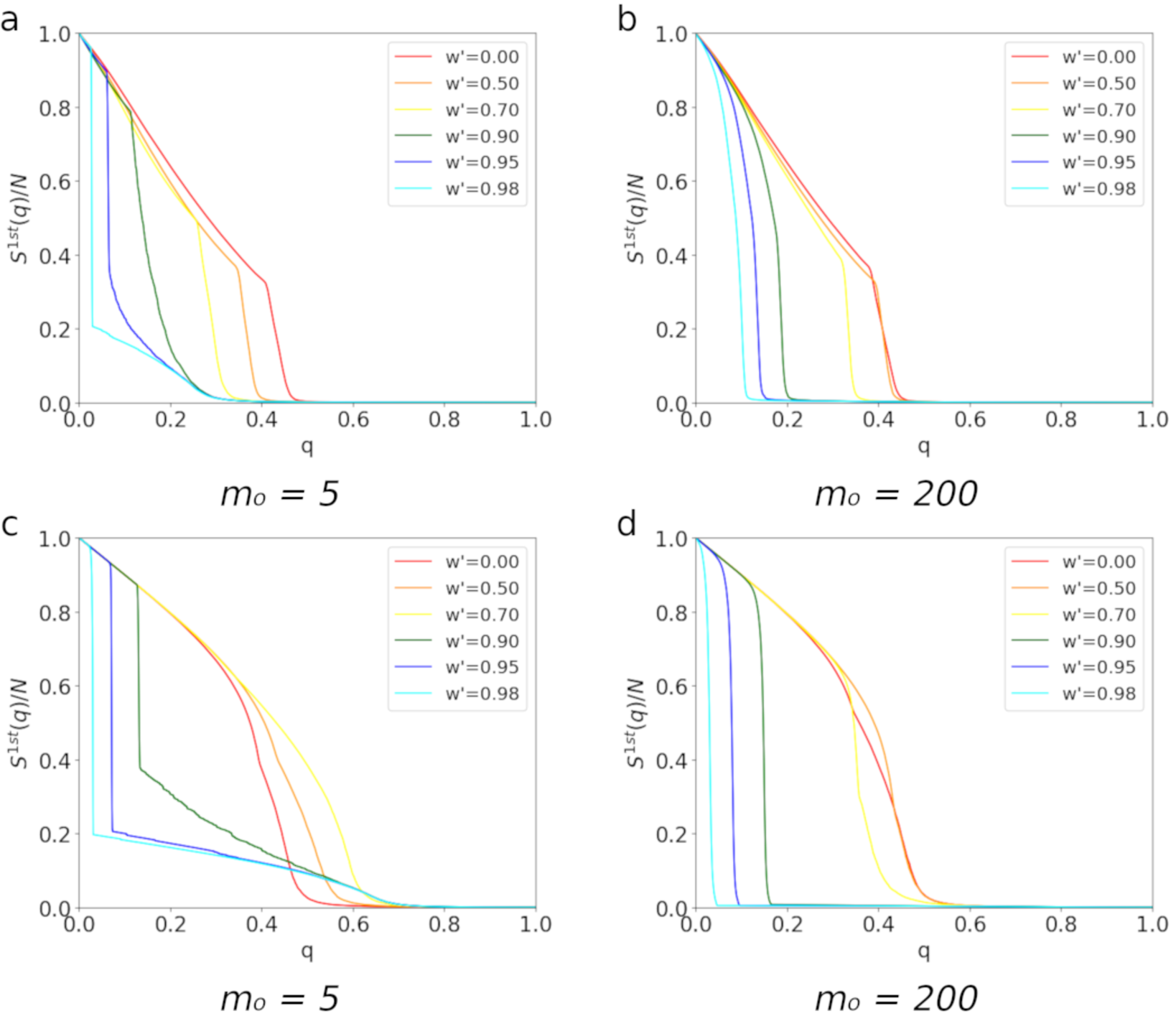}
\caption{Comparison of the areas under the curves 
represented as the robustness against MB attacks
in (a)(b) SF networks at $\nu = 1$
and (c)(d) nearly regular networks at $\nu = -100$
with (a)(c) $m_{o} = 5$ and (b)(d) $200$ modules.
The colored lines from red to light-blue
correspond to values of $w'$
for increasing the modularity $Q$.}
\label{fig2_q-vs-SN}
\end{figure}

The connectivity in a network 
is usually evaluated by the robustness index \cite{Schneider11}
\begin{equation}
 R \stackrel{\rm def}{=} \frac{1}{N}
 \left( \sum_{i=1}^{N-1} \frac{S^{1st}(q)}{N} \right), \;\;\;
 q = \frac{1}{N},
 \; \frac{2}{N}, \; \ldots, \; \frac{i}{N}, \; \ldots, \; \frac{N-1}{N},
\label{eq_def_R}
\end{equation}
where $S^{1st}(q)$ denote
the size of largest connected component (the 1st maximum or 
giant component) at the fraction $q$ of attacks or failures.
%The sum is taken for 
%$q = 1/N, 2/N, 3/N, \ldots, N-1/N$, and $N/N=1$.
We mainly consider the intentional 
module-based (MB) attacks \cite{Cunha15}.
Because they give larger damages to separate the modules by removing inter-links
with priority than the conventional initial degree (ID) attacks and
initial betweenness (IB) attacks without recalculations of degrees
and betweenness centralities (BC) \cite{Freeman1977} after node removals.
Here, the robustness against ID or IB attacks is typically studied 
and compared to that against MB attacks in considering that
end-nodes of inter-links tend to have large BC and to be also chosen by
IB attacks. However, it is also possible that end-nodes of 
intra-links have high BC.
In addition, large BC nodes may be hubs,
therefore some chosen nodes are coincident in these attacks
but others are not as the targets.
These differences will appear in the results as shown later 
(also see \href{https://}{{\it Supplement}} for more details).
In the following results,
we show the averaged values of $Q$, $S^{1st}(q)/N$, $R$,
and the eigenvalue of Laplacian matrix over $100$ realizations.
Figure \ref{fig2_q-vs-SN} shows that
the curves of $S^{1st}(q)/N$ against MB attacks are shifted to left 
for the larger rewiring rate $w'$ with stronger modular structure
from red, orange, yellow, green, blue, to light-blue lines
in SF (Fig. \ref{fig2_q-vs-SN}ab) and
nearly regular networks (Fig. \ref{fig2_q-vs-SN}cd)
at $\nu = 1$ and $-100$, respectively,  for $m_{o} = 5$ and $200$ modules.
Interpolated similar results between these figures 
are obtained for the intermediate
$m_{o} = 10, 20, 50,$ and $100$ modules and 
in other networks whose degree distributions are exponential,
nearly Poisson, and narrower ones for $1/2 \geq \nu \geq -5$
(see \href{https://}{{\it Supplement Fig. S1 $\sim$ S5}}).
In addition, 
although there are not many differences in comparing the 
lines of $S^{1st}(q)/N$ with the same color at a rewiring rate $w'$ 
for SF (Fig. \ref{fig2_q-vs-SN}ab)
and nearly regular networks (Fig. \ref{fig2_q-vs-SN}cd) against MB attacks,
there are remarkable differences in comparing the 
lines of $S^{1st}(q)/N$ with the same color for 
SF and nearly regular networks against ID or IB attacks
(see \href{https://}{{\it Supplement Fig. S6 $\sim$ S15}}).
Such differences are consistent with the previous results \cite{Chujyo22}.
From the definition in Eq. (\ref{eq_def_R}),
the value of $0 \leq R \leq 1/2$ represents 
the area under the curve of $S^{1st}(q)/N$,
a smaller area means more vulnerable against node removals.

\begin{figure}[htb]
\centering
\includegraphics[width=.8\linewidth]{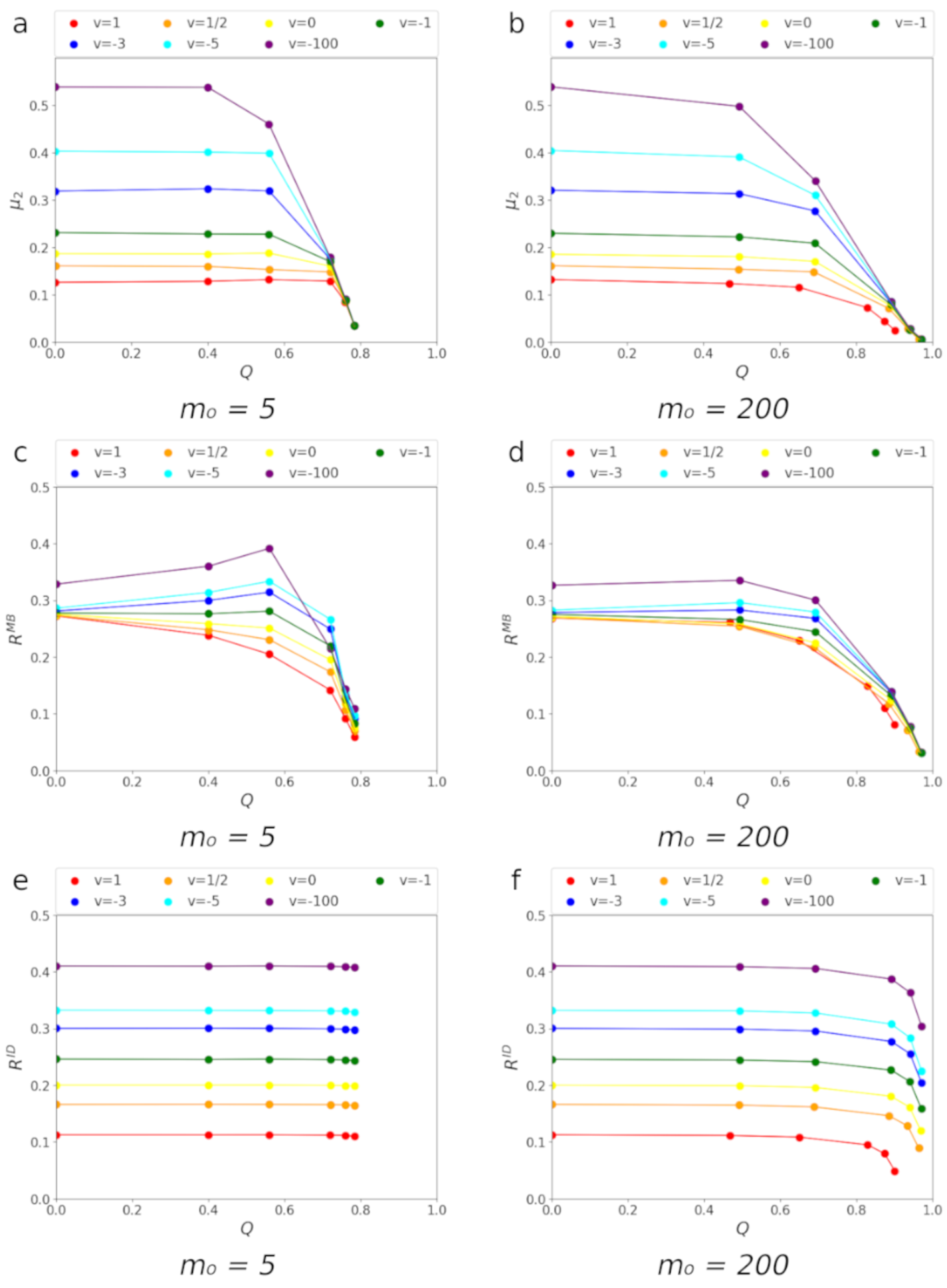}
\caption{Rapid decreasing of the vertical axis value:
(a)(b) the eigenvalue $\mu_{2}$ of Laplacian matrix 
and the robustness index $R$ against
(c)(d) MB and (e)(f) ID attacks
in modular networks measured by the modularity $Q$ of the horizontal axis value.
The degree distributions are continuously changed from
power-law (as SF networks at $\nu = 1$), power-law with exponential cut-off, 
exponential, nearly Poisson (as nearly ER random graphs at $\nu = -1$), and 
narrower ones (approaching regular networks at $\nu < -1$) by varying $\nu$
whose lines are rainbow colored from red to purple.
The modularity $Q$ is controlled by varying $w'$
in (a)(c)(e) $m_{o} = 5$ and (b)(d)(f) $200$ modules.}
\label{fig3_Q-vs-mu2_R}
\end{figure}

\subsection*{A new relation to the robustness}
To find  a new relation to the robustness, 
we also investigate the second smallest eigenvalue $\mu_{2}$ of Laplacian matrix
$L = A -D$,
where $A$ denotes the adjacency matrix
whose element is $A_{ij} = 1$ if nodes $i$ and $j$ are connected or 
$0$ otherwise, 
and $D$ denotes the diagonal matrix $\diag(k_{1}, k_{2}, \ldots, k_{N})$.
Note that $\mu_{2}$ is calculated 
for each of modular networks before attacks.
In Fig. \ref{fig3_Q-vs-mu2_R}abcd,
we emphasize the relation of $Q$ and $\mu_{2}$ or $R$ against MB attacks
in all of the networks denoted by colored lines
for $m_{o} = 5$ and $200$ modules.
Similar results are obtained for the intermediate
$m_{o} = 10, 20, 50,$ and $100$ modules.
In particular,
it is universal that both $\mu_{2}$ and $R$ are rapidly decreased 
for large $Q \gtrapprox 0.8$ as too strong modules or communities
with too dense intra-links
in all of these networks whose $P(k)$ are continuously
changed from power-law, exponential, near Poisson, to narrower ones
by varying $\nu$ denoted as different colored lines
(For more details, see \href{https://}{{\it Supplement Fig. S16,S17}}).
While, as shown in Fig. \ref{fig3_Q-vs-mu2_R}ef, 
the decreasing of $R$ against ID attacks for large $Q$ 
is not remarkable,
the results for IB attacks is the intermediate 
(For more details, see \href{https://}{{\it Supplement Fig. S18,S19}}).
However,
commonly in Fig. \ref{fig3_Q-vs-mu2_R}abcdef,
the decreasing becomes lower than the most gradual parts of red line
for almost non-modular SF networks known as the extremely vulnerable
\cite{Albert00,Callaway00}.
Moreover, we remark that higher values are shown 
from red to purple lines 
whose order corresponds to SF networks, 
ER random graphs, and regular networks.

\begin{figure}[htb]
\centering
\includegraphics[width=.4\linewidth]{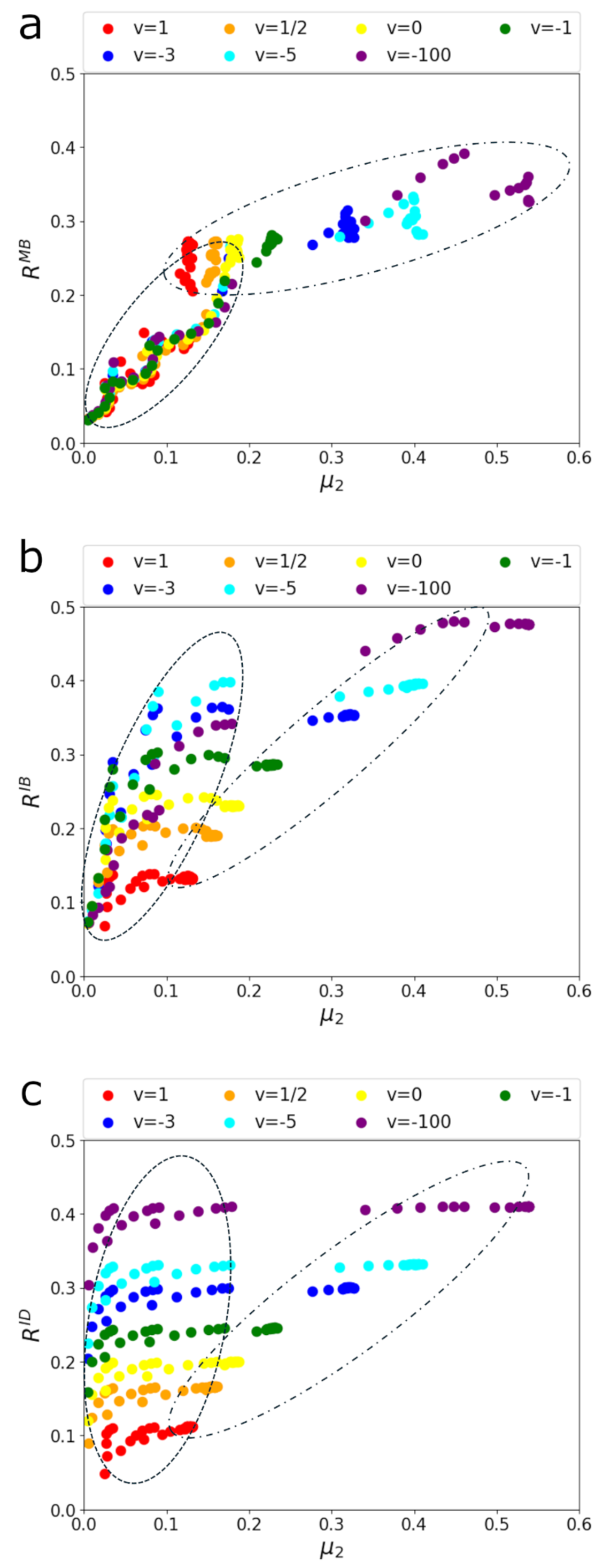}
\caption{Scatter plots of 
the eigenvalue $\mu_{2}$ of Laplacian matrix 
and the robustness index $R$
against (a) MB, (b) IB, (c) ID attacks in networks with
$m_{o} = 5, 10, 20, 50, 100$, and $200$  modules.
Rainbow colors for $\nu = 1, 1/2, 0, -1, -3, -5$ and $-100$ 
correspond to different $P(k)$ interpolated between 
power-law (as SF networks at $\nu = 1$), power-law with exponential cut-off, 
exponential, nearly Poisson (as ER random graphs at $\nu = -1$),
and narrower ones (approaching regular networks at $\nu < -1$).
Lower and upper dashed-oval parts are for $Q > 0.8$ and $Q < 0.8$,
respectively, in different $m_{o}$ modules 
and varying the rewiring rate $w'$.}
\label{fig4_mu2-vs-R}
\end{figure}

Figure \ref{fig4_mu2-vs-R} shows the correlations between
$\mu_{2}$ and $R$ against attacks in these modular networks
(see \href{https://}{{\it Supplement Fig. S20}}
as the detail results for 100 realizations of the networks by 
each parameter $\nu$ or $w'$).
Several superimposed points with a color represent
the results in networks for the different number $m_{o}$ of modules
and varying the rewiring rate $w'$ 
under the condition of the same $P(k)$ at each value of $\nu$
which corresponds to
power-law (as SF networks), exponential, near Poisson (as ER random graphs),
or narrower ones (in approaching regular networks).
It is common that the plots are located from down-left to top-right 
as decreasing both $\nu$ (from red to purple points) and 
$w'$ or its corresponding $Q$ to weaken the modular structure,
especially, the networks become stronger 
as $P(k)$ becomes narrower for smaller $\nu$.
The correlation coefficients are
$0.951$ for $Q > 0.8$ and $0.852$ for $Q < 0.8$
at the lower and the upper dashed-oval parts
in Fig. \ref{fig4_mu2-vs-R}a, respectively.
In other words, against MB attacks, 
small $\mu_{2}$ and $R$
appear clearly with higher correlations as the modularity $Q$ is larger.
In contrast, Fig. \ref{fig4_mu2-vs-R}bc shows that 
the correlation coefficients are 
$0.690$ for $Q > 0.8$ and $0.967$ for $Q < 0.8$ against IB attacks, and 
$0.198$ for $Q > 0.8$ and $0.963$ for $Q < 0.8$ against ID attacks.
Rather,
these higher correlations in the upper dashed-oval parts
represent the pure effect of $P(k)$ 
on the robustness against IB or ID attacks,
because of weak modular structure in $Q < 0.8$ 
(For more detail correlations by each value of $\nu$,
see \href{https://}{{\it Supplement Tables S1 and S2}}).

\section*{Discussion}
We briefly discusses what the maximization of $\mu_{2}$ means from 
the explanations to mathematical theories.
For a network
with the maximum and the minimum degrees $k_{max}$ and $k_{min}$, 
$\mu_{2}$ is dominant to be fast epidemic or information delivery,
since the solution of diffusion equation
$d \bm{x} / dt = - L \bm{x}$ is given by
\[
  \bm{x}(t) = \sum_{i = 1}^{N} c_{i} e^{- \mu_{i}} \bm{u}_{i}
  \approx c_{1} \bm{1} + c_{2} e^{- \mu_{2}} \bm{u}_{2},
\]
where $\bm{u}_{i}$ is the $i$-th eigenvector of Laplacian matrix $L$,
$\bm{u}_{1} = \bm{1} \stackrel{\rm def}{=} (1, 1, \ldots, 1)$, and 
the coefficients $c_{1}, c_{2}, \ldots, c_{N}$
are determined by the initial state
$\bm{x}(0) = \sum_{i = 1}^{N} c_{i} \bm{u}_{i}$.
Note that 
$e^{- \mu_{i}}$ is more quickly converged to zero than $e^{- \mu_{2}}$
for $i \geq 3$ 
because of $0 = \mu_{1} < \mu_{2} \leq \mu_{3} \leq \ldots \leq \mu_{N}$.
In other words,
a larger $\mu_{2}$ gives a more compact network with fast diffusion.
This property is related to the following theoretical results.
First,
the diameter $\rho$ (as the maximum length of the 
shortest path between nodes)
and the average length $\langle l_{ij} \rangle$ of the shortest paths
between nodes are upper bounded by
\begin{equation}
  O\left(\sqrt{\frac{k_{max}}{\mu_{2}}} \log N \right),
  \label{eq_u-bound_DL}
\end{equation}
for a connected network \cite{Mohar91,Meighem11},
where the length of the shortest path is counted by hops between nodes.
Thus, these smaller values of $\rho$ and $\langle l_{ij} \rangle$
give a compact network as $\mu_{2}$ is larger.
We remark that the construction of 
SF networks is based on efficiency 
and therefore compact with small $\langle l_{ij} \rangle$ 
even for $M = O(N)$ links \cite{Cancho01,Mathias01}
but vulnerable against hub attacks \cite{Albert00,Callaway00},
however the coexisting of compactness (or small-worldness)
and strong robustness is possible
in other networks especially as random regular.

Second,
the expanding constant
\[
  h(G) \stackrel{\rm def}{=} \min \left\{
  \frac{|\partial F|}{|F|}: \;\;\; F \subseteq V, \;\;
  0 < |F| < \frac{N}{2} \right\},
\]
is also bounded as
\begin{equation}
  \frac{\mu_{2}}{2} \leq h(G) \leq \sqrt{2 d \mu_{2}},
  \label{eq_expand}
\end{equation}
for a finite connected $d$-regular graph $G$ \cite{Davidoff03},
where $V$ denotes a set of nodes with the size $|V| = N$,
$\partial F$ denotes a set of links emanated from nodes in $F$ to
other nodes in $V - F$,
$\mu_{2}$ equals to the spectral gap
$\lambda_{N} - \lambda_{N-1}$,
$d = \lambda_{N} \geq \lambda_{N-1} \geq \ldots \geq \lambda_{1}$
as the eigenvalues of adjacency matrix $A$.
Thus,
$h(G)$ can be larger as the order of $\mu_{2}$ 
from the lower bound in the left-hand of Eq. (\ref{eq_expand}),
which means the strong connectivity 
with many inter-links in $\partial F$
against a bisection into $F$ and $V -F$ 
by these link removals \cite{Meighem11}.
Note that
$h(G)$ is also limited 
by the order of $\sqrt{\mu_{2}}$ 
from the upper bound in the right-hand of Eq. (\ref{eq_expand}).
In addition, for a network, 
$\mu_{2}$ can be larger, when $k_{min}$ is large from \cite{Meighem11}
\begin{equation}
 \mu_{2} \leq \frac{N}{N -1} k_{min}.
 \label{eq_bound-by-kmin}
\end{equation}
These Eqs. (\ref{eq_expand}) and (\ref{eq_bound-by-kmin})
become the bases for supporting our obtained result:
Weak modular
networks with narrower degree distributions (larger $k_{min}$)
have stronger connectivity against attacks, at the extreme case 
the nearly regular networks have the largest $R$ and $\mu_{2}$
in Figs. \ref{fig2_q-vs-SN}, \ref{fig3_Q-vs-mu2_R}, and \ref{fig4_mu2-vs-R}.
From Eqs. (\ref{eq_u-bound_DL}) and (\ref{eq_bound-by-kmin}),
smaller $k_{max}$ and larger $k_{min}$ are also desirable.

Third, in a $d$-regular graph,
a larger $\mu_{2}$ corresponds to a smaller $\lambda_{N-1}$.
However, $\lambda_{N-1}$ is greater than or equal to $2 \sqrt{d - 1}$
\cite{Lobotzky1994}.
The smallest case $\lambda_{N-1} = 2 \sqrt{d - 1}$ is the condition
whether or not a $d$-regular graph becomes a Ramanujan graph
known as good with small diameter $\rho$ and large expanding constant $h(G)$.
Strictly,
a Ramanujan graph has the constraints for the limited
numbers $N$ and $M = d \times N / 2$ of nodes and links, e.g.
$N = p^{r}$, $r \geq 2$, $q \geq 2 \sqrt{p}$, $d = (p^{r} - 1)/(p - 1)$,
where $p$ and $q$ are odd primes in an algebraic construction
\cite{Davidoff03,Lobotzky1994}.
Therefore it belongs to a very small subclass of $d$-regular graphs.
However, from the asymptotic property \cite{Friedman03},
the strong robustness of weak modular $d$-regular networks
may be corresponded to the goodness of Ramanujan graphs.
In more detail, it has been derived that \cite{Friedman03},
for any integer $d \geq 4$ and $\epsilon > 0$,
a random $d$-regular graph with sufficiently large $N$ asymptotically
approach a Ramanujan graph in the meaning that
the eigenvalues $\lambda_{i}$
of adjacency matrix except for $\lambda_{N} = d$ satisfy
$|\lambda_{i}| \leq 2 \sqrt{d - 1} + \epsilon$, $i = 1, \ldots, N-1$,
as $\epsilon \rightarrow 0$ for $N \rightarrow \infty$.
Note that, for $d=4$, the maximum spectral gap is
$d - 2 \sqrt{d - 1} \approx 0.53$ as nearly as vertical
axis intercepts of purple lines in Fig. \ref{fig3_Q-vs-mu2_R}ab.
In addition,
since a Ramanujan graph has links determined 
by operations on the finite group
of projective general linear group $PGL_{2}(q)$
or projective special linear group $PSL_{2}(q)$
using an equivalence class based on modulo
\cite{Davidoff03,Lobotzky1994},
neighbor nodes of two end-nodes of a link seem to be far away
on the paths except for the link, 
when $N$ is large.
This is also supported by that "{\it the frequency of times spent by
simple random walk in a nontrivial cycle is almost surely $0$ on
every infinite Ramanujan graph}", where "{\it nontrivial cycle
is not purely a backtracking cycle}" \cite{Lyons15}.
Note that short cycles exist in the local module structure.
Thus, 
Ramanujan graphs are probably non-modular 
$d$-regular graphs, and close to random ones for a large $N$.
As further studies with deeper insights, 
there remain theoretical issues or more careful numerical checks,
which may also be  useful from an application points of view.

\section*{Conclusion}
We have comprehensively investigated the robustness of connectivity
in modular networks with continuously changing degree distributions,
which include power-law (SF networks), nearly Poisson (ER random graphs),
and nearly unimodal (regular networks) for interpolating the previous
analyses \cite{Shai15,Nguyen21,Kim23} at only three types
of typical networks.
Although tremendous efforts 
\cite{Albert00,Callaway00,Schneider11,Tanizawa12,Holme11,Dorogovtsev04,Cohen10}
has been devoted to slow progress
research on the robustness in complex networks,
we have additionally revealed the universal property that
strong modular networks, even with any degree distributions, 
are extremely vulnerable against intentional attacks.
Moreover,
between the statistical robustness of connectivity \cite{Schneider11,Tanizawa12}
and the algebraic connectivity \cite{Meighem11,Davidoff03,Mohar91}, 
we have found a new relation that
both the robustness index $R$ and
the second smallest eigenvalue $\mu_{2}$ of Laplacian matrix 
become smaller as the modularity $Q$ is larger.
In particular,
the correlations between $R$ and $\mu_{2}$ are high,
when MB attacks are given in a strong modular structure with $Q > 0.8$,
while ID or IB attacks are given
in a weak modular structure with $Q < 0.8$.
The later case as the pure effect of degree distributions 
is also supported by that
the robustness becomes stronger as networks approach regular
from SF \cite{Chujyo22}, 
and that $\mu_{2}$ can become larger in the upper and lower bounds
\cite{Mohar91,Meighem11,Davidoff03}
as $k_{min}$ is larger and $k_{max}$ is smaller
with narrower degree distributions.
Therefore
too dense modules or communities should be avoided
for maintaining the essential network function of connectivity
against intentional attacks or disasters.
In other words,
the excessive numbers of links to neighbor nodes counted by hops are unsuitable
for not only technological infrastructures or biochemical reactions 
but also social meanings
in increasing contemporary threats of military conflicts or large disasters, 
though dense unions as connecting nodes seem to give a positive impression.
In more details, 
future studies are remained to elucidate the mechanism in slightly 
different behavior for varying the number $m_{o}$ of modules
and the robustness against other more destructive attacks \cite{Artime24},
e.g. attacks on collective influencers \cite{Makse15} or
candidates of FVS \cite{Zhou16},
in taking into account that
the worst node removals is CND problem
\cite{Santos18} as NP-hard.

%\begin{figure}[htb]
%  \includegraphics[width=.4\textwidth]{fig_direction.eps}
%
%  \includegraphics[width=.4\textwidth]{fig_neighbors.eps}
%\caption{} \label{fig_mapping}
%\end{figure}

\section*{Data availability}
The data analyzed in this study are available from the corresponding author on reasonable request.
%The datasets used and/or analysed during the current study available from the corresponding author on reasonable request.

\section*{Abbreviations}
SF: scale-free, CND: critical node detection, FVS: feedback vertex set,
NP: non-deterministic polynomial time, ER: Erd\"{o}s-R\'{e}nyi,
GN: growing network, IPA: inverse preferential attachment,
MB: module-based, ID: initial degree, IB: initial betweenness,
BC: betweenness centralities.

%\section*{Acknowledgments}
%The authors express appreciation to anonymous reviewers 
%for their valuable comments to improve the manuscript.

\section*{Author Information}
\begin{description}
  \item[Affiliation] Graduate School of 
    Advanced Institute of Science and Technology, 
    Japan Advanced Institute of Science and Technology, 
    Ishikawa 923-1292 Japan\\
    Yukio Hayashi, and Taishi Ogawa
  \item[Contributions]
    Y.H. conceived and designed the research, 
    analyzed data and discuss results, 
    and wrote the manuscript. T.O. implemented and performed 
    numerical experiments.
    %H.K. preprocessed the data of networks.
  \item[Competing Interests] 
    The authors declare no competing interests.
  \item[Corresponding Author] Yukio Hayashi
\end{description}

\newpage
\section*{Supplementary Materials}
\renewcommand{\thefigure}{S\arabic{figure}}
\renewcommand{\thetable}{S\arabic{table}}
\setcounter{figure}{0}
\setcounter{table}{0}

We set the parameters
$\nu = 1, 1/2, 0, -1, -3, -5$, and $-100$
in the attachment probabilities proportional to $k_{i}^{\nu}$,
the numbers $m_{o} = 5, 10, 20, 50. 100$, and $200$ of modules,
the number $m = 2$ of attached links per time-step in growing,
and the size $N = 10^{4}$.
In the following results,
we show the averaged values over $100$ realizations.

\begin{figure}[htb]
  \begin{minipage}{.48\textwidth}
    \includegraphics[width=.9\textwidth]{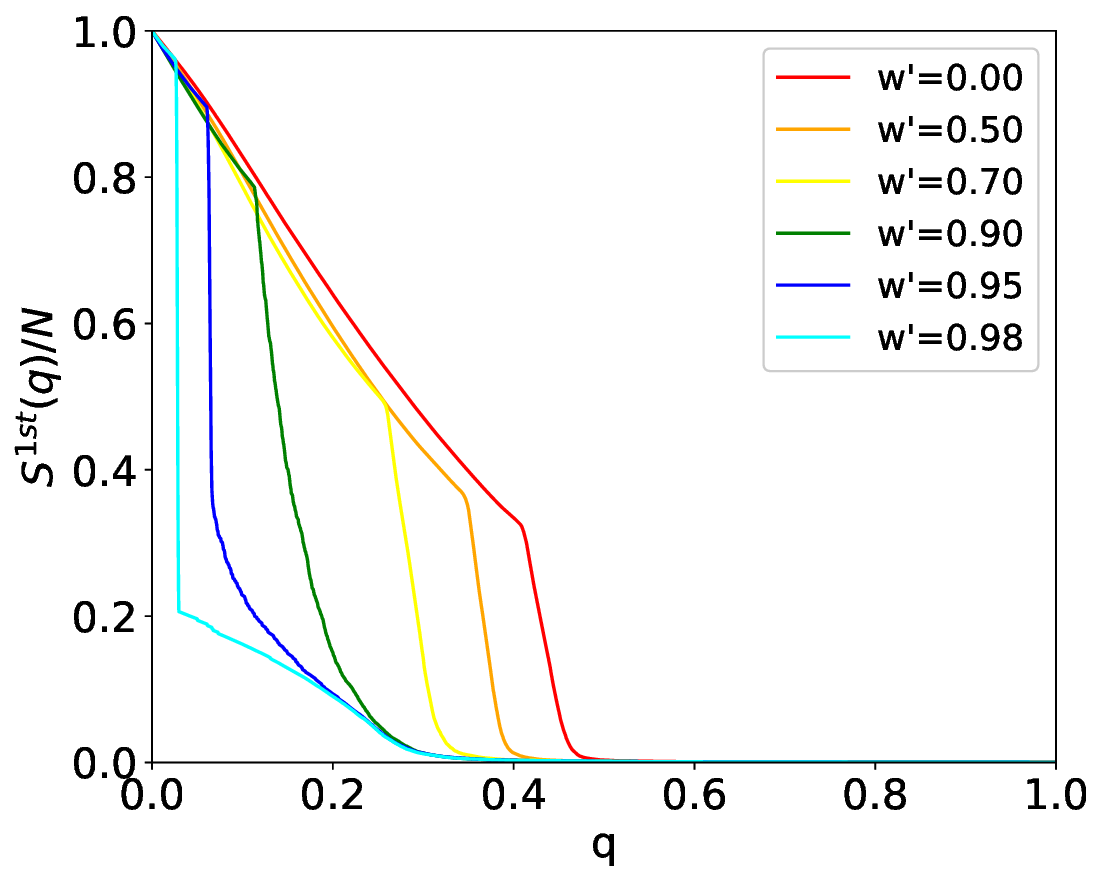}
    \begin{center} (a) $m_{o} = 5$ \end{center}  
  \end{minipage}
  \hfill  
  \begin{minipage}{.48\textwidth}
    \includegraphics[width=.9\textwidth]{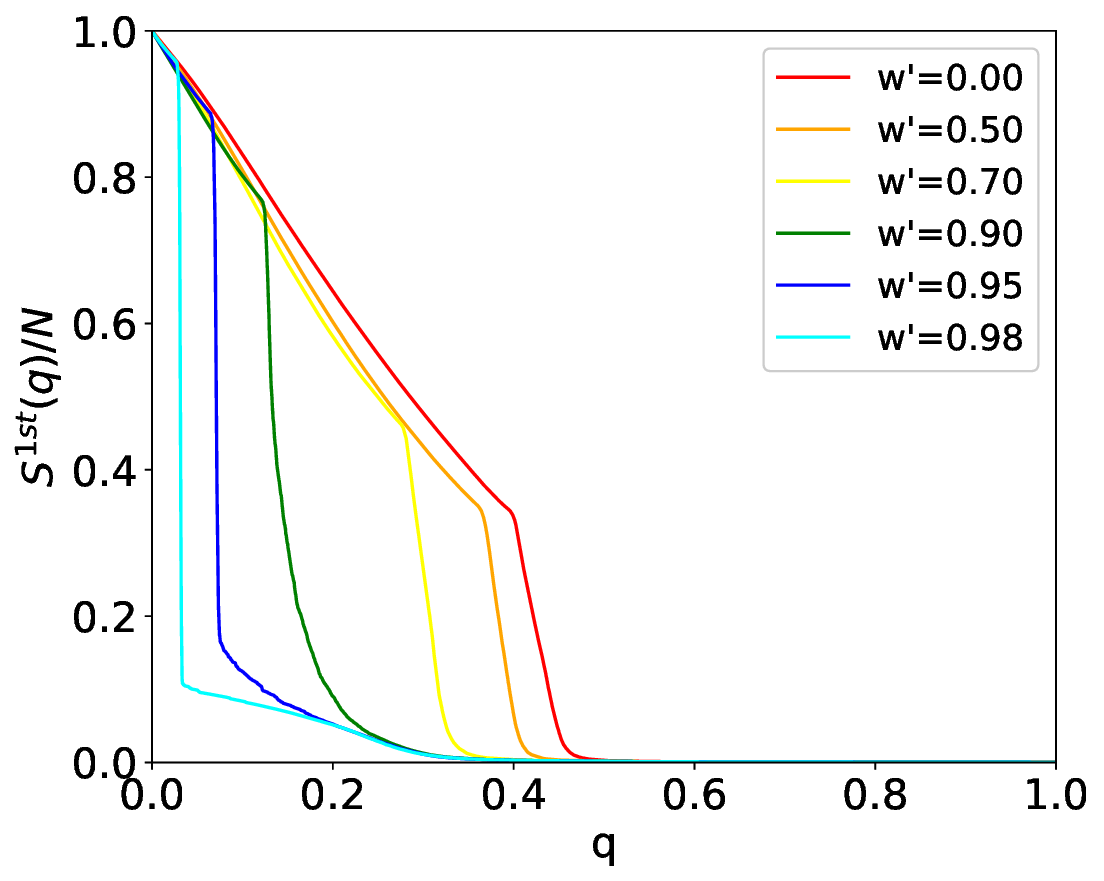}
    \begin{center} (b) $m_{o} = 10$ \end{center}
  \end{minipage}    
  \hfill
  \begin{minipage}{.48\textwidth}
    \includegraphics[width=.9\textwidth]{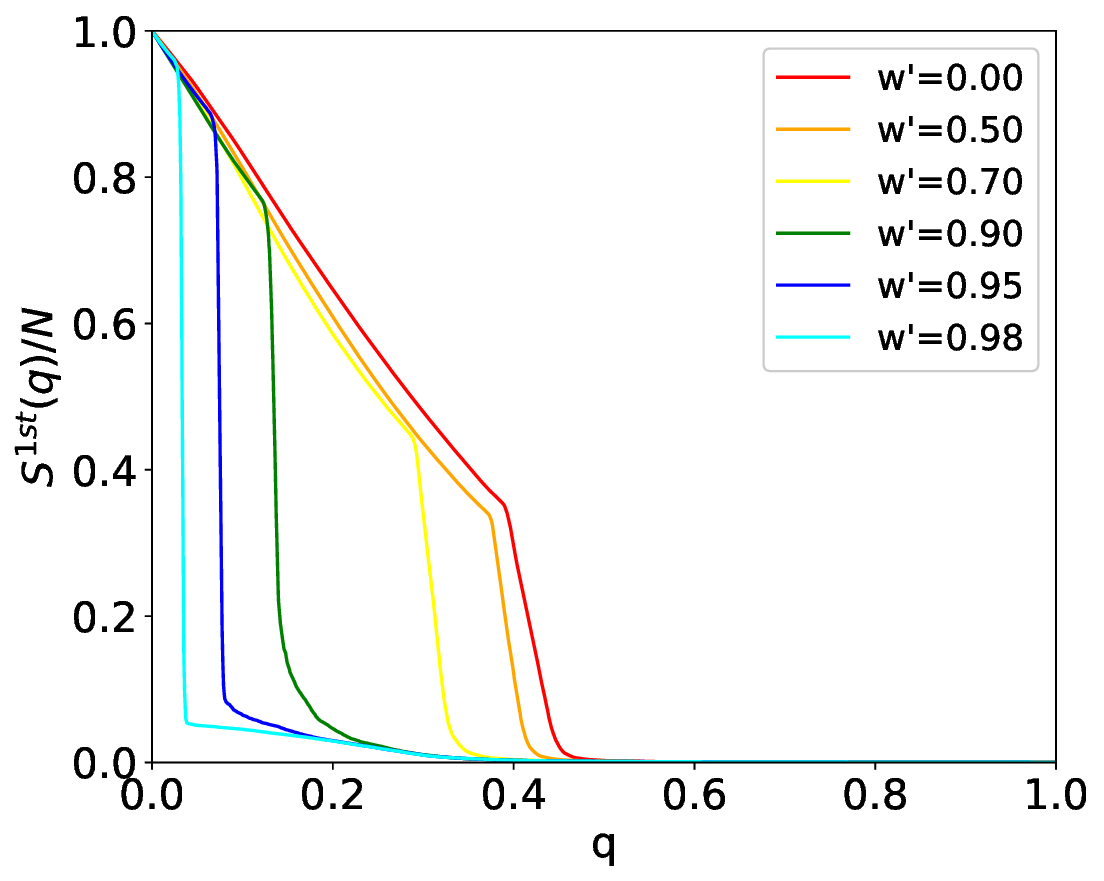}
    \begin{center} (c) $m_{o} = 20$ \end{center}
  \end{minipage}
  \hfill  
  \begin{minipage}{.48\textwidth}
    \includegraphics[width=.9\textwidth]{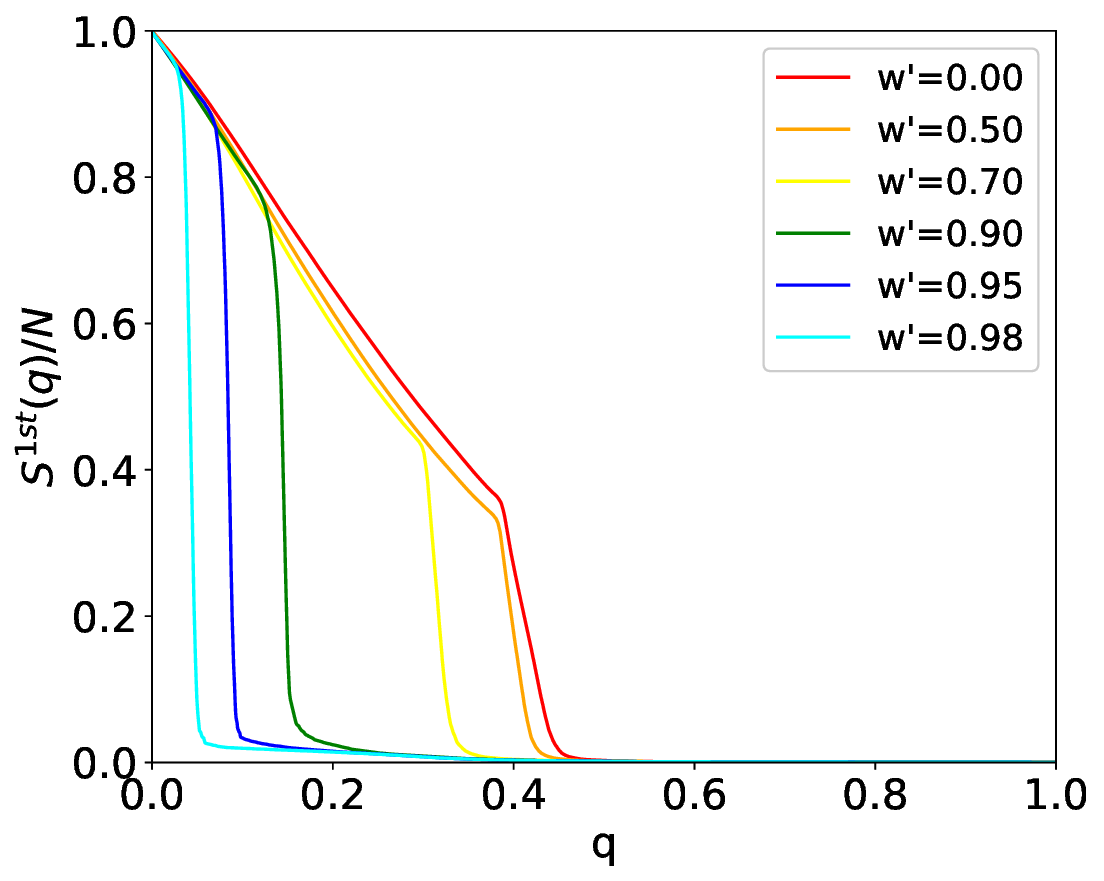}
    \begin{center} (d) $m_{o} = 50$ \end{center}
  \end{minipage}     
  \hfill 
  \begin{minipage}{.48\textwidth}
    \includegraphics[width=.9\textwidth]{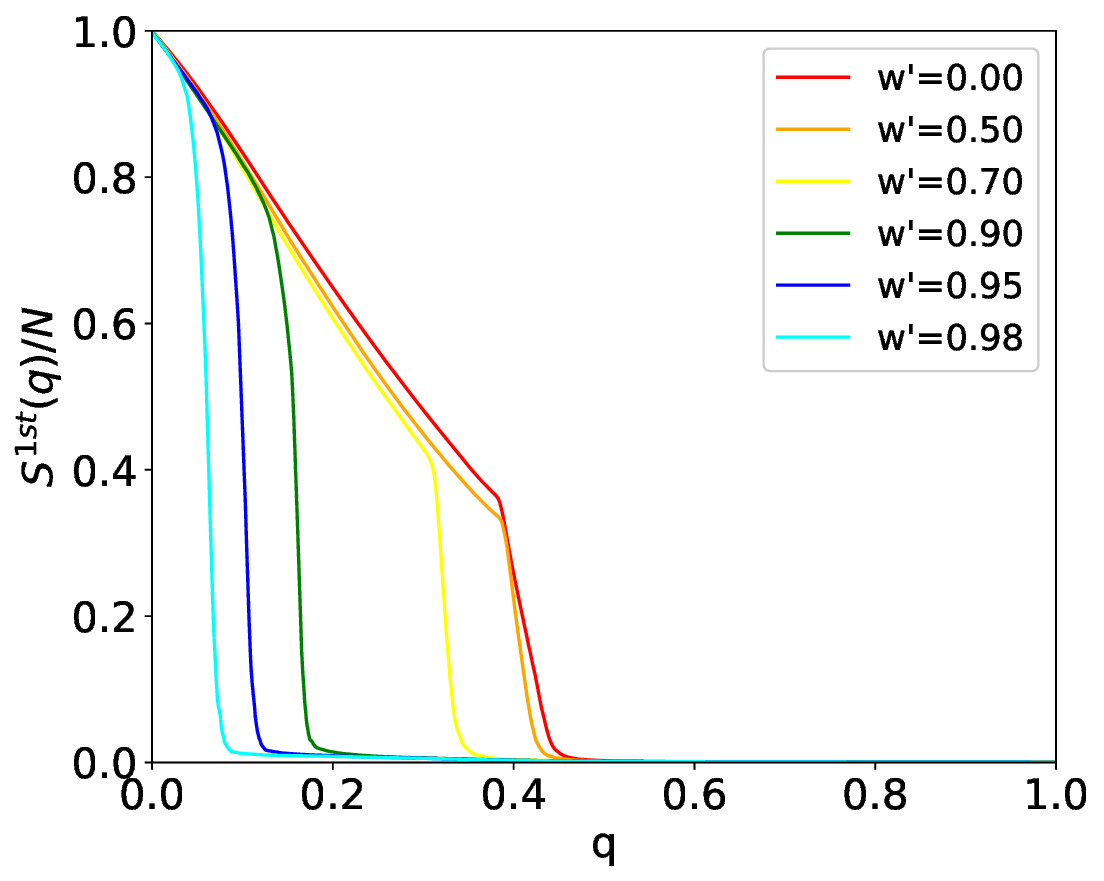}
    \begin{center} (e) $m_{o} = 100$ \end{center}
  \end{minipage}
  \hfill  
  \begin{minipage}{.48\textwidth}
    \includegraphics[width=.9\textwidth]{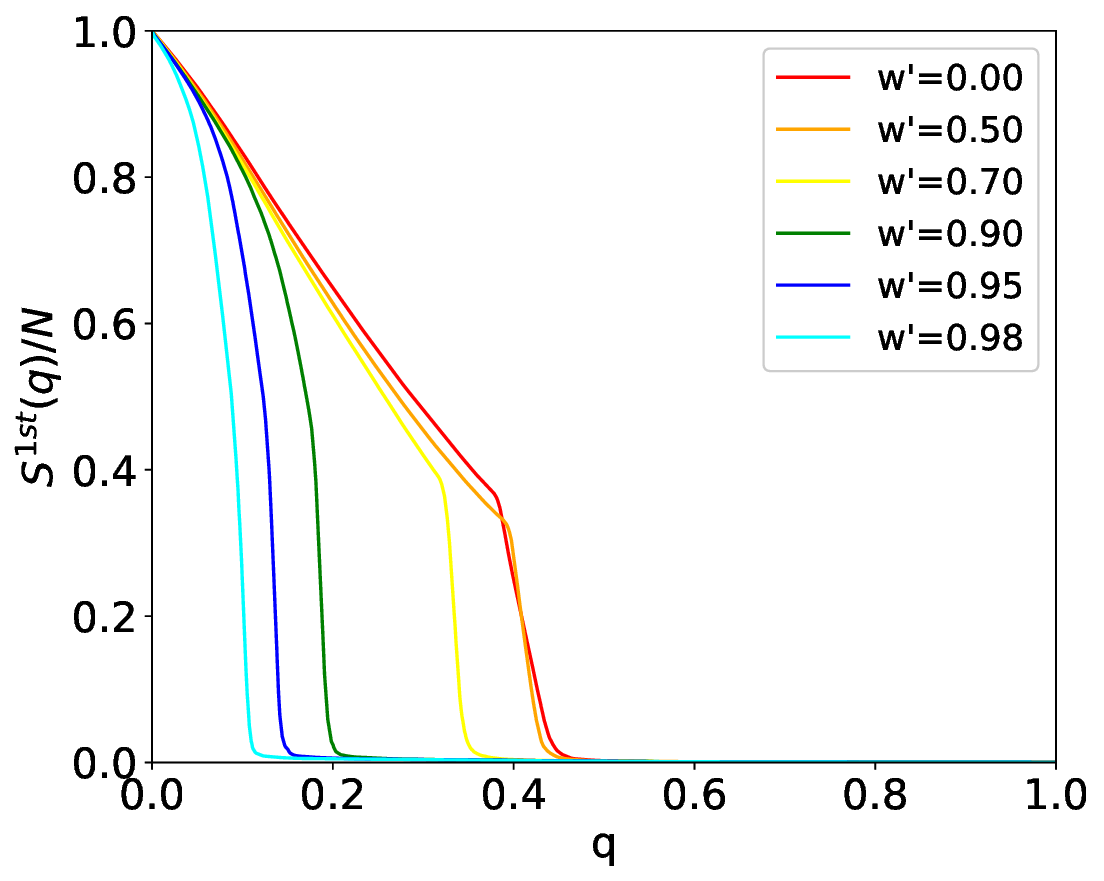}
    \begin{center} (f) $m_{o} = 200$ \end{center}
  \end{minipage}     
%\centering
%\includegraphics[width=.8\textwidth]{resize_figS2.eps}
\caption{Comparison of the areas under the curves 
represented as the robustness against MB attacks
in SF networks at $\nu = 1$ with $m_{o}$ modules.}
\label{fig_MB_nu1}
\end{figure}

In Figs. \ref{fig_MB_nu1} $\sim$ \ref{fig_ID_nu-100},
it is common that 
the curves of $S(q)/N$ against attacks are shifted to left 
for larger rewiring rate $w'$ with stronger modular structure
colored from red, orange, yellow, green, blue, to light-blue.
However, in comparing these figures, 
the areas under the curves are larger as $\nu$ is smaller 
in approaching regular from SF networks.

In Figs. \ref{fig_Q-mu2} $\sim$ \ref{fig_Q-R^ID},
it is common that higher values for each of the vertical axes
are shown from red to purple lines,
which correspond to $P(k)$ changed from 
power-law (as SF networks), power-law with exponential cut-off, 
exponential, nearly Poisson (as ER random graphs), and 
narrower ones (approaching regular networks) by varying $\nu$, 
although 
the intervals of lines and decreasing behavior are slightly different
in comparing these figures.

\newpage
\begin{table}[htb]\centering
\caption{The correlation coefficients between $\mu_{2}$ and $R$
against attacks in Fig. 5 for modular networks,
whose $P(k)$ are controlled by $\nu$ as the attachment probabilities
proportional to $k_{i}^{\nu}$.}

\begin{scriptsize}
\begin{center} (a) MB attacks \end{center}
\begin{tabular}{ccc} \hline
  $\nu$ & $Q < 0.8$ & $Q > 0.8$ \\ \hline\hline
  1     & 0.111 & 0.822 \\
  1/2   & 0.661 & 0.977 \\
  0     & 0.555 & 0.977 \\
  -1    & 0.909 & 0.975 \\
  -3    & 0.300 & 0.968 \\
  -5    & 0.097 & 0.960 \\
  -100  & -0.095 & 0.955 \\ \hline
  Total & 0.852 & 0.951 \\ \hline
\end{tabular}

\begin{center} (b) IB attacks \end{center}
\begin{tabular}{ccc} \hline
  $\nu$ & $Q < 0.8$ & $Q > 0.8$ \\ \hline\hline
  1     & 0.357 & 0.562 \\
  1/2   & 0.375 & 0.675 \\
  0     & 0.303 & 0.689 \\
  -1    & 0.758 & 0.749 \\
  -3    & 0.933 & 0.841 \\
  -5    & 0.982 & 0.898 \\
  -100  & 0.729 & 0.960 \\ \hline
  Total & 0.967 & 0.690 \\ \hline
\end{tabular}

\begin{center} (c) ID attacks \end{center}
\begin{tabular}{ccc} \hline
$\nu$ & $Q < 0.8$ & $Q > 0.8$ \\ \hline\hline
  1     & 0.776 & 0.576 \\
  1/2   & 0.775 & 0.619 \\
  0     & 0.961 & 0.609 \\
  -1    & 0.933 & 0.596 \\
  -3    & 0.929 & 0.605 \\
  -5    & 0.961 & 0.593 \\
  -100  & 0.841 & 0.588 \\ \hline
  Total & 0.963 & 0.198 \\ \hline
\end{tabular}
\end{scriptsize}
\label{table_detail}
\end{table}

\newpage
\begin{table}[htb]\centering
\caption{The correlation coefficients between $\mu_{2}$ and $R$
against attacks in Fig. S20 for 100 realizations of the networks,
whose $P(k)$ are controlled by $\nu$ as the attachment probabilities
proportional to $k_{i}^{\nu}$.}

\begin{scriptsize}
\begin{center} (a) MB attacks \end{center}
\begin{tabular}{ccc} \hline
  $\nu$ & $Q < 0.8$ & $Q > 0.8$ \\ \hline\hline
  1     & 0.550 & 0.755 \\
  1/2   & 0.685 & 0.953 \\
  0     & 0.766 & 0.963 \\
  -1    & 0.840 & 0.966 \\
  -3    & 0.861 & 0.968 \\
  -5    & 0.849 & 0.963 \\
  -100  & 0.861 & 0.966 \\ \hline
  Total & 0.775 & 0.945 \\ \hline
\end{tabular}

\begin{center} (b) IB attacks \end{center}
\begin{tabular}{ccc} \hline
  $\nu$ & $Q < 0.8$ & $Q > 0.8$ \\ \hline\hline
  1     & -0.316 & 0.571 \\
  1/2   & -0.498 & 0.712 \\
  0     & -0.517 & 0.736 \\
  -1    & -0.189 & 0.799 \\
  -3    &  0.492 & 0.878 \\
  -5    &  0.660 & 0.920 \\
  -100  &  0.929 & 0.957 \\ \hline
  Total &  0.862 & 0.724 \\ \hline
\end{tabular}

\begin{center} (c) ID attacks \end{center}
\begin{tabular}{ccc} \hline
$\nu$ & $Q < 0.8$ & $Q > 0.8$ \\ \hline\hline
  1     & 0.127 & 0.573 \\
  1/2   & 0.114 & 0.632 \\
  0     & 0.115 & 0.630 \\
  -1    & 0.147 & 0.620 \\
  -3    & 0.167 & 0.634 \\
  -5    & 0.235 & 0.621 \\
  -100  & 0.302 & 0.615 \\ \hline
  Total & 0.775 & 0.200 \\ \hline
\end{tabular}
\end{scriptsize}
\label{table_100reallization}
\end{table}

\newpage

\begin{figure}[htb]
  \begin{minipage}{.48\textwidth}
    \includegraphics[width=.9\textwidth]{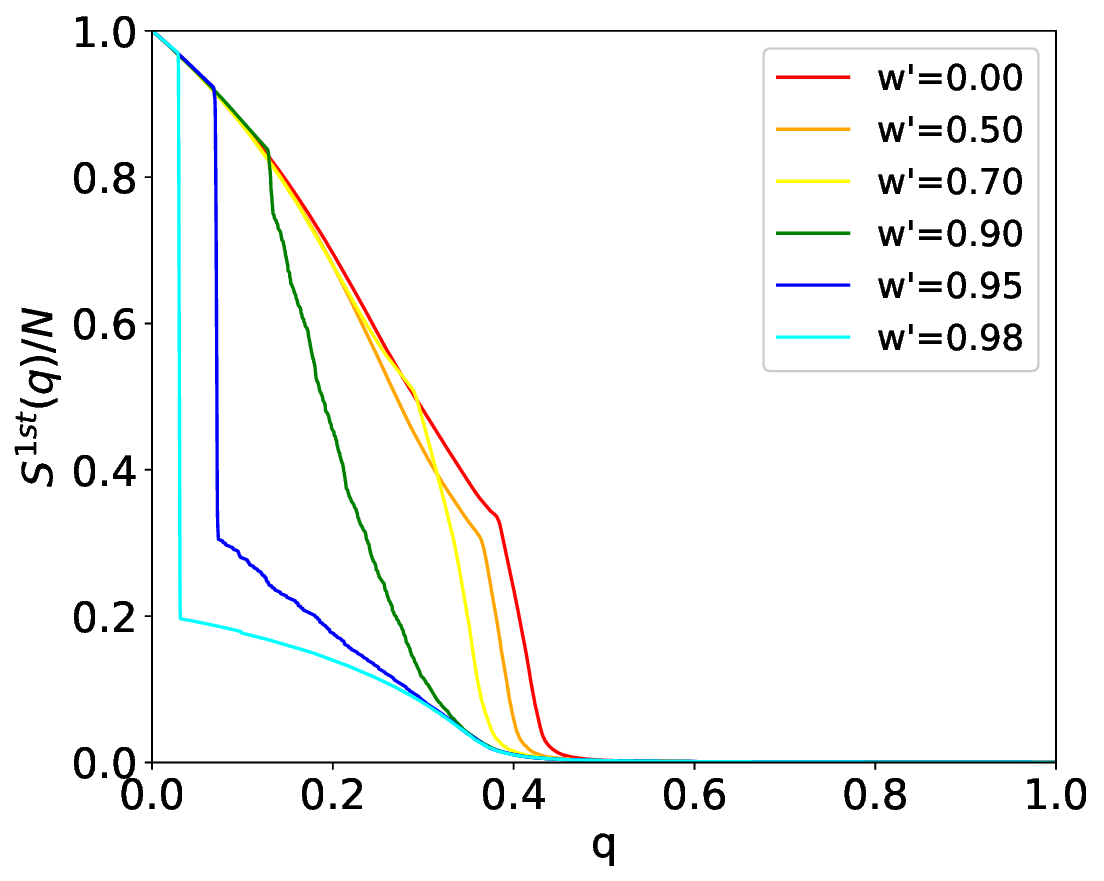}
    \begin{center} (a) $m_{o} = 5$ \end{center}  
  \end{minipage}
  \hfill  
  \begin{minipage}{.48\textwidth}
    \includegraphics[width=.9\textwidth]{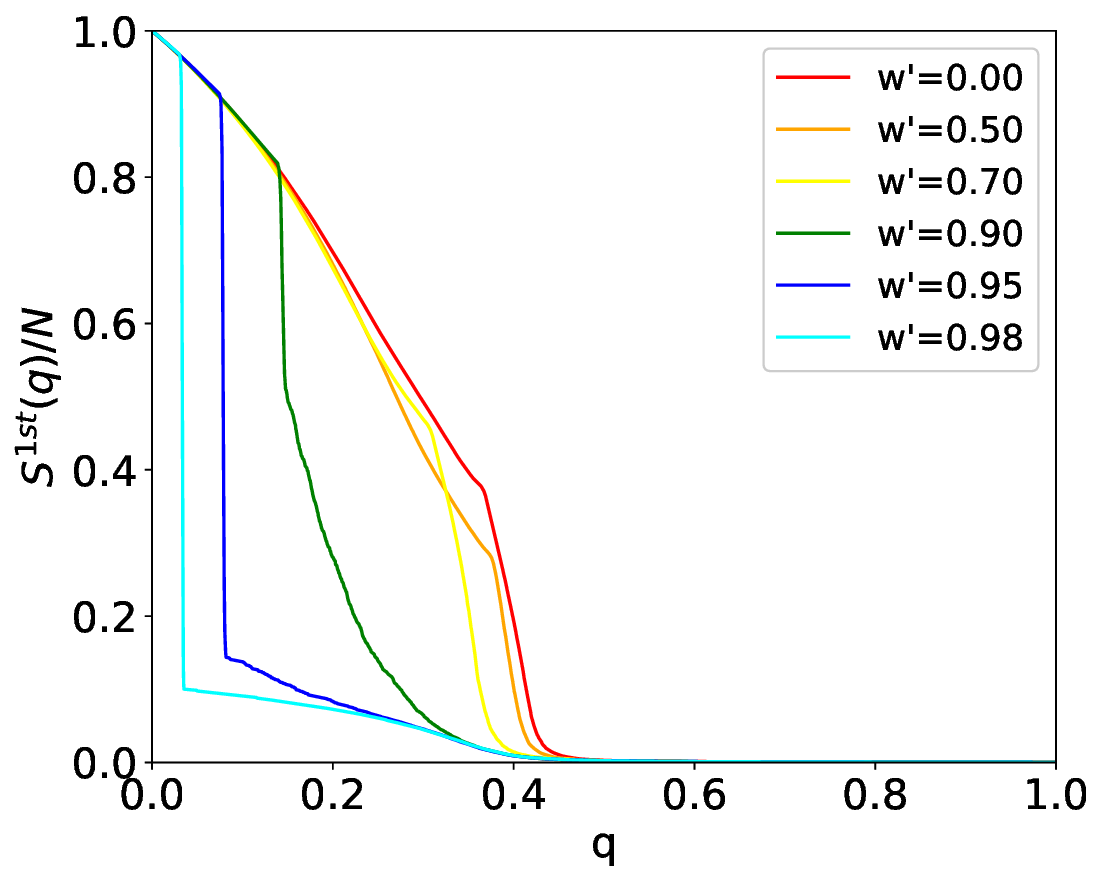}
    \begin{center} (b) $m_{o} = 10$ \end{center}
  \end{minipage}    
  \hfill
  \begin{minipage}{.48\textwidth}
    \includegraphics[width=.9\textwidth]{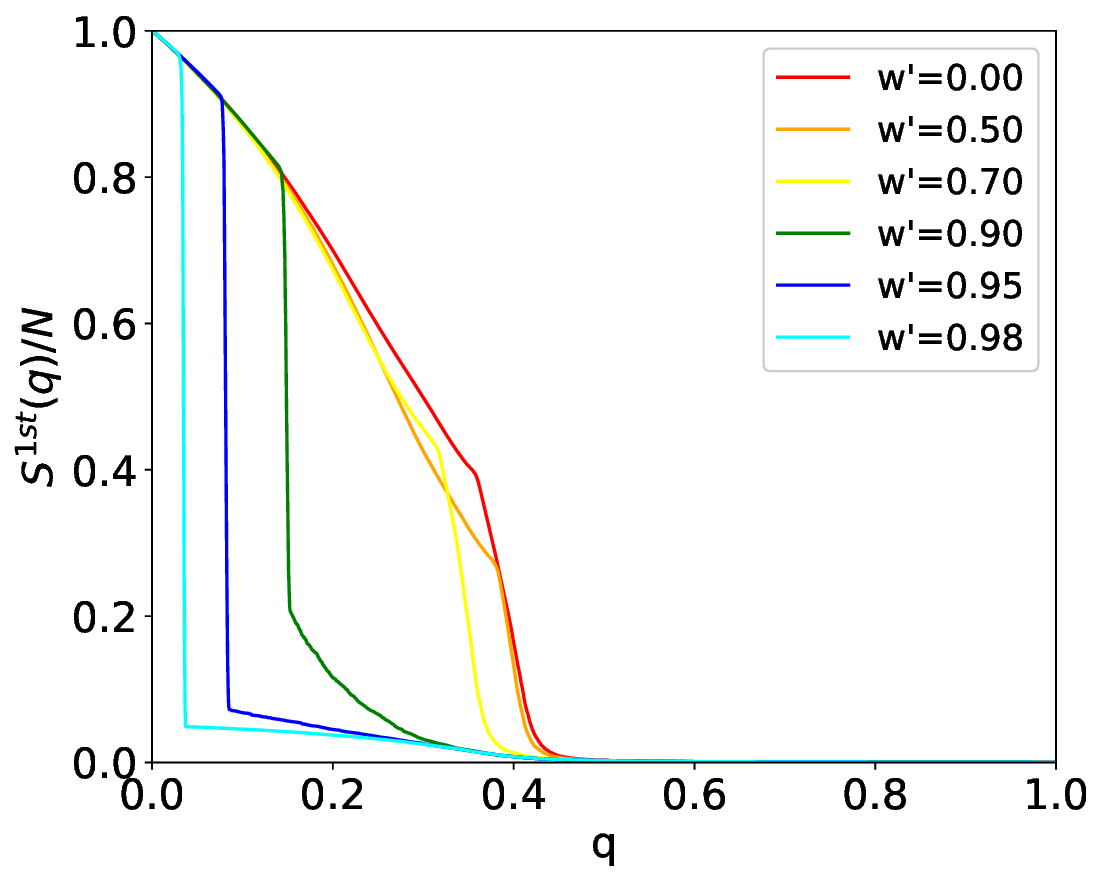}
    \begin{center} (c) $m_{o} = 20$ \end{center}
  \end{minipage}
  \hfill  
  \begin{minipage}{.48\textwidth}
    \includegraphics[width=.9\textwidth]{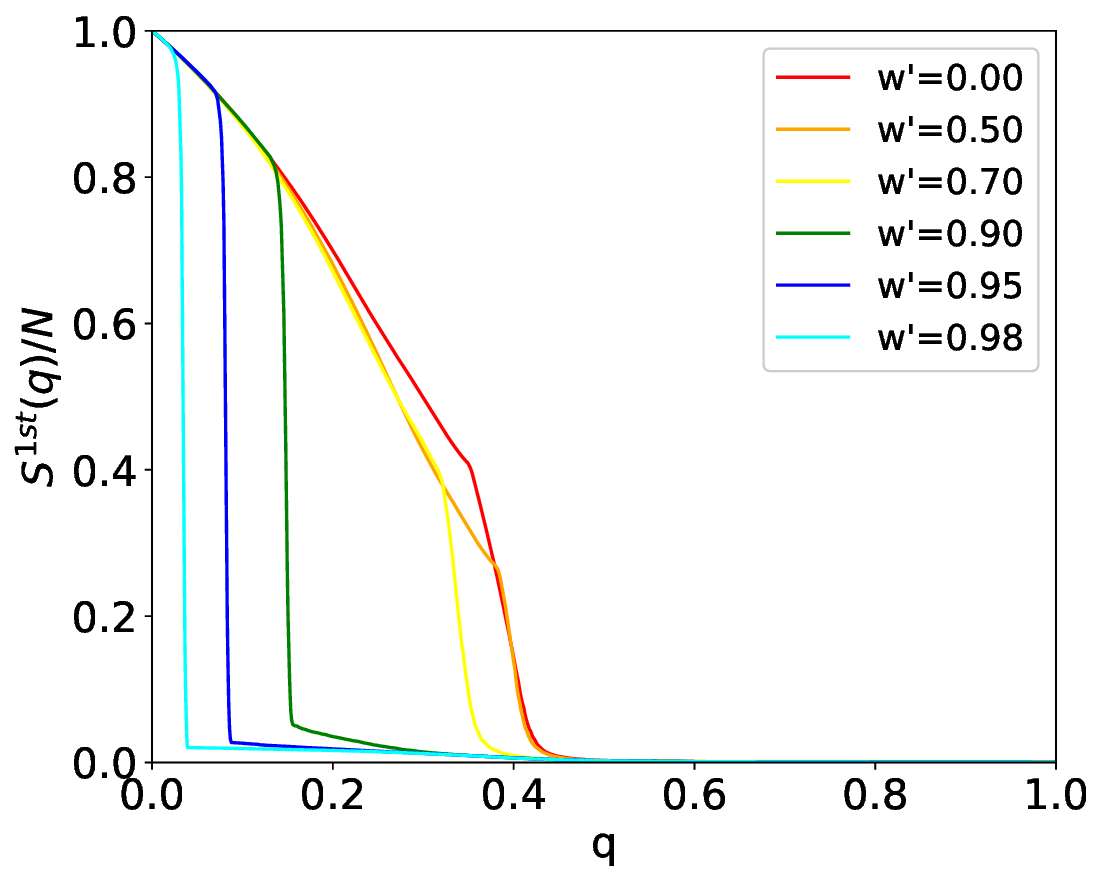}
    \begin{center} (d) $m_{o} = 50$ \end{center}
  \end{minipage}     
  \hfill 
  \begin{minipage}{.48\textwidth}
    \includegraphics[width=.9\textwidth]{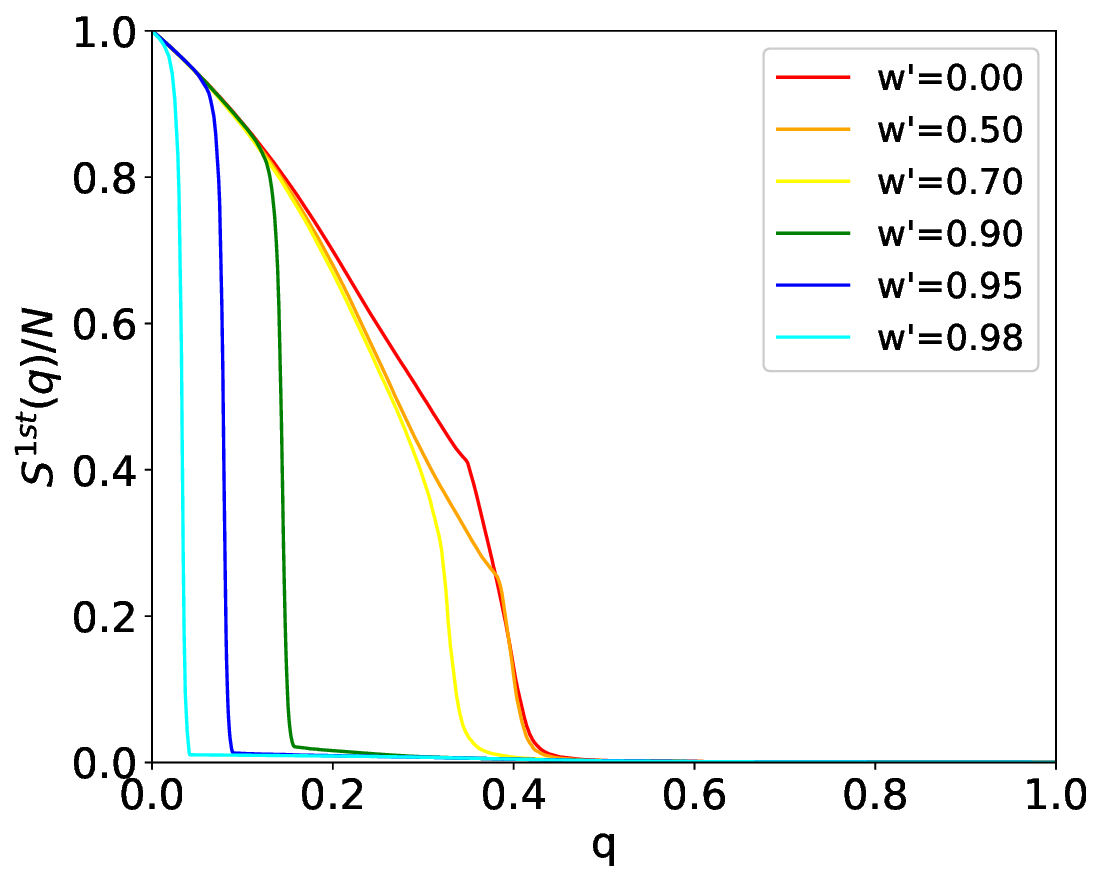}
    \begin{center} (e) $m_{o} = 100$ \end{center}
  \end{minipage}
  \hfill  
  \begin{minipage}{.48\textwidth}
    \includegraphics[width=.9\textwidth]{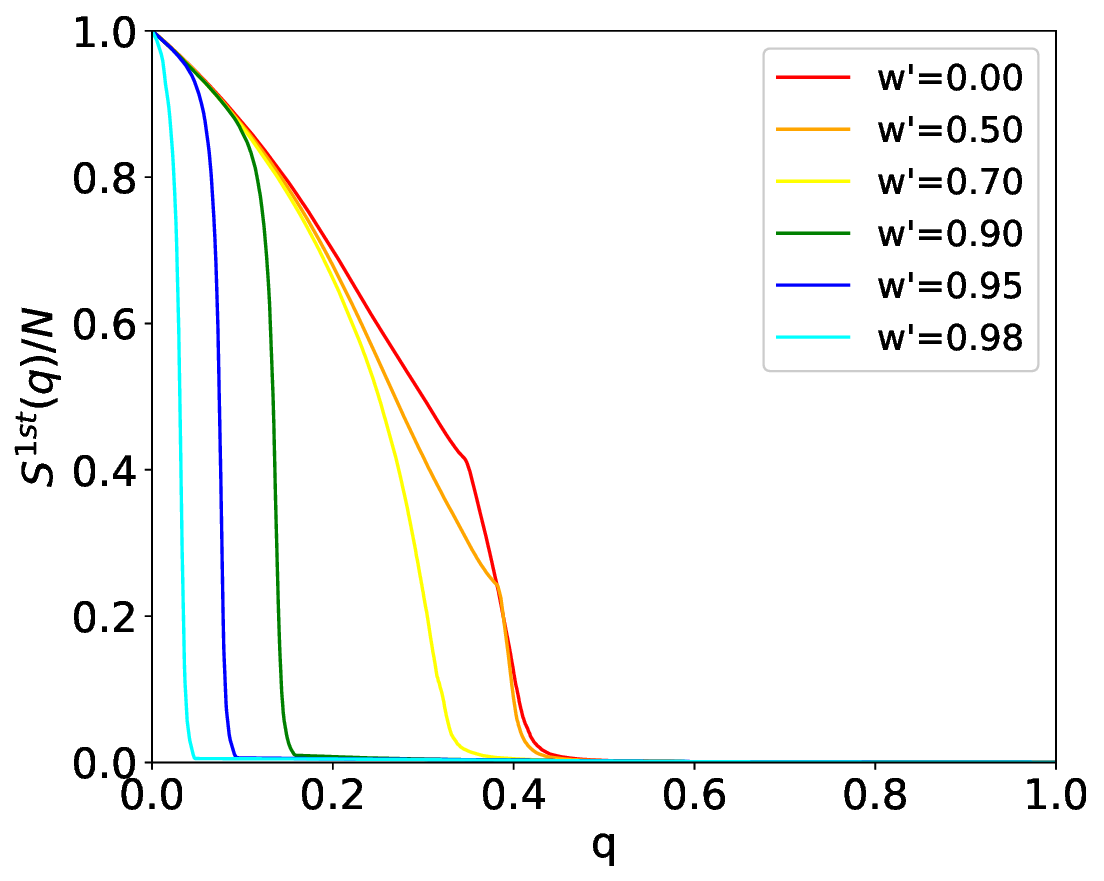}
    \begin{center} (f) $m_{o} = 200$ \end{center}
  \end{minipage}       
%\centering
%\includegraphics[width=.8\textwidth]{resize_figS3.eps}
\caption{Comparison of the areas under the curves 
represented as the robustness against MB attacks
in randomly attached networks at $\nu = 0$ with $m_{o}$ modules.}
\label{fig_MB_nu0}
\end{figure}

\begin{figure}[htb]
  \begin{minipage}{.48\textwidth}
    \includegraphics[width=.9\textwidth]{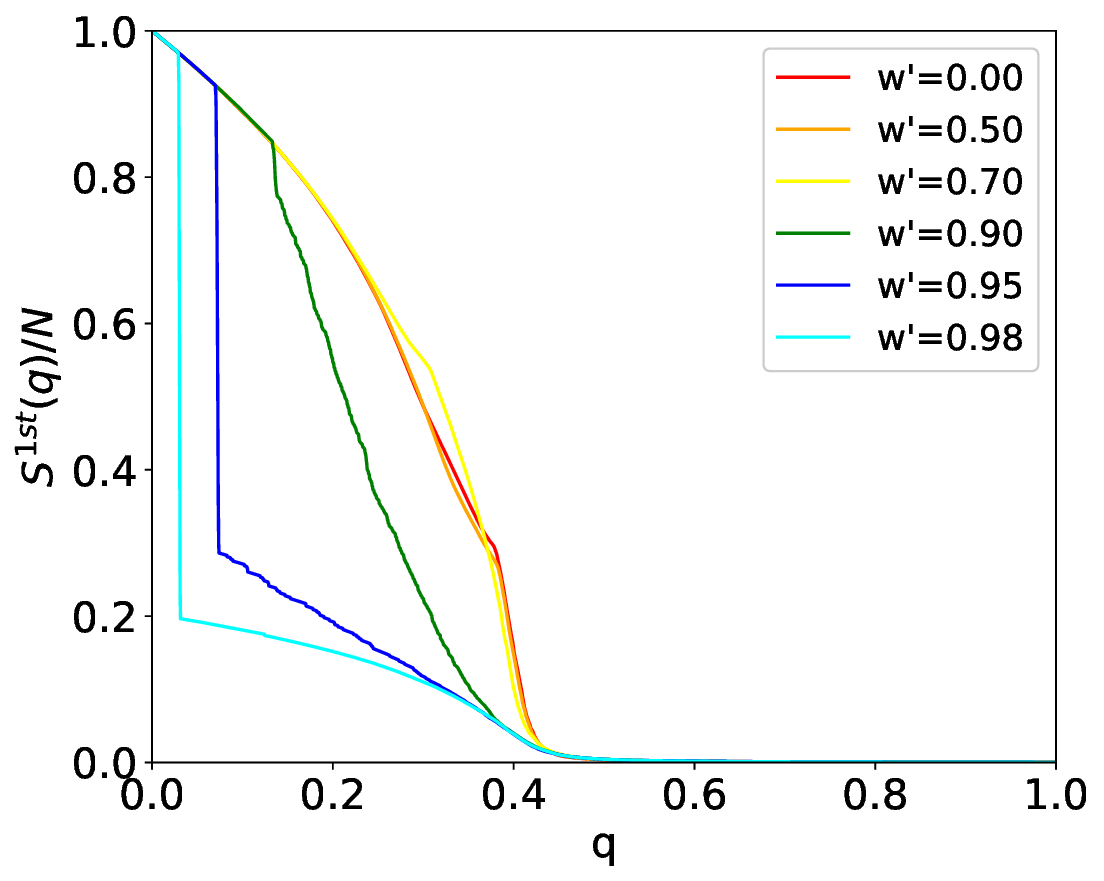}
    \begin{center} (a) $m_{o} = 5$ \end{center}  
  \end{minipage}
  \hfill  
  \begin{minipage}{.48\textwidth}
    \includegraphics[width=.9\textwidth]{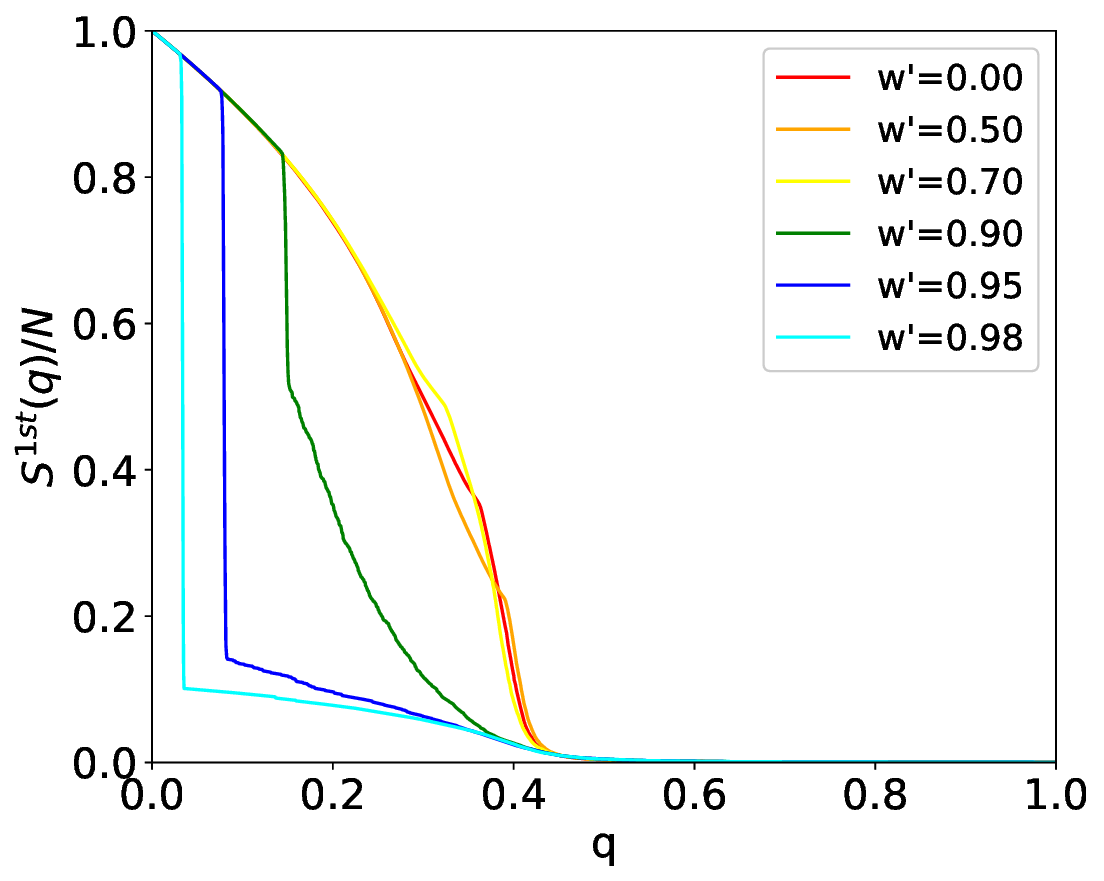}
    \begin{center} (b) $m_{o} = 10$ \end{center}
  \end{minipage}    
  \hfill
  \begin{minipage}{.48\textwidth}
    \includegraphics[width=.9\textwidth]{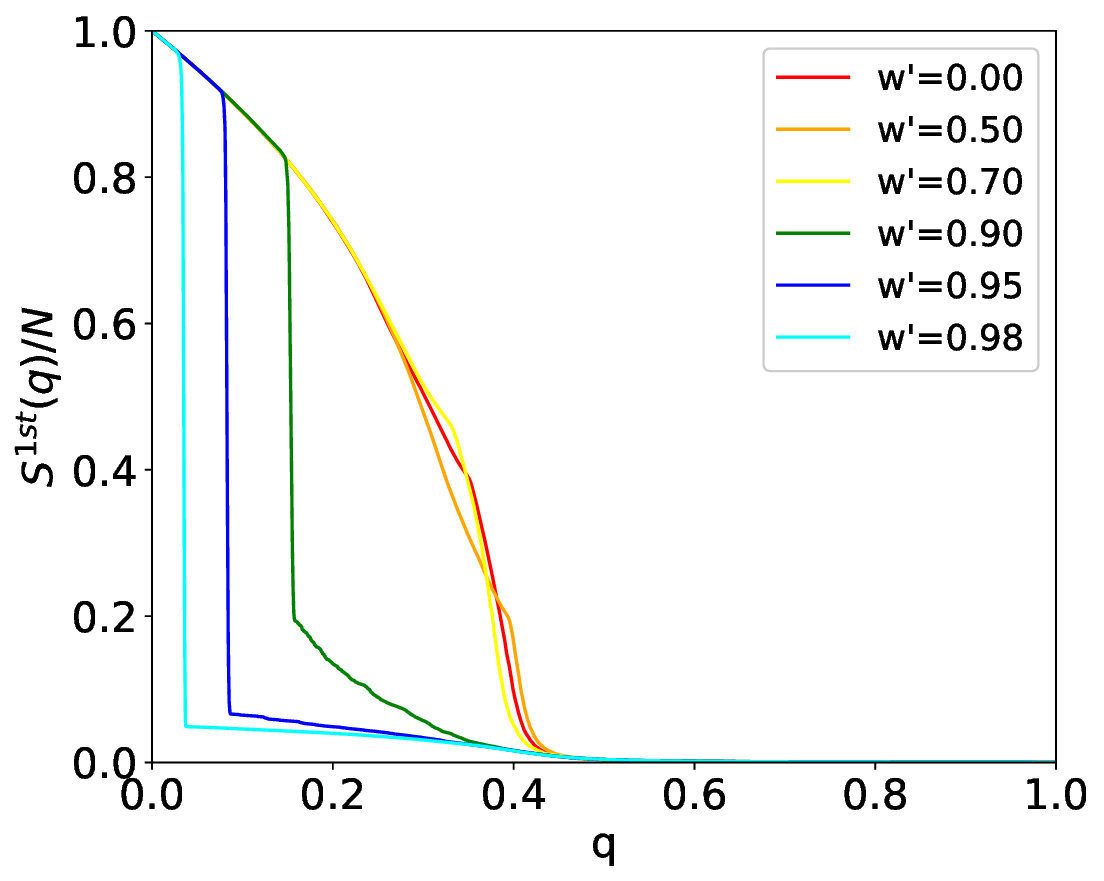}
    \begin{center} (c) $m_{o} = 20$ \end{center}
  \end{minipage}
  \hfill  
  \begin{minipage}{.48\textwidth}
    \includegraphics[width=.9\textwidth]{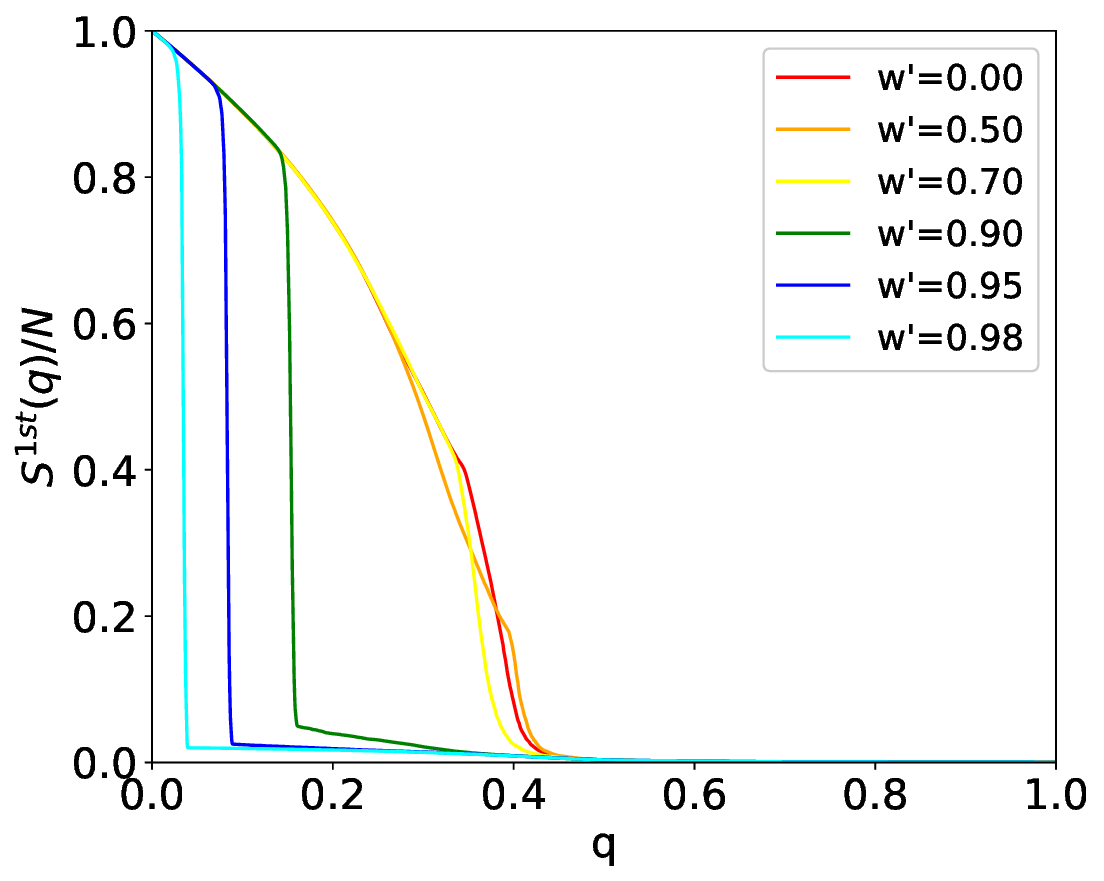}
    \begin{center} (d) $m_{o} = 50$ \end{center}
  \end{minipage}     
  \hfill 
  \begin{minipage}{.48\textwidth}
    \includegraphics[width=.9\textwidth]{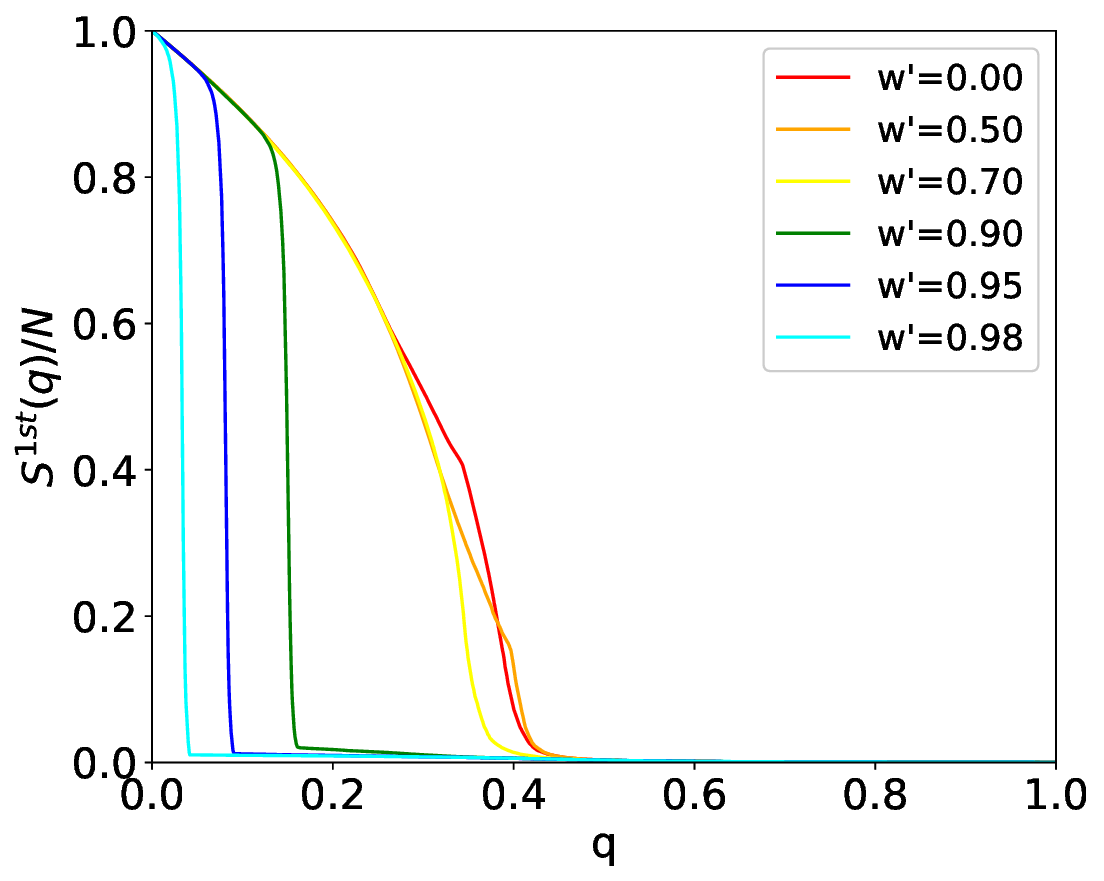}
    \begin{center} (e) $m_{o} = 100$ \end{center}
  \end{minipage}
  \hfill  
  \begin{minipage}{.48\textwidth}
    \includegraphics[width=.9\textwidth]{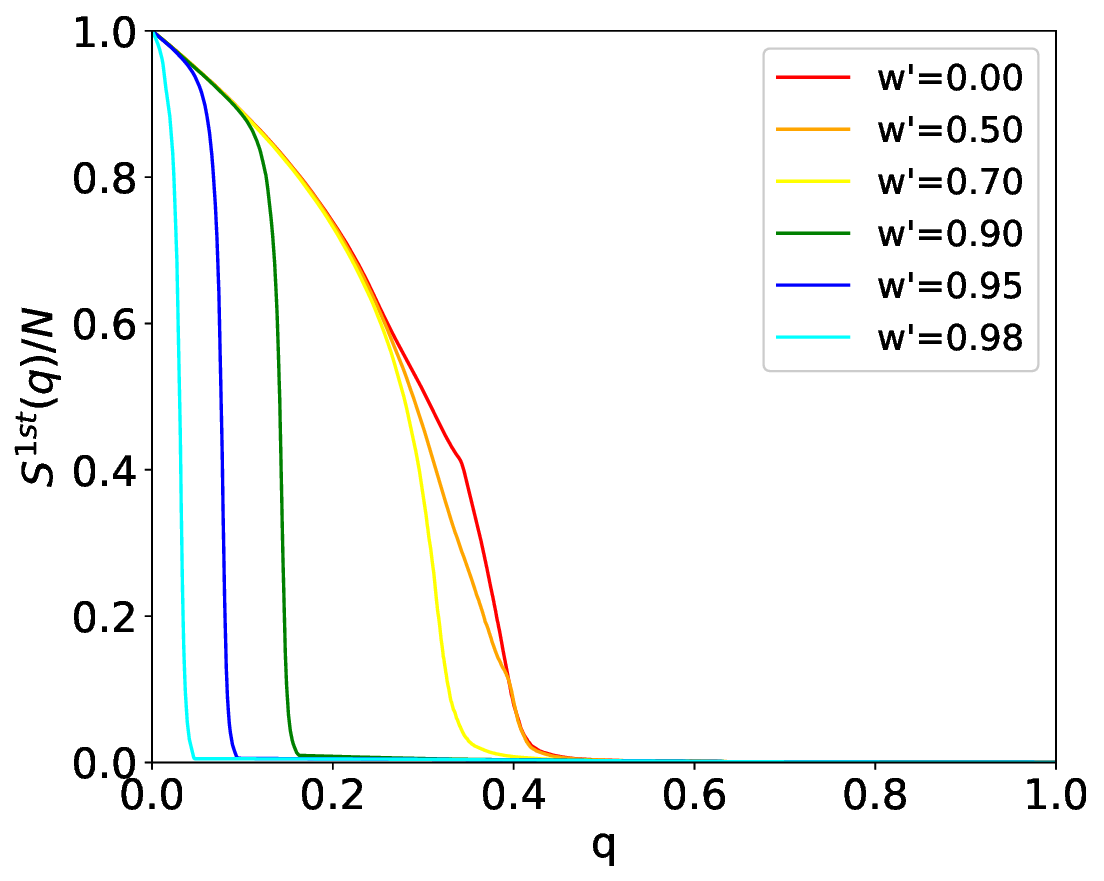}
    \begin{center} (f) $m_{o} = 200$ \end{center}
  \end{minipage}       
%\centering
%\includegraphics[width=.8\textwidth]{resize_figS4.eps}
\caption{Comparison of the areas under the curves 
represented as the robustness against MB attacks
in nearly ER random graphs at $\nu = -1$ with $m_{o}$ modules.}
\label{fig_MB_nu-1}
\end{figure}

\begin{figure}[htb]
  \begin{minipage}{.48\textwidth}
    \includegraphics[width=.9\textwidth]{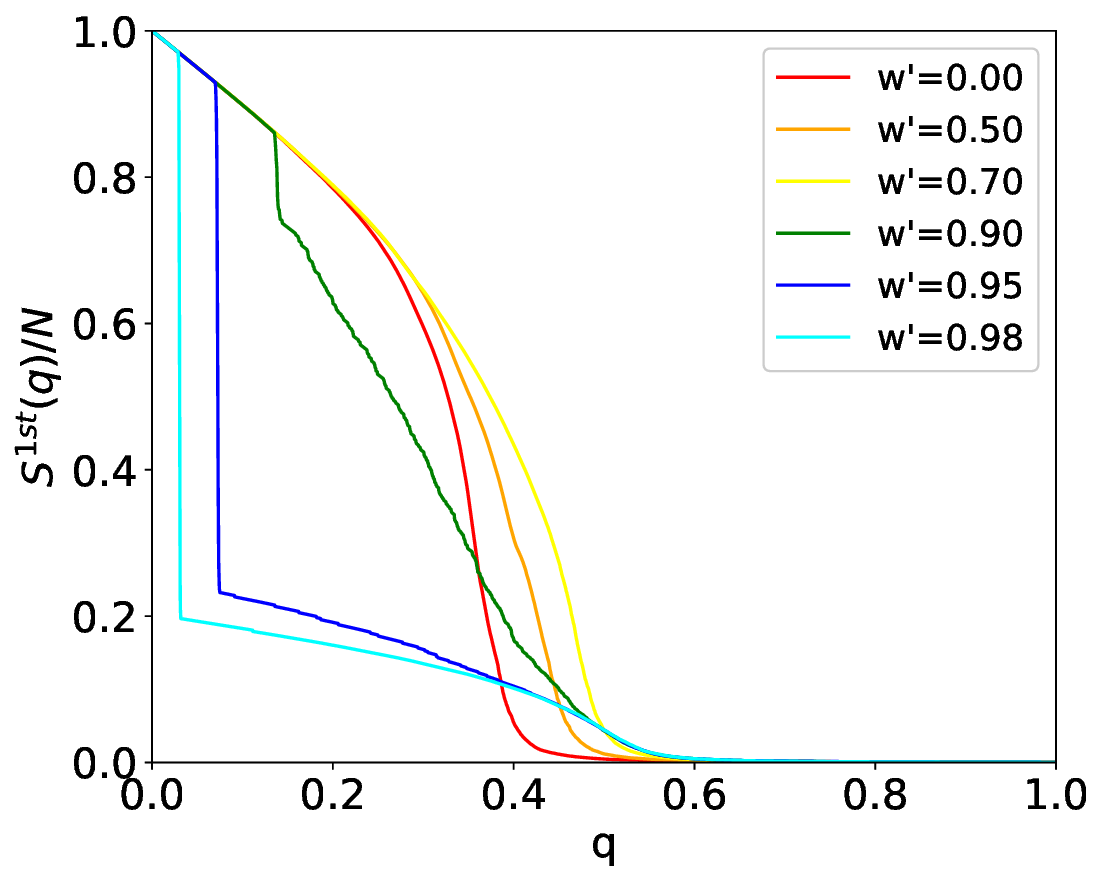}
    \begin{center} (a) $m_{o} = 5$ \end{center}  
  \end{minipage}
  \hfill  
  \begin{minipage}{.48\textwidth}
    \includegraphics[width=.9\textwidth]{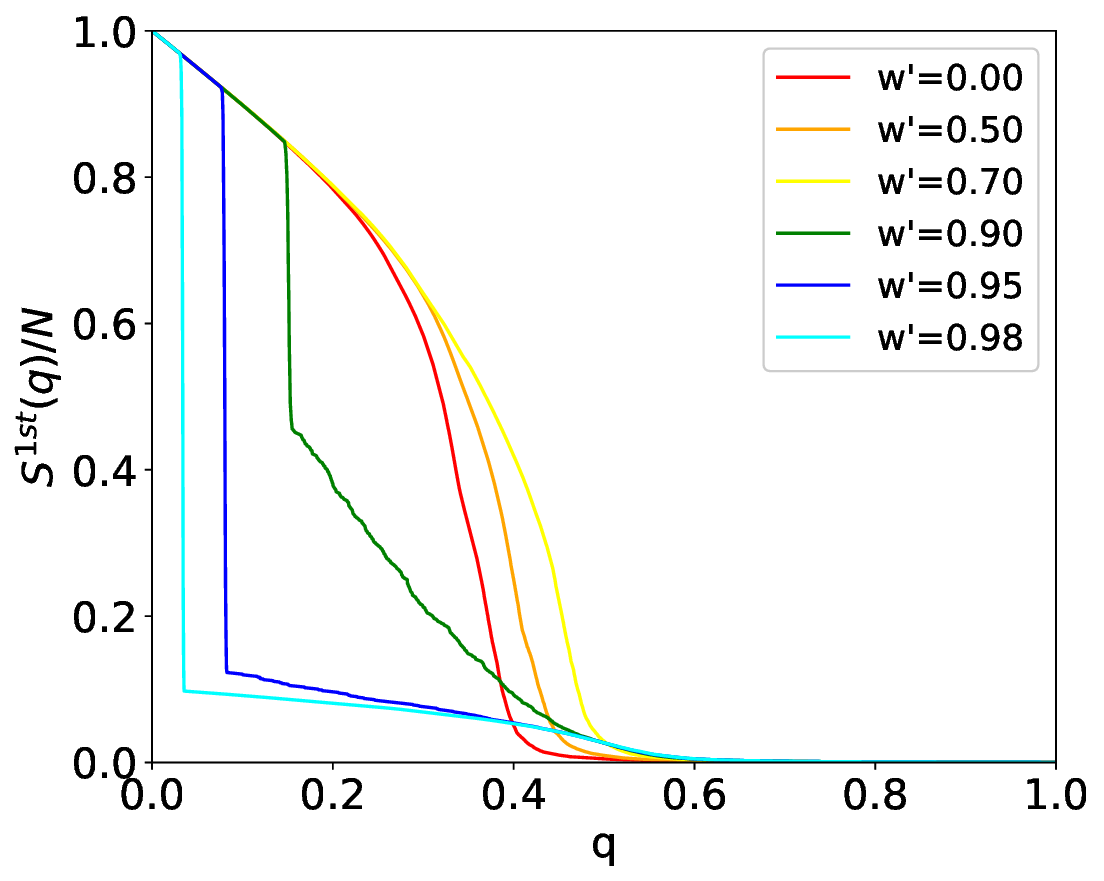}
    \begin{center} (b) $m_{o} = 10$ \end{center}
  \end{minipage}    
  \hfill
  \begin{minipage}{.48\textwidth}
    \includegraphics[width=.9\textwidth]{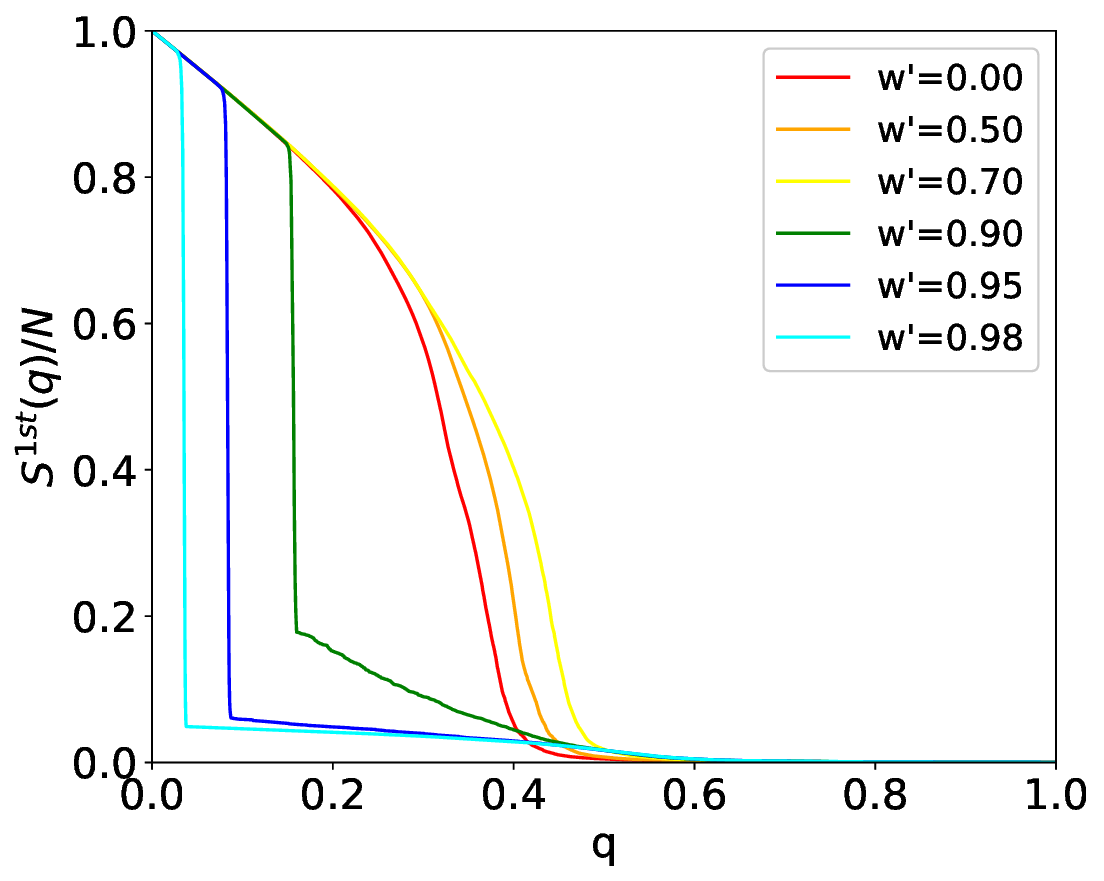}
    \begin{center} (c) $m_{o} = 20$ \end{center}
  \end{minipage}
  \hfill  
  \begin{minipage}{.48\textwidth}
    \includegraphics[width=.9\textwidth]{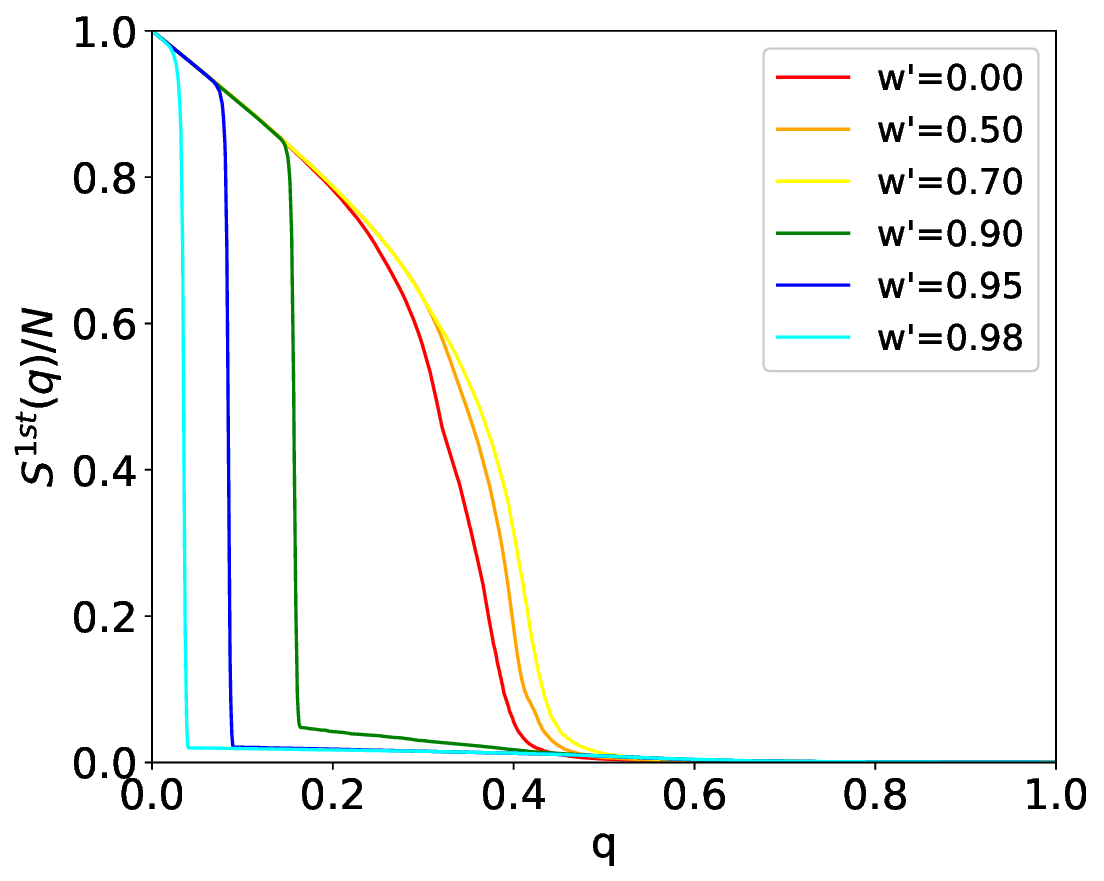}
    \begin{center} (d) $m_{o} = 50$ \end{center}
  \end{minipage}     
  \hfill 
  \begin{minipage}{.48\textwidth}
    \includegraphics[width=.9\textwidth]{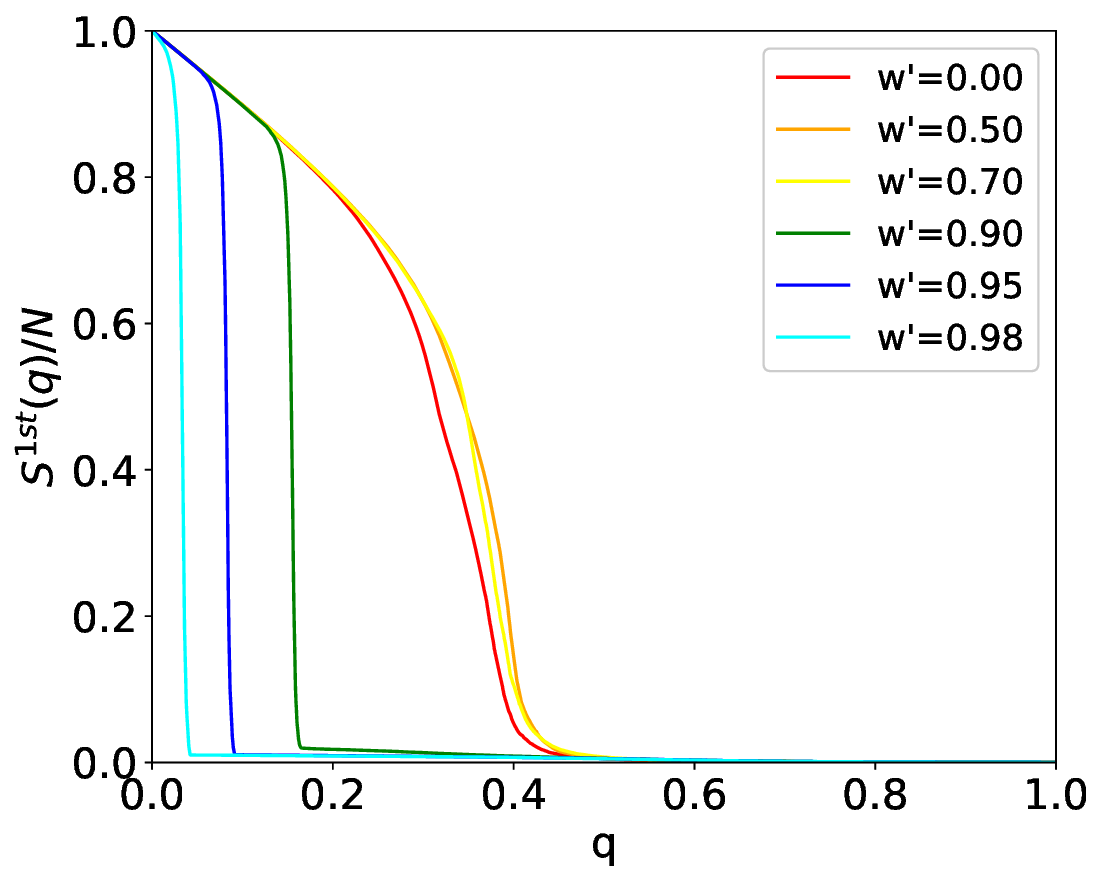}
    \begin{center} (e) $m_{o} = 100$ \end{center}
  \end{minipage}
  \hfill  
  \begin{minipage}{.48\textwidth}
    \includegraphics[width=.9\textwidth]{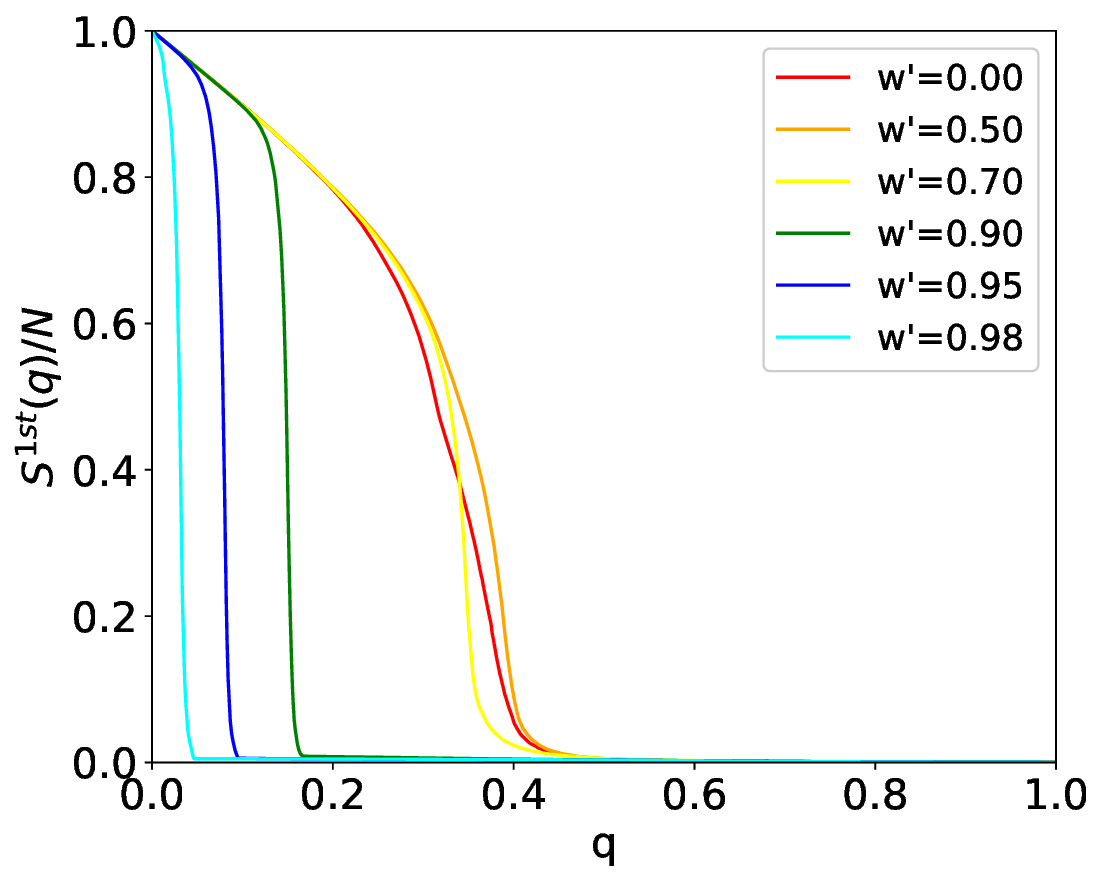}
    \begin{center} (f) $m_{o} = 200$ \end{center}
  \end{minipage}       
%\centering
%\includegraphics[width=.8\textwidth]{resize_figS5.eps}
\caption{Comparison of the areas under the curves 
represented as the robustness against MB attacks
in networks of narrower $P(k)$ at $\nu = -5$ with $m_{o}$ modules.}
\label{fig_MB_nu-5}
\end{figure}

\begin{figure}[htb]
  \begin{minipage}{.48\textwidth}
    \includegraphics[width=.9\textwidth]{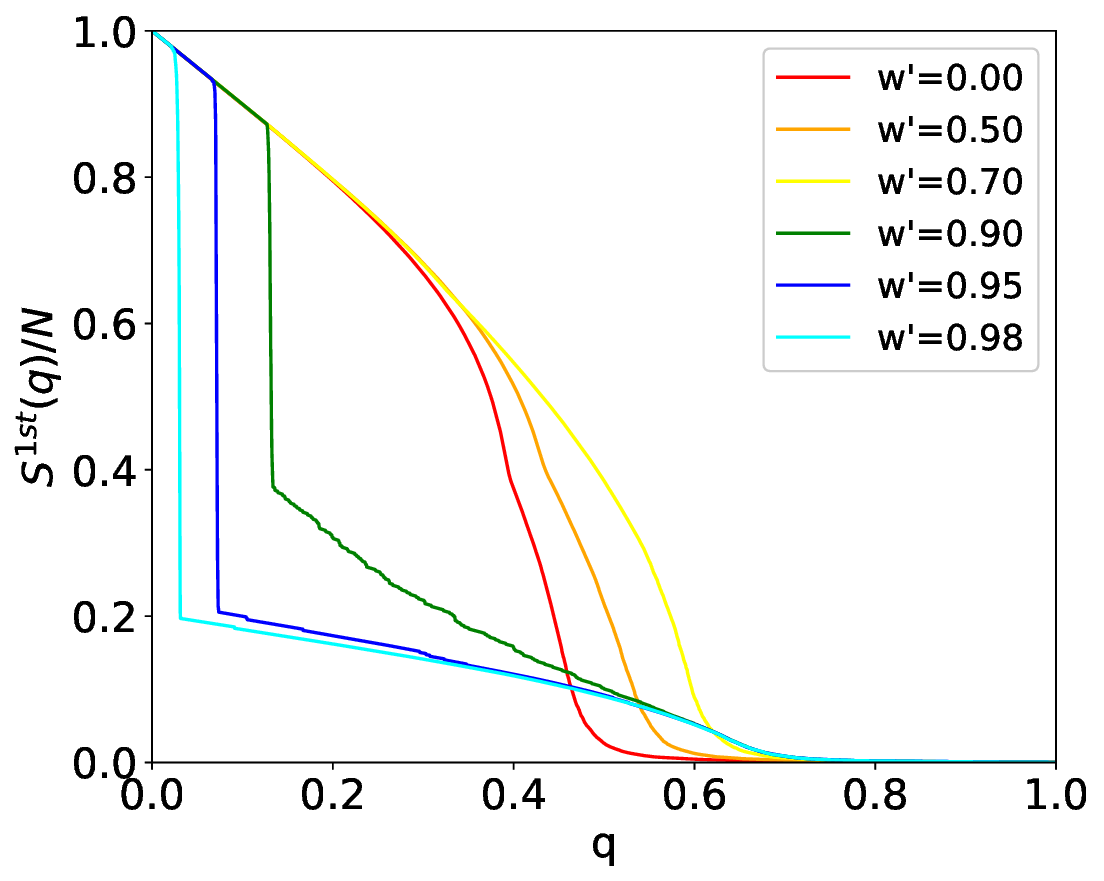}
    \begin{center} (a) $m_{o} = 5$ \end{center}  
  \end{minipage}
  \hfill  
  \begin{minipage}{.48\textwidth}
    \includegraphics[width=.9\textwidth]{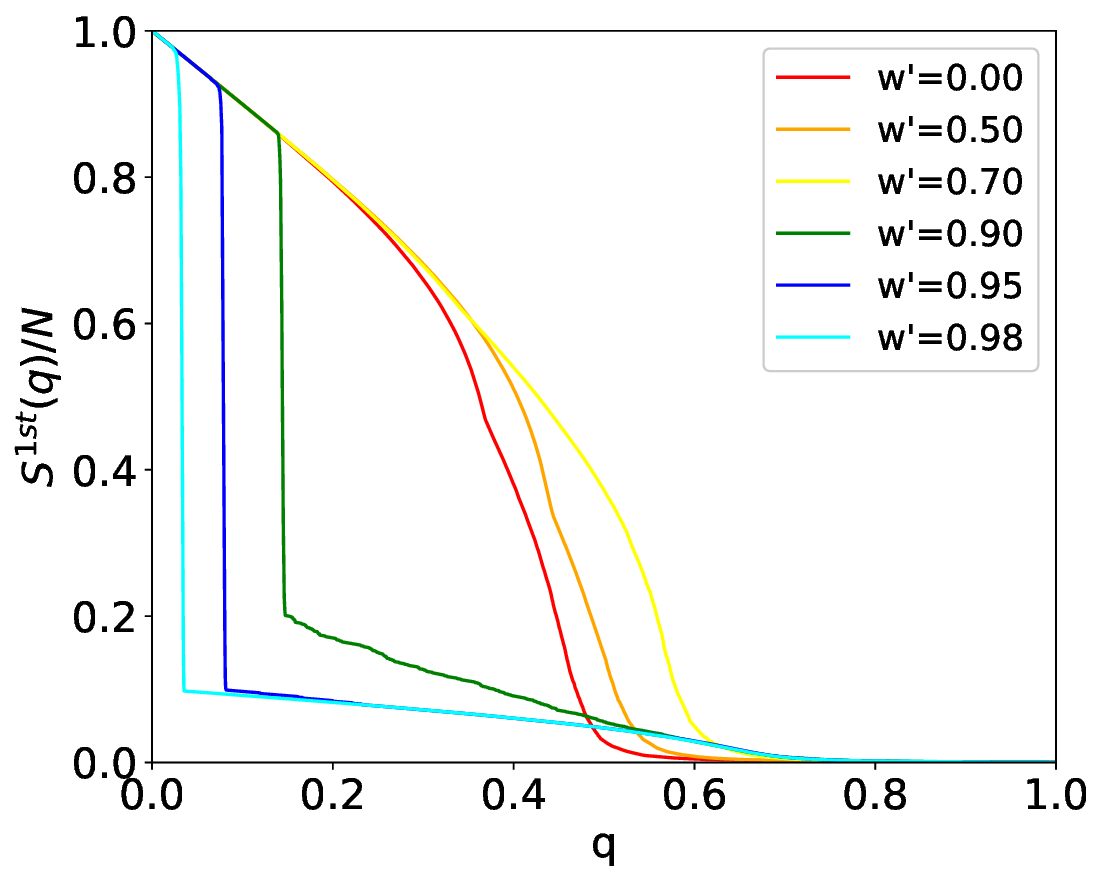}
    \begin{center} (b) $m_{o} = 10$ \end{center}
  \end{minipage}    
  \hfill
  \begin{minipage}{.48\textwidth}
    \includegraphics[width=.9\textwidth]{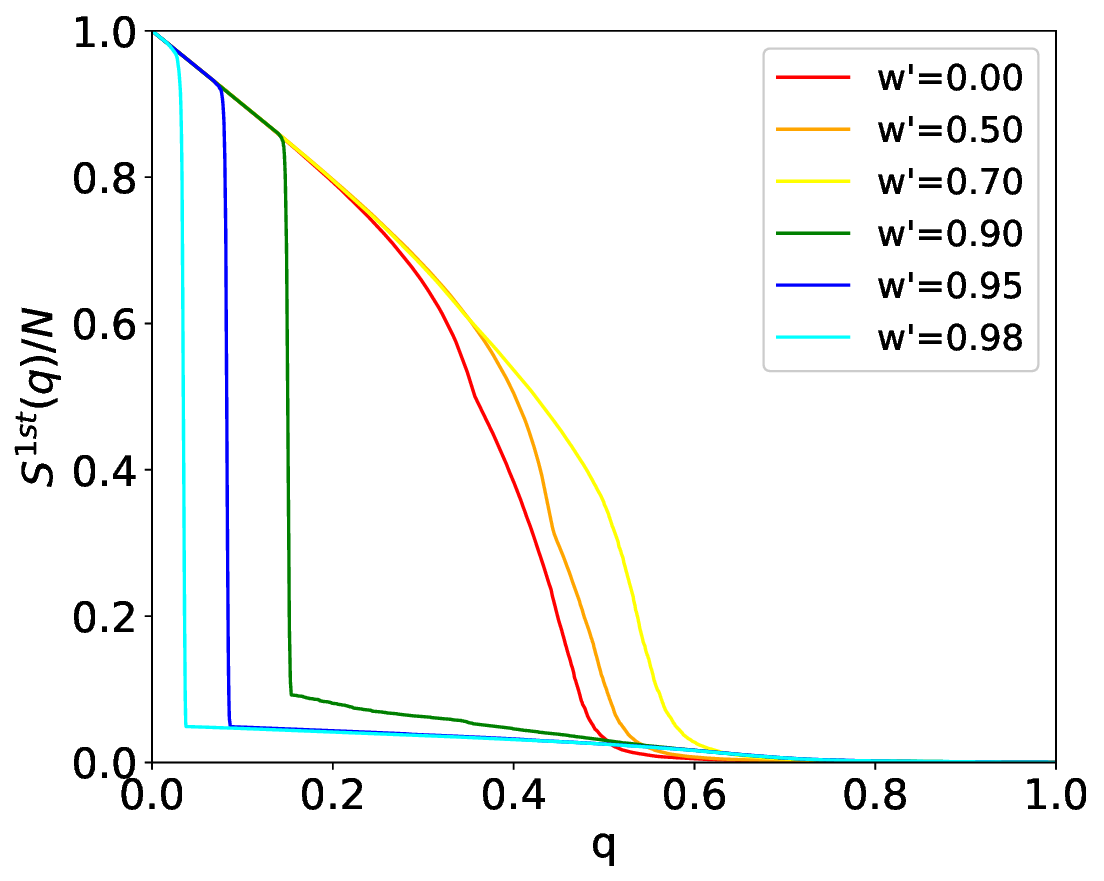}
    \begin{center} (c) $m_{o} = 20$ \end{center}
  \end{minipage}
  \hfill  
  \begin{minipage}{.48\textwidth}
    \includegraphics[width=.9\textwidth]{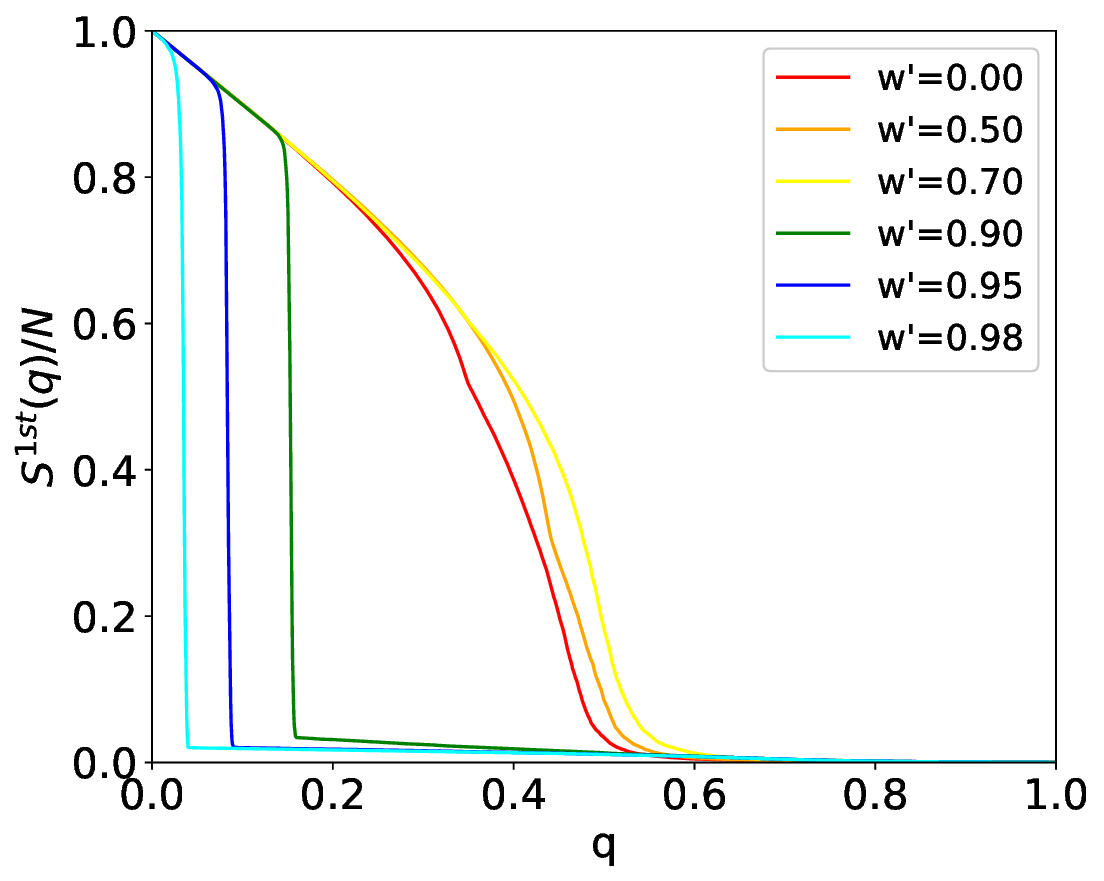}
    \begin{center} (d) $m_{o} = 50$ \end{center}
  \end{minipage}     
  \hfill 
  \begin{minipage}{.48\textwidth}
    \includegraphics[width=.9\textwidth]{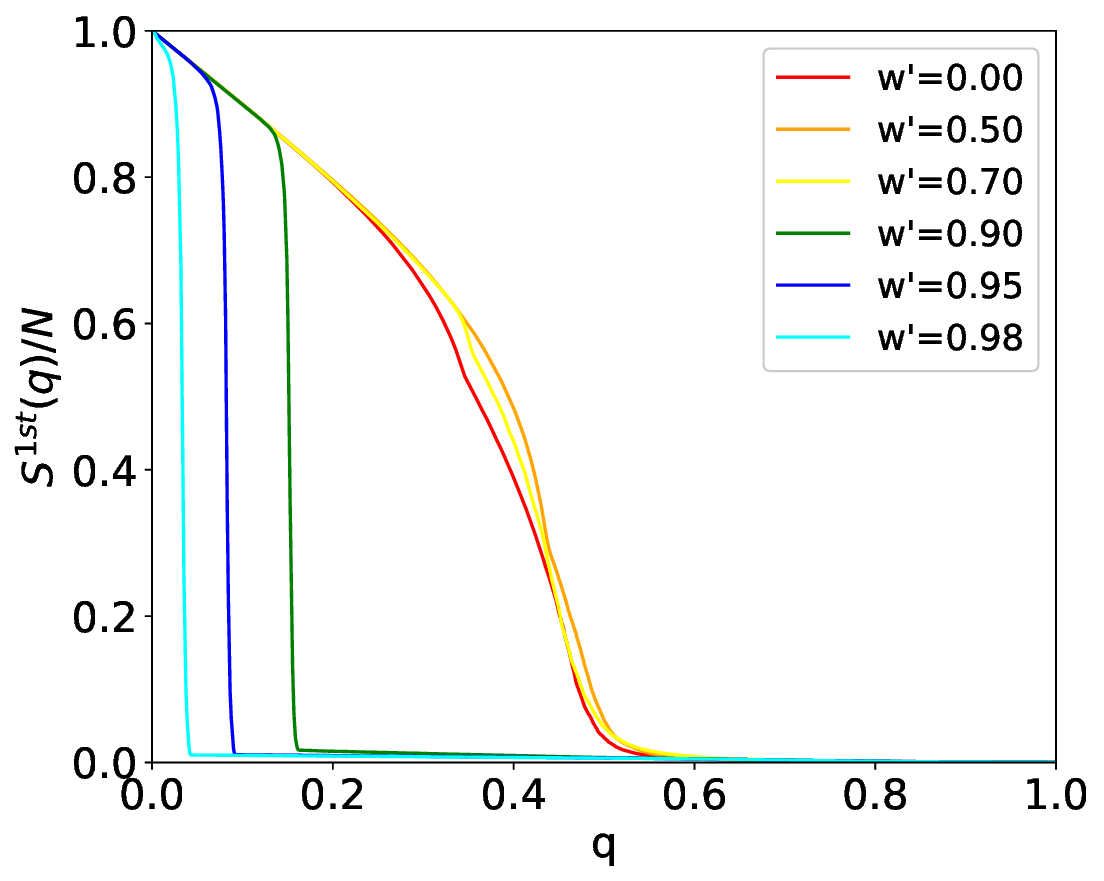}
    \begin{center} (e) $m_{o} = 100$ \end{center}
  \end{minipage}
  \hfill  
  \begin{minipage}{.48\textwidth}
    \includegraphics[width=.9\textwidth]{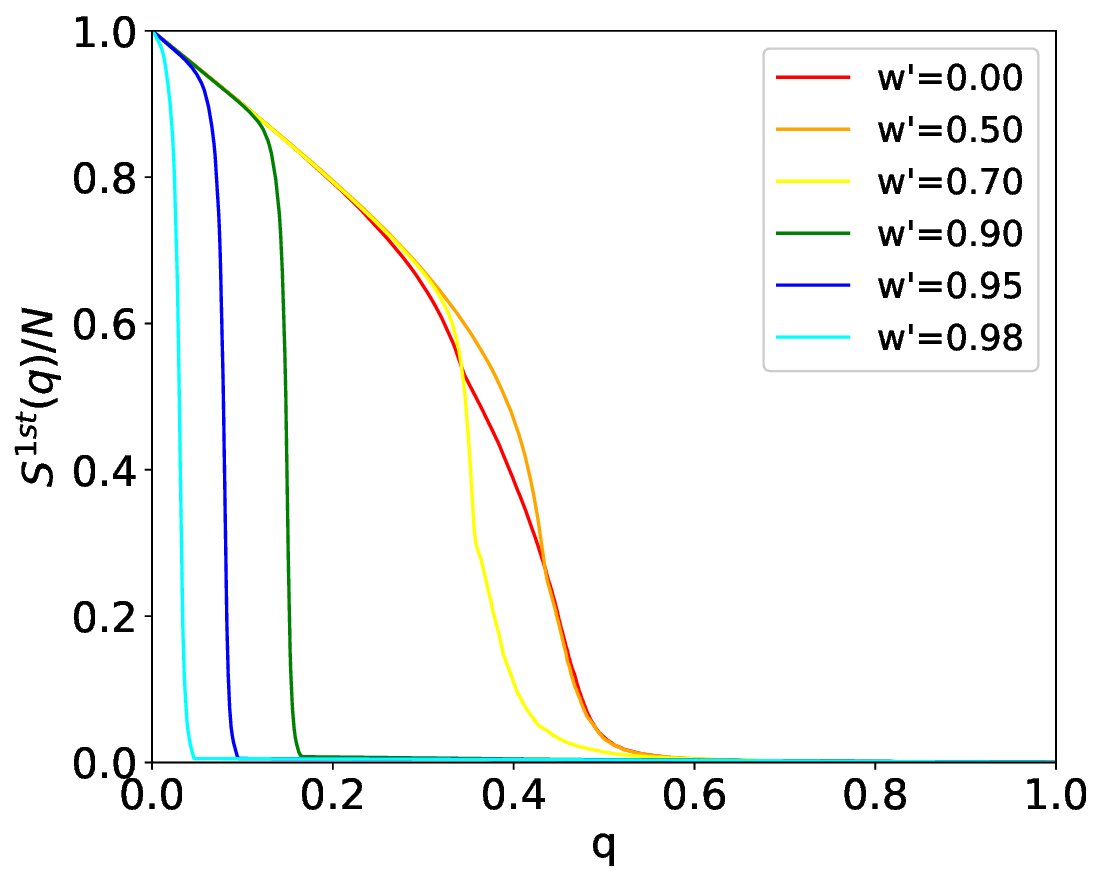}
    \begin{center} (f) $m_{o} = 200$ \end{center}
  \end{minipage}       
%\centering
%\includegraphics[width=.8\textwidth]{resize_figS6.eps}
\caption{Comparison of the areas under the curves 
represented as the robustness against MB attacks
in nearly regular networks at $\nu = -100$ with $m_{o}$ modules.}
\label{fig_MB_nu-100}
\end{figure}

\begin{figure}[htb]
  \begin{minipage}{.48\textwidth}
    \includegraphics[width=.9\textwidth]{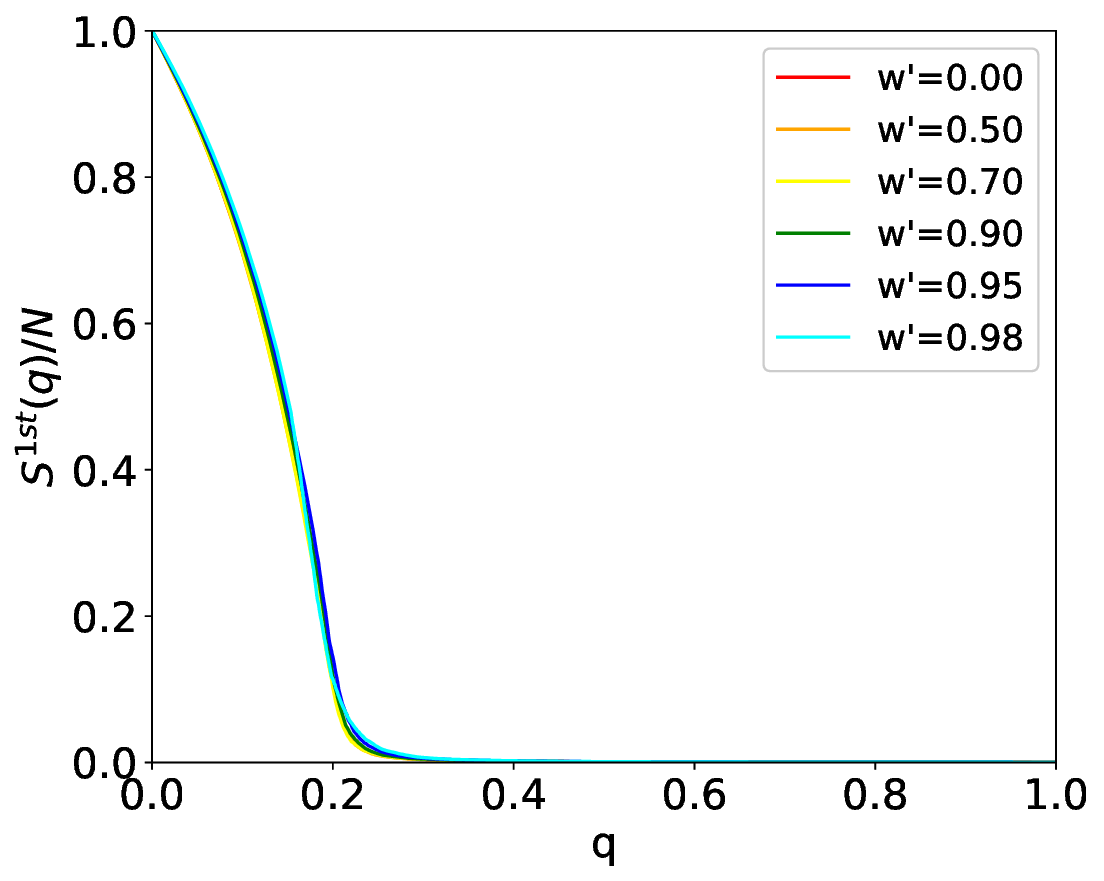}
    \begin{center} (a) $m_{o} = 5$ \end{center}  
  \end{minipage}
  \hfill  
  \begin{minipage}{.48\textwidth}
    \includegraphics[width=.9\textwidth]{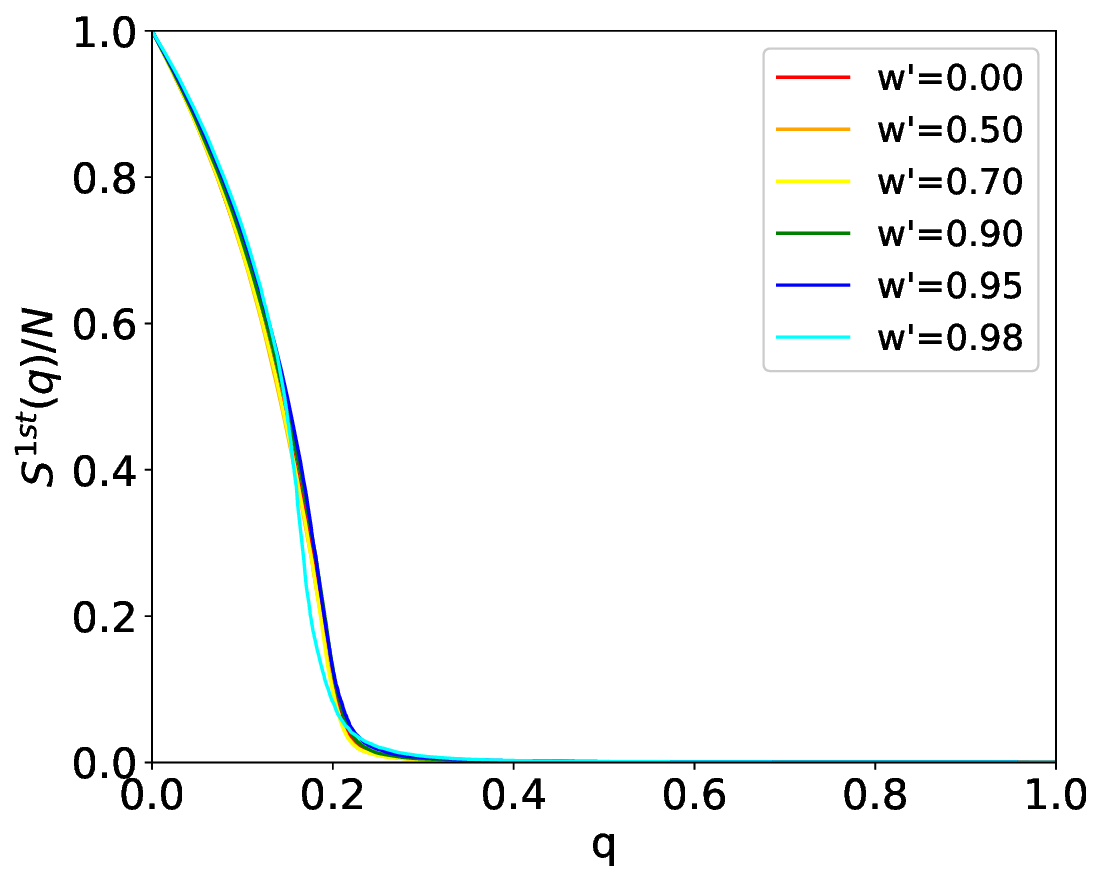}
    \begin{center} (b) $m_{o} = 10$ \end{center}
  \end{minipage}    
  \hfill
  \begin{minipage}{.48\textwidth}
    \includegraphics[width=.9\textwidth]{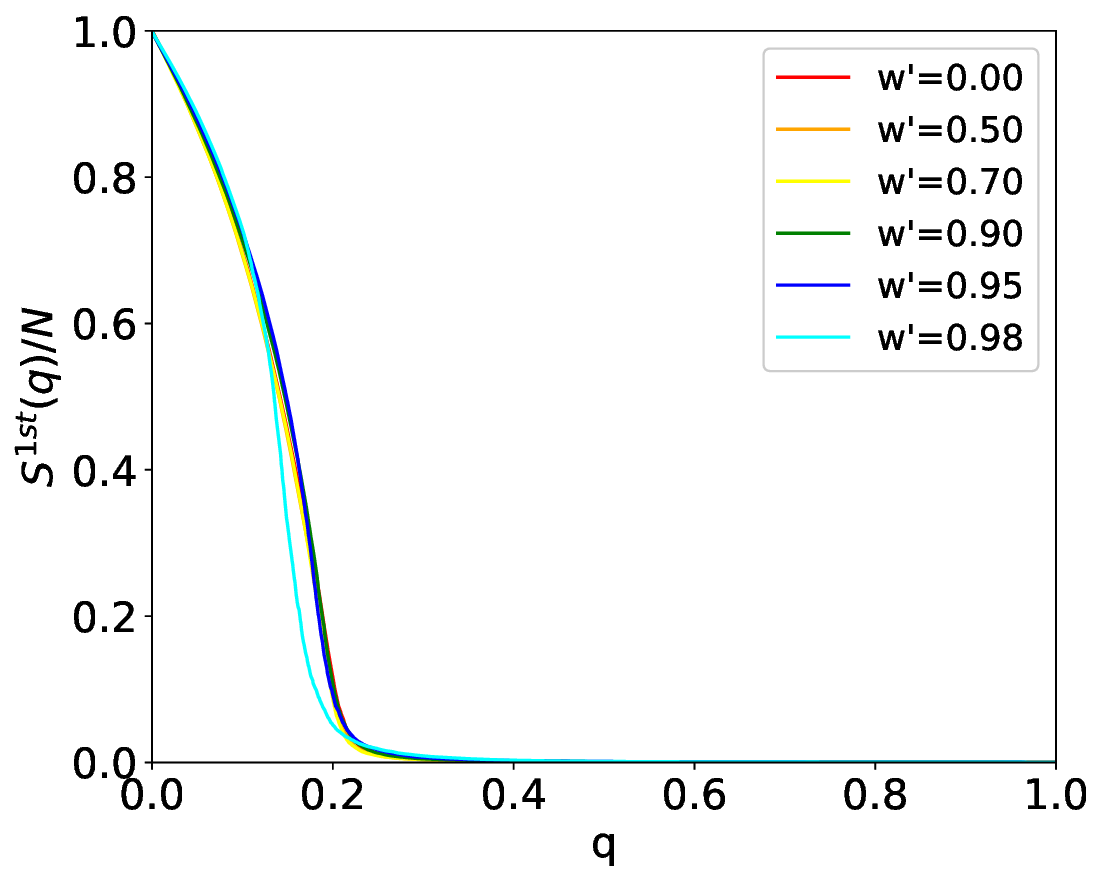}
    \begin{center} (c) $m_{o} = 20$ \end{center}
  \end{minipage}
  \hfill  
  \begin{minipage}{.48\textwidth}
    \includegraphics[width=.9\textwidth]{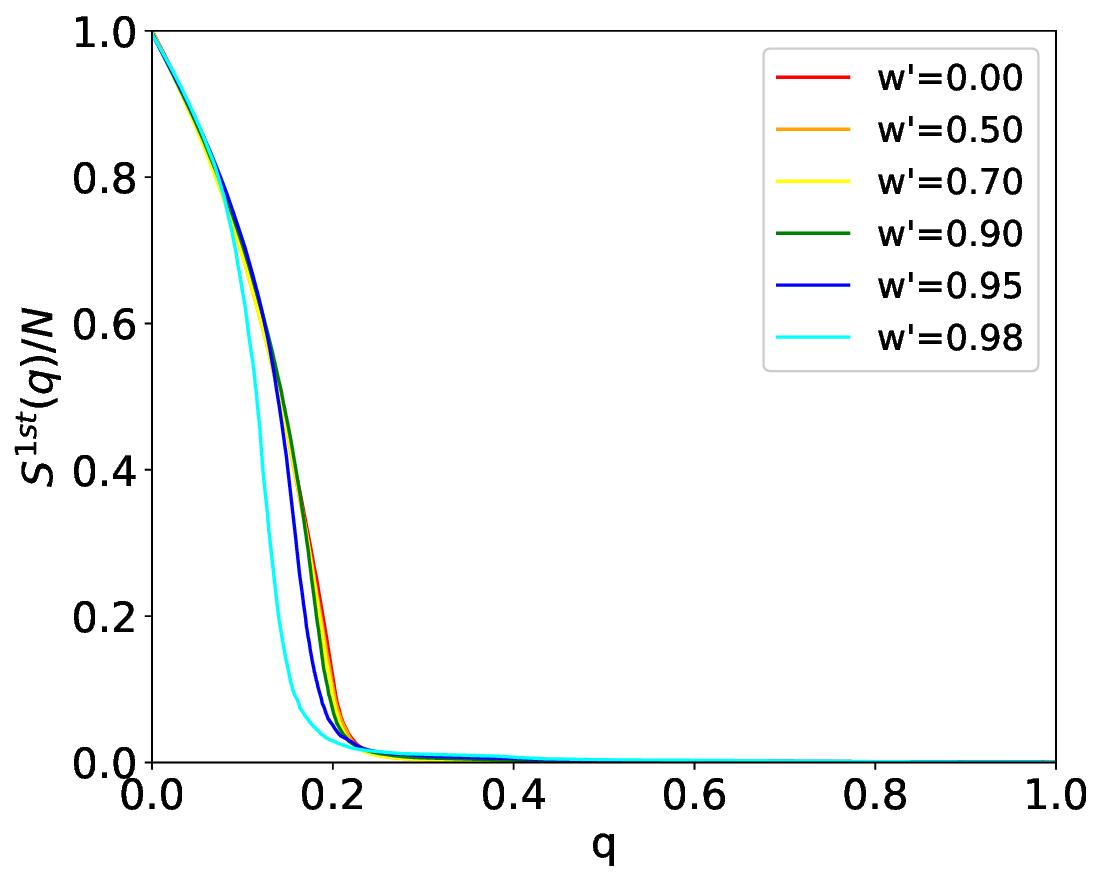}
    \begin{center} (d) $m_{o} = 50$ \end{center}
  \end{minipage}     
  \hfill 
  \begin{minipage}{.48\textwidth}
    \includegraphics[width=.9\textwidth]{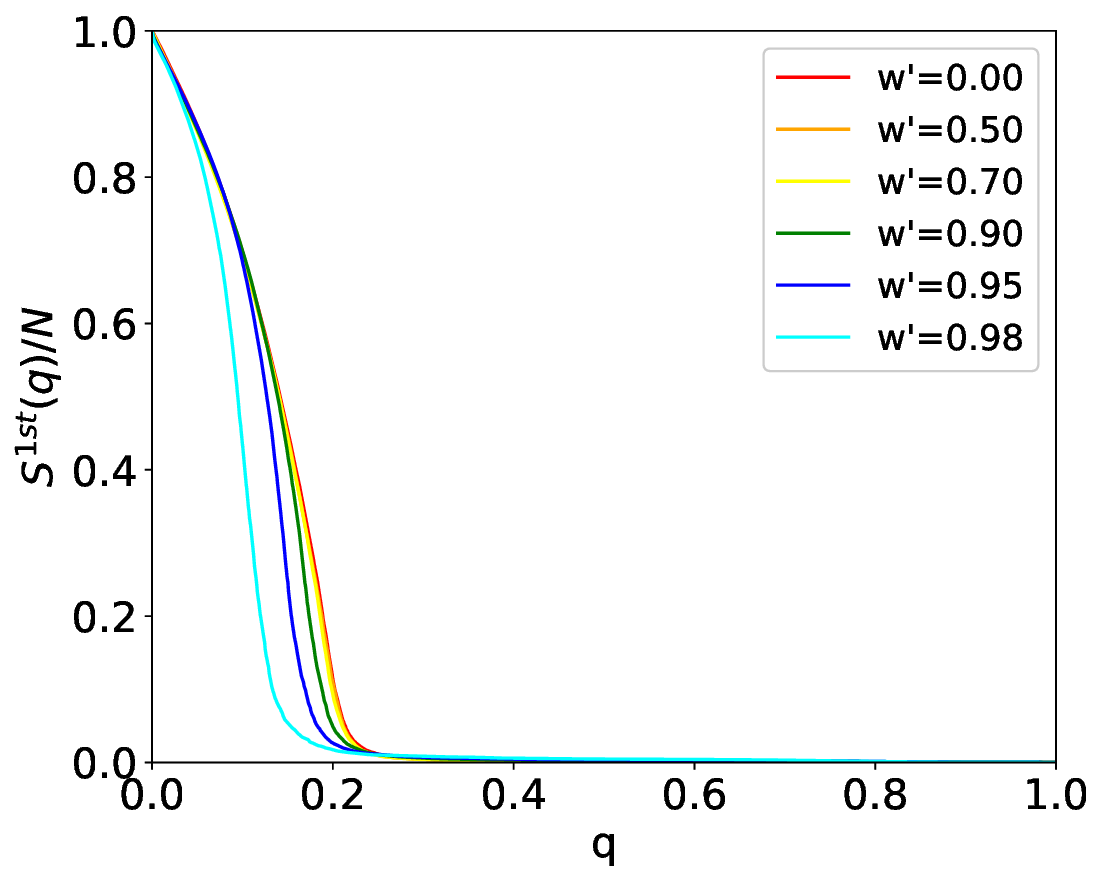}
    \begin{center} (e) $m_{o} = 100$ \end{center}
  \end{minipage}
  \hfill  
  \begin{minipage}{.48\textwidth}
    \includegraphics[width=.9\textwidth]{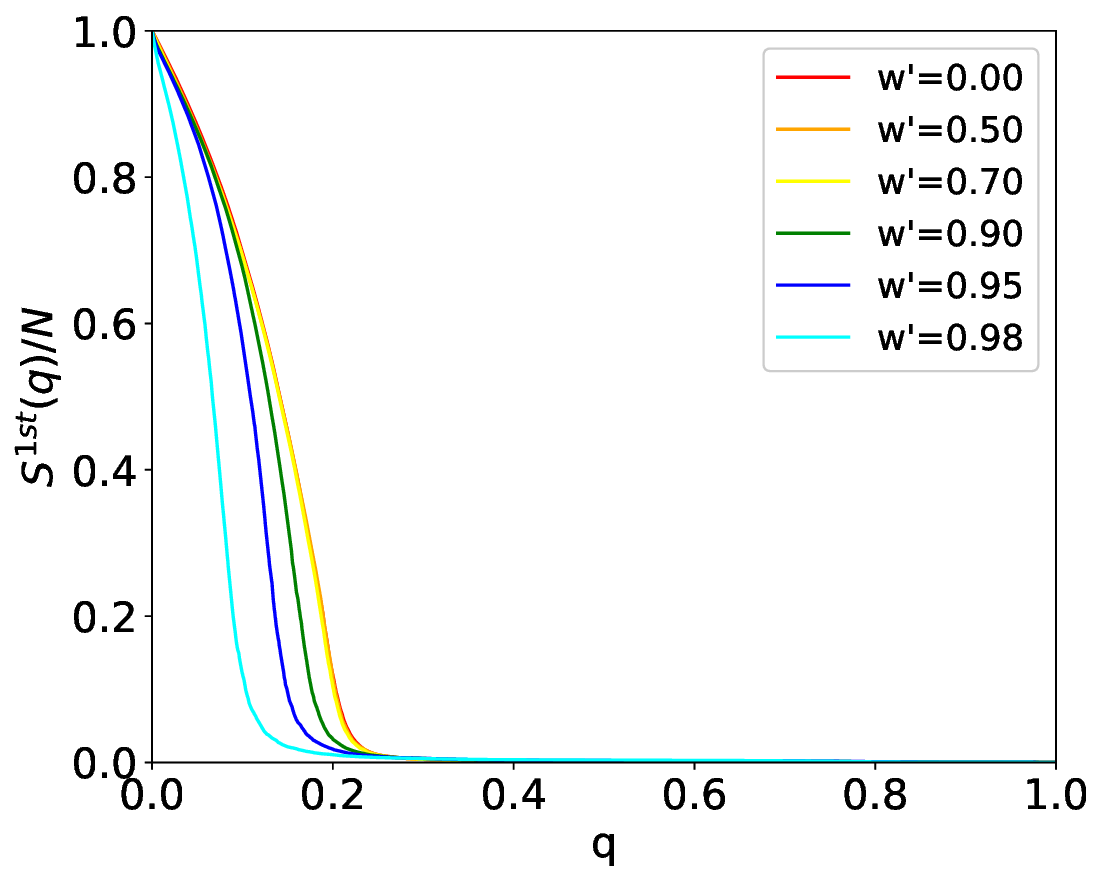}
    \begin{center} (f) $m_{o} = 200$ \end{center}
  \end{minipage}       
%\centering
%\includegraphics[width=.8\textwidth]{resize_figS7.eps}
\caption{Comparison of the areas under the curves 
represented as the robustness against IB attacks
in SF networks at $\nu = 1$ with $m_{o}$ modules.}
\label{fig_IB_nu1}
\end{figure}

\begin{figure}[htb]
  \begin{minipage}{.48\textwidth}
    \includegraphics[width=.9\textwidth]{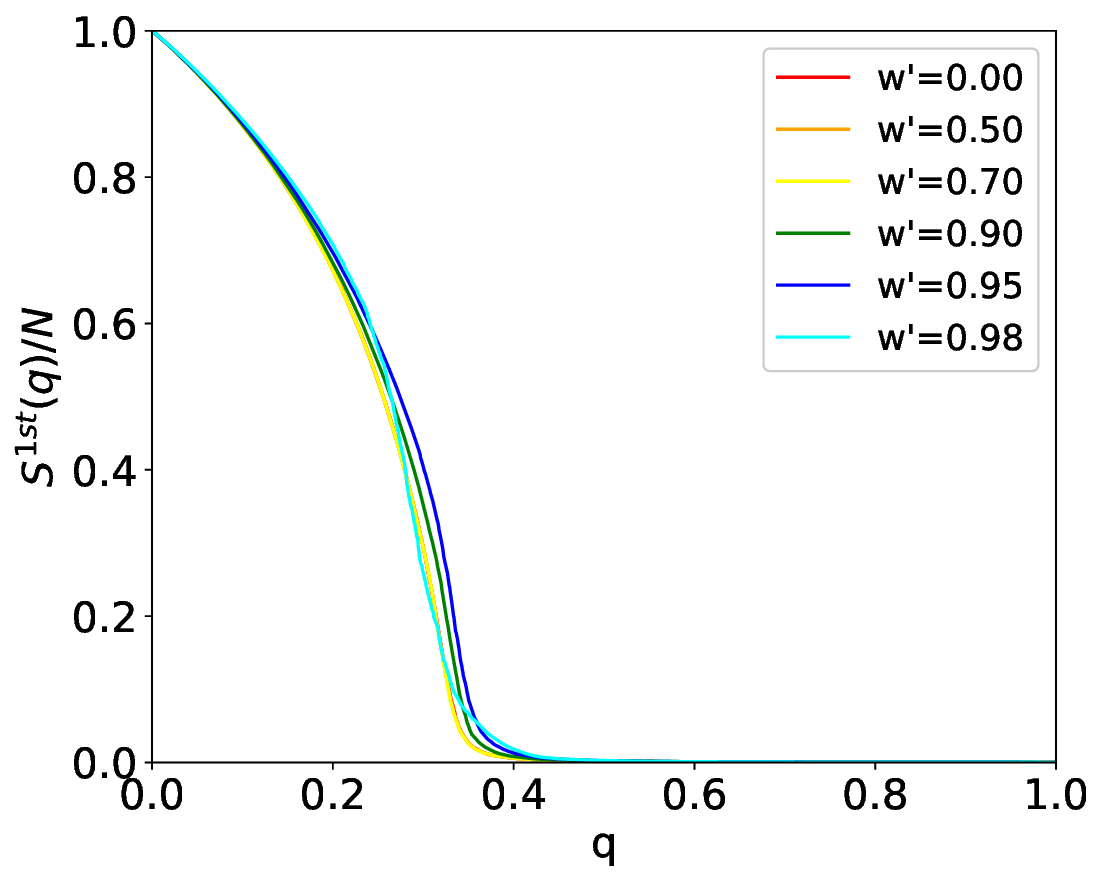}
    \begin{center} (a) $m_{o} = 5$ \end{center}  
  \end{minipage}
  \hfill  
  \begin{minipage}{.48\textwidth}
    \includegraphics[width=.9\textwidth]{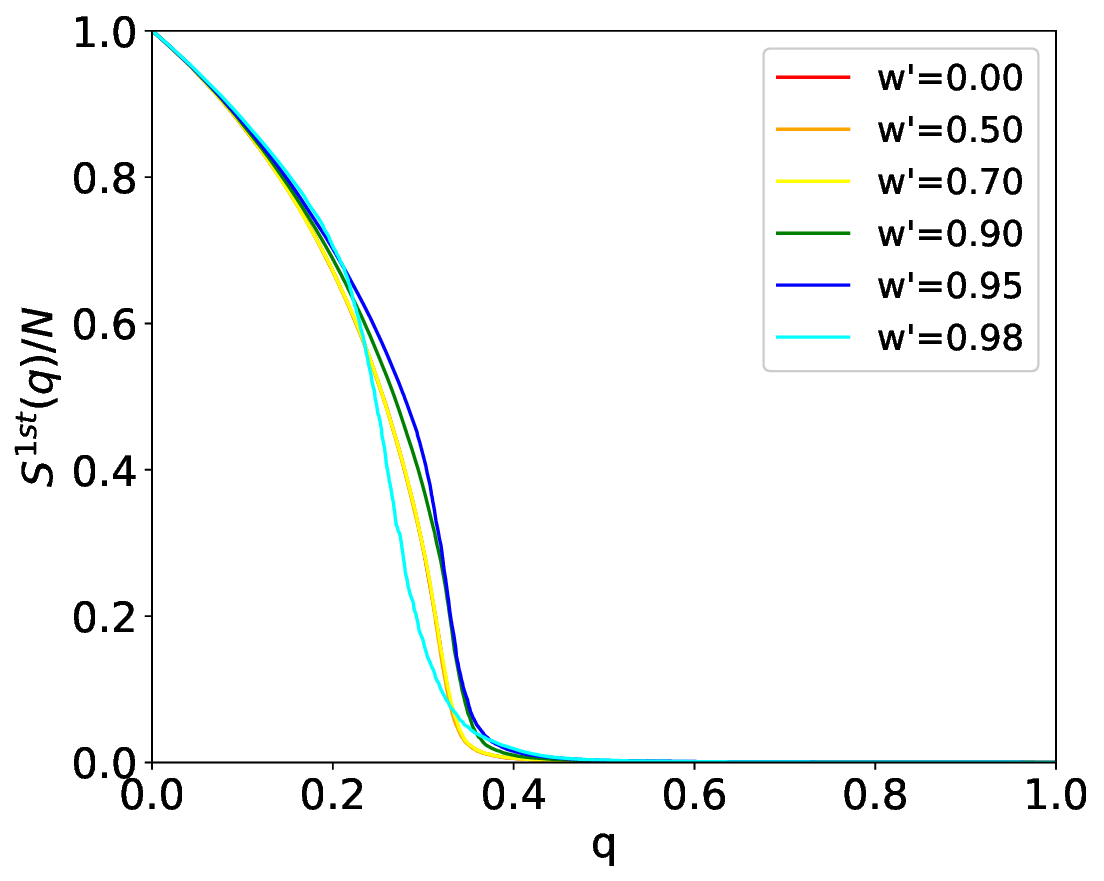}
    \begin{center} (b) $m_{o} = 10$ \end{center}
  \end{minipage}    
  \hfill
  \begin{minipage}{.48\textwidth}
    \includegraphics[width=.9\textwidth]{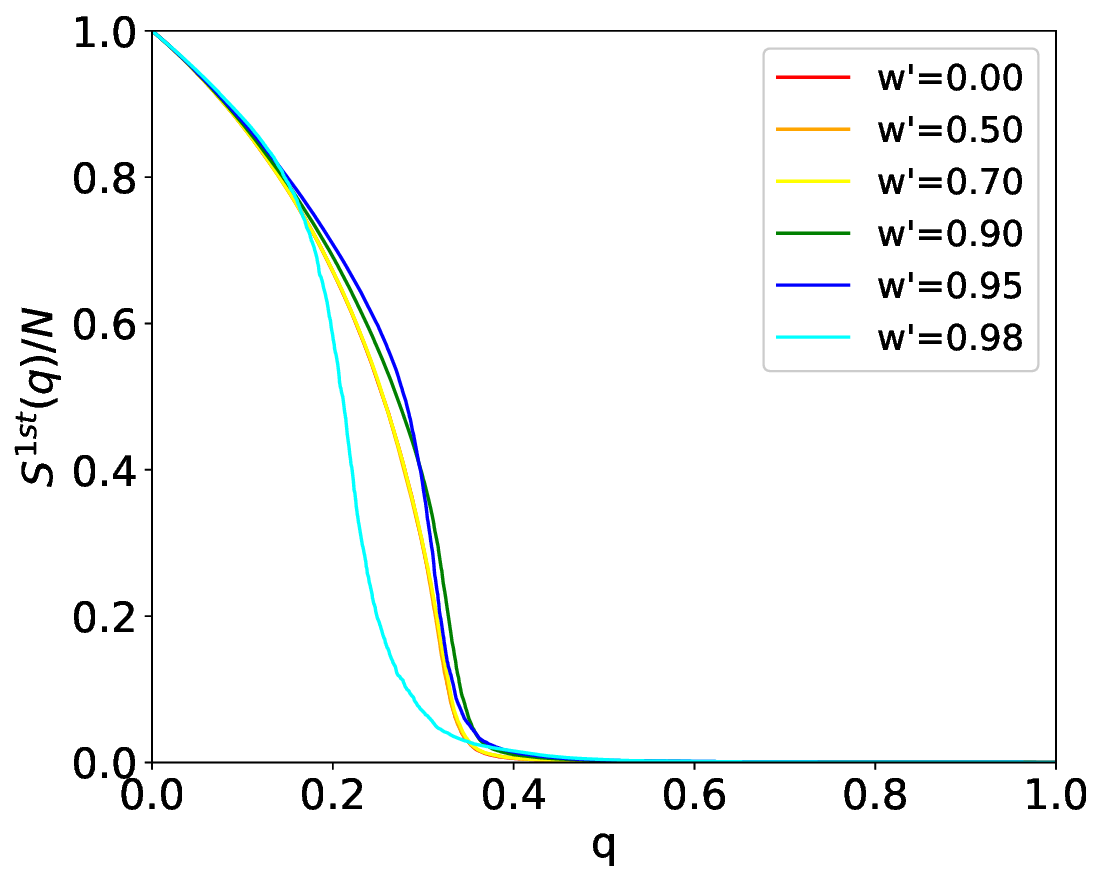}
    \begin{center} (c) $m_{o} = 20$ \end{center}
  \end{minipage}
  \hfill  
  \begin{minipage}{.48\textwidth}
    \includegraphics[width=.9\textwidth]{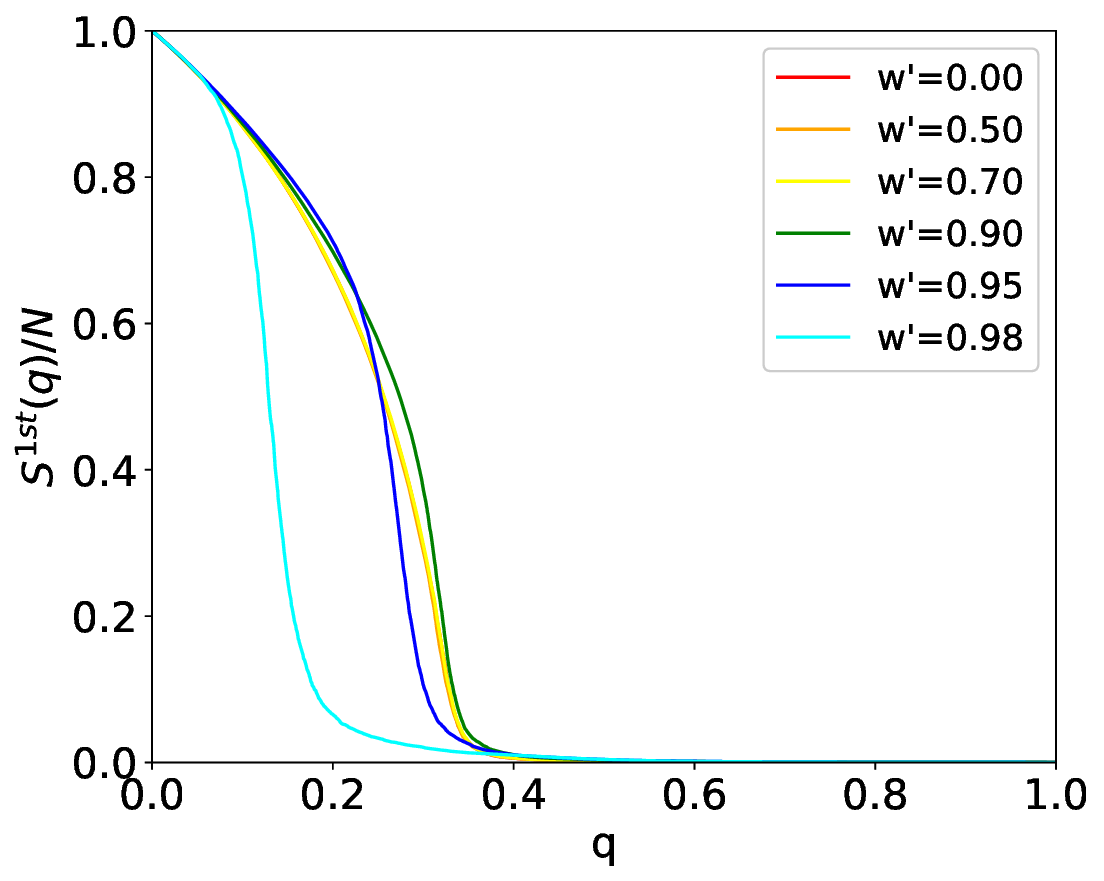}
    \begin{center} (d) $m_{o} = 50$ \end{center}
  \end{minipage}     
  \hfill 
  \begin{minipage}{.48\textwidth}
    \includegraphics[width=.9\textwidth]{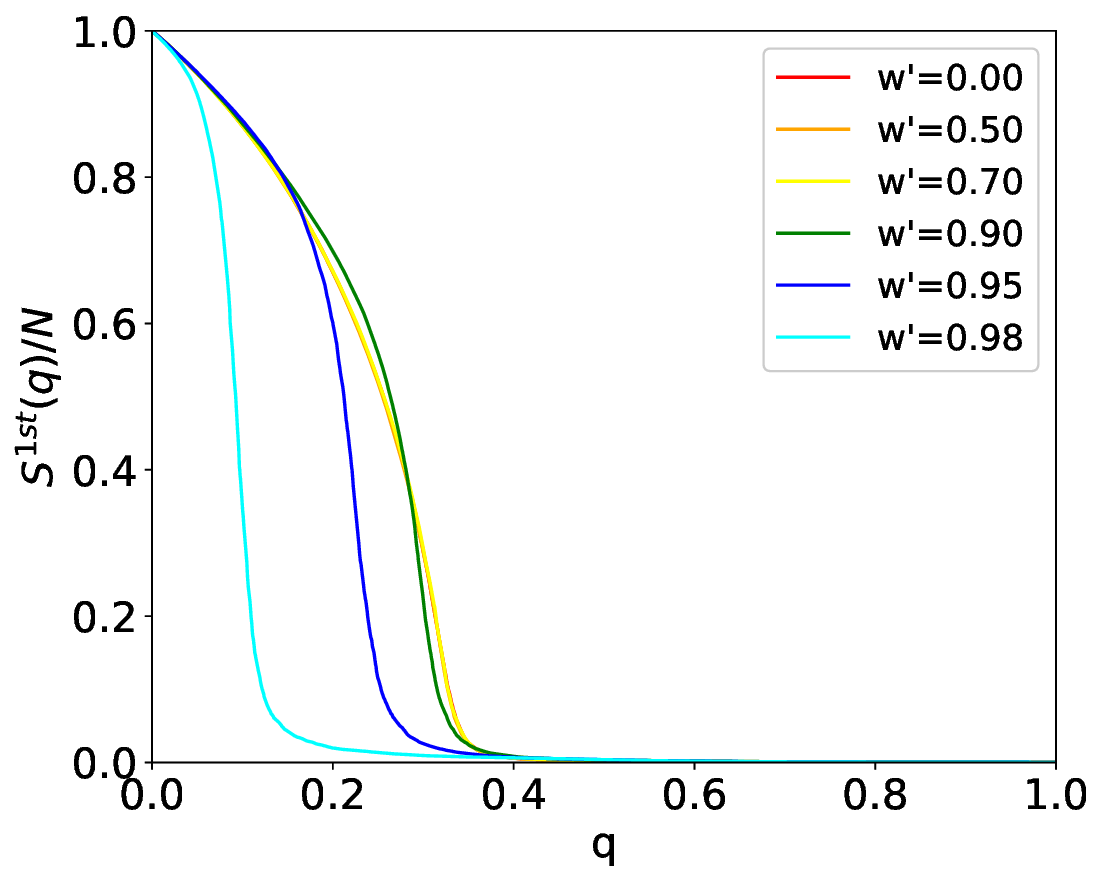}
    \begin{center} (e) $m_{o} = 100$ \end{center}
  \end{minipage}
  \hfill  
  \begin{minipage}{.48\textwidth}
    \includegraphics[width=.9\textwidth]{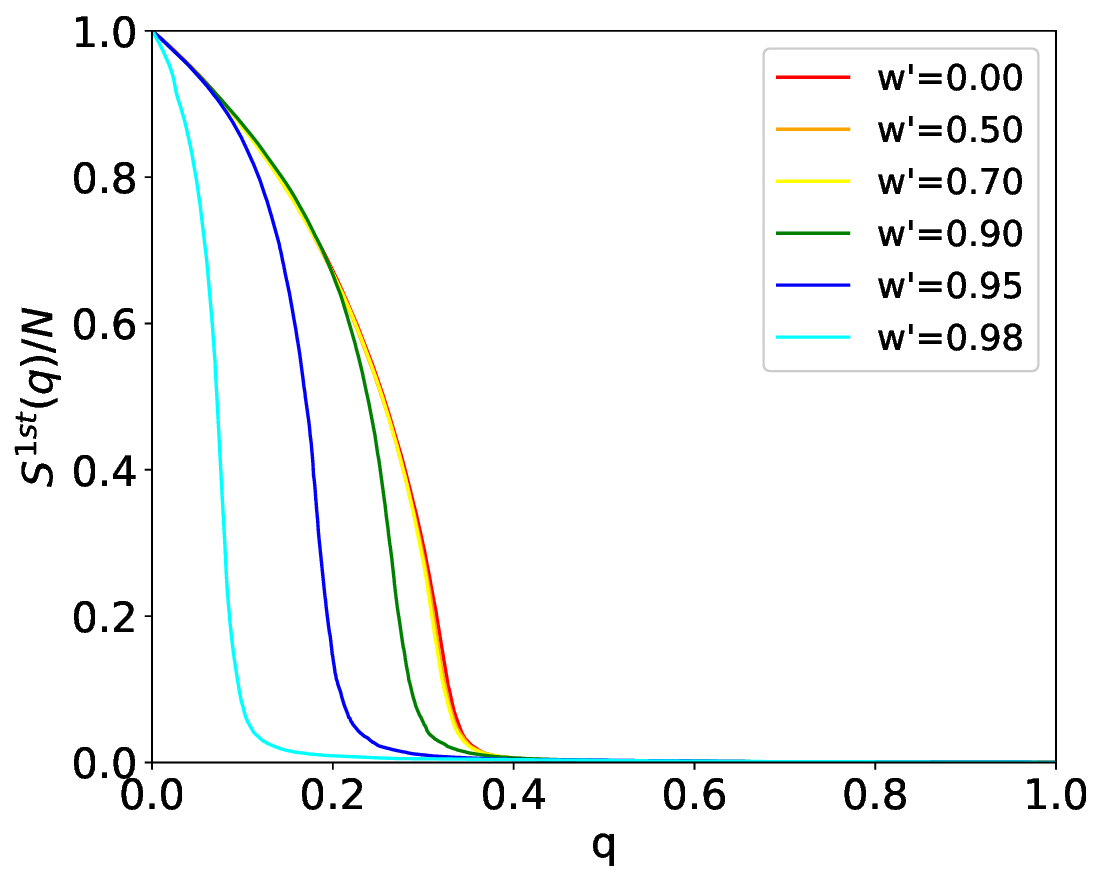}
    \begin{center} (f) $m_{o} = 200$ \end{center}
  \end{minipage}       
%\centering
%\includegraphics[width=.8\textwidth]{resize_figS8.eps}
\caption{Comparison of the areas under the curves 
represented as the robustness against IB attacks
in randomly attached networks at $\nu = 0$ with $m_{o}$ modules.}
\label{fig_IB_nu0}
\end{figure}

\begin{figure}[htb]
  \begin{minipage}{.48\textwidth}
    \includegraphics[width=.9\textwidth]{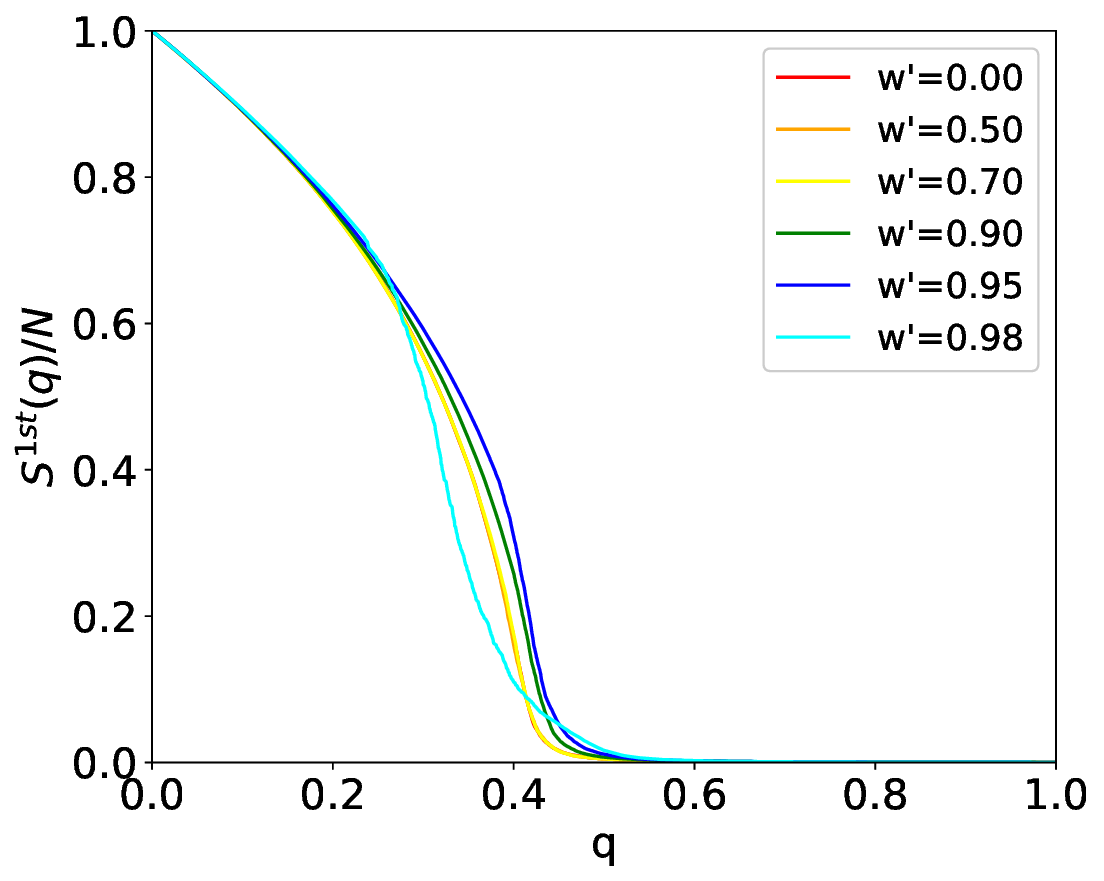}
    \begin{center} (a) $m_{o} = 5$ \end{center}  
  \end{minipage}
  \hfill  
  \begin{minipage}{.48\textwidth}
    \includegraphics[width=.9\textwidth]{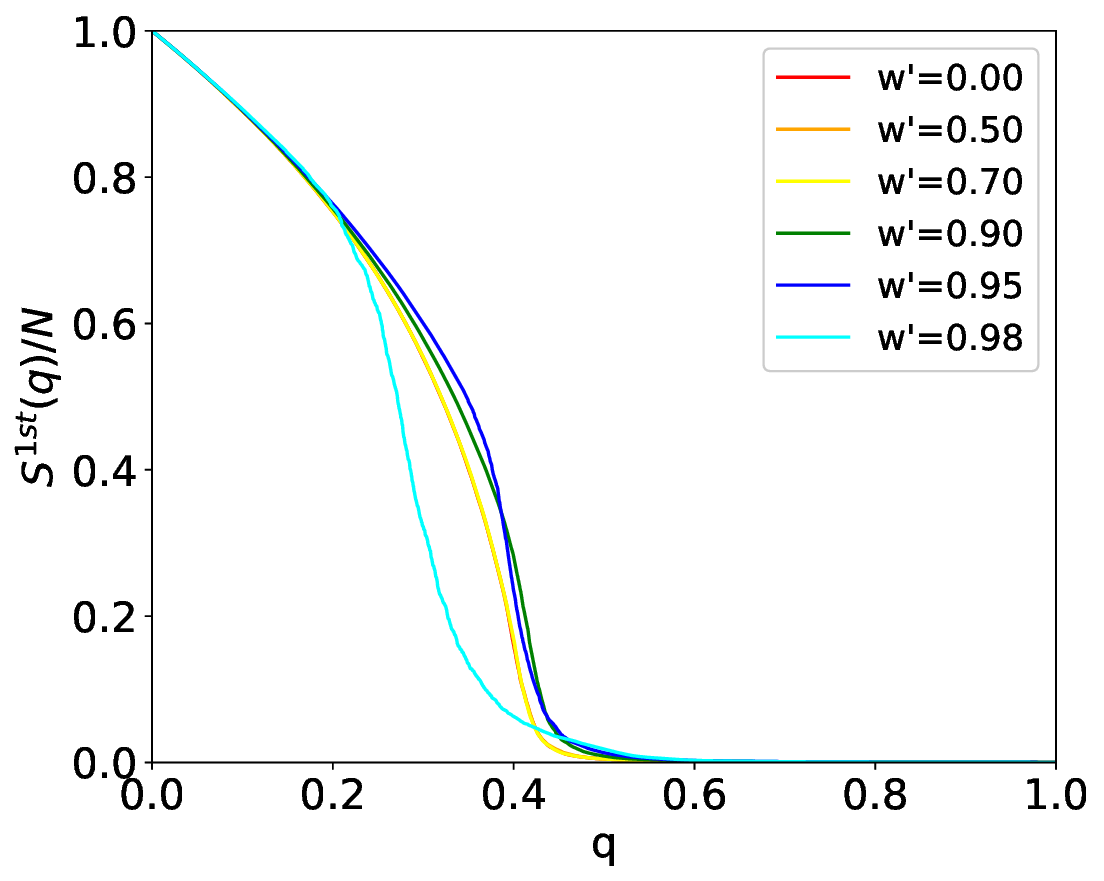}
    \begin{center} (b) $m_{o} = 10$ \end{center}
  \end{minipage}    
  \hfill
  \begin{minipage}{.48\textwidth}
    \includegraphics[width=.9\textwidth]{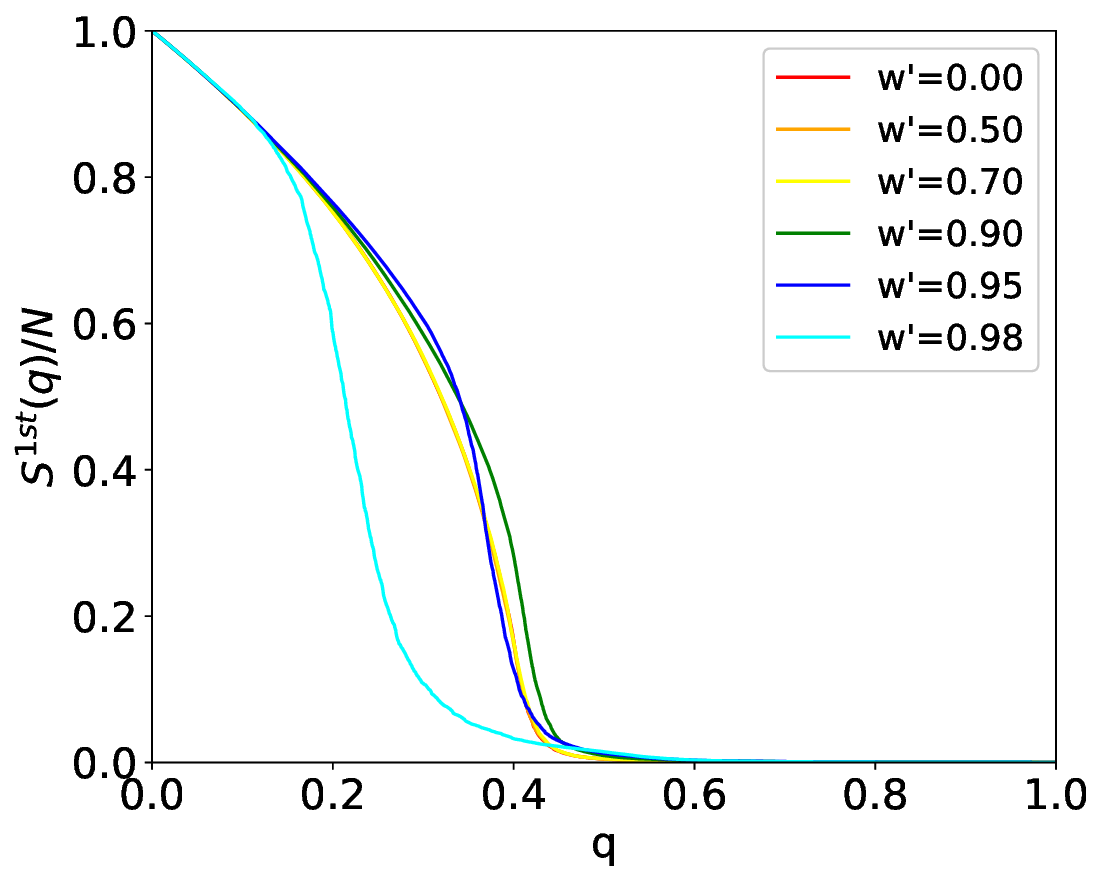}
    \begin{center} (c) $m_{o} = 20$ \end{center}
  \end{minipage}
  \hfill  
  \begin{minipage}{.48\textwidth}
    \includegraphics[width=.9\textwidth]{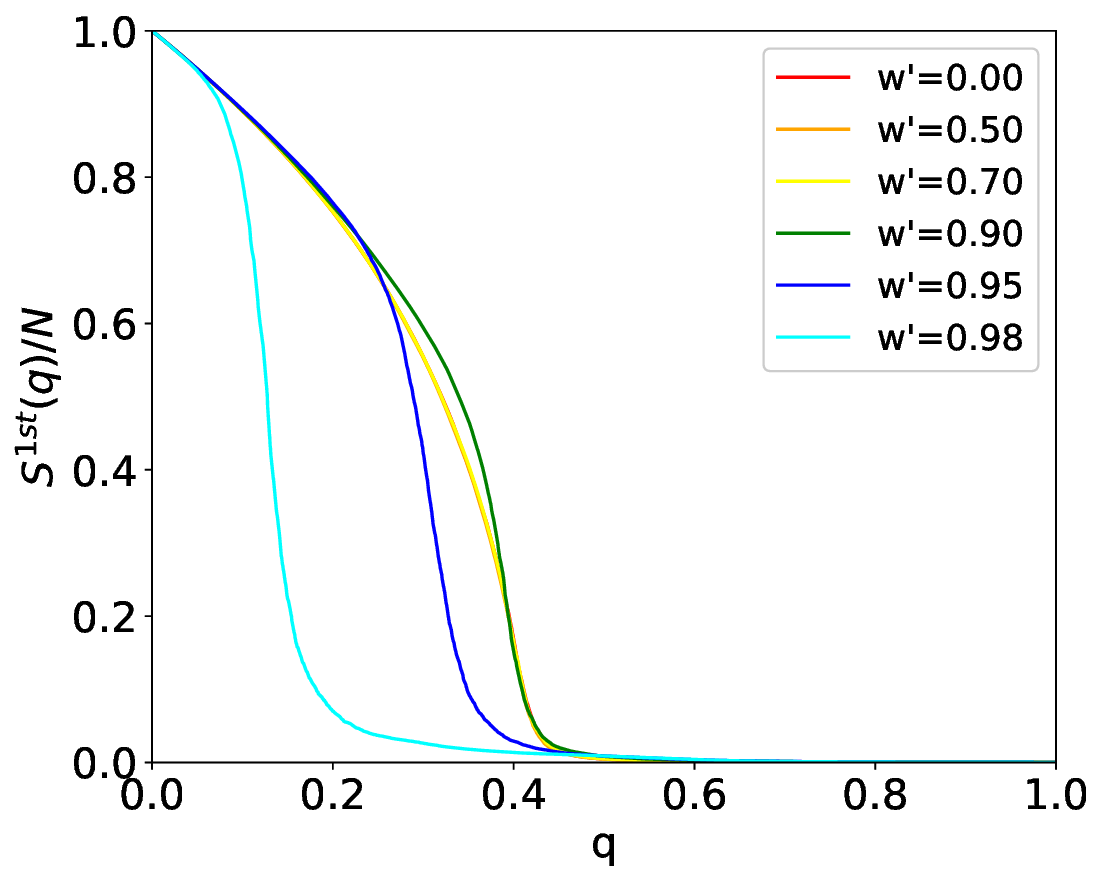}
    \begin{center} (d) $m_{o} = 50$ \end{center}
  \end{minipage}     
  \hfill 
  \begin{minipage}{.48\textwidth}
    \includegraphics[width=.9\textwidth]{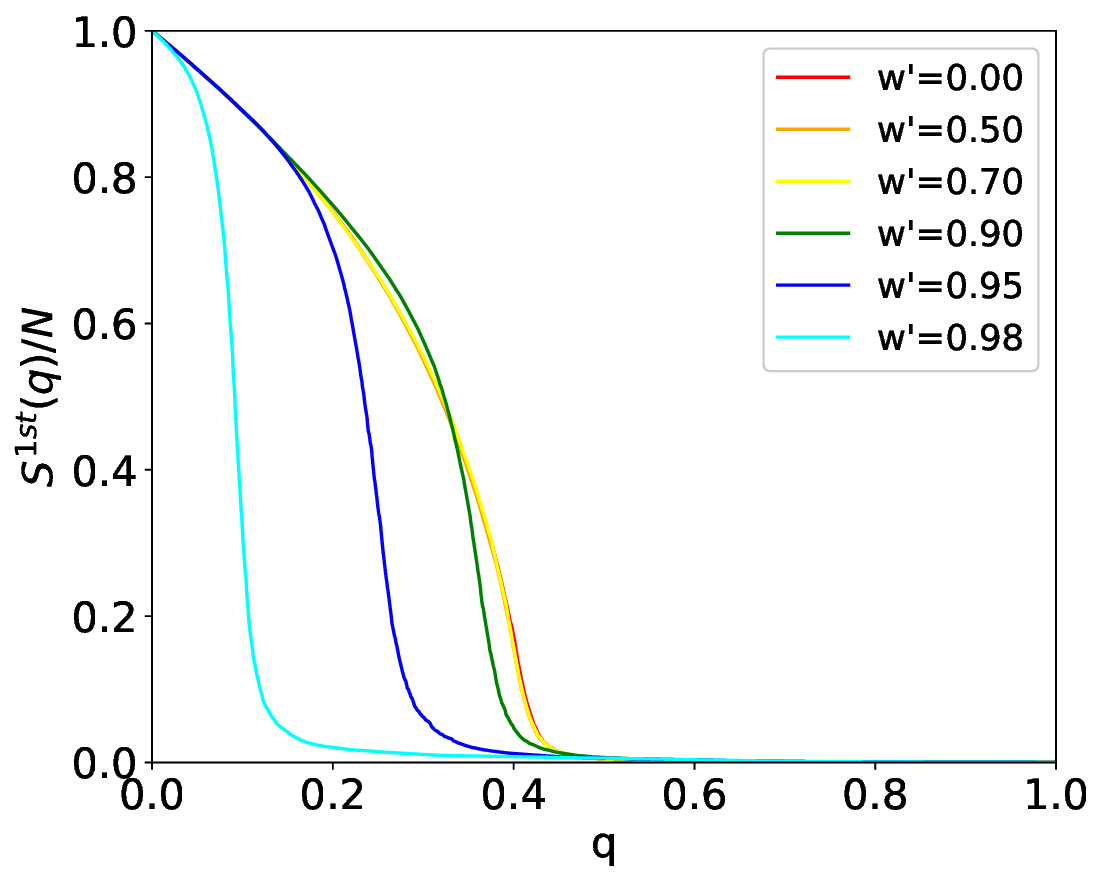}
    \begin{center} (e) $m_{o} = 100$ \end{center}
  \end{minipage}
  \hfill  
  \begin{minipage}{.48\textwidth}
    \includegraphics[width=.9\textwidth]{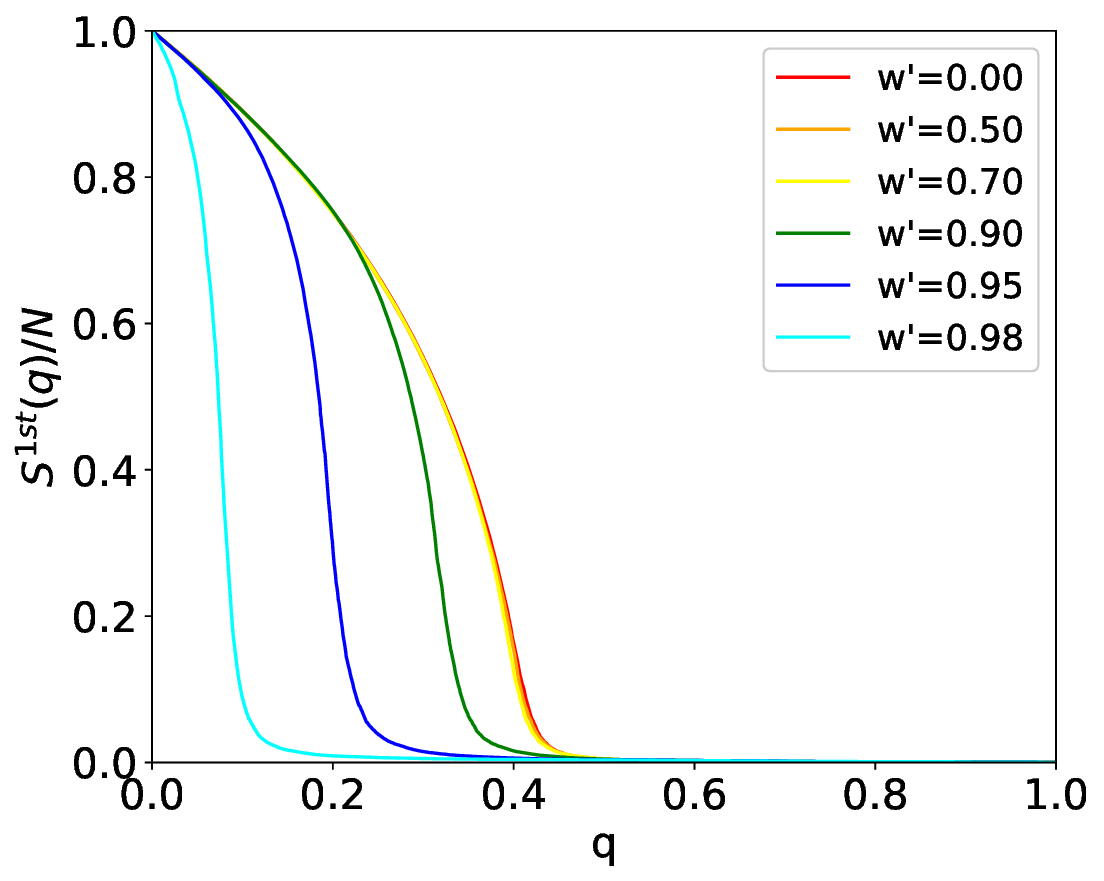}
    \begin{center} (f) $m_{o} = 200$ \end{center}
  \end{minipage}       
%\centering
%\includegraphics[width=.8\textwidth]{resize_figS9.eps}
\caption{Comparison of the areas under the curves 
represented as the robustness against IB attacks
in nearly ER random graphs at $\nu = -1$ with $m_{o}$ modules.}
\label{fig_IB_nu-1}
\end{figure}

\begin{figure}[htb]
  \begin{minipage}{.48\textwidth}
    \includegraphics[width=.9\textwidth]{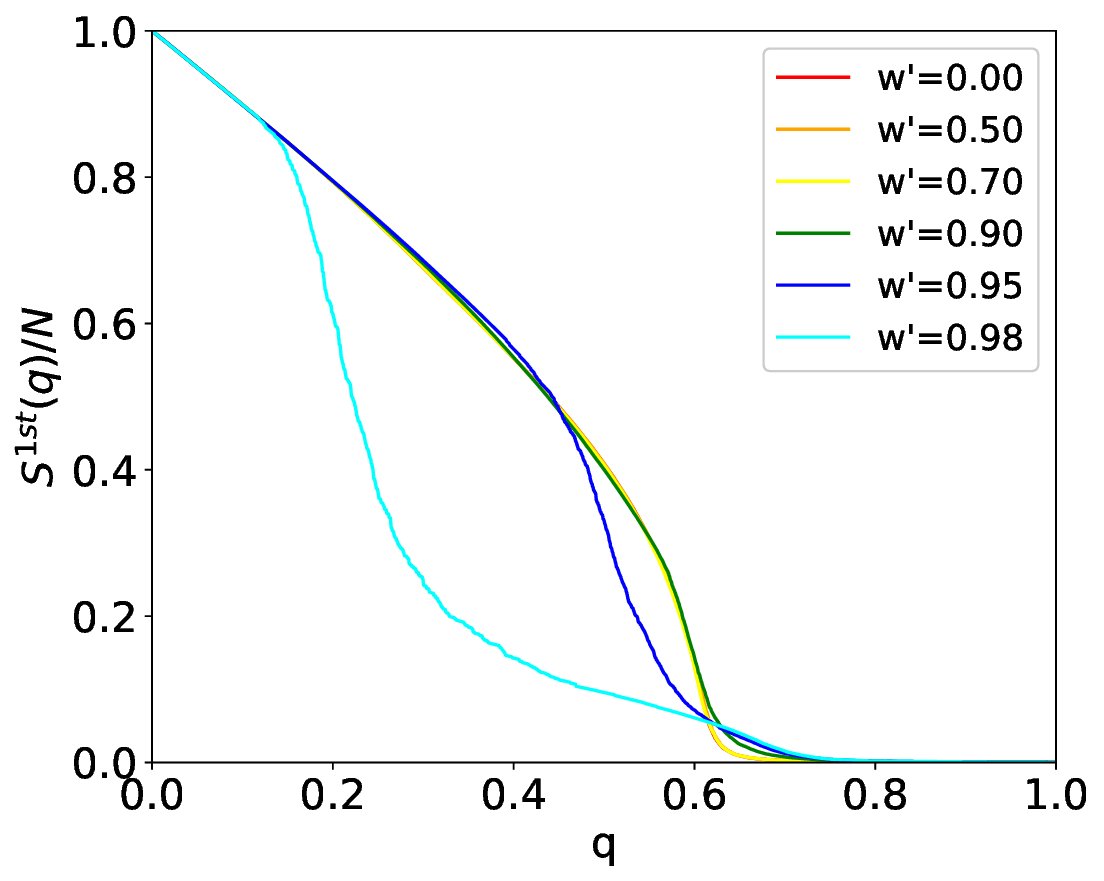}
    \begin{center} (a) $m_{o} = 5$ \end{center}  
  \end{minipage}
  \hfill  
  \begin{minipage}{.48\textwidth}
    \includegraphics[width=.9\textwidth]{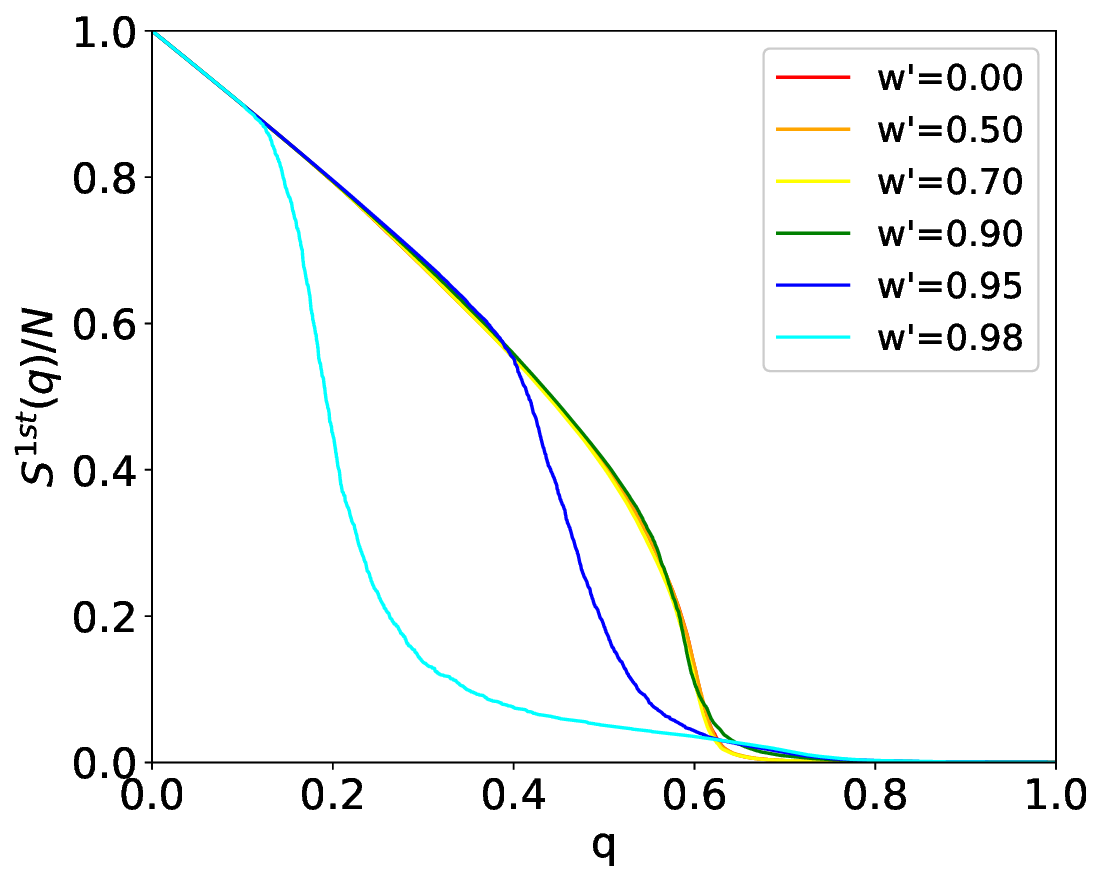}
    \begin{center} (b) $m_{o} = 10$ \end{center}
  \end{minipage}    
  \hfill
  \begin{minipage}{.48\textwidth}
    \includegraphics[width=.9\textwidth]{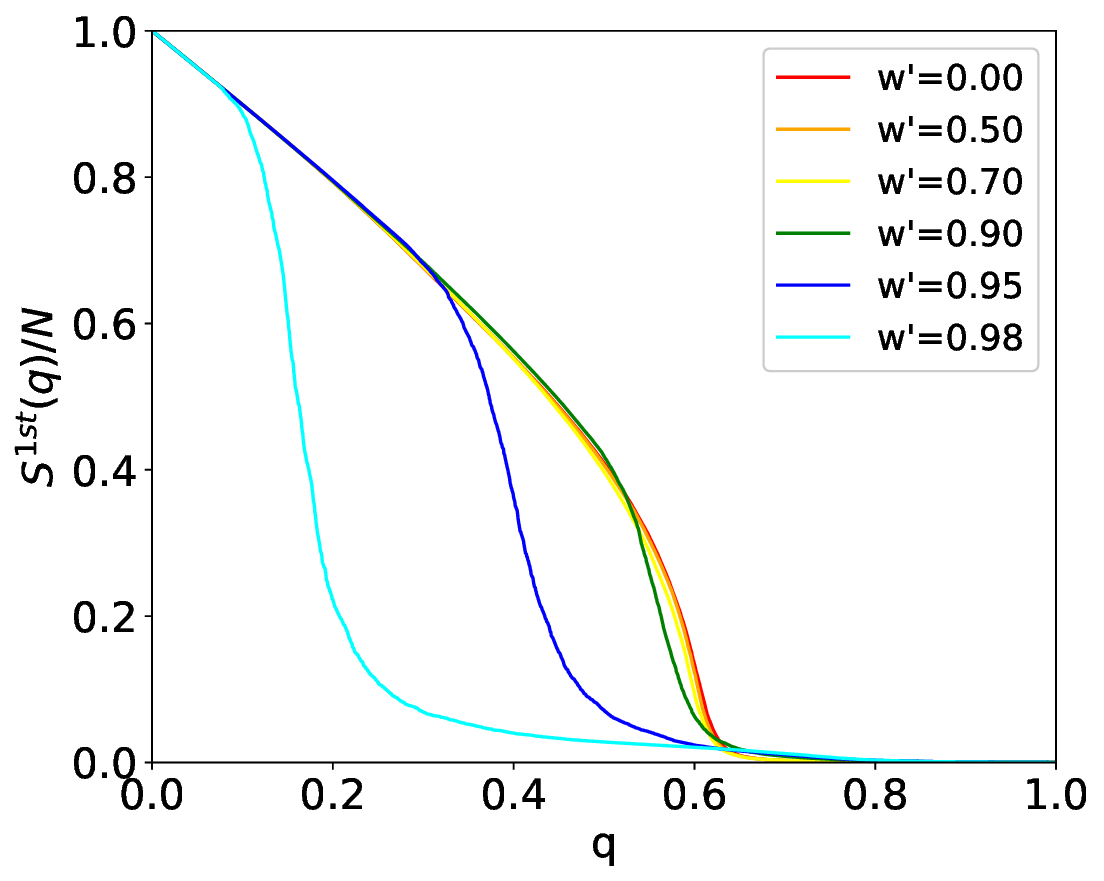}
    \begin{center} (c) $m_{o} = 20$ \end{center}
  \end{minipage}
  \hfill  
  \begin{minipage}{.48\textwidth}
    \includegraphics[width=.9\textwidth]{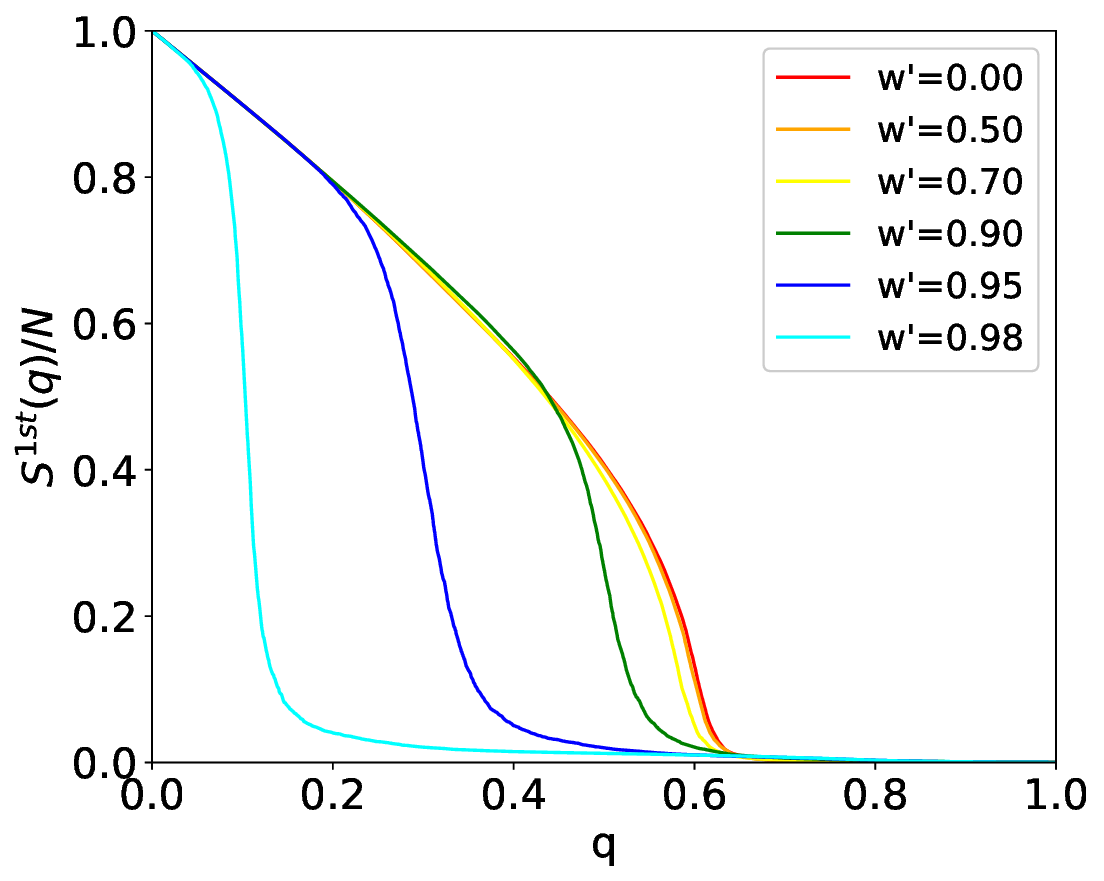}
    \begin{center} (d) $m_{o} = 50$ \end{center}
  \end{minipage}     
  \hfill 
  \begin{minipage}{.48\textwidth}
    \includegraphics[width=.9\textwidth]{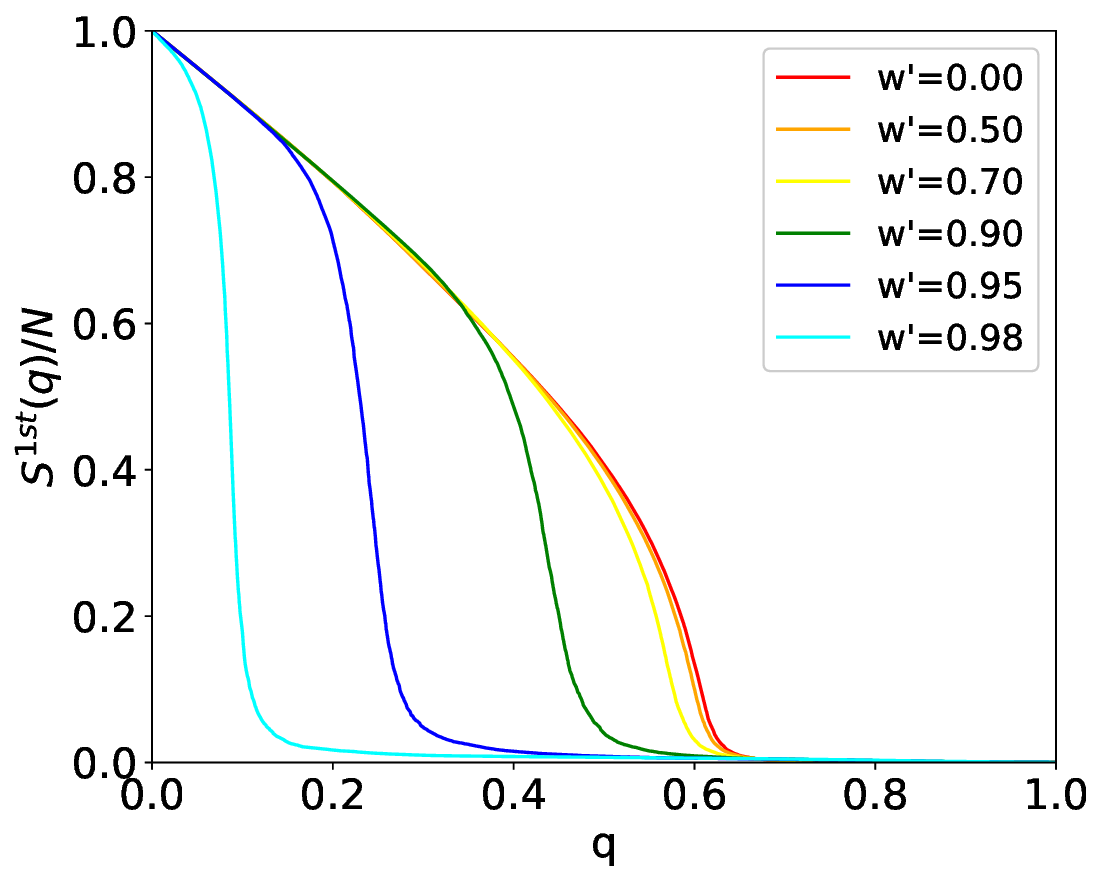}
    \begin{center} (e) $m_{o} = 100$ \end{center}
  \end{minipage}
  \hfill  
  \begin{minipage}{.48\textwidth}
    \includegraphics[width=.9\textwidth]{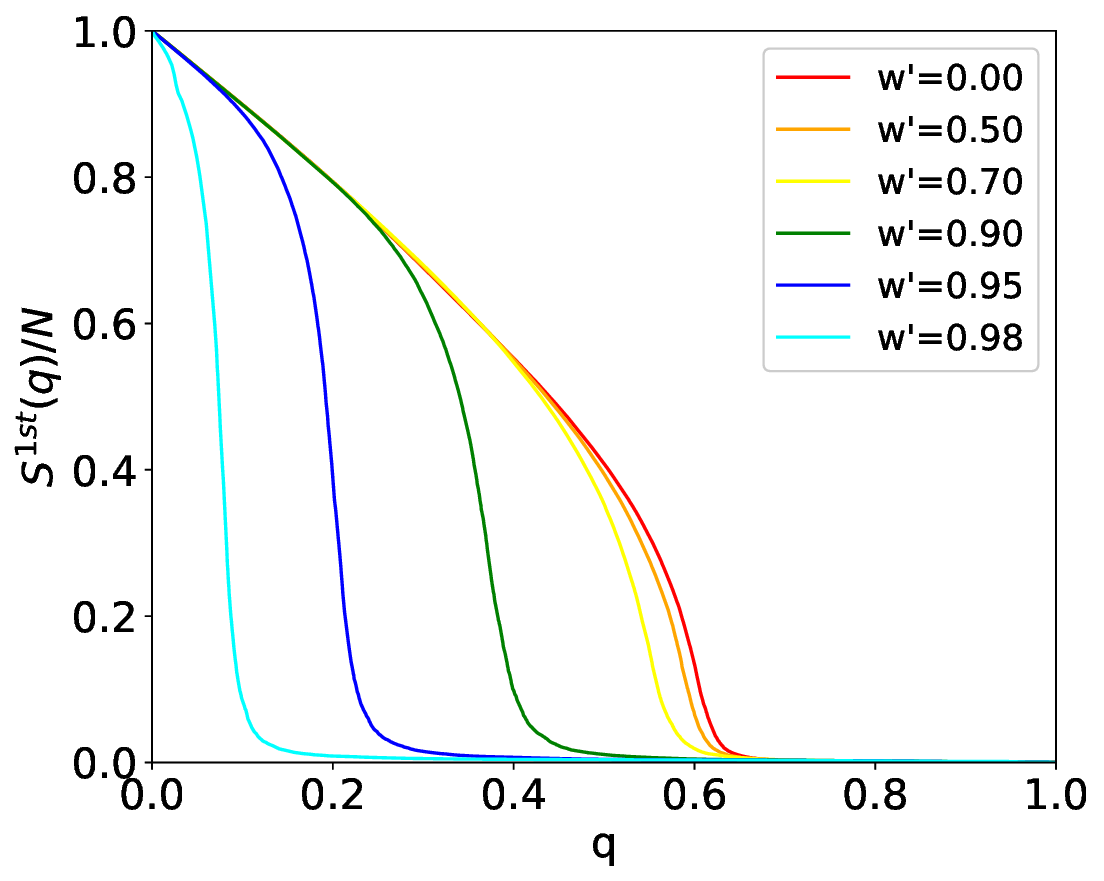}
    \begin{center} (f) $m_{o} = 200$ \end{center}
  \end{minipage}       
%\centering
%\includegraphics[width=.8\textwidth]{resize_figS10.eps}
\caption{Comparison of the areas under the curves 
represented as the robustness against IB attacks
in networks of narrower $P(k)$ at $\nu = -5$ with $m_{o}$ modules.}
\label{fig_IB_nu-5}
\end{figure}

\begin{figure}[htb]
  \begin{minipage}{.48\textwidth}
    \includegraphics[width=.9\textwidth]{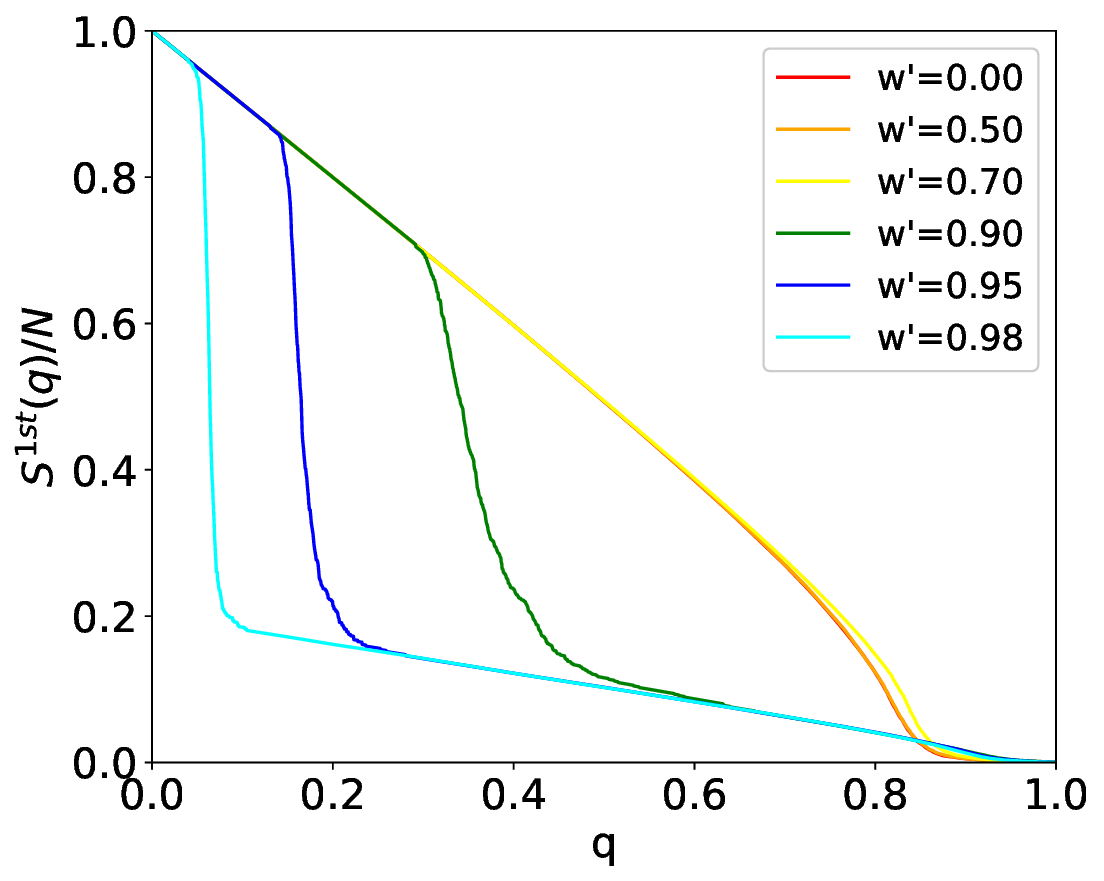}
    \begin{center} (a) $m_{o} = 5$ \end{center}  
  \end{minipage}
  \hfill  
  \begin{minipage}{.48\textwidth}
    \includegraphics[width=.9\textwidth]{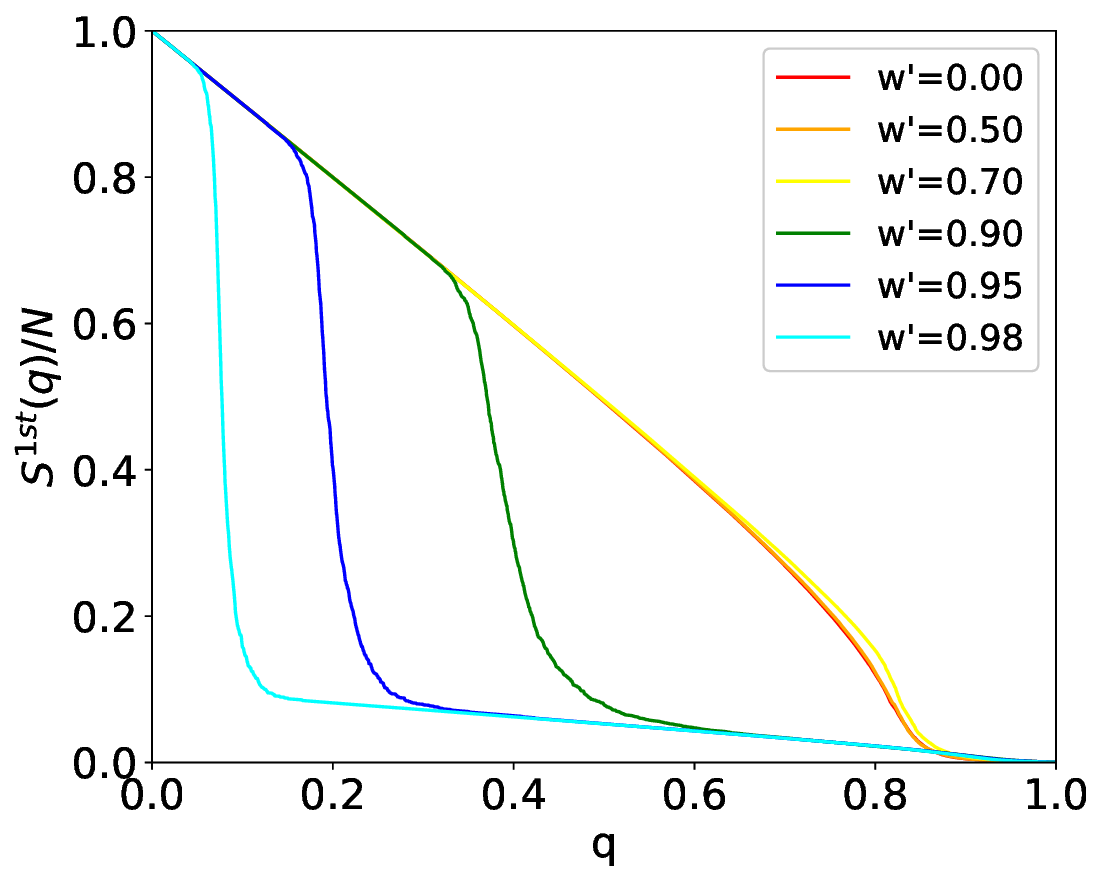}
    \begin{center} (b) $m_{o} = 10$ \end{center}
  \end{minipage}    
  \hfill
  \begin{minipage}{.48\textwidth}
    \includegraphics[width=.9\textwidth]{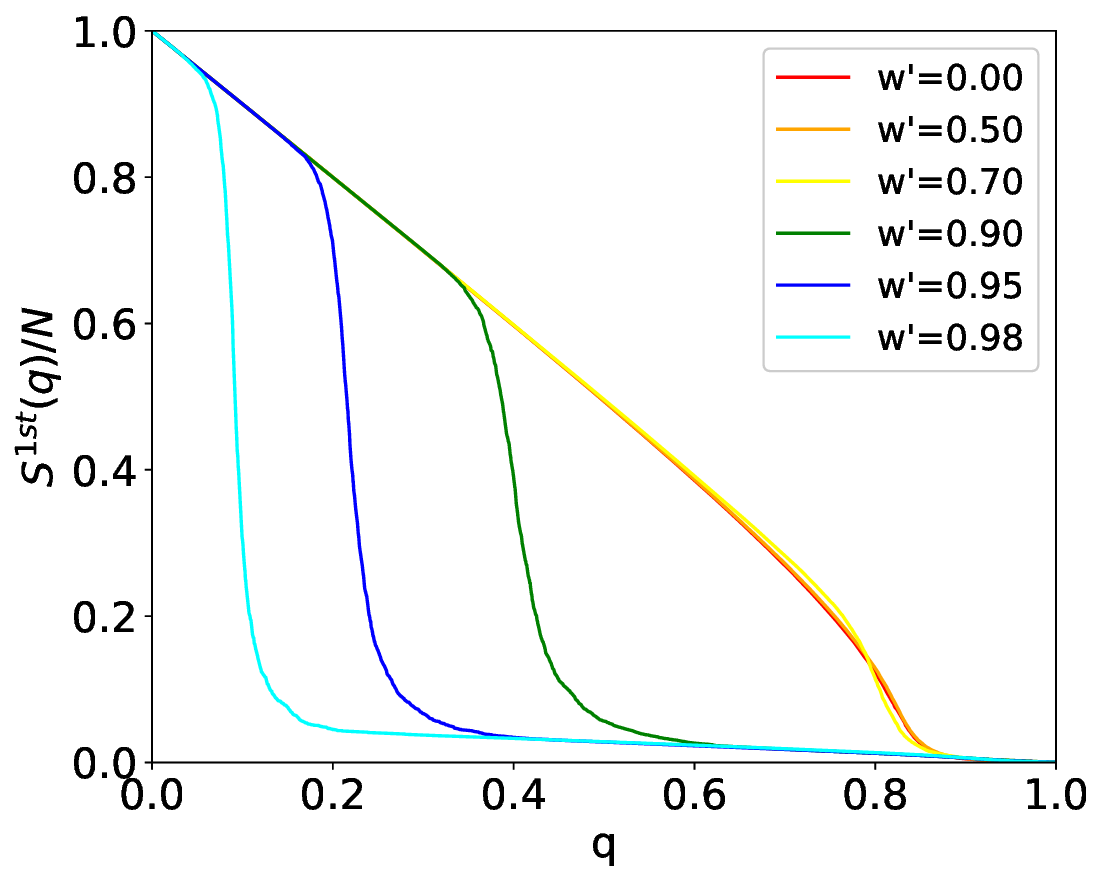}
    \begin{center} (c) $m_{o} = 20$ \end{center}
  \end{minipage}
  \hfill  
  \begin{minipage}{.48\textwidth}
    \includegraphics[width=.9\textwidth]{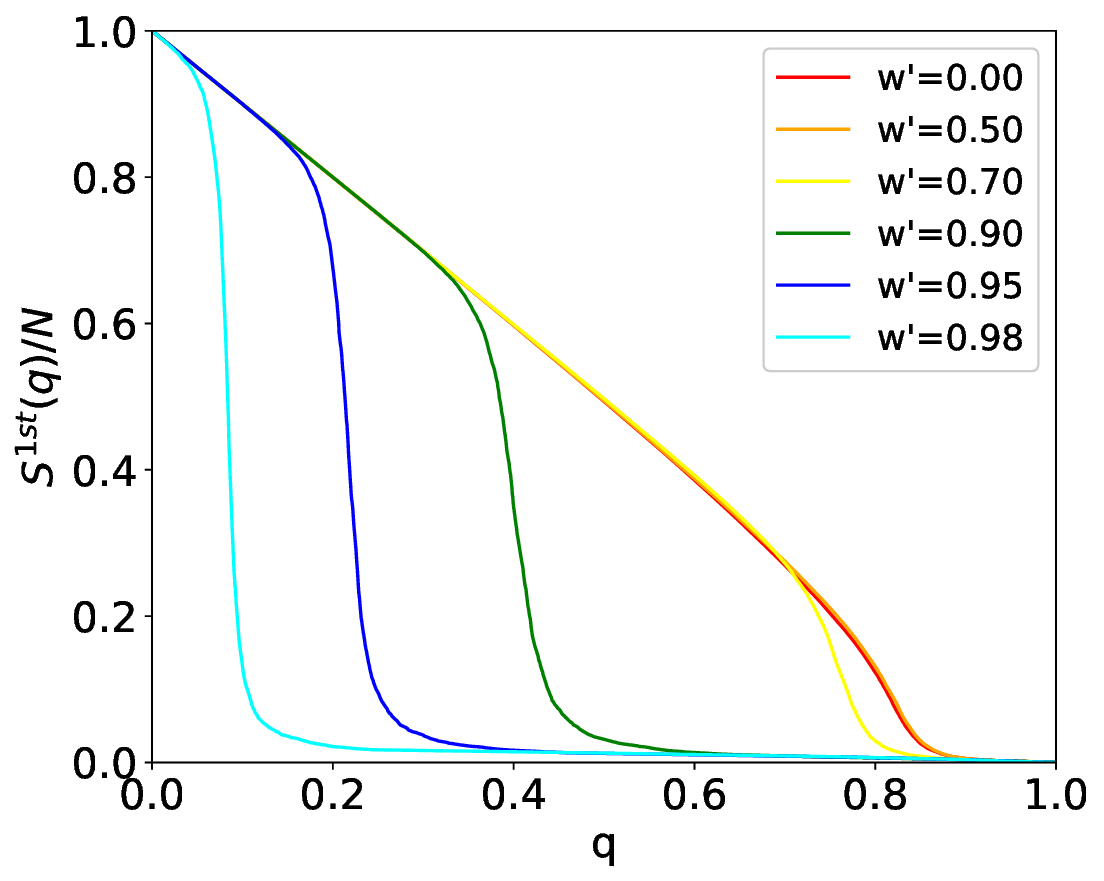}
    \begin{center} (d) $m_{o} = 50$ \end{center}
  \end{minipage}     
  \hfill 
  \begin{minipage}{.48\textwidth}
    \includegraphics[width=.9\textwidth]{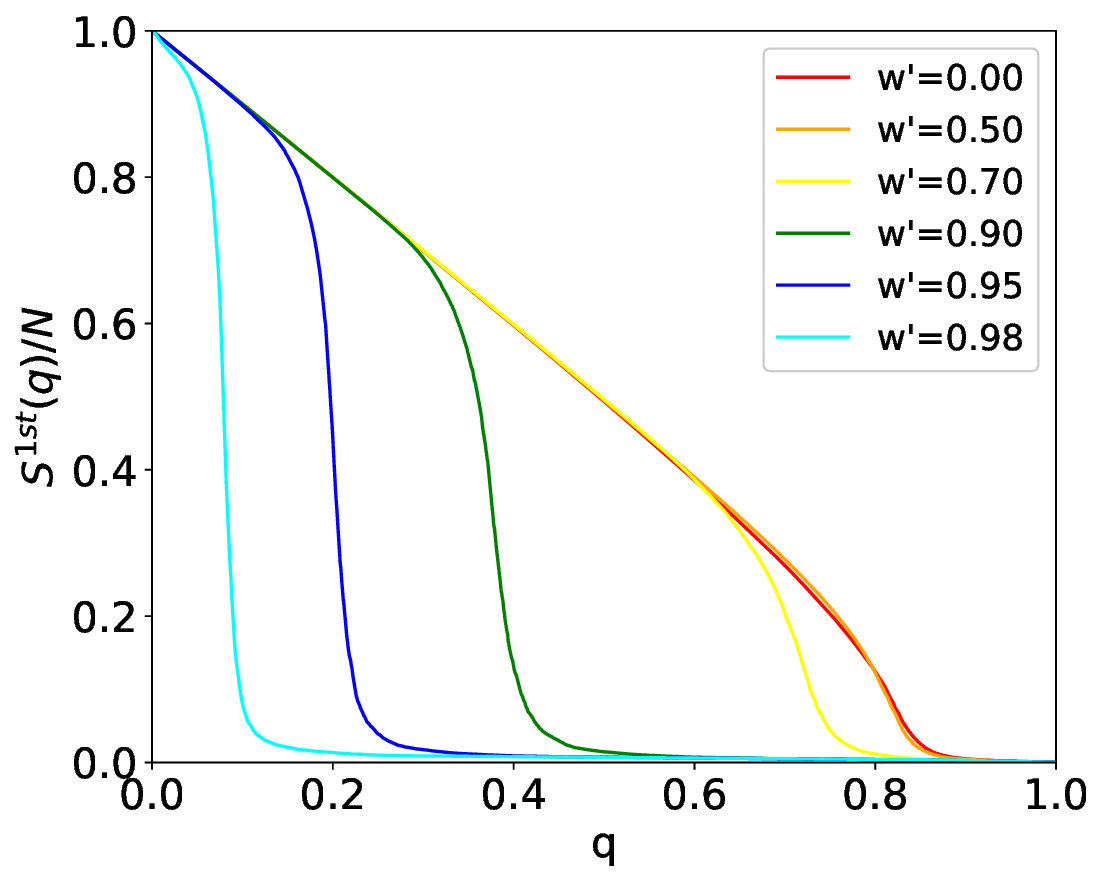}
    \begin{center} (e) $m_{o} = 100$ \end{center}
  \end{minipage}
  \hfill  
  \begin{minipage}{.48\textwidth}
    \includegraphics[width=.9\textwidth]{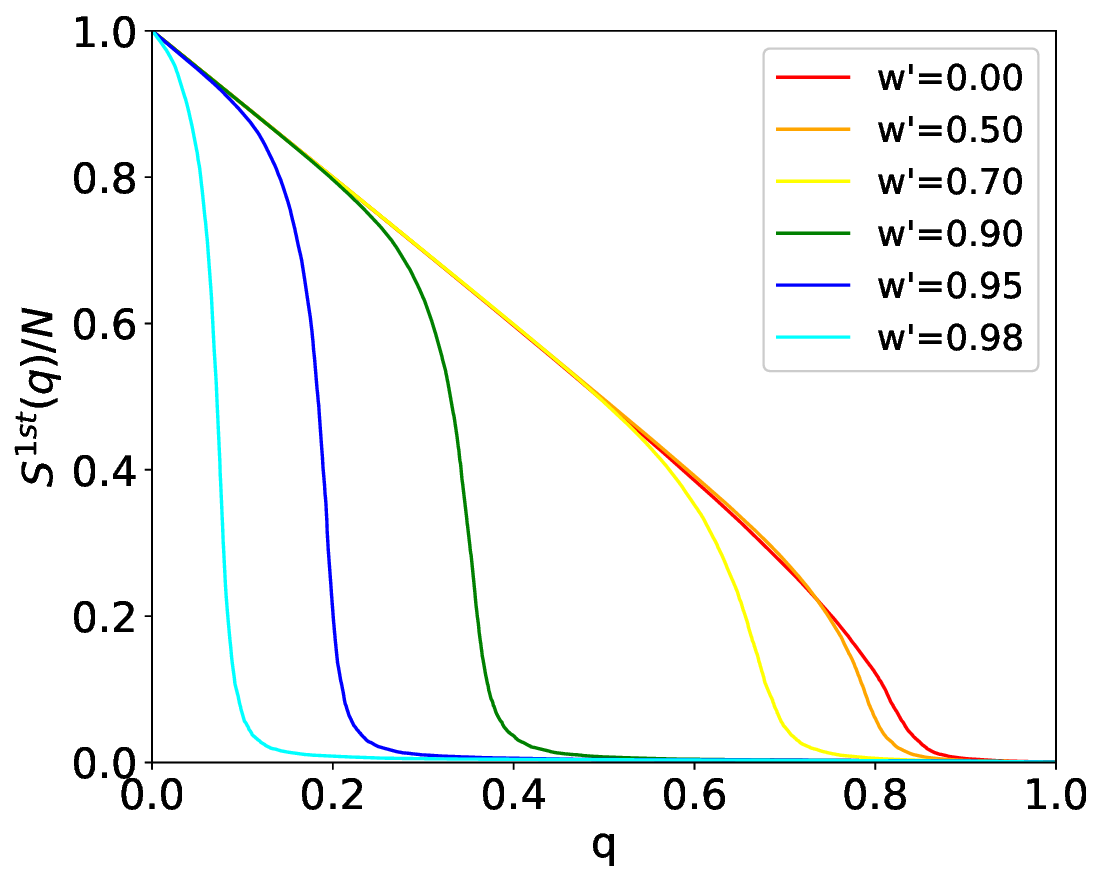}
    \begin{center} (f) $m_{o} = 200$ \end{center}
  \end{minipage}       
%\centering
%\includegraphics[width=.8\textwidth]{resize_figS11.eps}
\caption{Comparison of the areas under the curves 
represented as the robustness against IB attacks
in nearly regular networks at $\nu = -100$ with $m_{o}$ modules.}
\label{fig_IB_nu-100}
\end{figure}

\begin{figure}[htb]
  \begin{minipage}{.48\textwidth}
    \includegraphics[width=.9\textwidth]{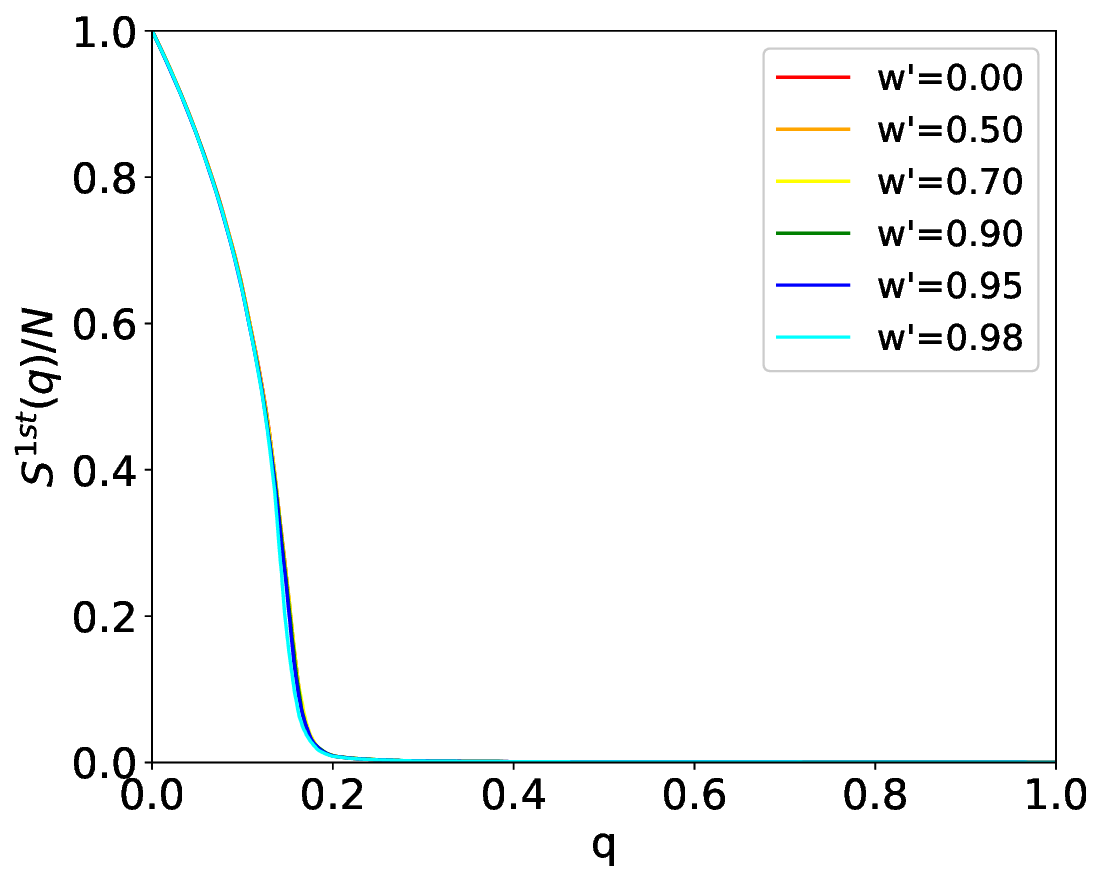}
    \begin{center} (a) $m_{o} = 5$ \end{center}  
  \end{minipage}
  \hfill  
  \begin{minipage}{.48\textwidth}
    \includegraphics[width=.9\textwidth]{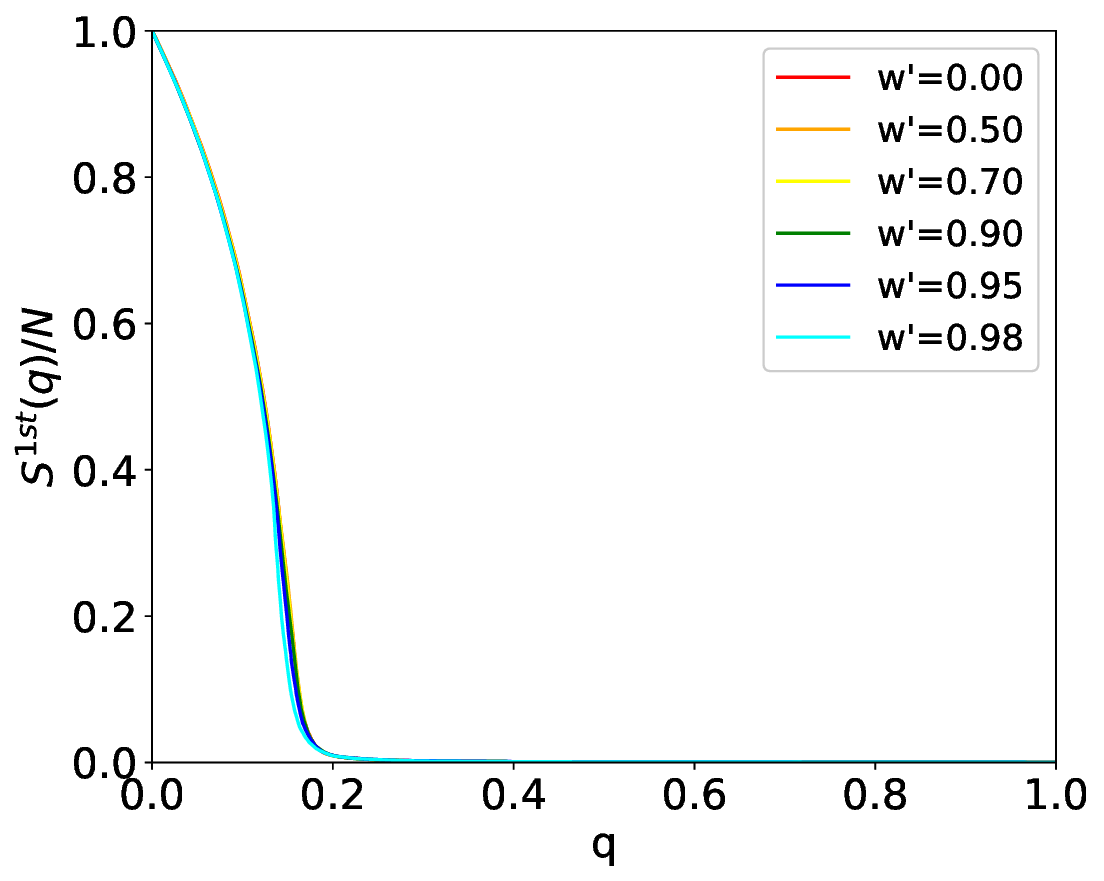}
    \begin{center} (b) $m_{o} = 10$ \end{center}
  \end{minipage}    
  \hfill
  \begin{minipage}{.48\textwidth}
    \includegraphics[width=.9\textwidth]{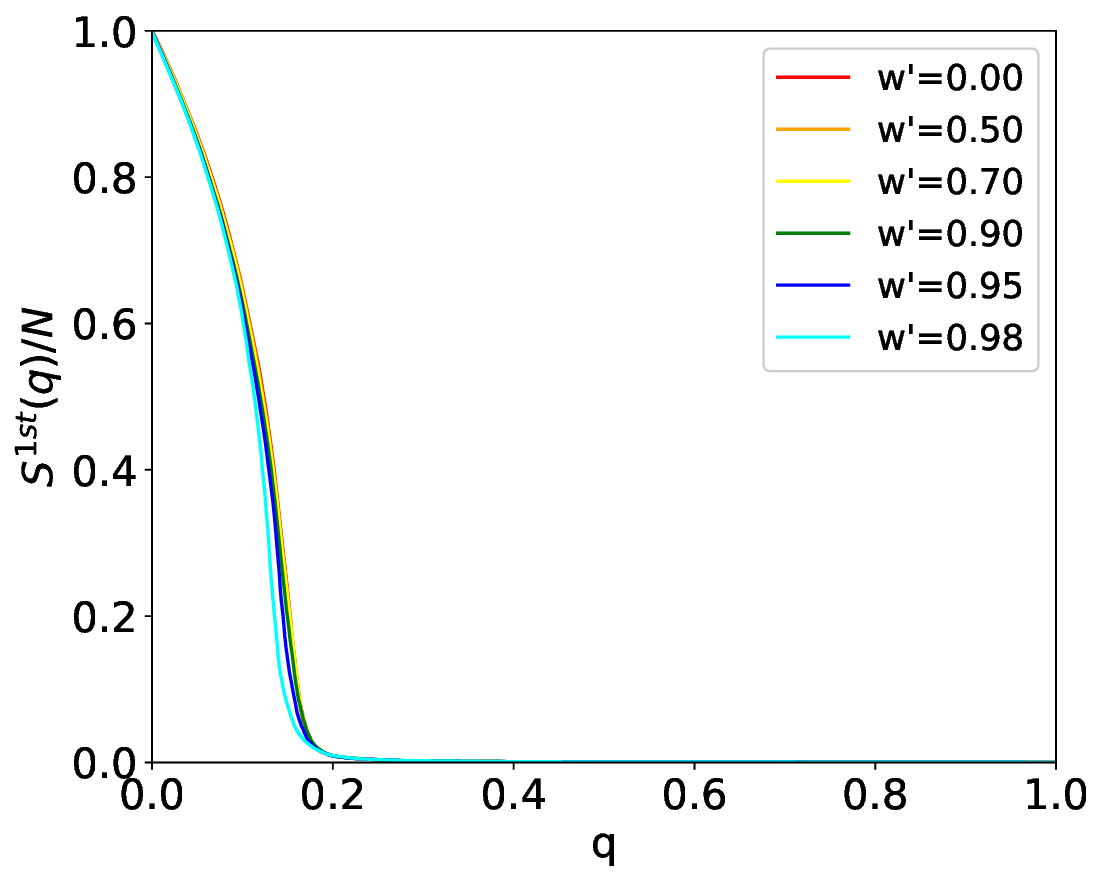}
    \begin{center} (c) $m_{o} = 20$ \end{center}
  \end{minipage}
  \hfill  
  \begin{minipage}{.48\textwidth}
    \includegraphics[width=.9\textwidth]{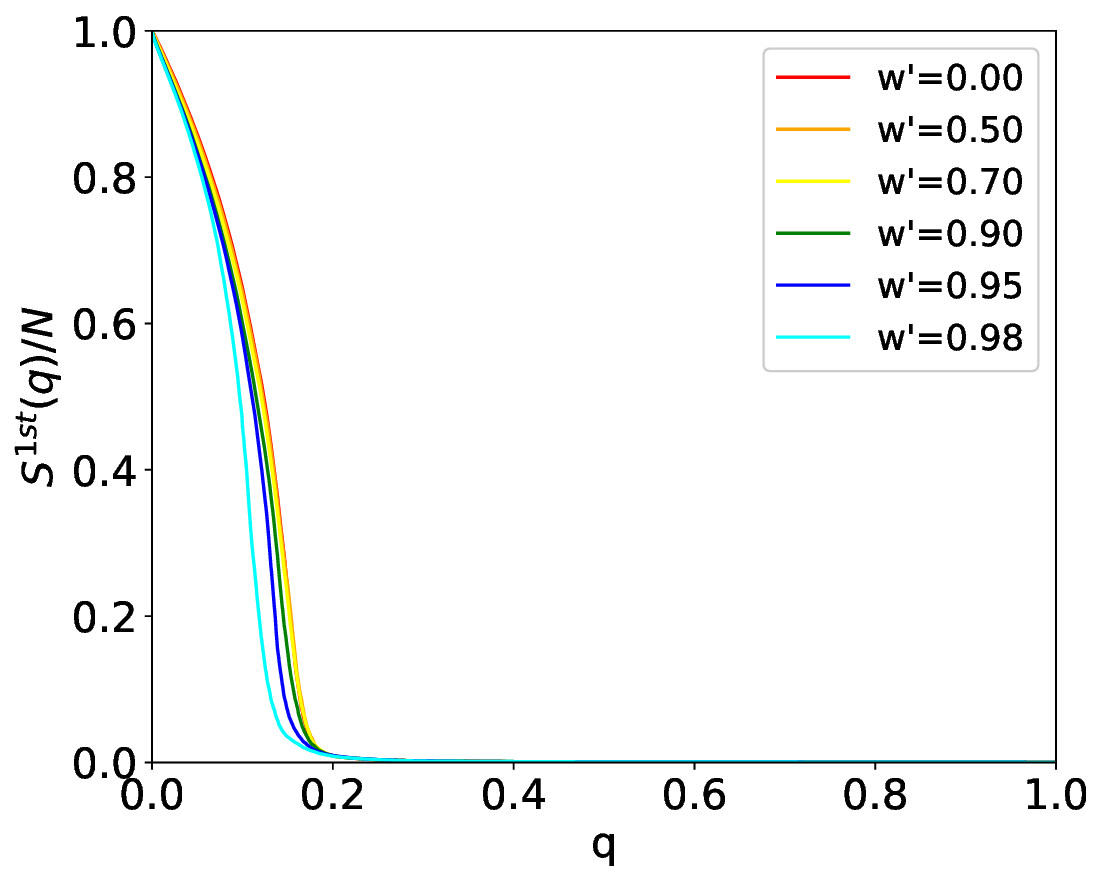}
    \begin{center} (d) $m_{o} = 50$ \end{center}
  \end{minipage}     
  \hfill 
  \begin{minipage}{.48\textwidth}
    \includegraphics[width=.9\textwidth]{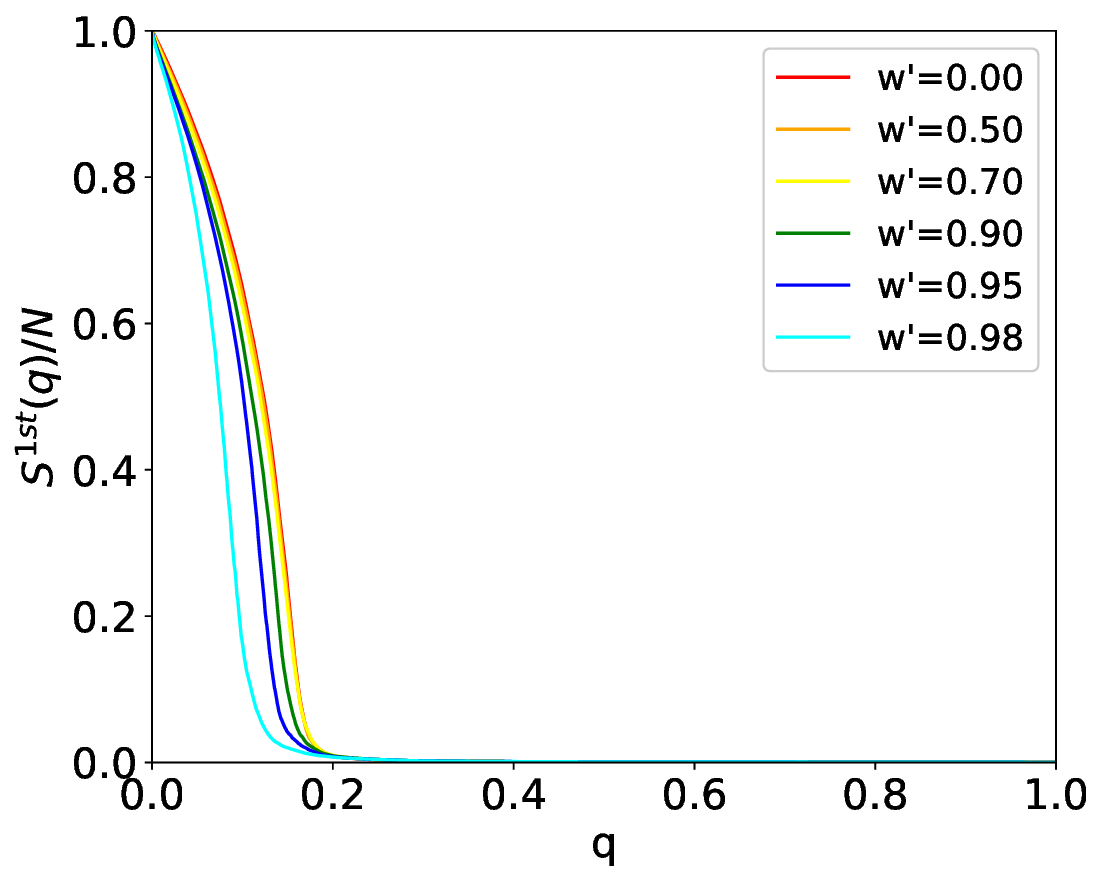}
    \begin{center} (e) $m_{o} = 100$ \end{center}
  \end{minipage}
  \hfill  
  \begin{minipage}{.48\textwidth}
    \includegraphics[width=.9\textwidth]{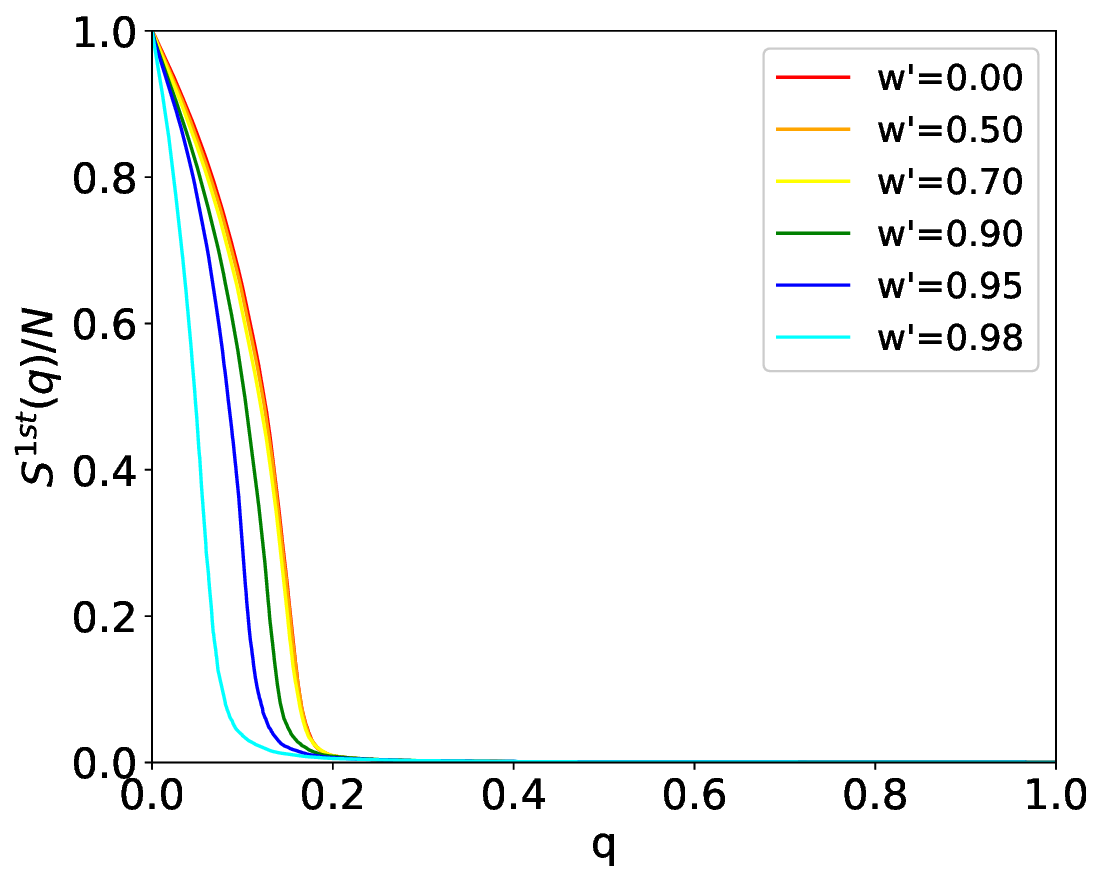}
    \begin{center} (f) $m_{o} = 200$ \end{center}
  \end{minipage}       
%\centering
%\includegraphics[width=.8\textwidth]{resize_figS12.eps}
\caption{Comparison of the areas under the curves 
represented as the robustness against ID attacks
in SF networks at $\nu = 1$ with $m_{o}$ modules.}
\label{fig_ID_nu1}
\end{figure}

\begin{figure}[htb]
  \begin{minipage}{.48\textwidth}
    \includegraphics[width=.9\textwidth]{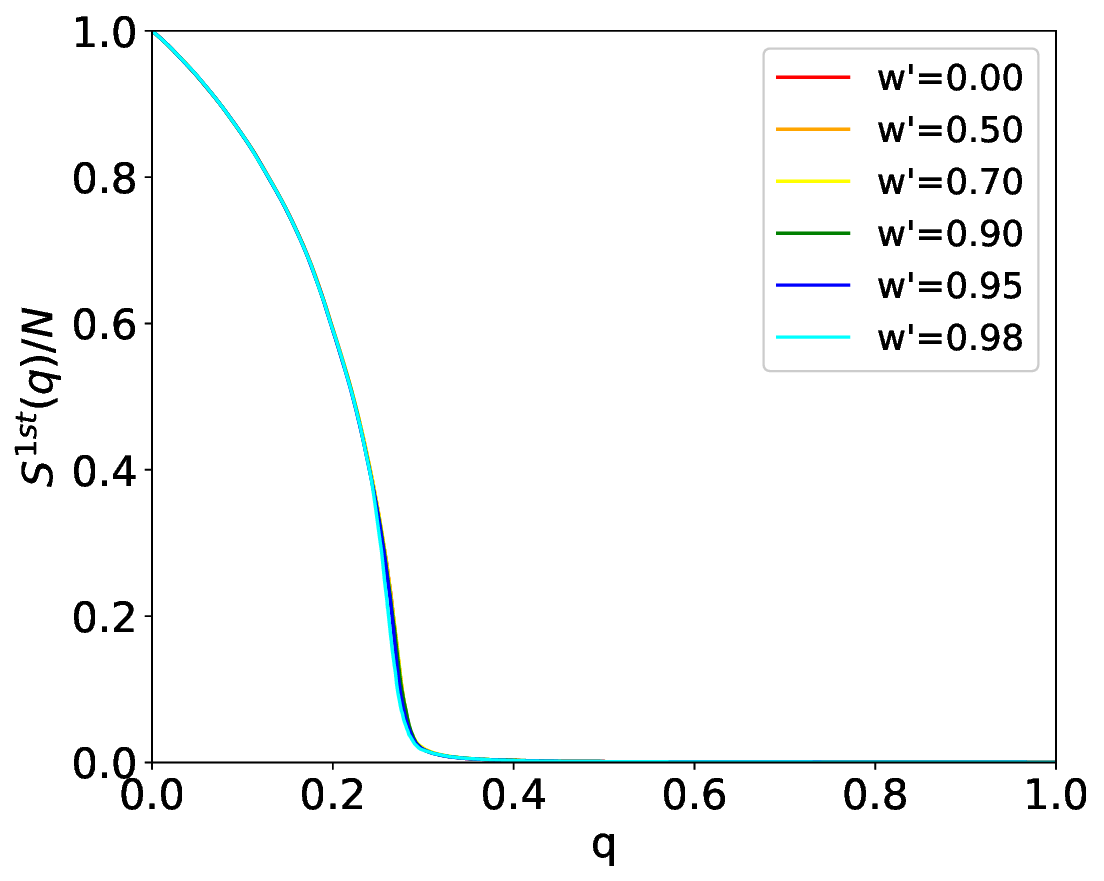}
    \begin{center} (a) $m_{o} = 5$ \end{center}  
  \end{minipage}
  \hfill  
  \begin{minipage}{.48\textwidth}
    \includegraphics[width=.9\textwidth]{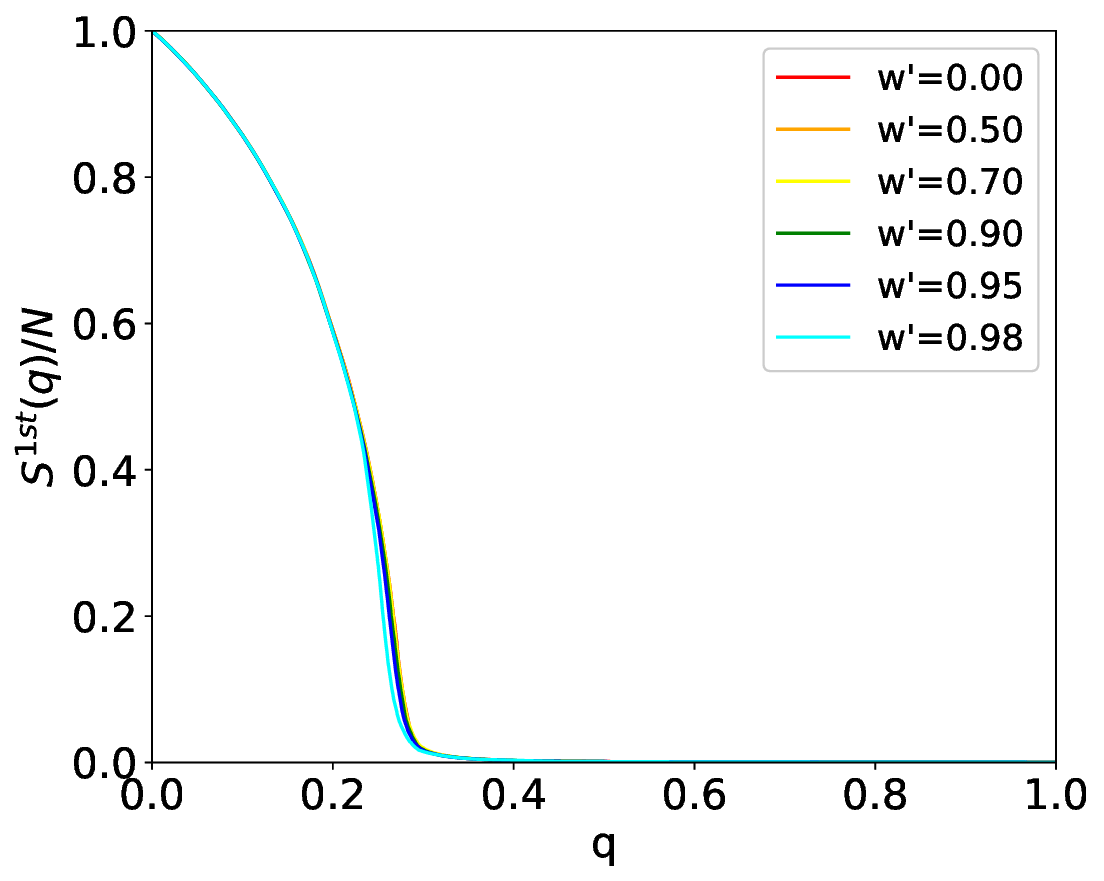}
    \begin{center} (b) $m_{o} = 10$ \end{center}
  \end{minipage}    
  \hfill
  \begin{minipage}{.48\textwidth}
    \includegraphics[width=.9\textwidth]{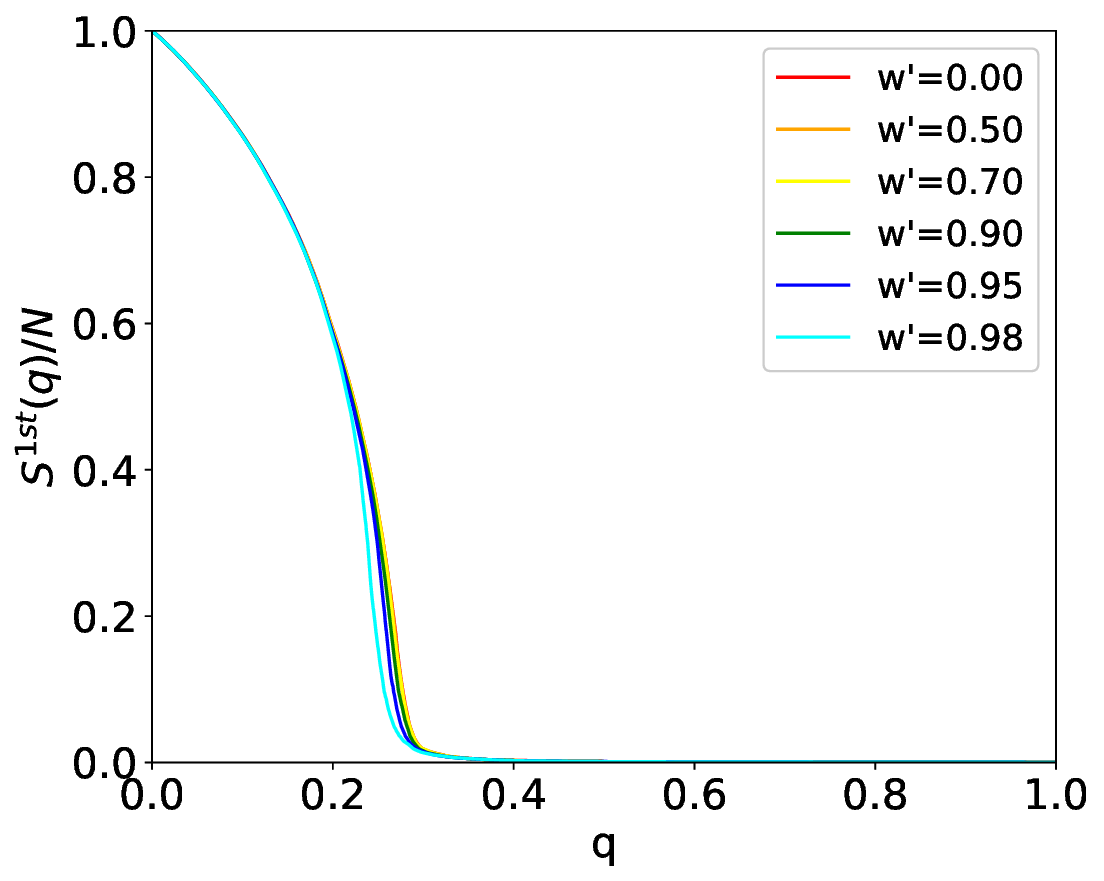}
    \begin{center} (c) $m_{o} = 20$ \end{center}
  \end{minipage}
  \hfill  
  \begin{minipage}{.48\textwidth}
    \includegraphics[width=.9\textwidth]{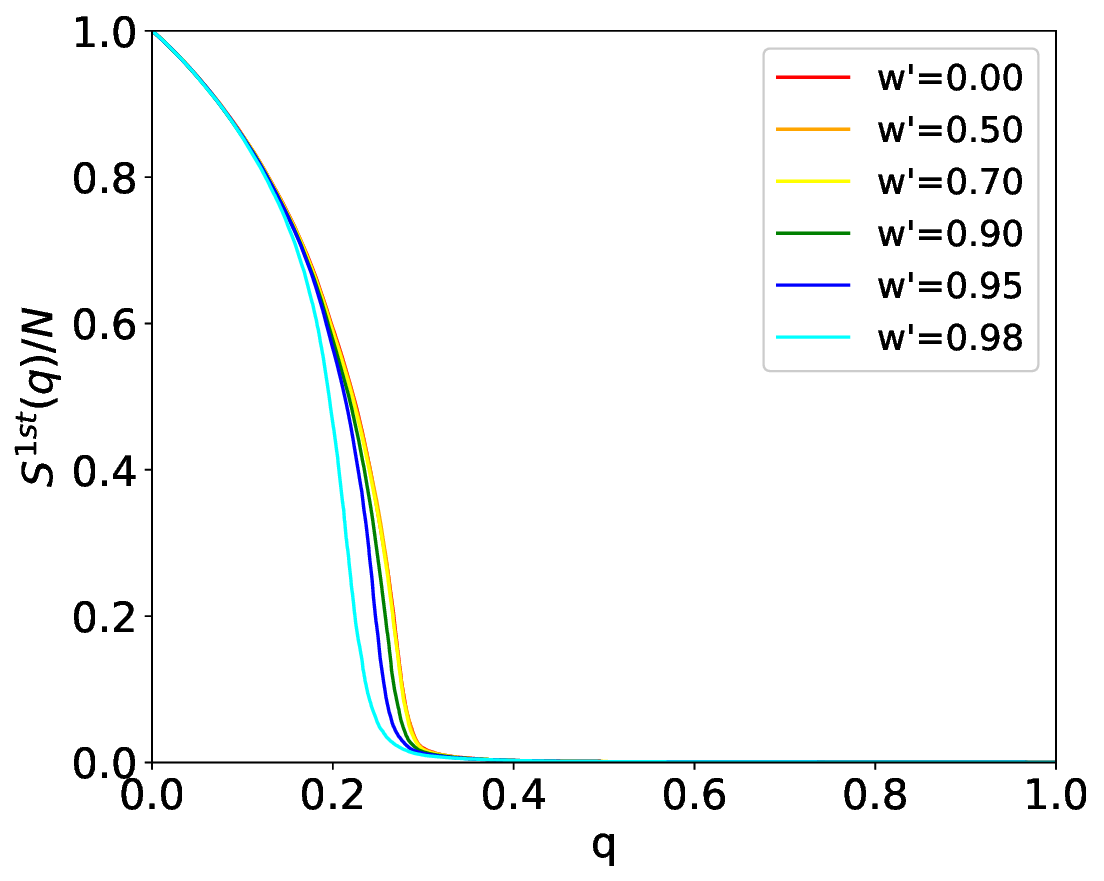}
    \begin{center} (d) $m_{o} = 50$ \end{center}
  \end{minipage}     
  \hfill 
  \begin{minipage}{.48\textwidth}
    \includegraphics[width=.9\textwidth]{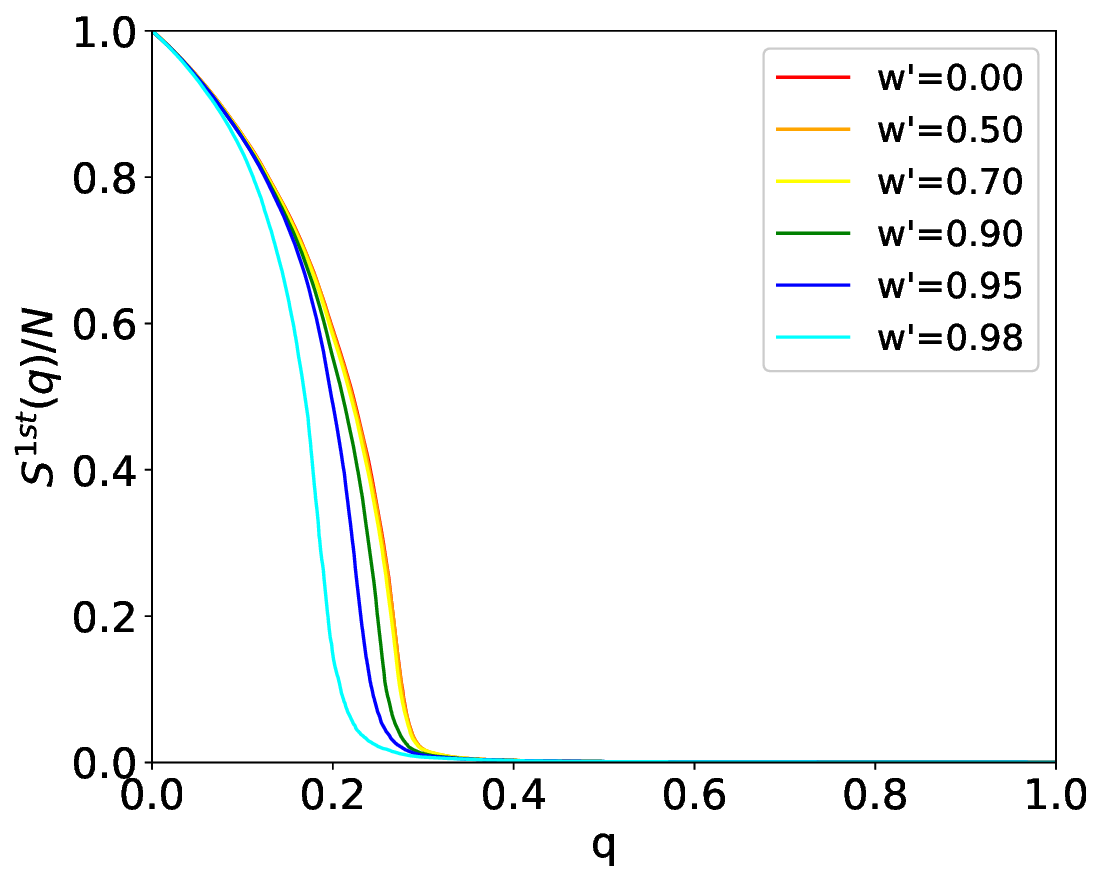}
    \begin{center} (e) $m_{o} = 100$ \end{center}
  \end{minipage}
  \hfill  
  \begin{minipage}{.48\textwidth}
    \includegraphics[width=.9\textwidth]{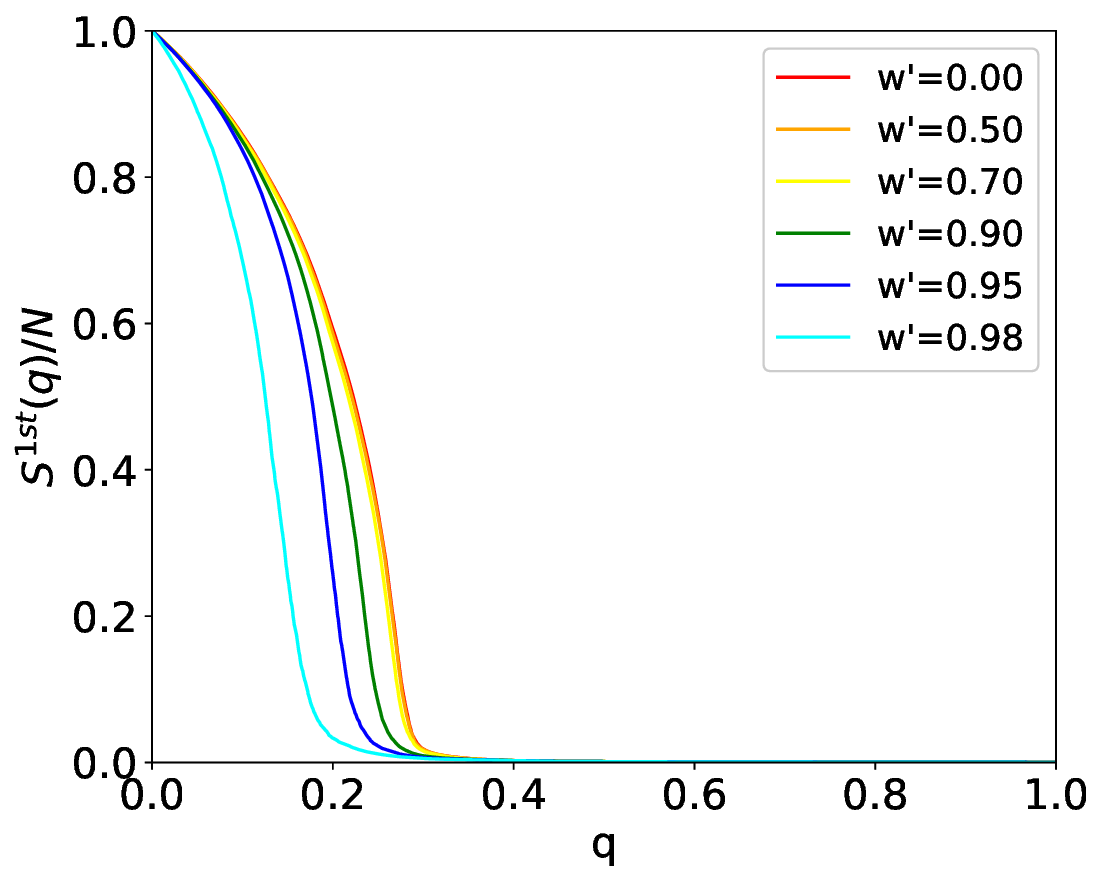}
    \begin{center} (f) $m_{o} = 200$ \end{center}
  \end{minipage}       
%\centering
%\includegraphics[width=.8\textwidth]{resize_figS13.eps}
\caption{Comparison of the areas under the curves 
represented as the robustness against ID attacks
in randomly attached networks at $\nu = 0$ with $m_{o}$ modules.}
\label{fig_ID_nu0}
\end{figure}

\begin{figure}[htb]
  \begin{minipage}{.48\textwidth}
    \includegraphics[width=.9\textwidth]{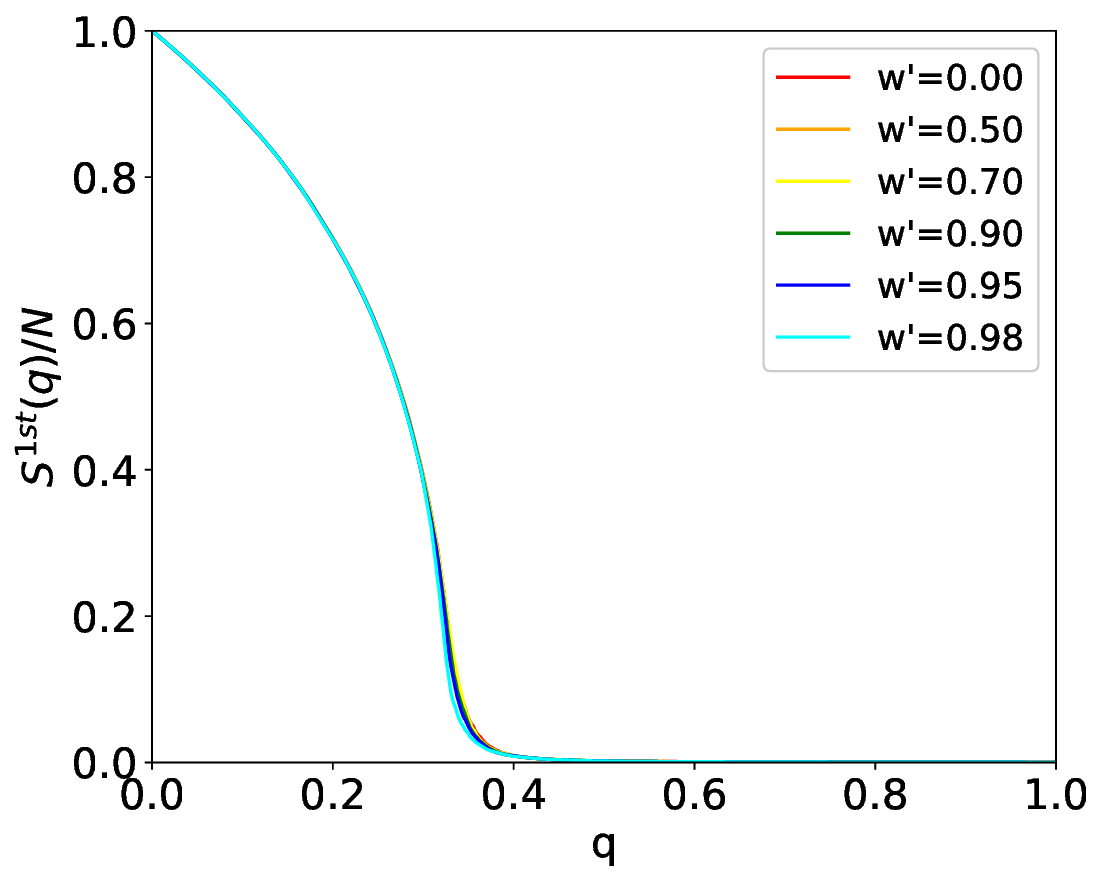}
    \begin{center} (a) $m_{o} = 5$ \end{center}  
  \end{minipage}
  \hfill  
  \begin{minipage}{.48\textwidth}
    \includegraphics[width=.9\textwidth]{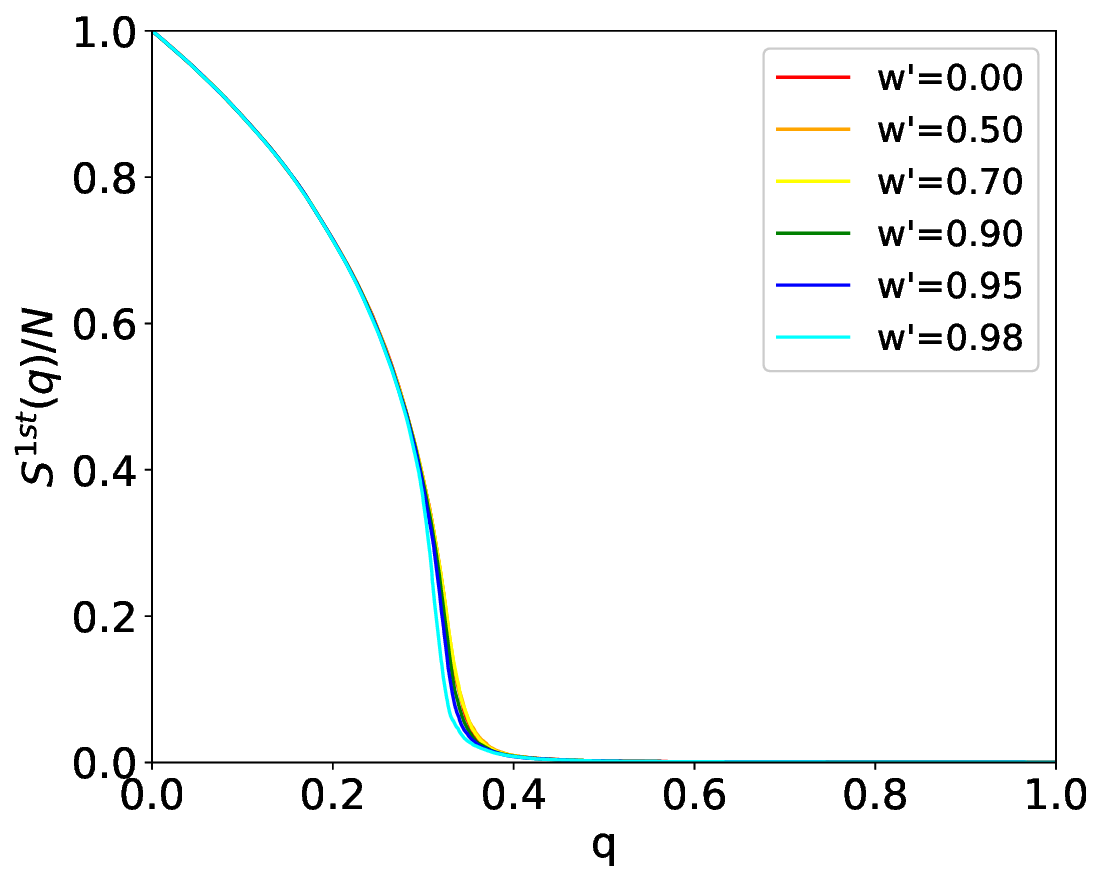}
    \begin{center} (b) $m_{o} = 10$ \end{center}
  \end{minipage}    
  \hfill
  \begin{minipage}{.48\textwidth}
    \includegraphics[width=.9\textwidth]{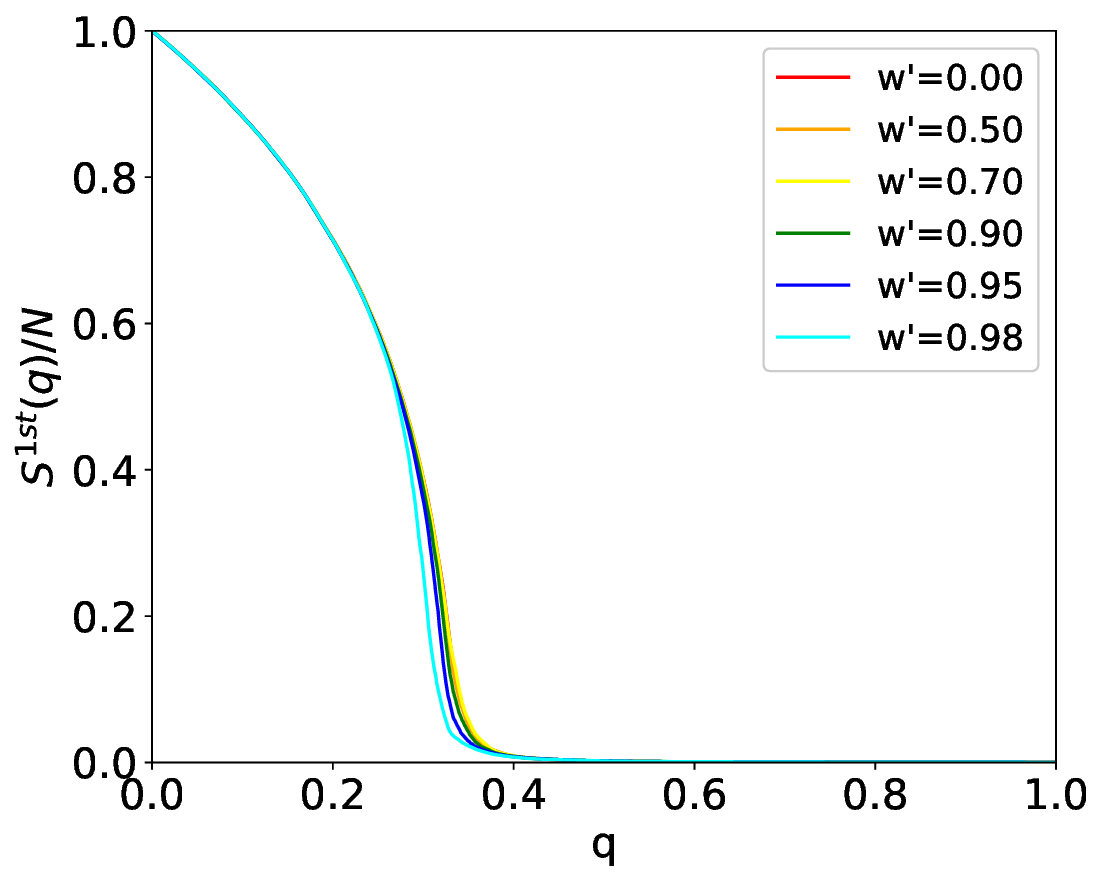}
    \begin{center} (c) $m_{o} = 20$ \end{center}
  \end{minipage}
  \hfill  
  \begin{minipage}{.48\textwidth}
    \includegraphics[width=.9\textwidth]{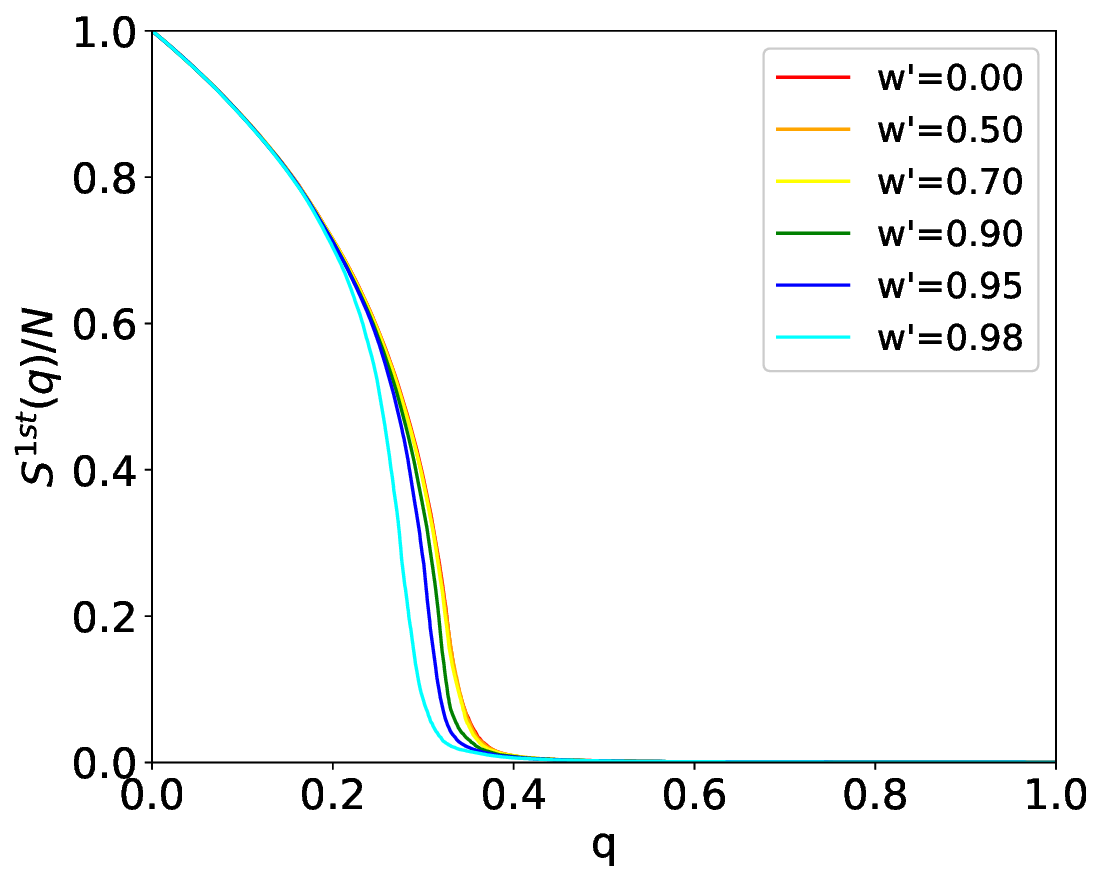}
    \begin{center} (d) $m_{o} = 50$ \end{center}
  \end{minipage}     
  \hfill 
  \begin{minipage}{.48\textwidth}
    \includegraphics[width=.9\textwidth]{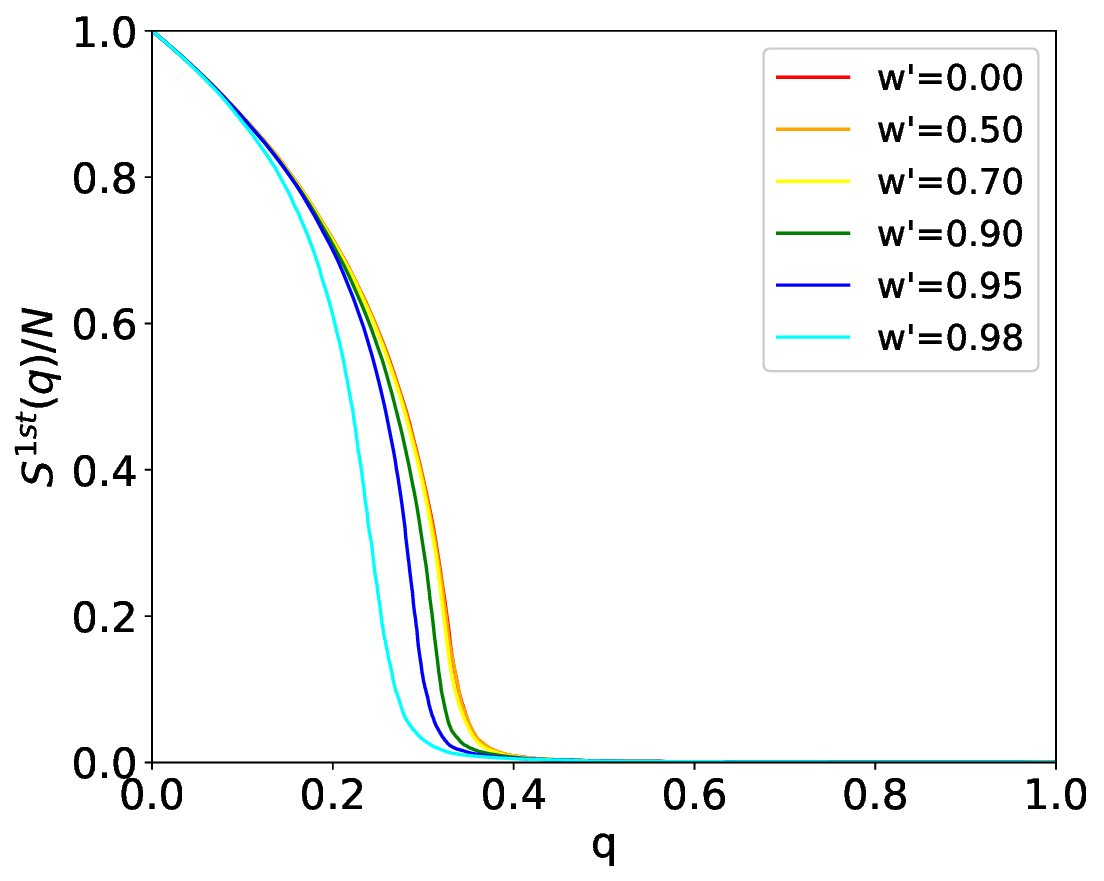}
    \begin{center} (e) $m_{o} = 100$ \end{center}
  \end{minipage}
  \hfill  
  \begin{minipage}{.48\textwidth}
    \includegraphics[width=.9\textwidth]{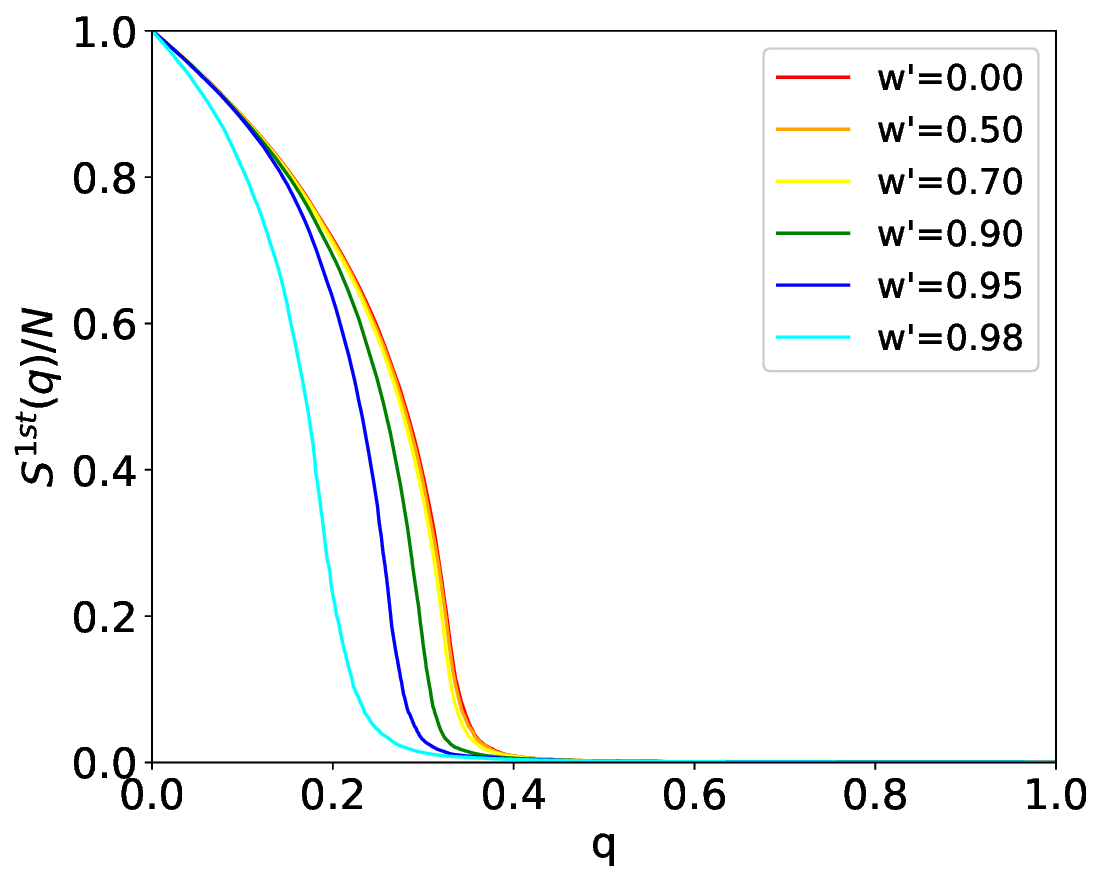}
    \begin{center} (f) $m_{o} = 200$ \end{center}
  \end{minipage}       
%\centering
%\includegraphics[width=.8\textwidth]{resize_figS14.eps}
\caption{Comparison of the areas under the curves 
represented as the robustness against ID attacks
in nearly ER random graphs at $\nu = -1$ with $m_{o}$ modules.}
\label{fig_ID_nu-1}
\end{figure}

\begin{figure}[htb]
  \begin{minipage}{.48\textwidth}
    \includegraphics[width=.9\textwidth]{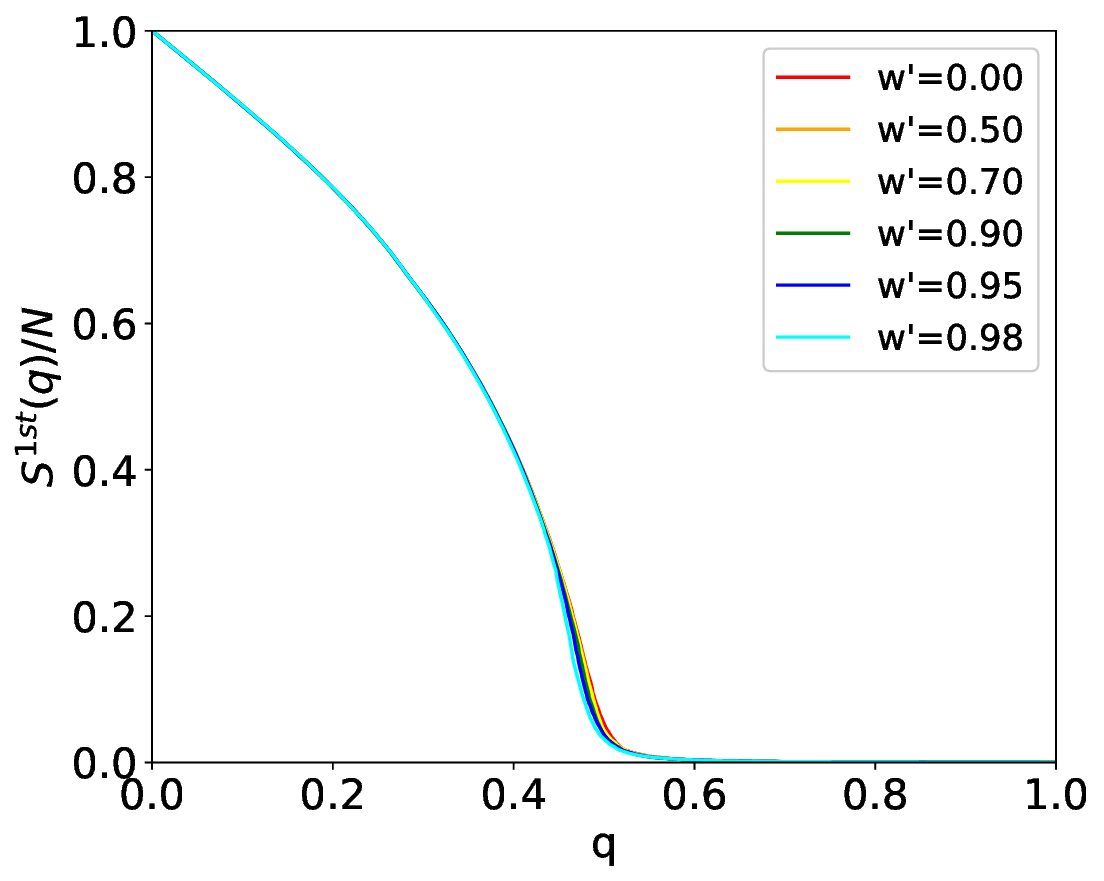}
    \begin{center} (a) $m_{o} = 5$ \end{center}  
  \end{minipage}
  \hfill  
  \begin{minipage}{.48\textwidth}
    \includegraphics[width=.9\textwidth]{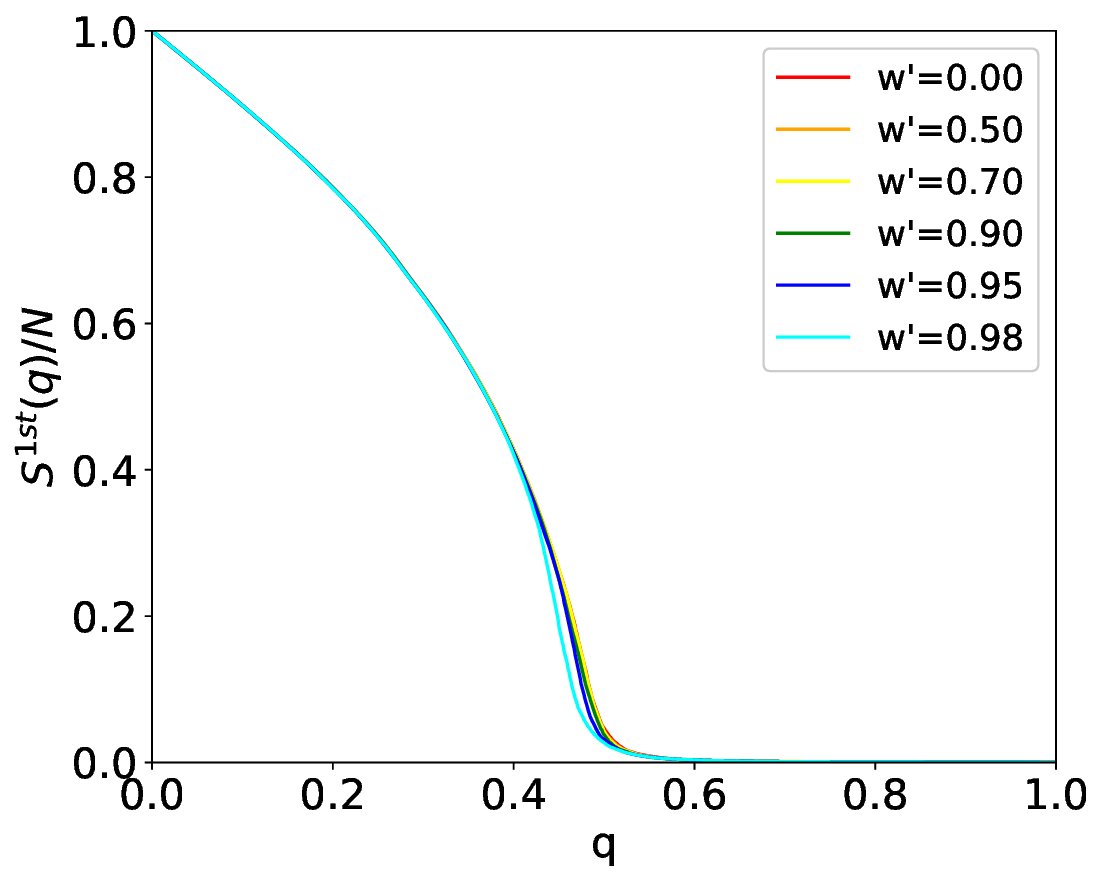}
    \begin{center} (b) $m_{o} = 10$ \end{center}
  \end{minipage}    
  \hfill
  \begin{minipage}{.48\textwidth}
    \includegraphics[width=.9\textwidth]{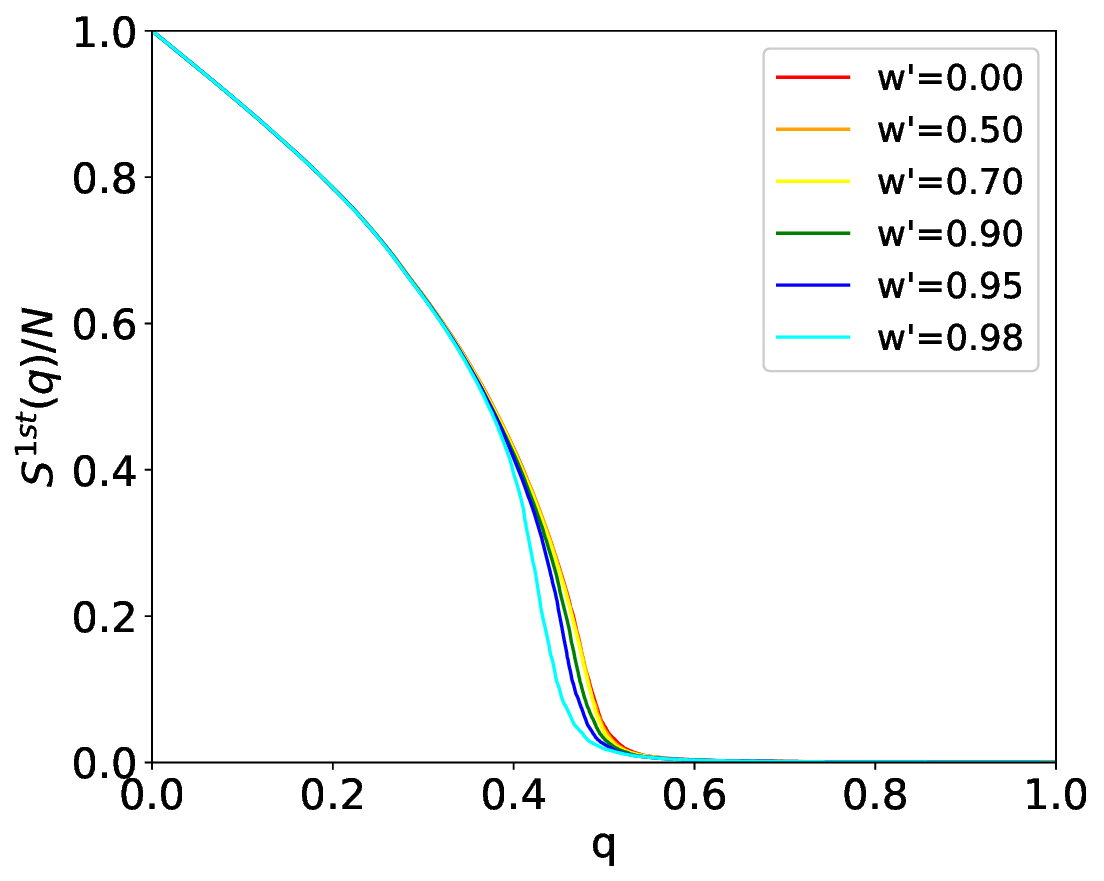}
    \begin{center} (c) $m_{o} = 20$ \end{center}
  \end{minipage}
  \hfill  
  \begin{minipage}{.48\textwidth}
    \includegraphics[width=.9\textwidth]{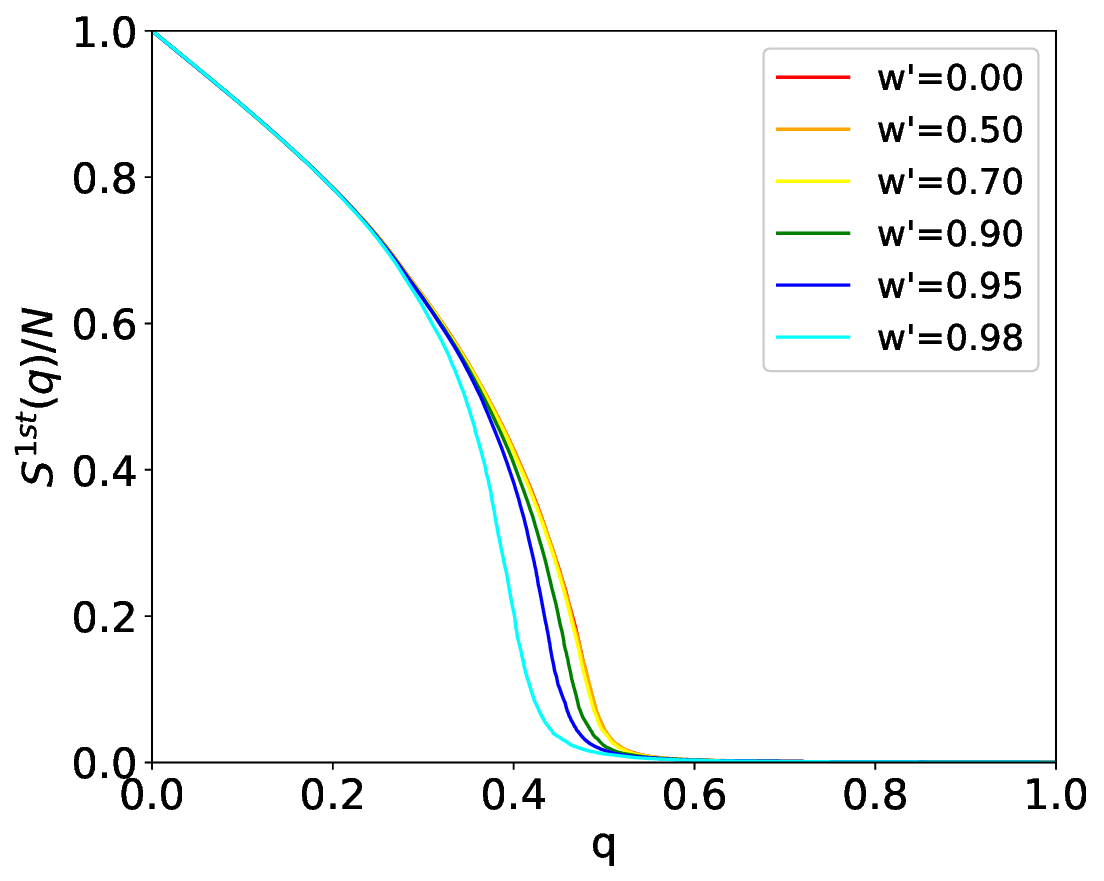}
    \begin{center} (d) $m_{o} = 50$ \end{center}
  \end{minipage}     
  \hfill 
  \begin{minipage}{.48\textwidth}
    \includegraphics[width=.9\textwidth]{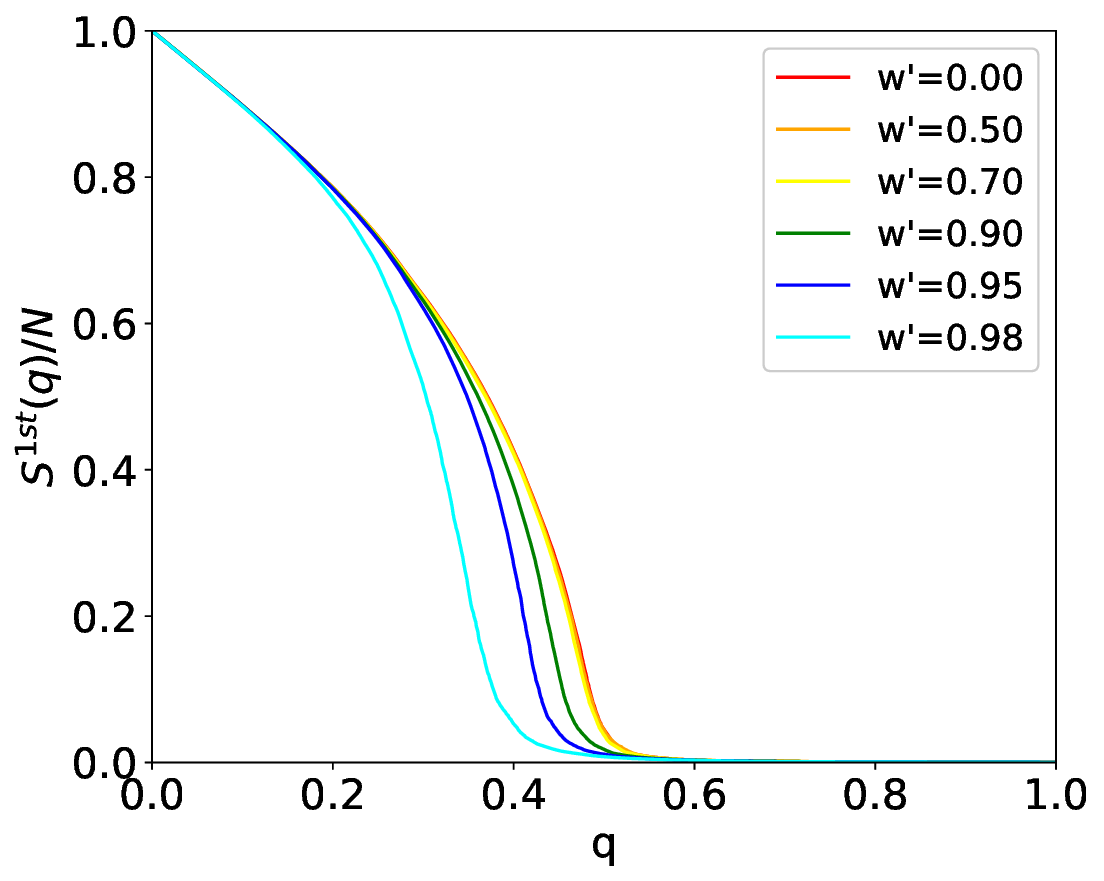}
    \begin{center} (e) $m_{o} = 100$ \end{center}
  \end{minipage}
  \hfill  
  \begin{minipage}{.48\textwidth}
    \includegraphics[width=.9\textwidth]{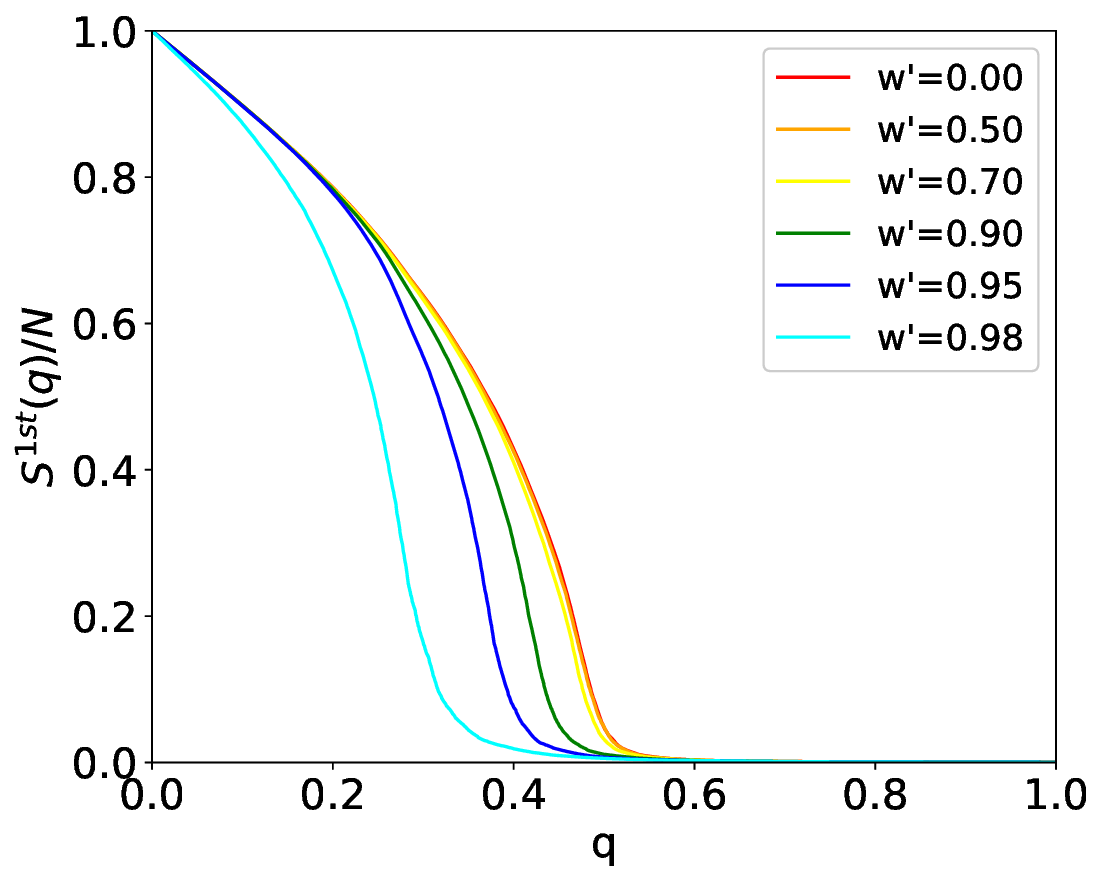}
    \begin{center} (f) $m_{o} = 200$ \end{center}
  \end{minipage}       
%\centering
%\includegraphics[width=.8\textwidth]{resize_figS15.eps}
\caption{Comparison of the areas under the curves 
represented as the robustness against ID attacks
in networks of narrower $P(k)$ at $\nu = -5$ with $m_{o}$ modules.}
\label{fig_ID_nu-5}
\end{figure}

\begin{figure}[htb]
  \begin{minipage}{.48\textwidth}
    \includegraphics[width=.9\textwidth]{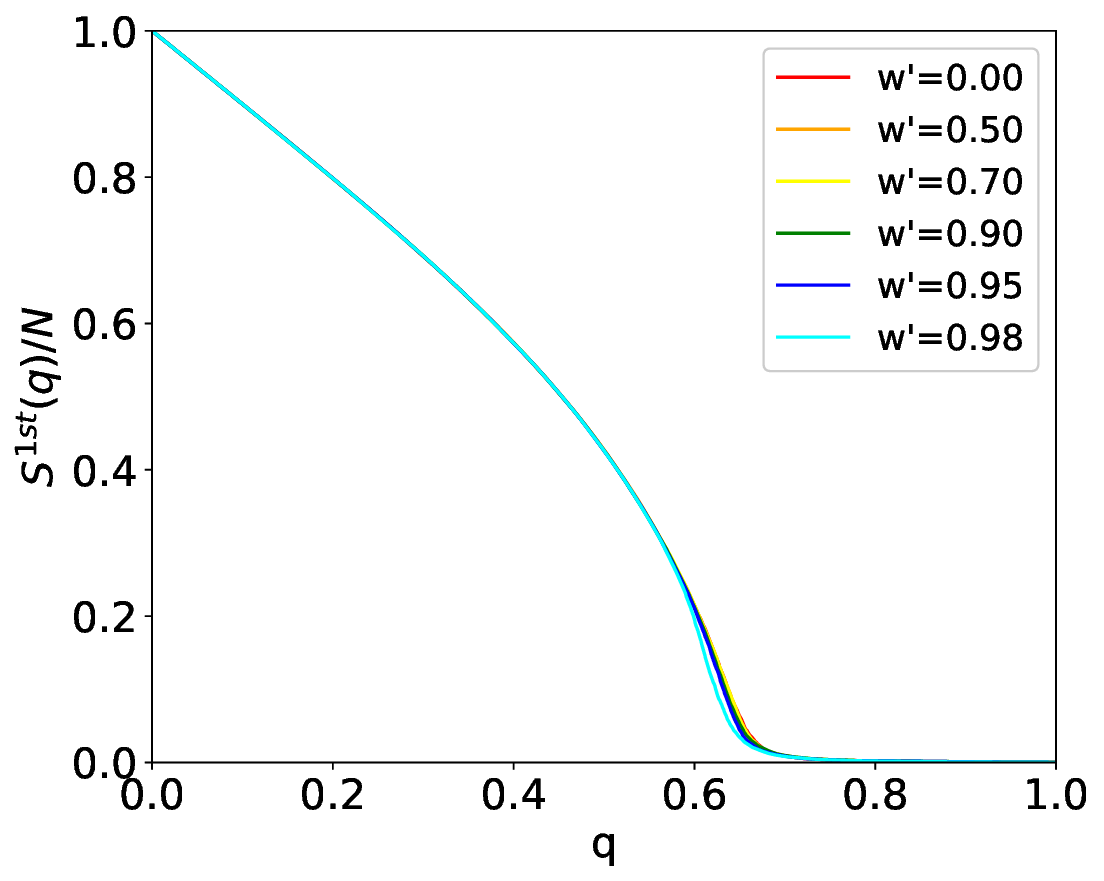}
    \begin{center} (a) $m_{o} = 5$ \end{center}  
  \end{minipage}
  \hfill  
  \begin{minipage}{.48\textwidth}
    \includegraphics[width=.9\textwidth]{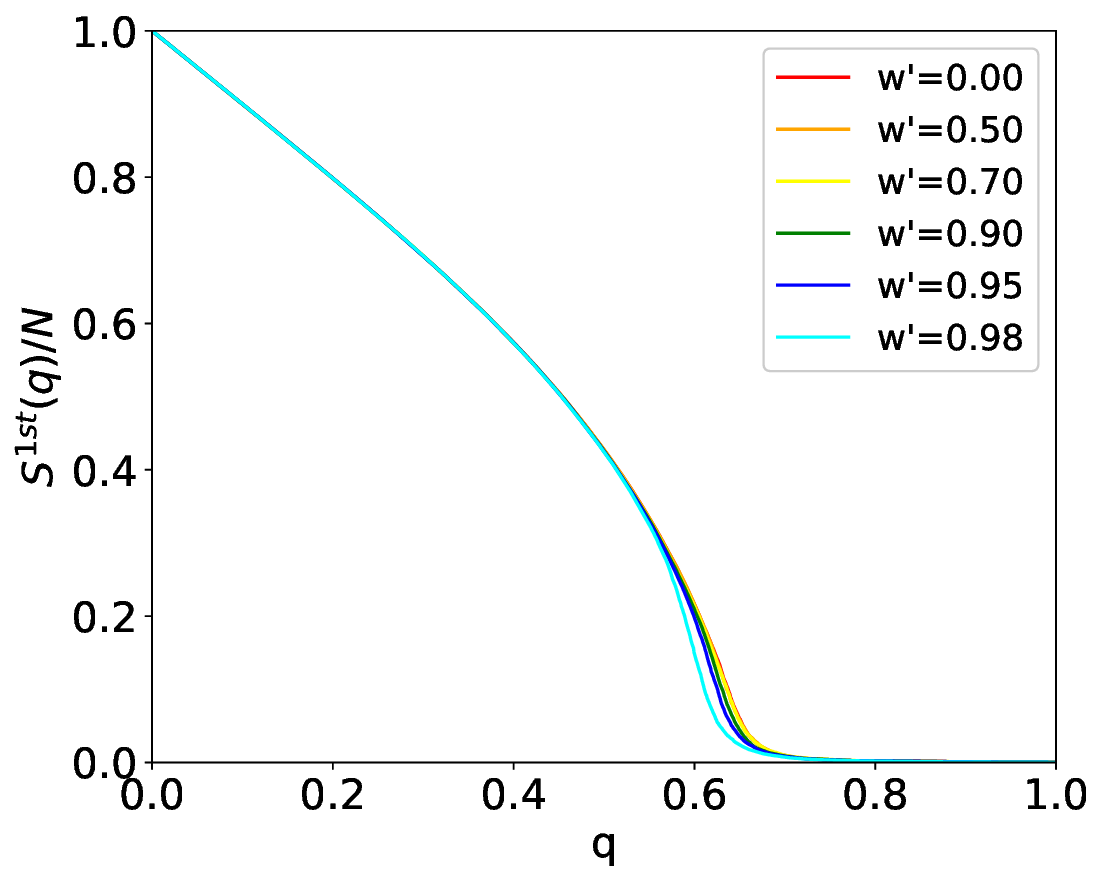}
    \begin{center} (b) $m_{o} = 10$ \end{center}
  \end{minipage}    
  \hfill
  \begin{minipage}{.48\textwidth}
    \includegraphics[width=.9\textwidth]{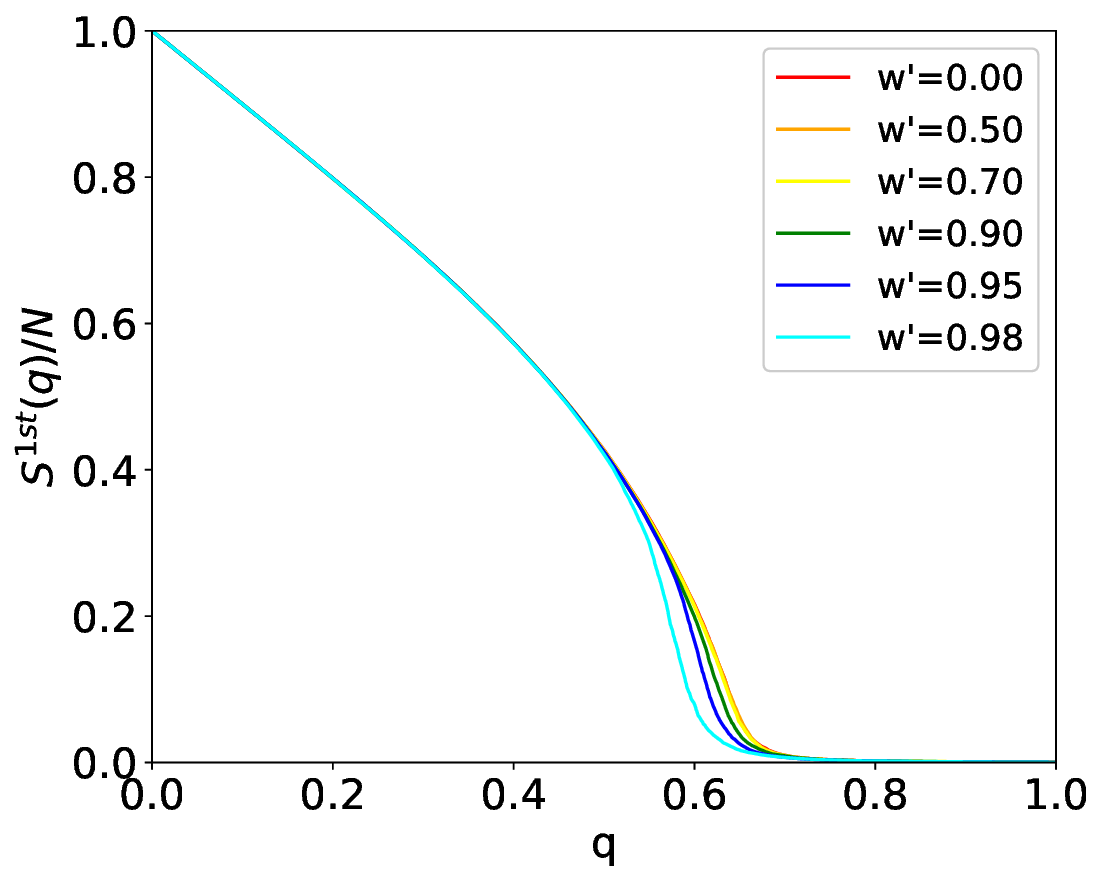}
    \begin{center} (c) $m_{o} = 20$ \end{center}
  \end{minipage}
  \hfill  
  \begin{minipage}{.48\textwidth}
    \includegraphics[width=.9\textwidth]{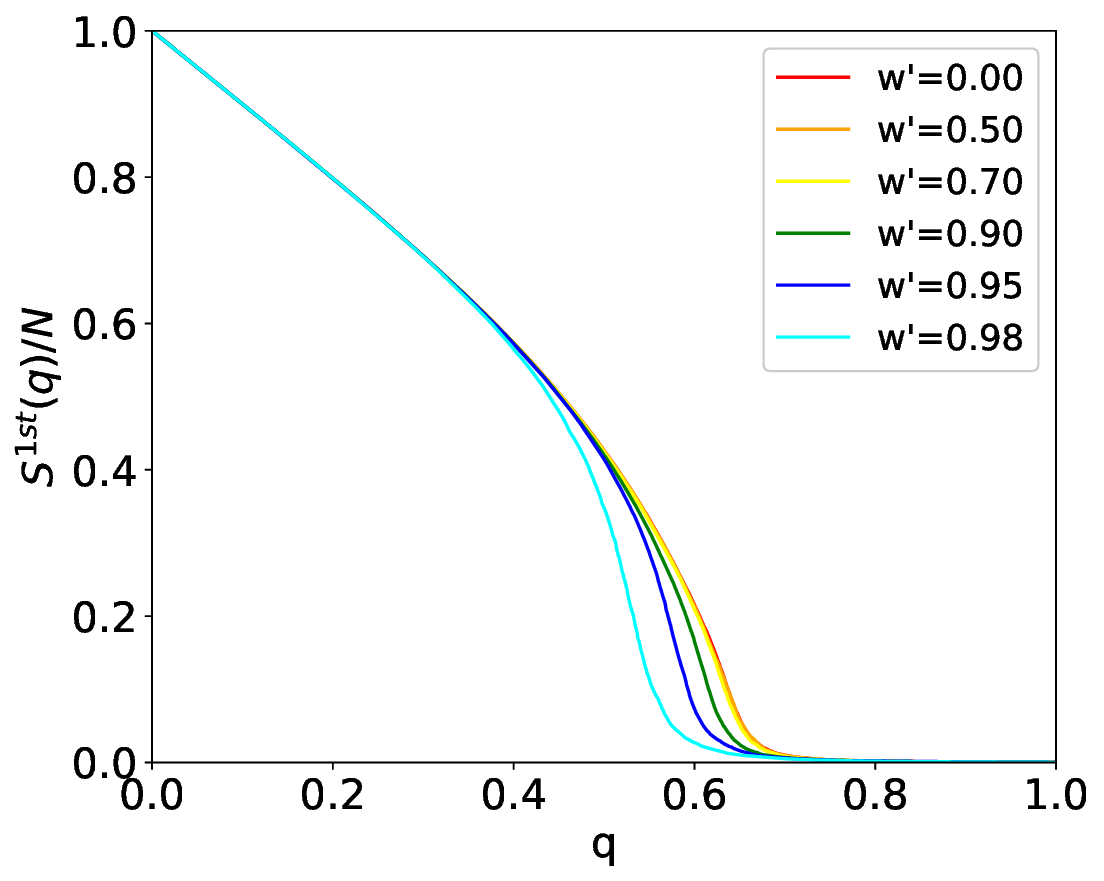}
    \begin{center} (d) $m_{o} = 50$ \end{center}
  \end{minipage}     
  \hfill 
  \begin{minipage}{.48\textwidth}
    \includegraphics[width=.9\textwidth]{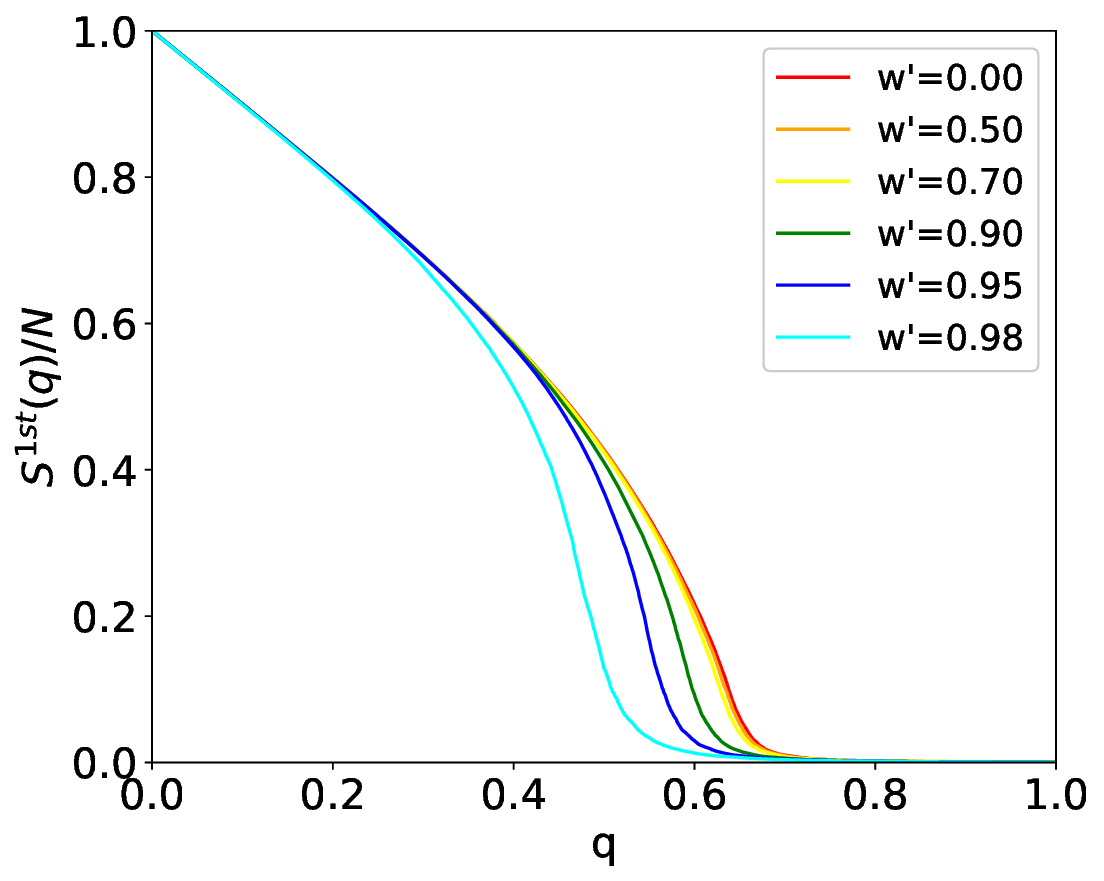}
    \begin{center} (e) $m_{o} = 100$ \end{center}
  \end{minipage}
  \hfill  
  \begin{minipage}{.48\textwidth}
    \includegraphics[width=.9\textwidth]{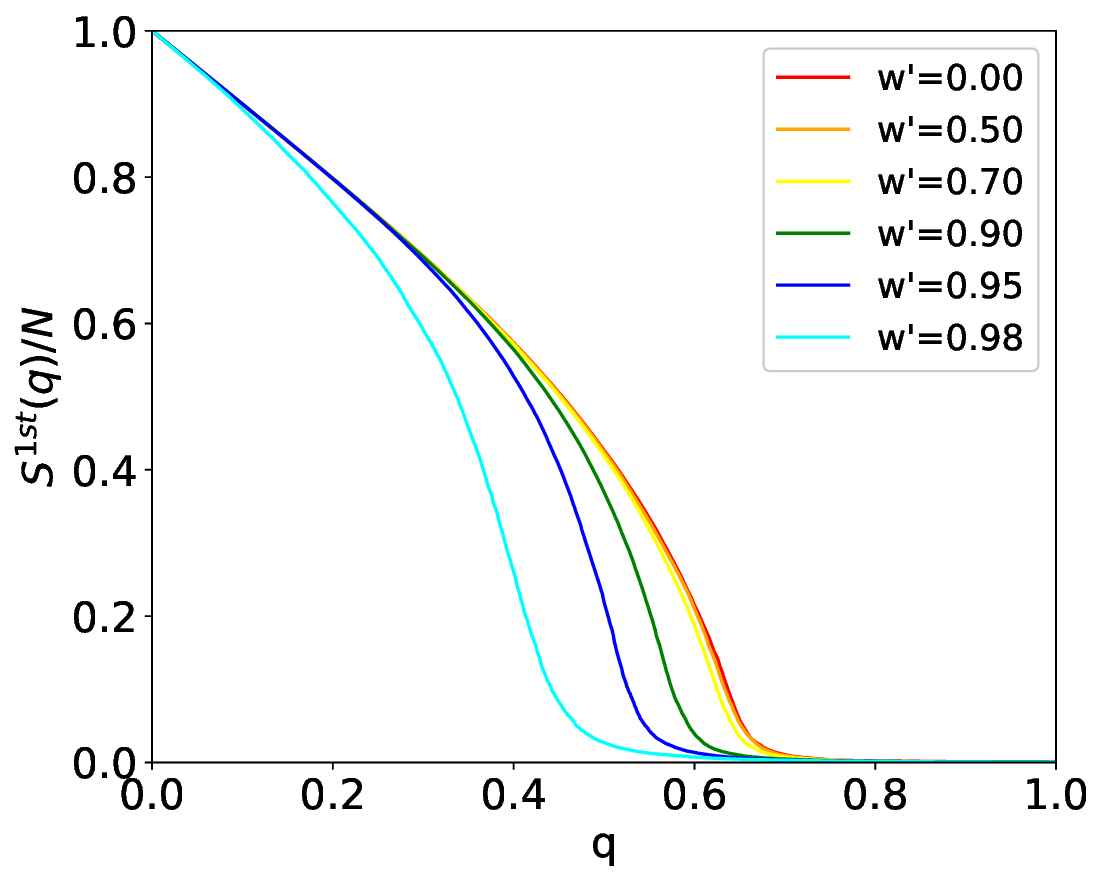}
    \begin{center} (f) $m_{o} = 200$ \end{center}
  \end{minipage}       
%\centering
%\includegraphics[width=.8\textwidth]{resize_figS16.eps}
\caption{Comparison of the areas under the curves 
represented as the robustness against ID attacks
in nearly regular networks at $\nu = -100$ with $m_{o}$ modules.}
\label{fig_ID_nu-100}
\end{figure}

\begin{figure}[htb]
  \begin{minipage}{.48\textwidth}
    \includegraphics[width=.9\textwidth]{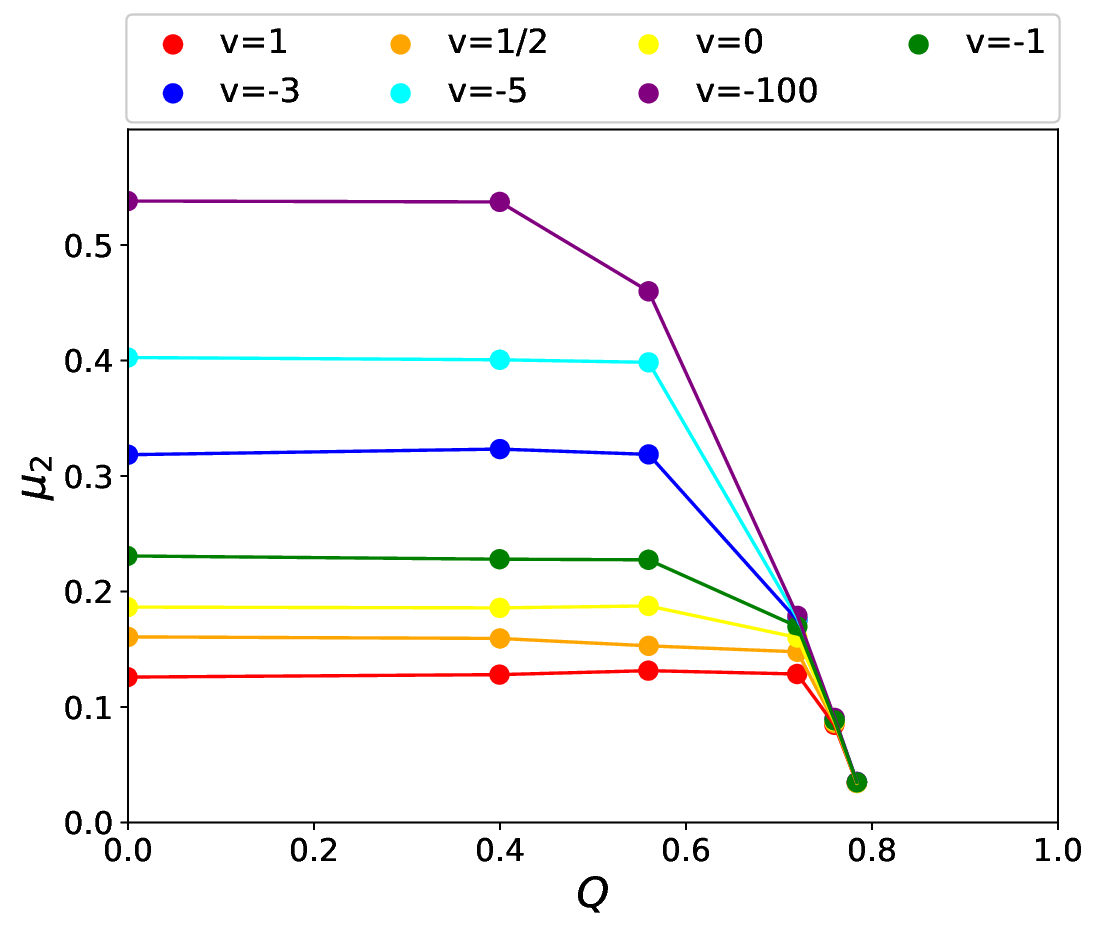}
    \begin{center} (a) $m_{o} = 5$ \end{center}  
  \end{minipage}
  \hfill  
  \begin{minipage}{.48\textwidth}
    \includegraphics[width=.9\textwidth]{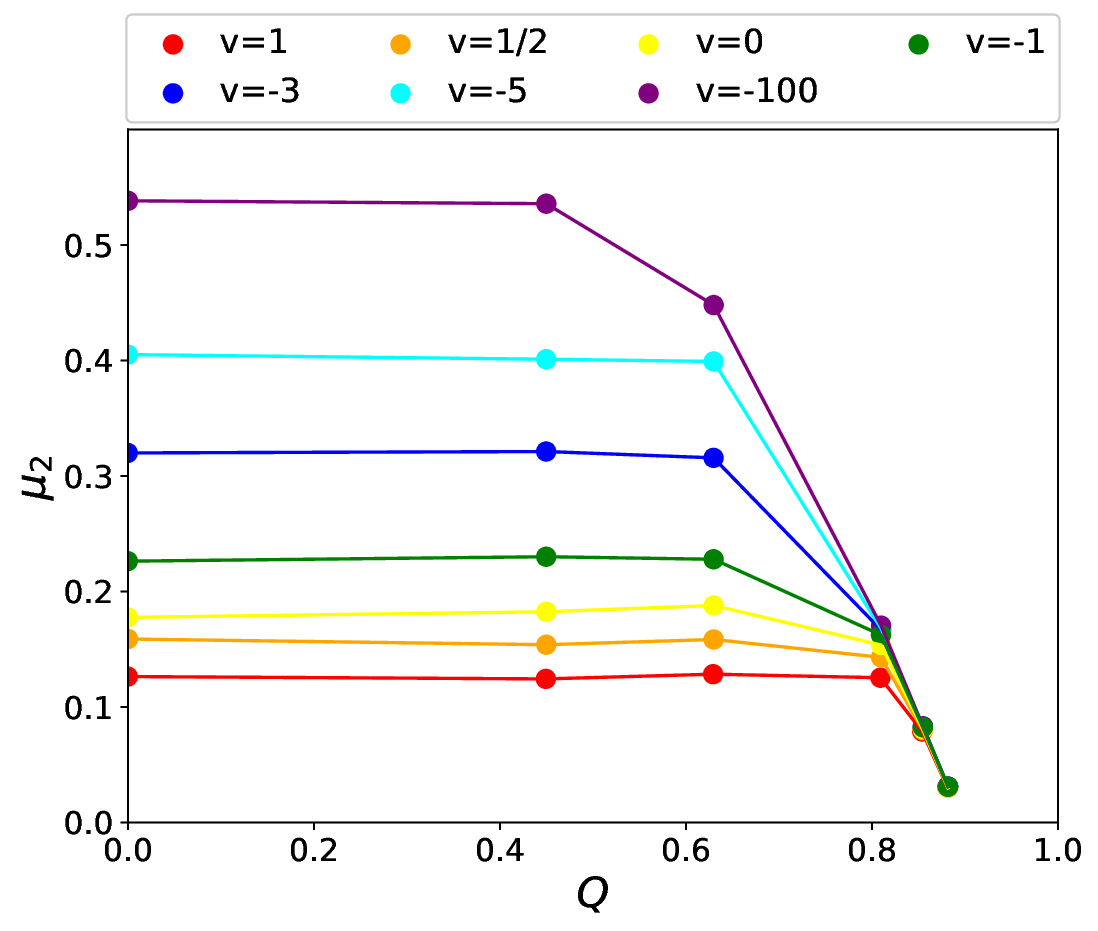}
    \begin{center} (b) $m_{o} = 10$ \end{center}
  \end{minipage}    
  \hfill
  \begin{minipage}{.48\textwidth}
    \includegraphics[width=.9\textwidth]{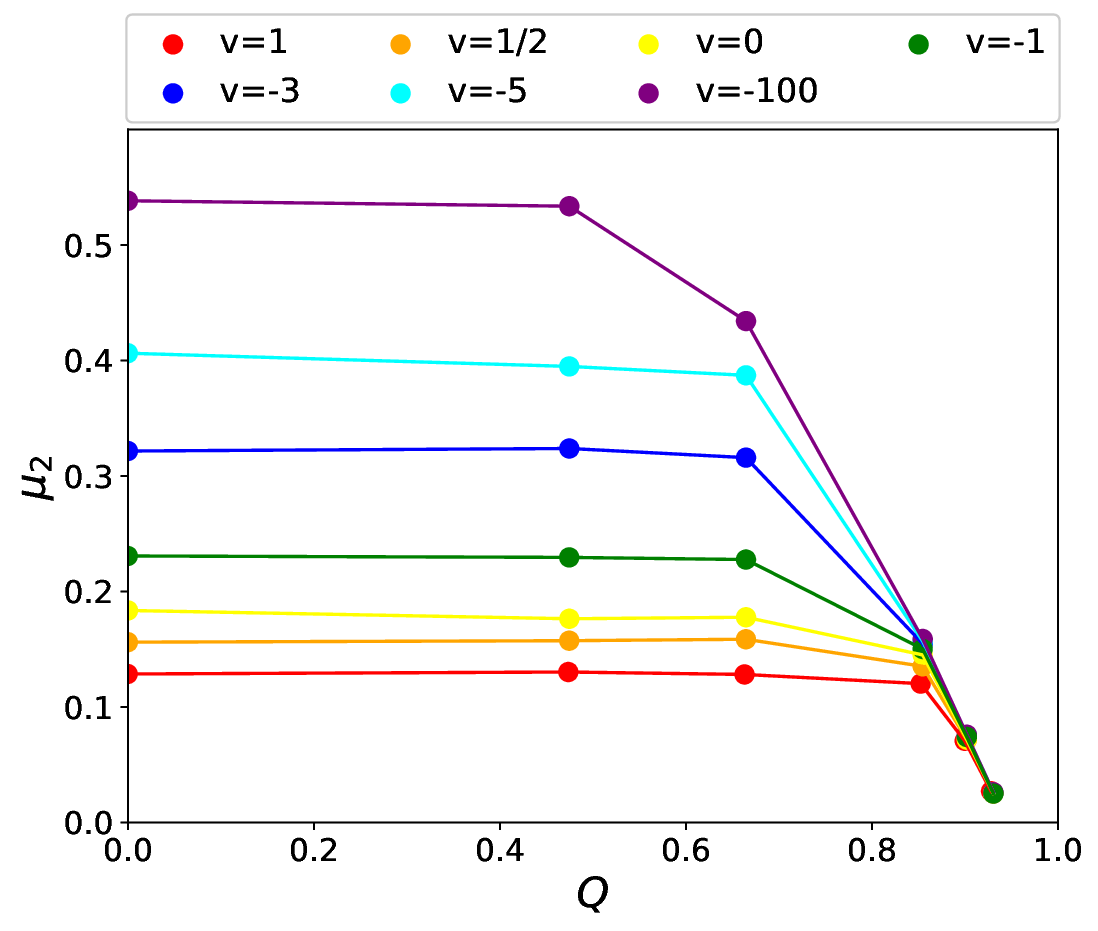}
    \begin{center} (c) $m_{o} = 20$ \end{center}
  \end{minipage}
  \hfill  
  \begin{minipage}{.48\textwidth}
    \includegraphics[width=.9\textwidth]{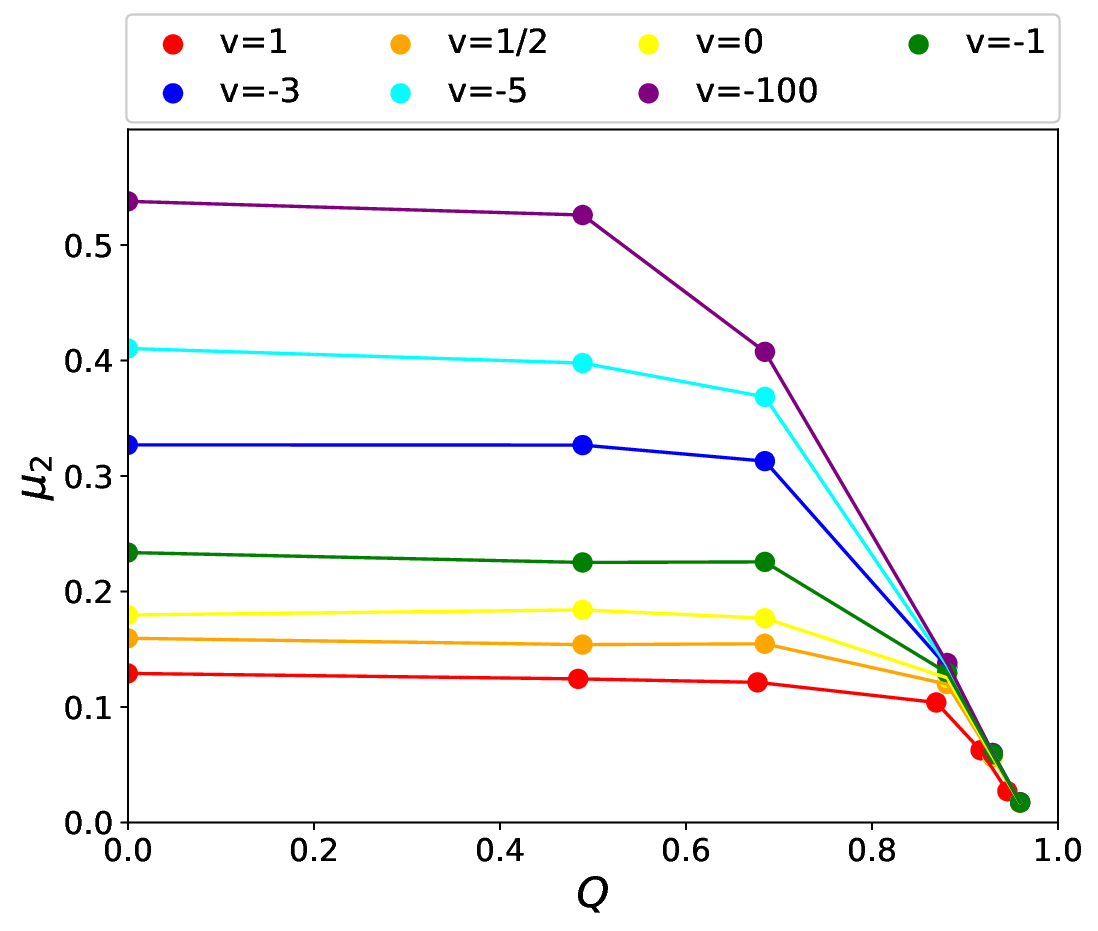}
    \begin{center} (d) $m_{o} = 50$ \end{center}
  \end{minipage}     
  \hfill 
  \begin{minipage}{.48\textwidth}
    \includegraphics[width=.9\textwidth]{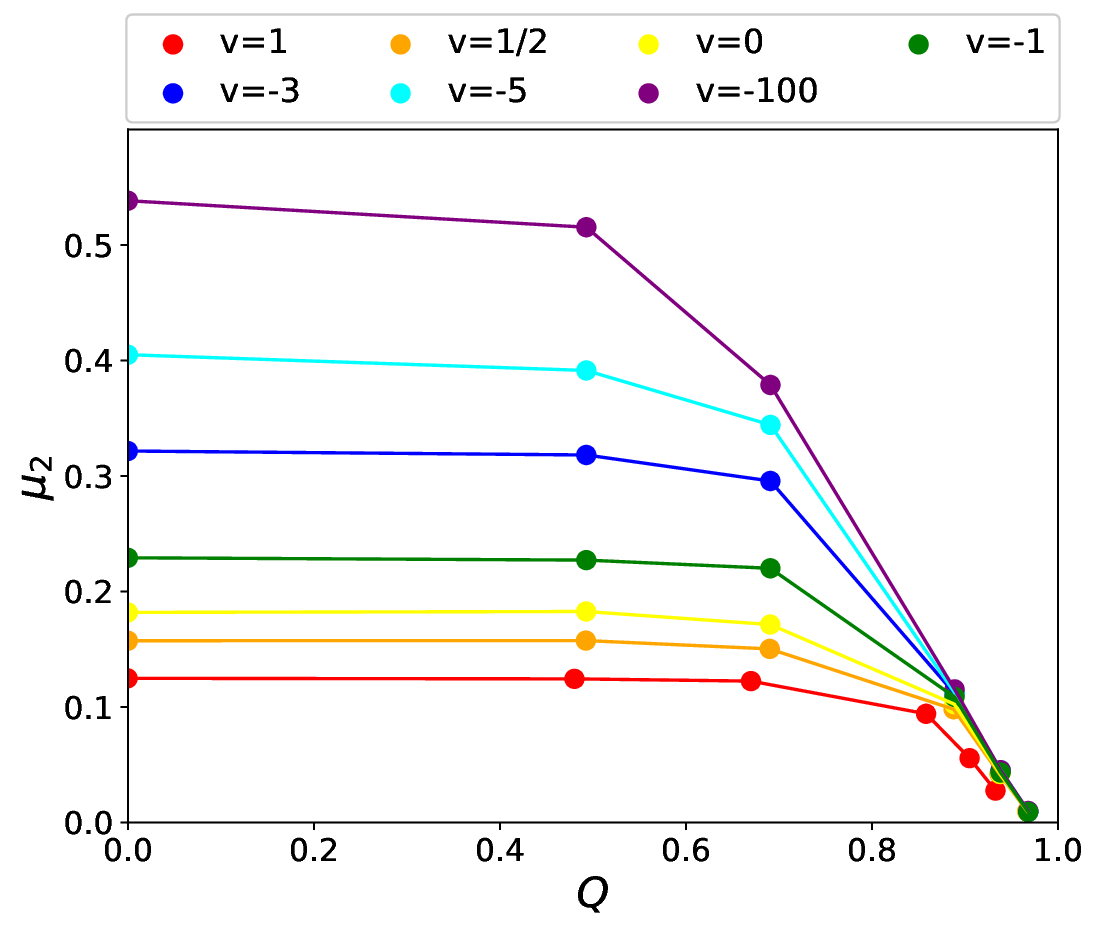}
    \begin{center} (e) $m_{o} = 100$ \end{center}
  \end{minipage}
  \hfill  
  \begin{minipage}{.48\textwidth}
    \includegraphics[width=.9\textwidth]{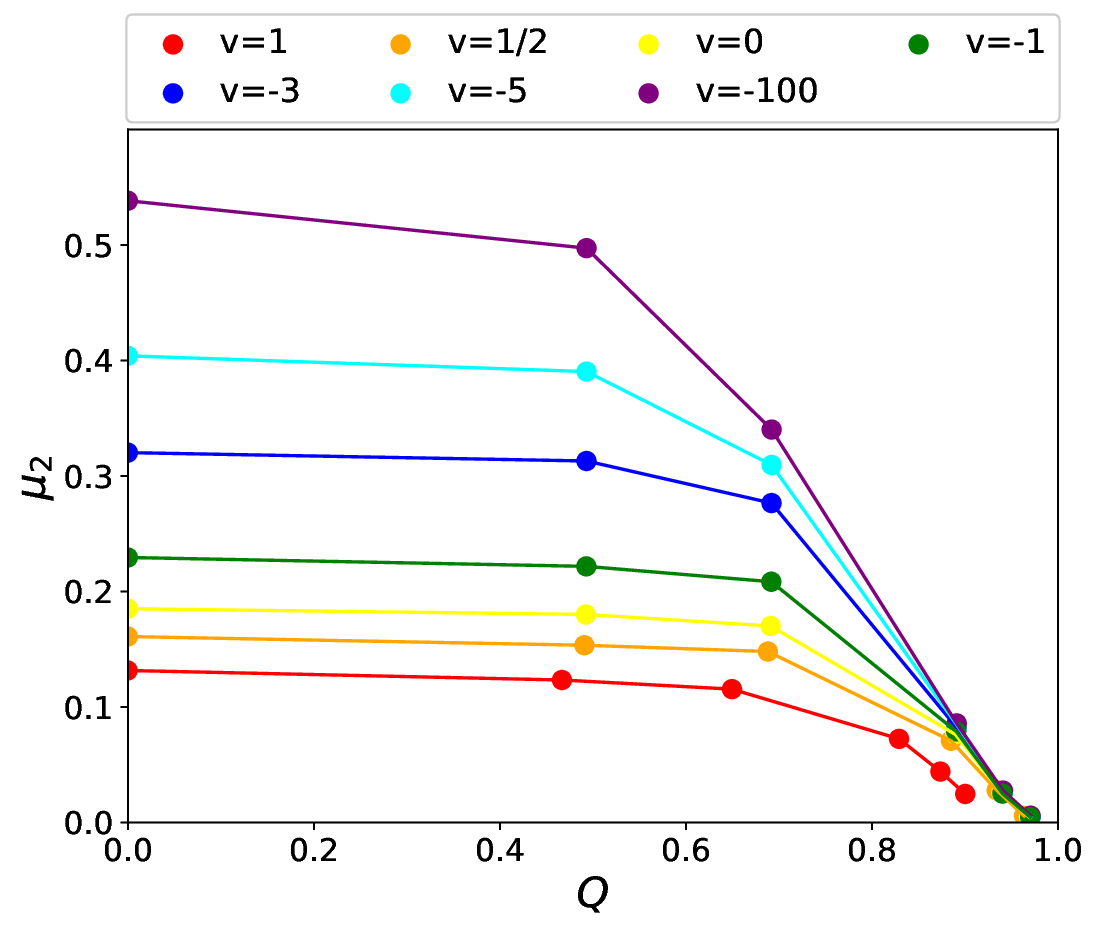}
    \begin{center} (f) $m_{o} = 200$ \end{center}
  \end{minipage}       
%\centering
%\includegraphics[width=.8\textwidth]{resize_figS17.eps}
\caption{Rapid decreasing of 
the eigenvalue $\mu_{2}$ of Laplacian matrix
in modular networks with larger $Q$.}
\label{fig_Q-mu2}
\end{figure}

\begin{figure}[htb]
  \begin{minipage}{.48\textwidth}
    \includegraphics[width=.9\textwidth]{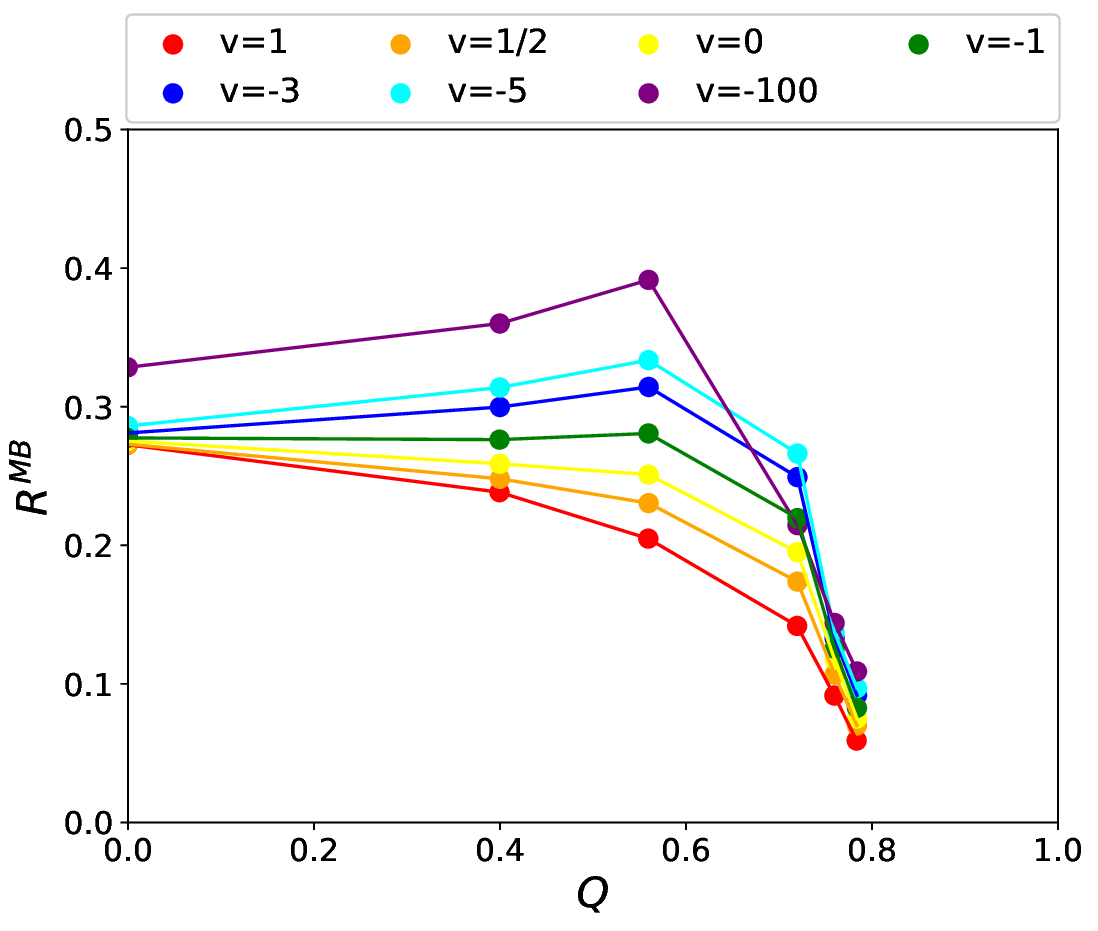}
    \begin{center} (a) $m_{o} = 5$ \end{center}  
  \end{minipage}
  \hfill  
  \begin{minipage}{.48\textwidth}
    \includegraphics[width=.9\textwidth]{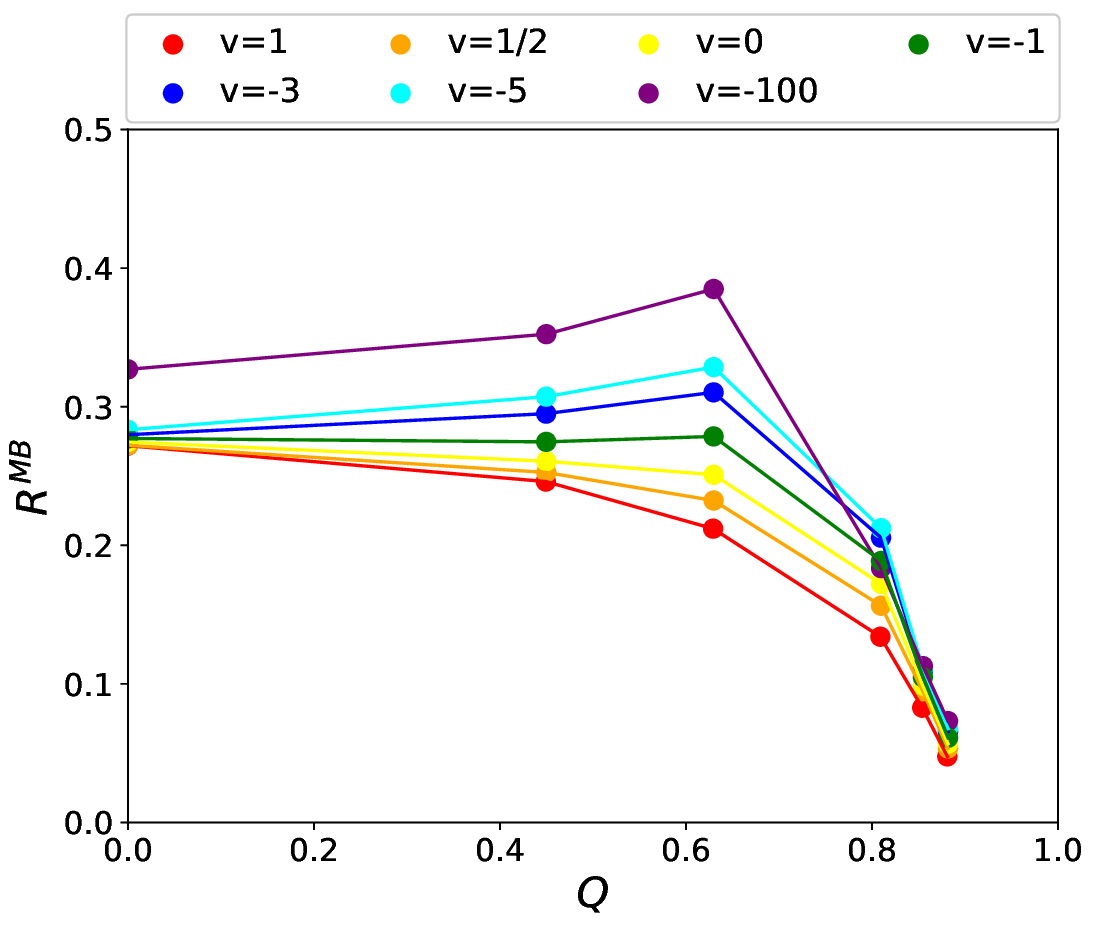}
    \begin{center} (b) $m_{o} = 10$ \end{center}
  \end{minipage}    
  \hfill
  \begin{minipage}{.48\textwidth}
    \includegraphics[width=.9\textwidth]{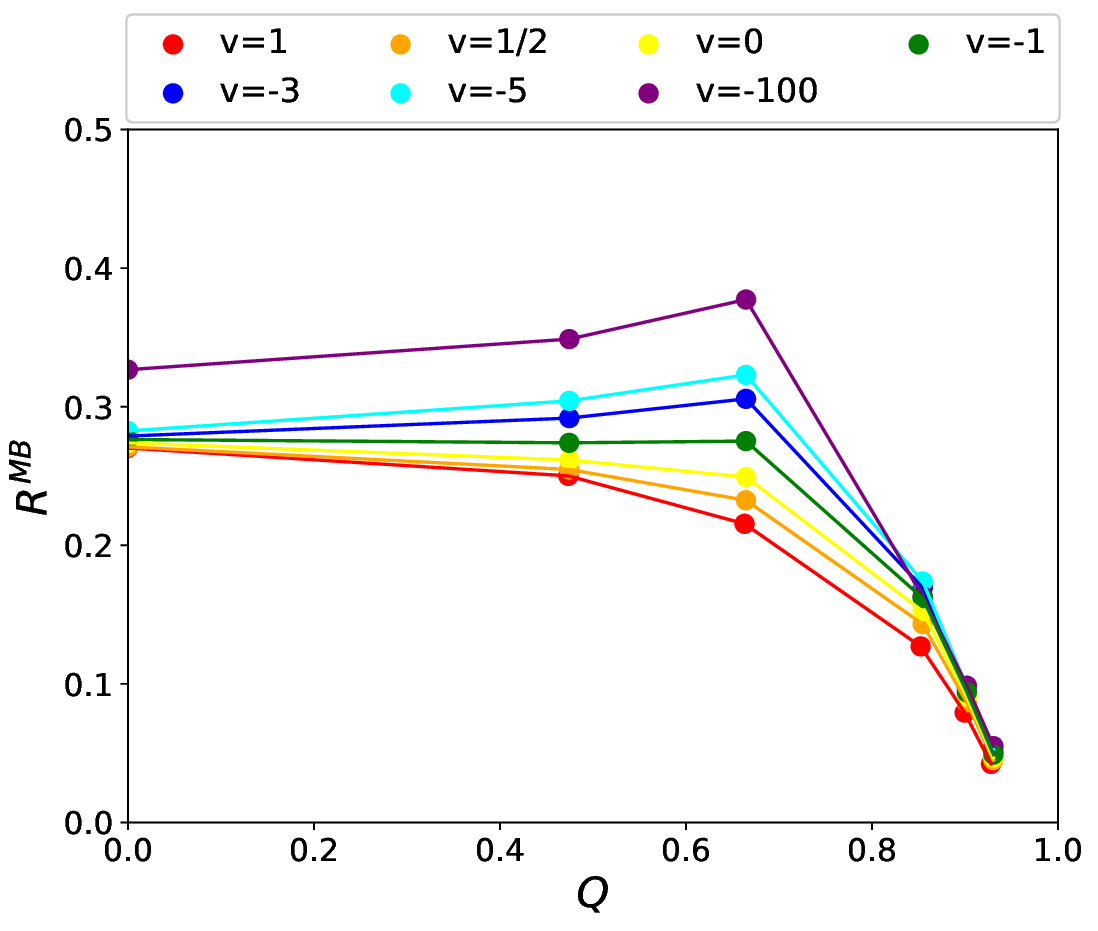}
    \begin{center} (c) $m_{o} = 20$ \end{center}
  \end{minipage}
  \hfill  
  \begin{minipage}{.48\textwidth}
    \includegraphics[width=.9\textwidth]{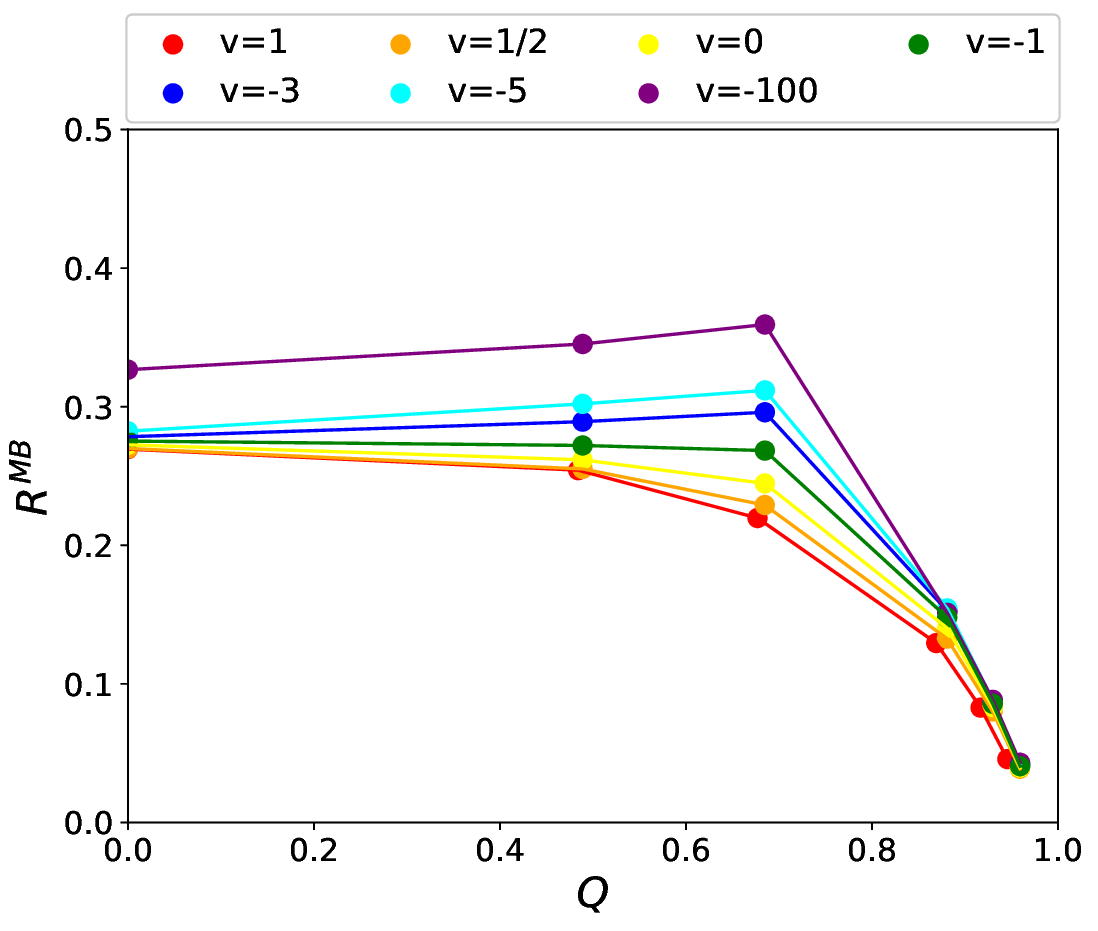}
    \begin{center} (d) $m_{o} = 50$ \end{center}
  \end{minipage}     
  \hfill 
  \begin{minipage}{.48\textwidth}
    \includegraphics[width=.9\textwidth]{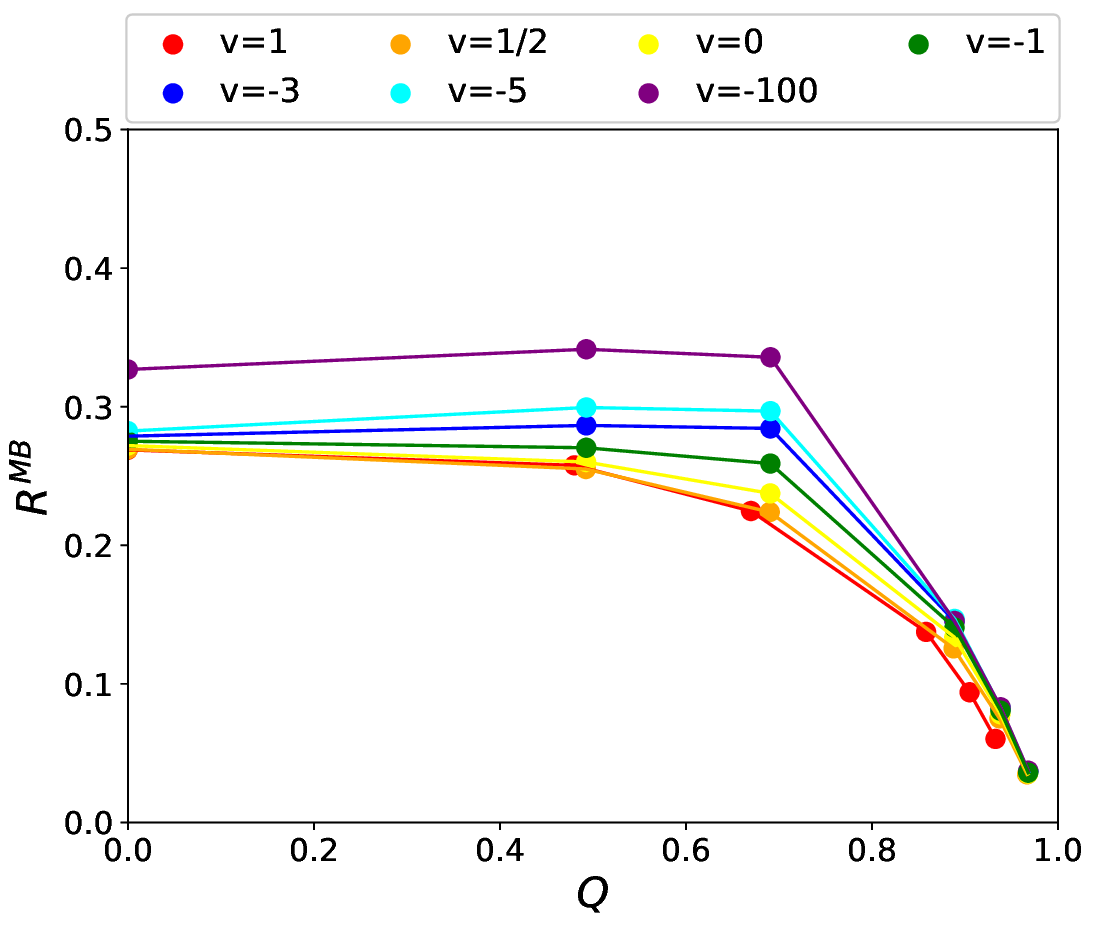}
    \begin{center} (e) $m_{o} = 100$ \end{center}
  \end{minipage}
  \hfill  
  \begin{minipage}{.48\textwidth}
    \includegraphics[width=.9\textwidth]{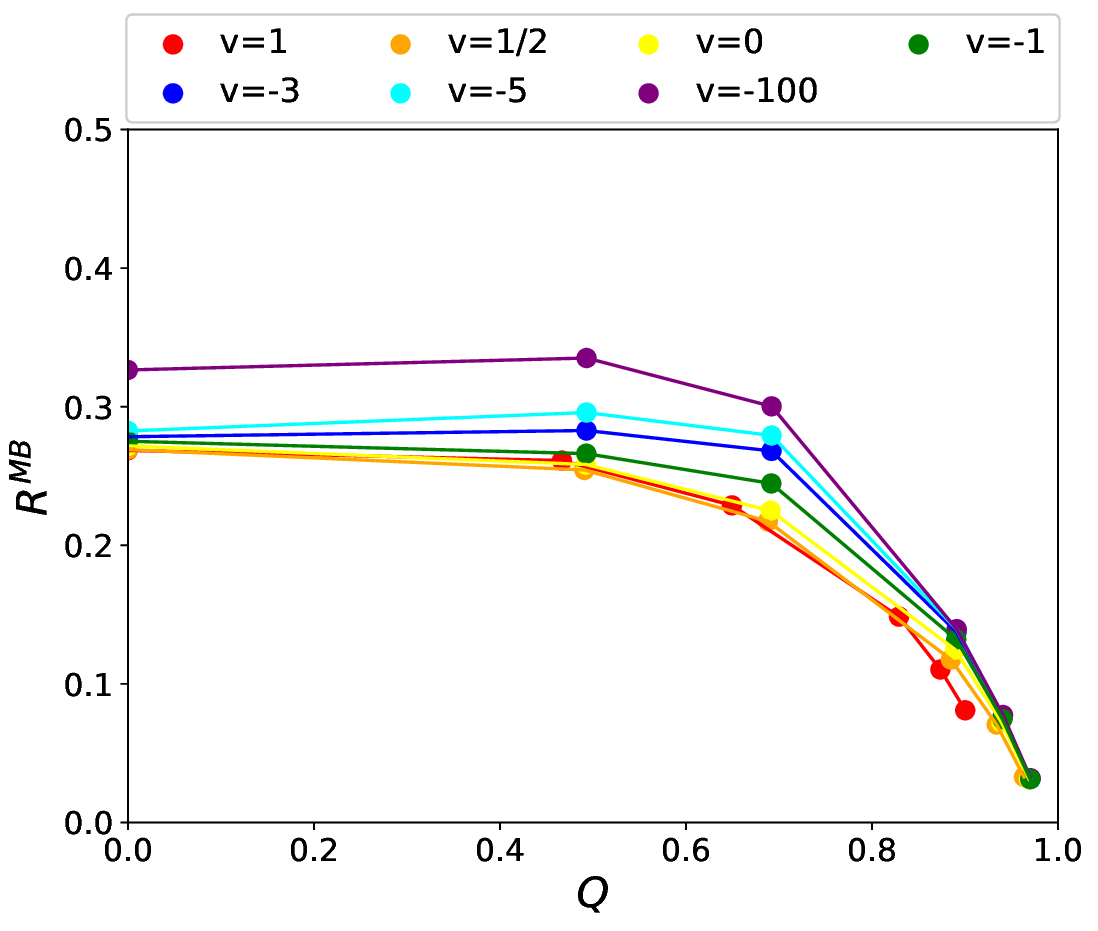}
    \begin{center} (f) $m_{o} = 200$ \end{center}
  \end{minipage}       
%\centering
%\includegraphics[width=.8\textwidth]{resize_figS18.eps}
\caption{Rapid decreasing of
the robustness index $R$ against MB attacks  
in modular networks with larger $Q$.}
\label{fig_Q-R^MB}
\end{figure}

\begin{figure}[htb]
  \begin{minipage}{.48\textwidth}
    \includegraphics[width=.9\textwidth]{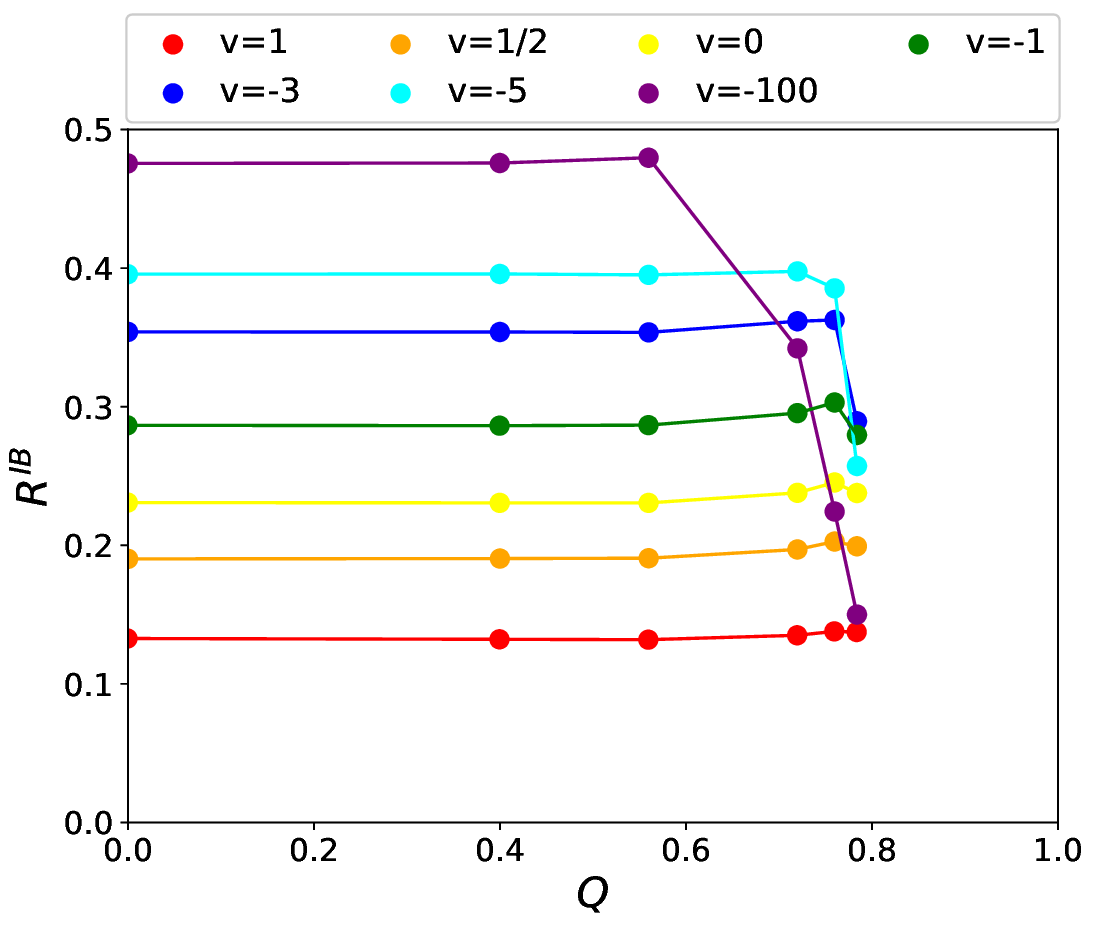}
    \begin{center} (a) $m_{o} = 5$ \end{center}  
  \end{minipage}
  \hfill  
  \begin{minipage}{.48\textwidth}
    \includegraphics[width=.9\textwidth]{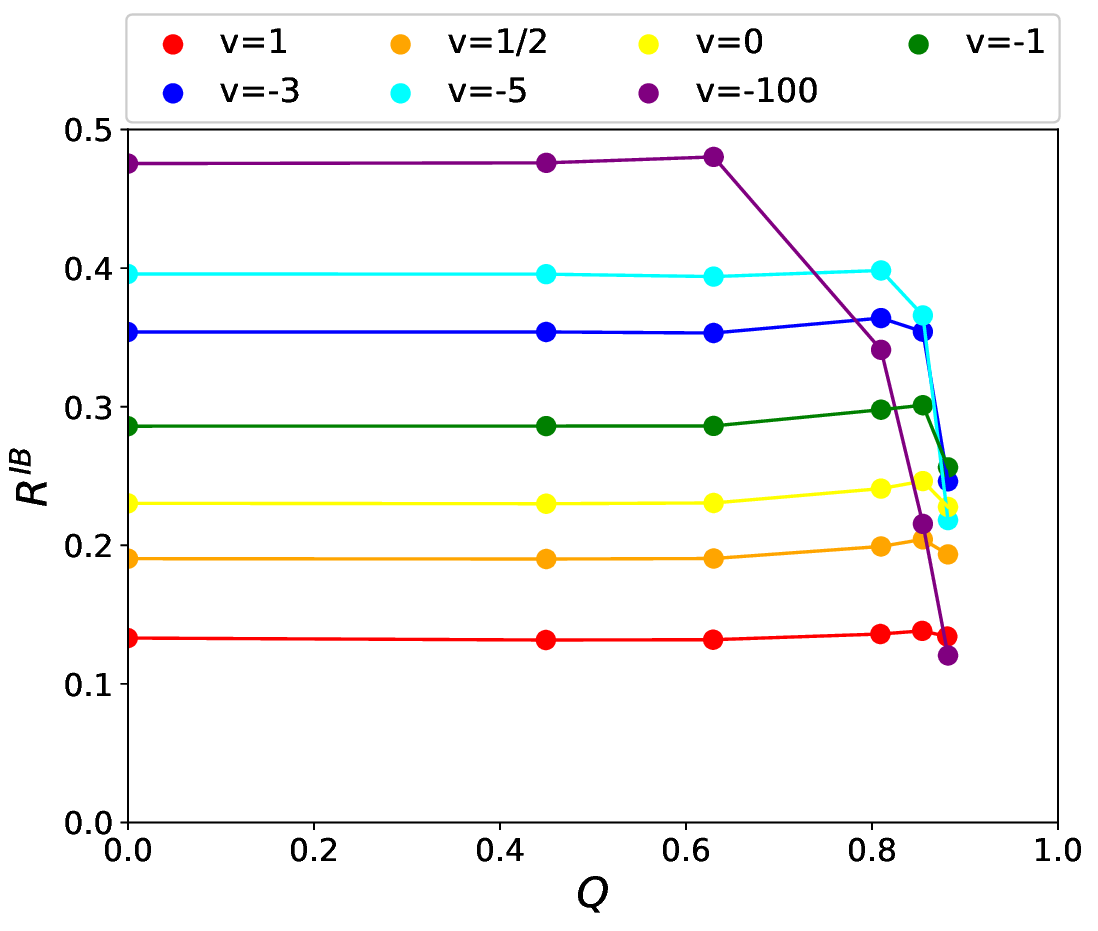}
    \begin{center} (b) $m_{o} = 10$ \end{center}
  \end{minipage}    
  \hfill
  \begin{minipage}{.48\textwidth}
    \includegraphics[width=.9\textwidth]{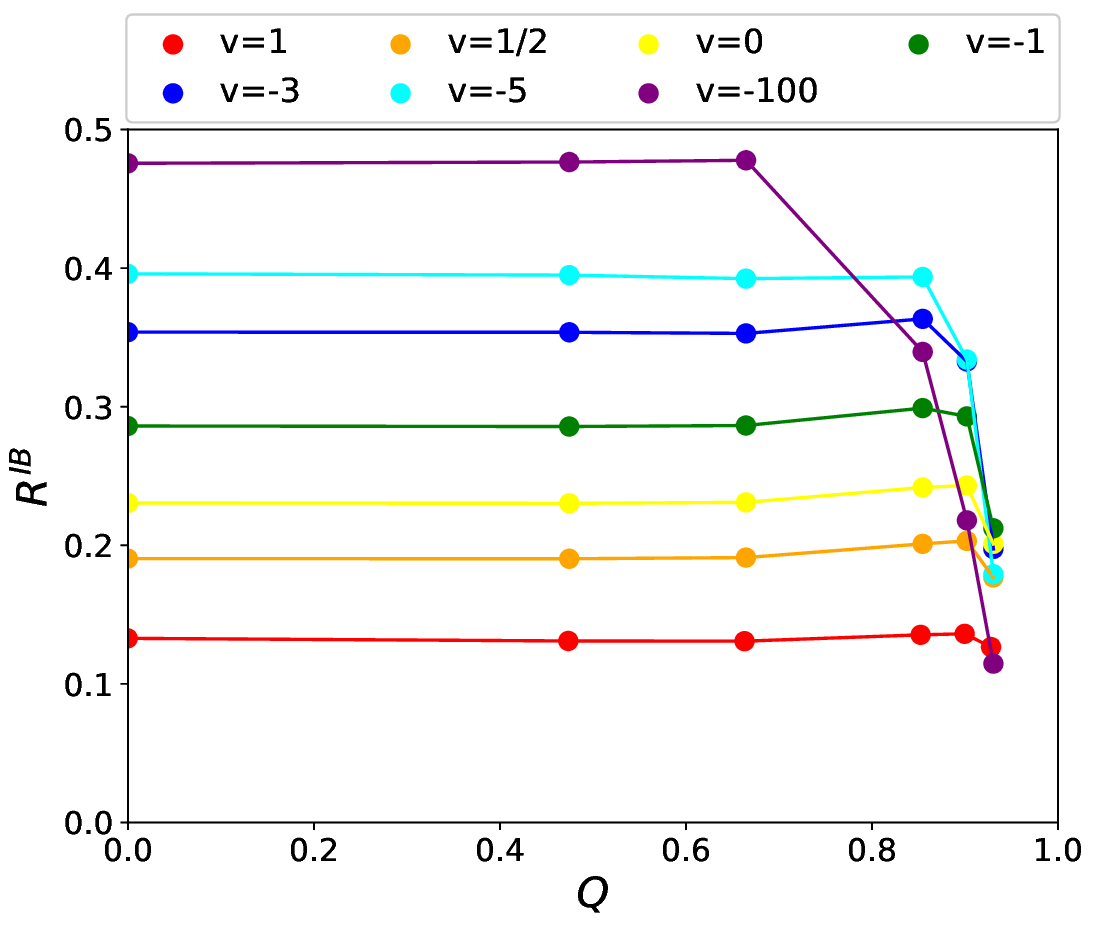}
    \begin{center} (c) $m_{o} = 20$ \end{center}
  \end{minipage}
  \hfill  
  \begin{minipage}{.48\textwidth}
    \includegraphics[width=.9\textwidth]{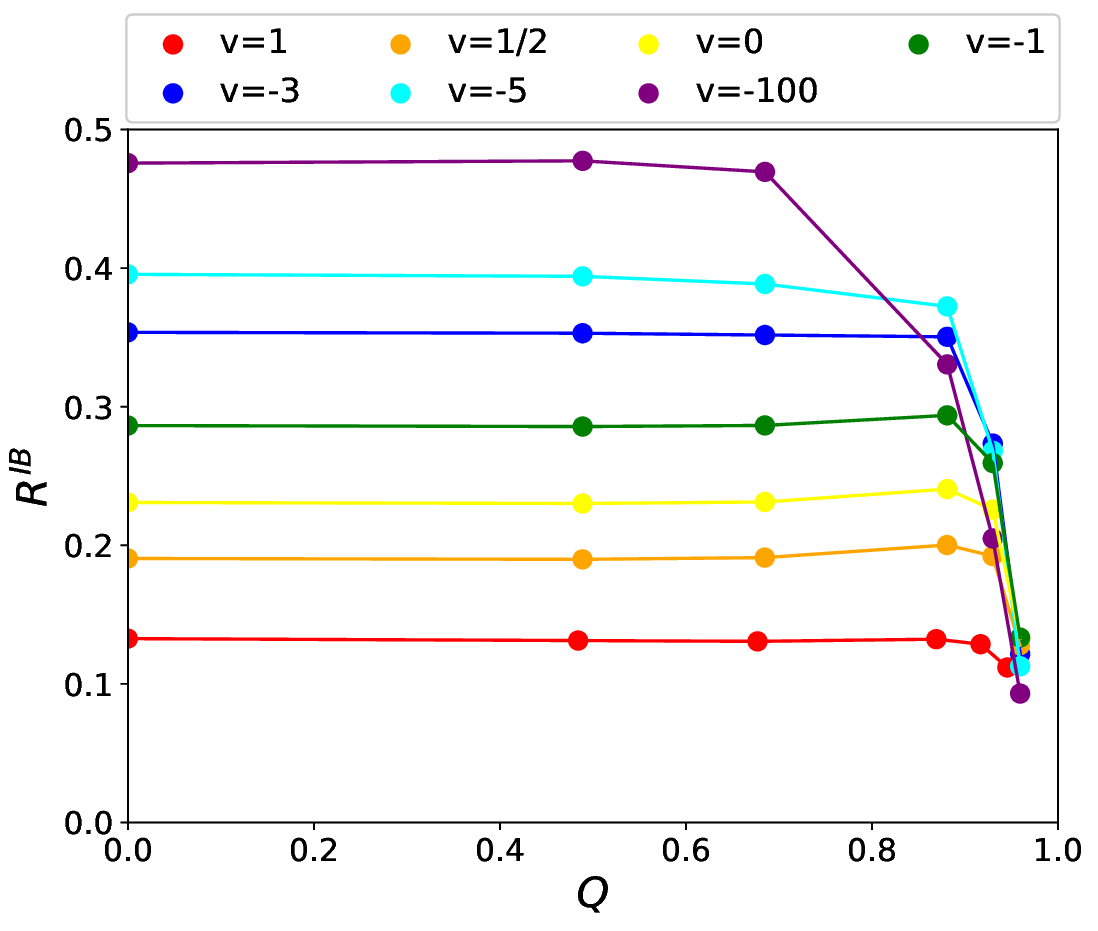}
    \begin{center} (d) $m_{o} = 50$ \end{center}
  \end{minipage}     
  \hfill 
  \begin{minipage}{.48\textwidth}
    \includegraphics[width=.9\textwidth]{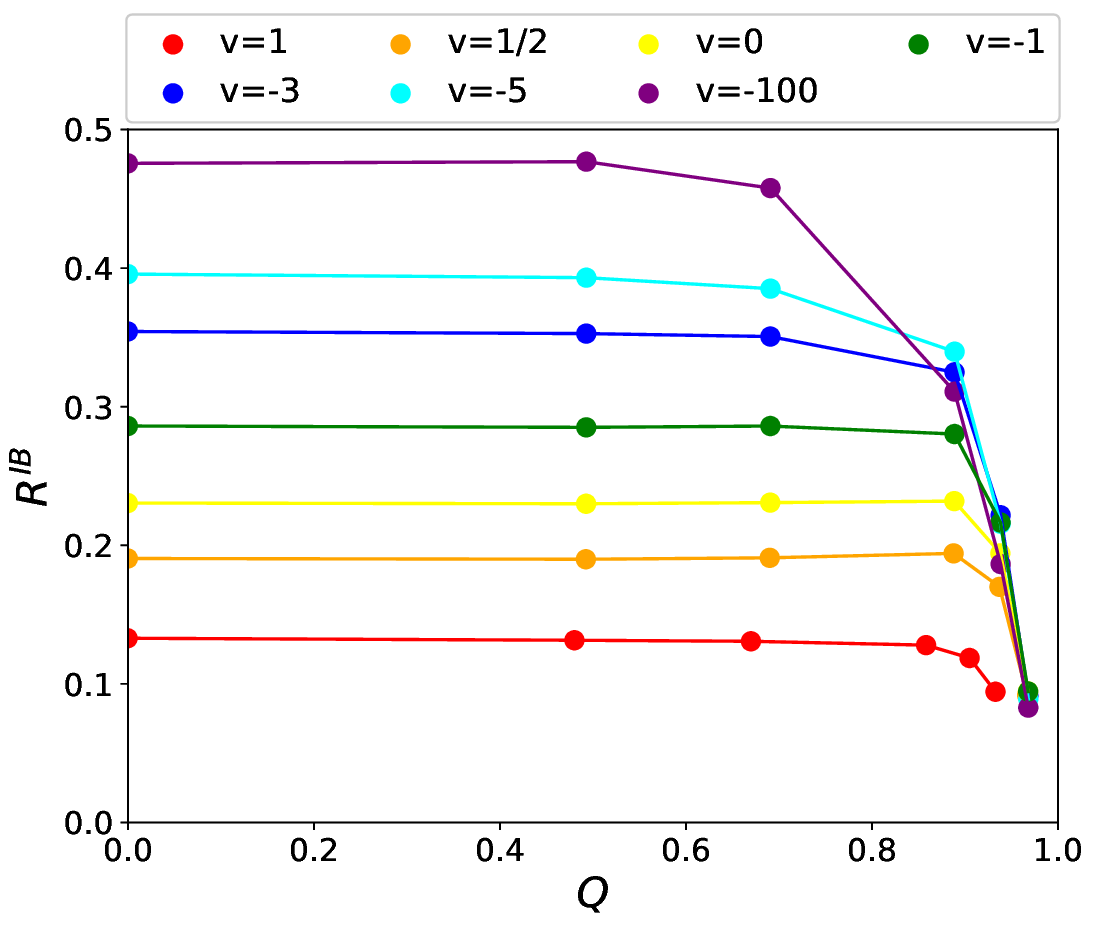}
    \begin{center} (e) $m_{o} = 100$ \end{center}
  \end{minipage}
  \hfill  
  \begin{minipage}{.48\textwidth}
    \includegraphics[width=.9\textwidth]{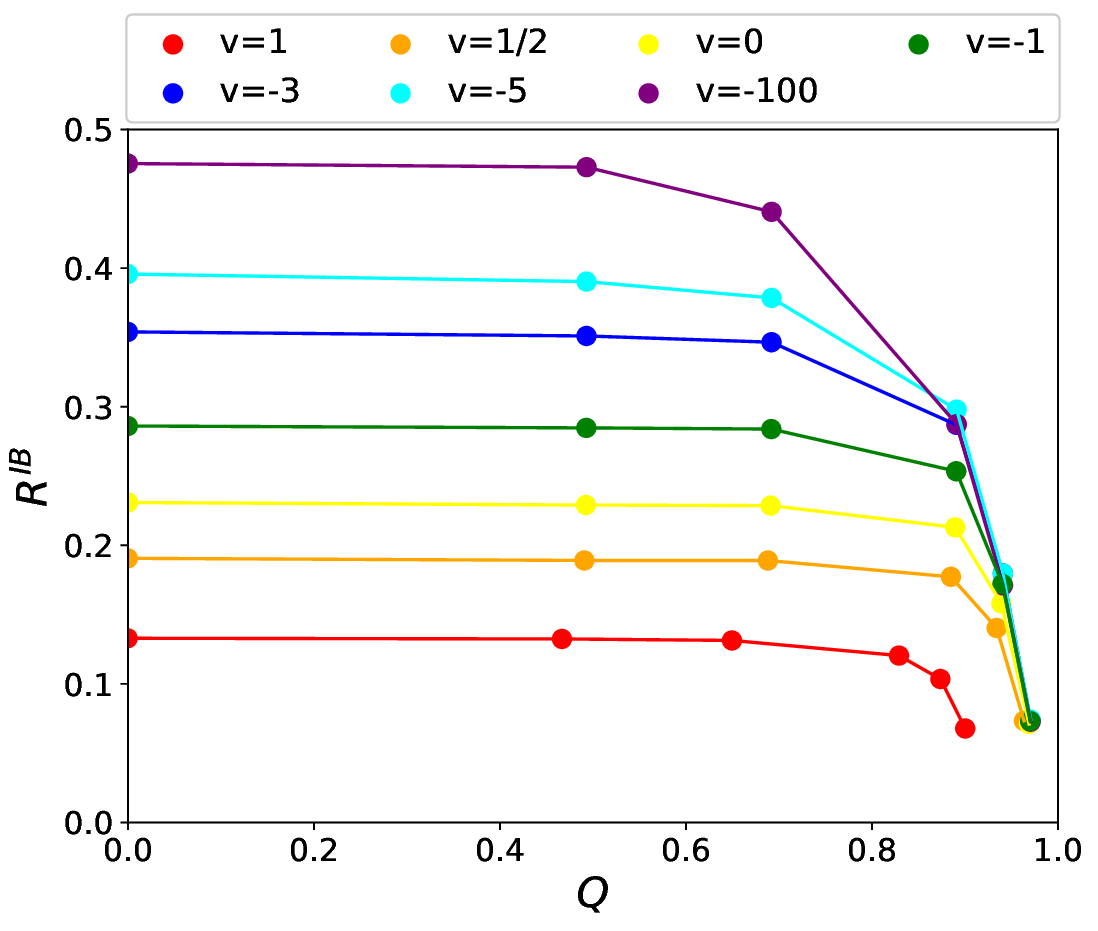}
    \begin{center} (f) $m_{o} = 200$ \end{center}
  \end{minipage}       
%\centering
%\includegraphics[width=.8\textwidth]{resize_figS19.eps}
\caption{Rapid decreasing of
the robustness index $R$ against IB attacks  
in modular networks with larger $Q$.}
\label{fig_Q-R^IB}
\end{figure}

\begin{figure}[htb]
  \begin{minipage}{.48\textwidth}
    \includegraphics[width=.9\textwidth]{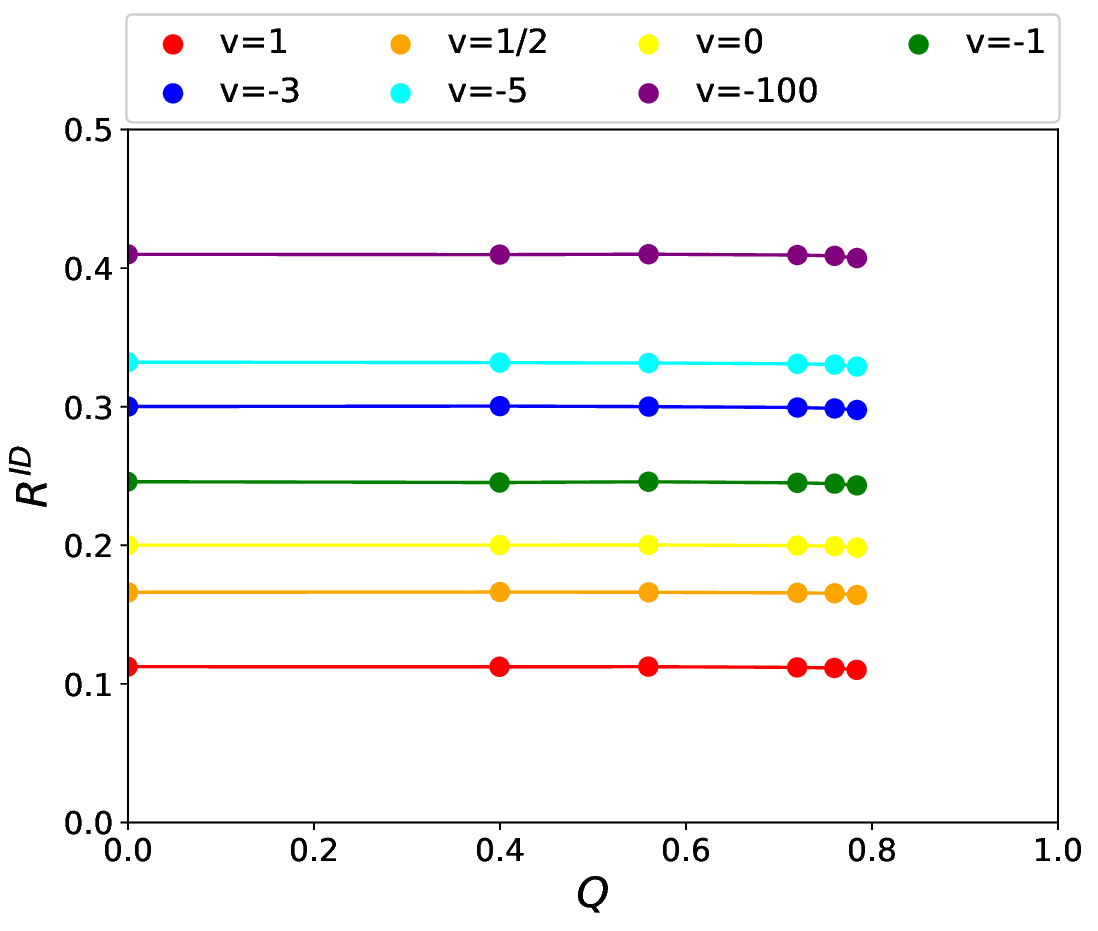}
    \begin{center} (a) $m_{o} = 5$ \end{center}  
  \end{minipage}
  \hfill  
  \begin{minipage}{.48\textwidth}
    \includegraphics[width=.9\textwidth]{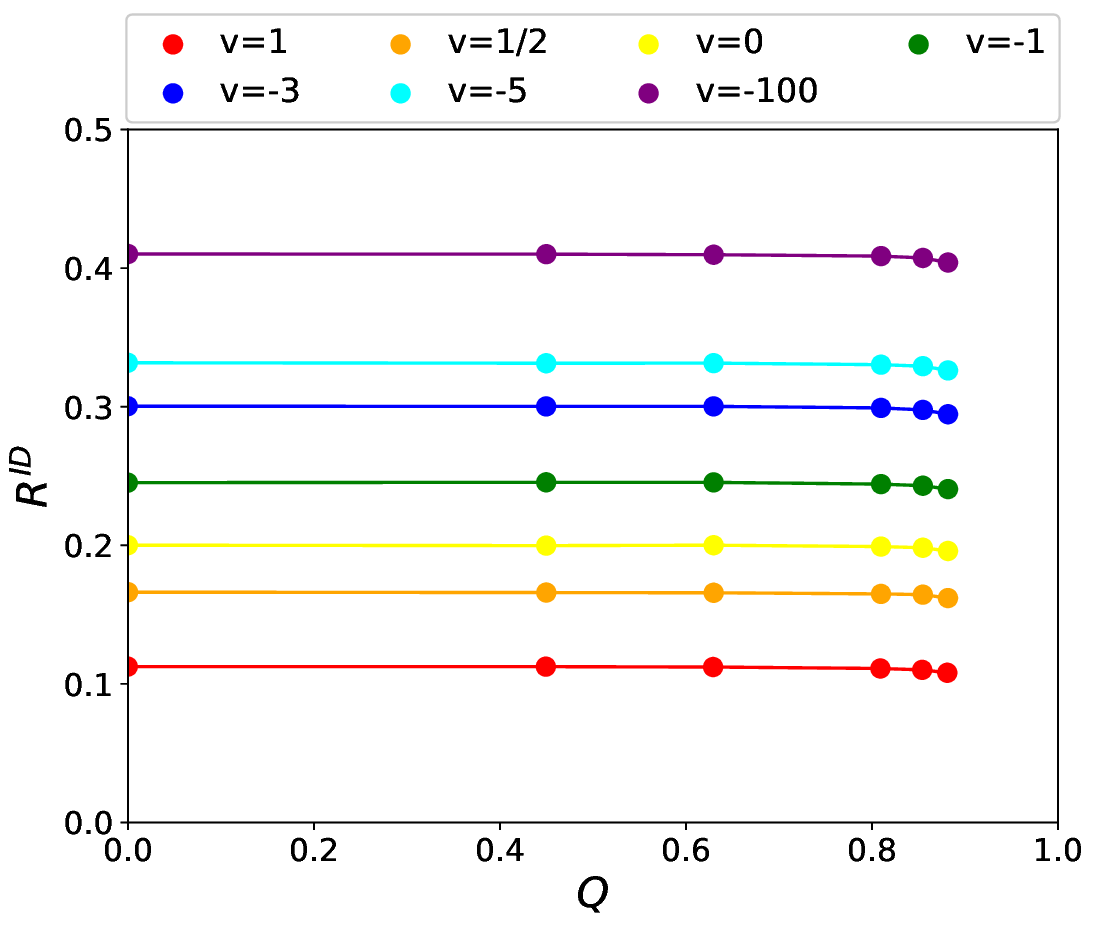}
    \begin{center} (b) $m_{o} = 10$ \end{center}
  \end{minipage}    
  \hfill
  \begin{minipage}{.48\textwidth}
    \includegraphics[width=.9\textwidth]{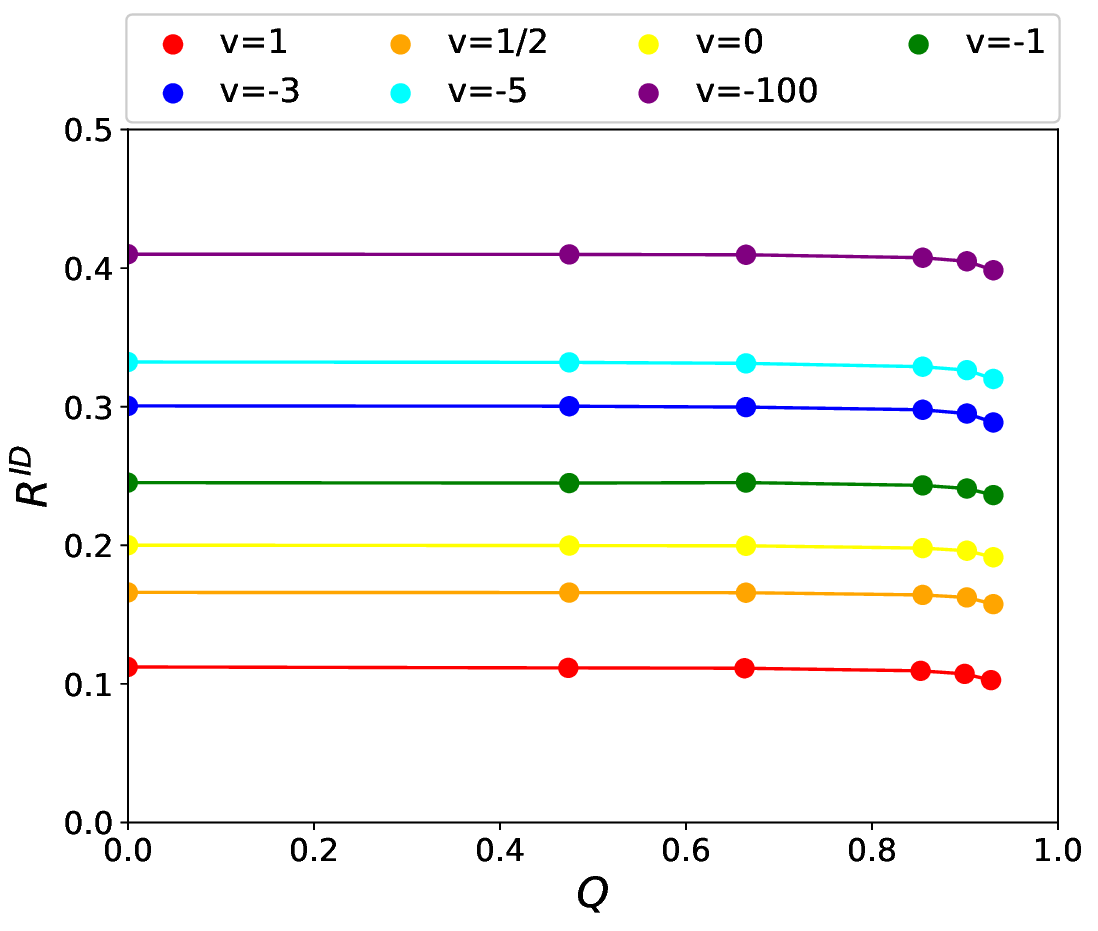}
    \begin{center} (c) $m_{o} = 20$ \end{center}
  \end{minipage}
  \hfill  
  \begin{minipage}{.48\textwidth}
    \includegraphics[width=.9\textwidth]{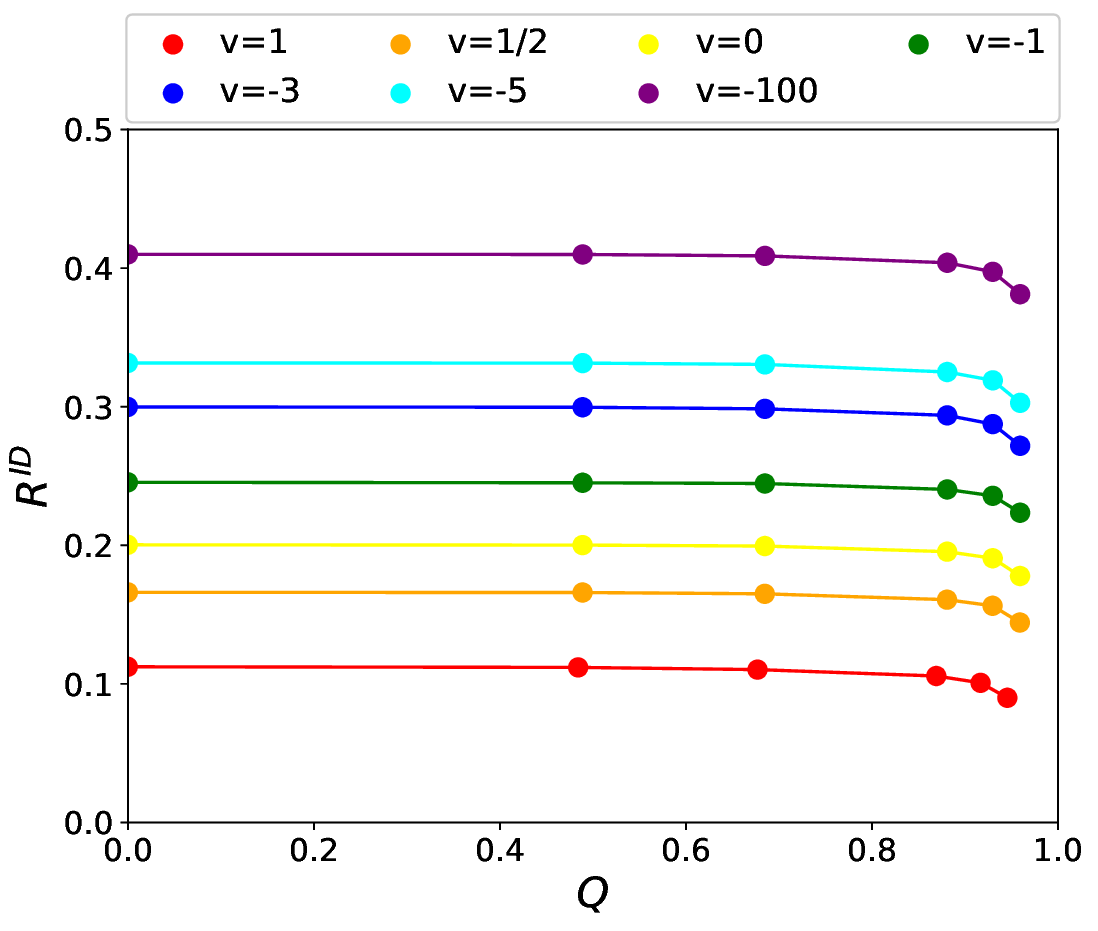}
    \begin{center} (d) $m_{o} = 50$ \end{center}
  \end{minipage}     
  \hfill 
  \begin{minipage}{.48\textwidth}
    \includegraphics[width=.9\textwidth]{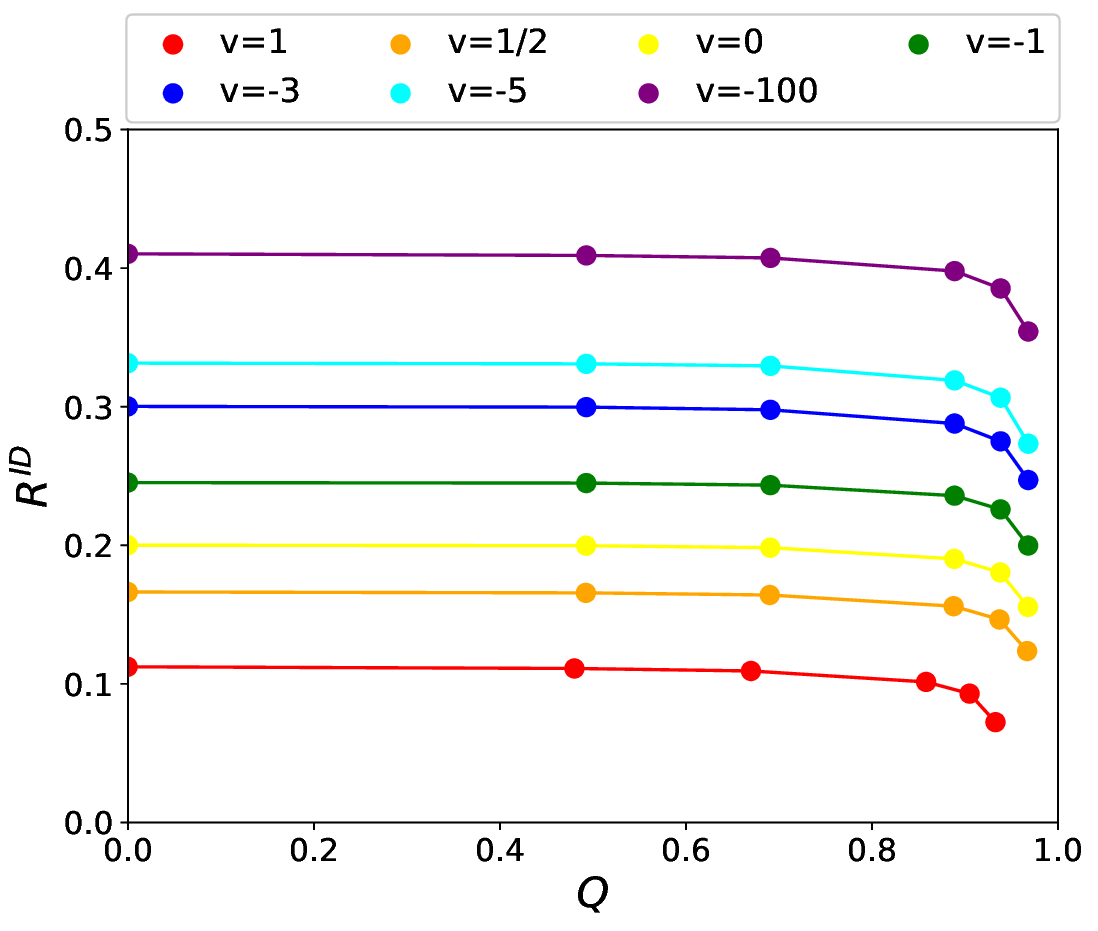}
    \begin{center} (e) $m_{o} = 100$ \end{center}
  \end{minipage}
  \hfill  
  \begin{minipage}{.48\textwidth}
    \includegraphics[width=.9\textwidth]{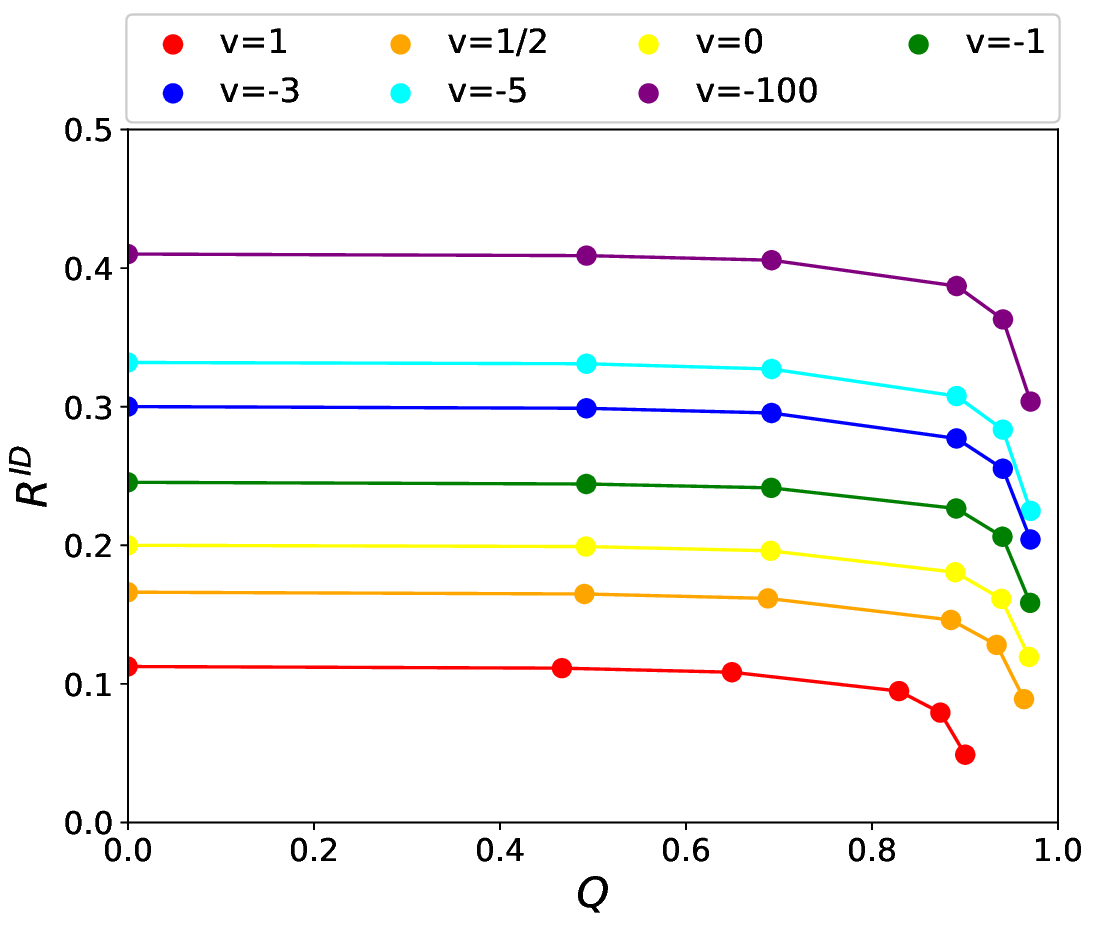}
    \begin{center} (f) $m_{o} = 200$ \end{center}
  \end{minipage}       
%\centering
%\includegraphics[width=.8\textwidth]{resize_figS20.eps}
\caption{Rapid decreasing of
the robustness index $R$ against ID attacks  
in modular networks with larger $Q$.}
\label{fig_Q-R^ID}
\end{figure}

\begin{figure}[htb]
  \begin{minipage}{.48\textwidth}
    \includegraphics[width=.9\textwidth]{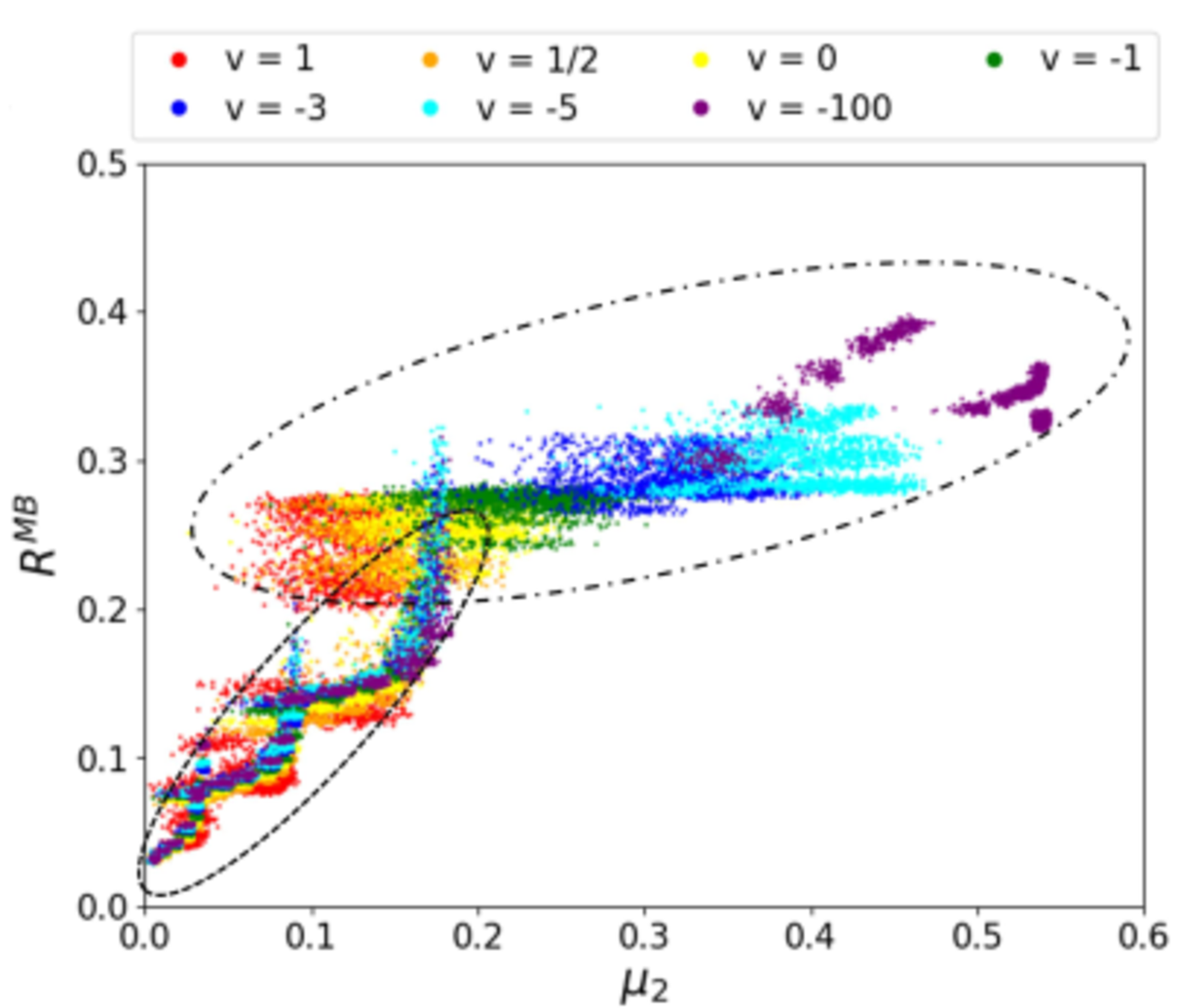}
    \begin{center} (a) \end{center}  
  \end{minipage}
  \hfill  
  \begin{minipage}{.48\textwidth}
    \includegraphics[width=.9\textwidth]{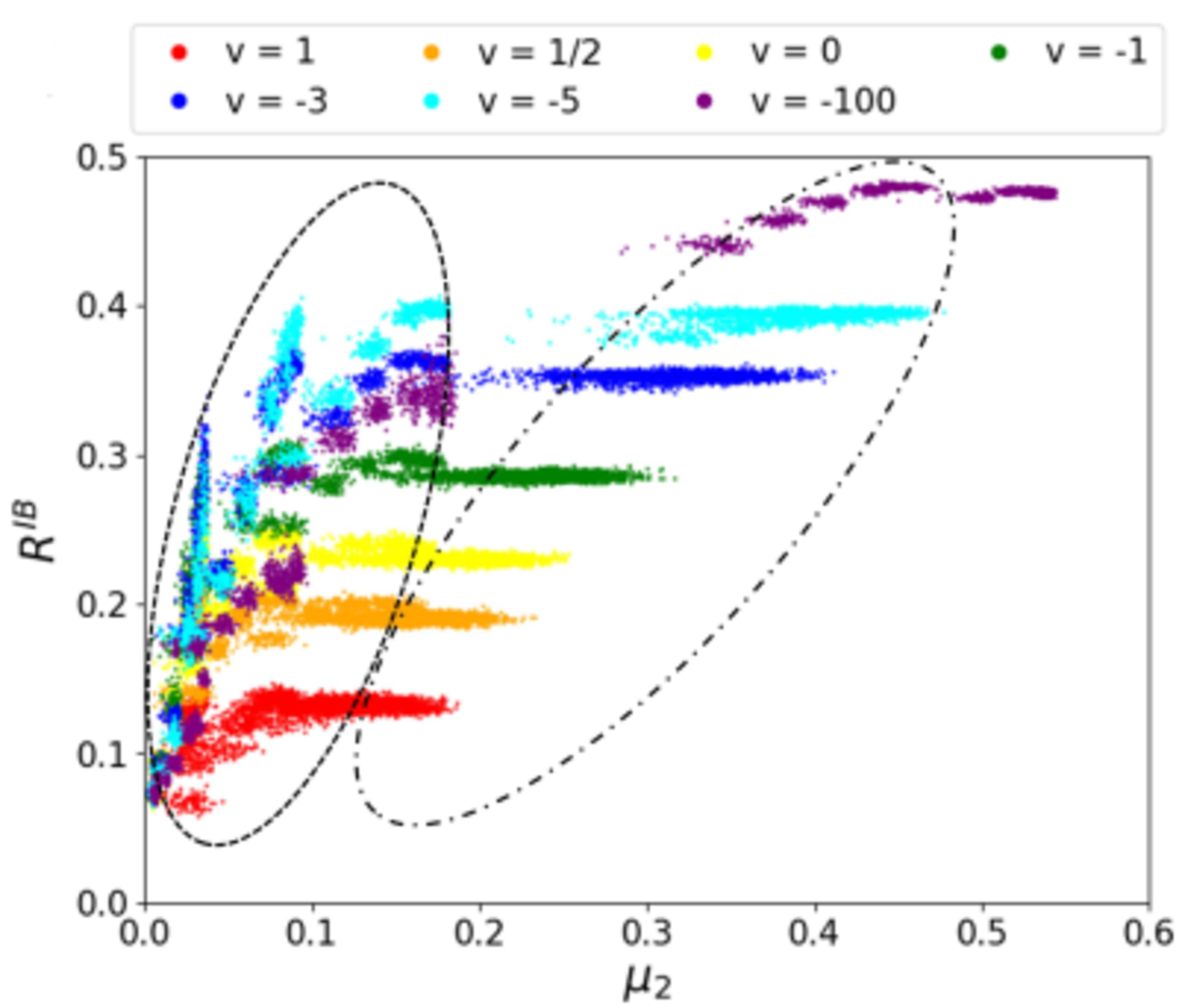}
    \begin{center} (b) \end{center}
  \end{minipage}    
  \hfill
  \begin{minipage}{.48\textwidth}
    \includegraphics[width=.9\textwidth]{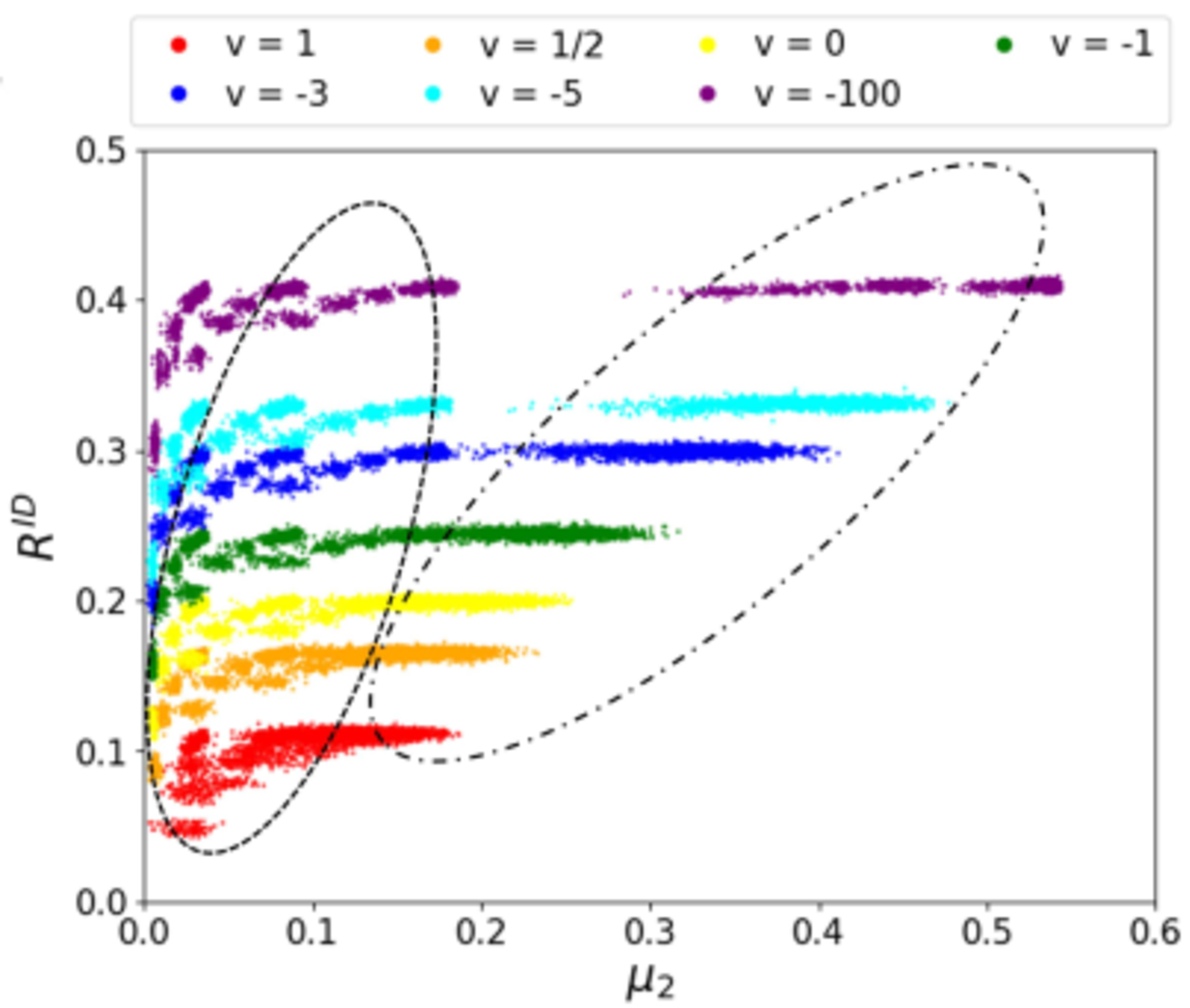}
    \begin{center} (c) \end{center}
  \end{minipage}
%\centering
%\includegraphics[width=.4\linewidth]{resize_figS21.eps}
\caption{Scatter plots of 
the eigenvalue $\mu_{2}$ of Laplacian matrix 
and the robustness index $R$
against (a) MB, (b) IB, (c) ID attacks in
100 realizations of the networks with
$m_{o} = 5, 10, 20, 50, 100$, and $200$  modules.
Rainbow colors for $\nu = 1, 1/2, 0, -1, -3, -5$ and $-100$ 
correspond to different $P(k)$ interpolated from 
power-law (as SF networks), power-law with exponential cut-off, 
exponential, nearly Poisson (as ER random graphs), and narrower ones
(approaching regular networks).
Lower and upper dashed-oval parts are for $Q > 0.8$ and $Q < 0.8$,
respectively, in different $m_{o}$ modules 
and varying the rewiring rate $w'$.}
\label{figS21_mu2-vs-R}
\end{figure}

\end{document}